\documentclass[longauth]{aa} 

\usepackage{graphicx}
\usepackage{txfonts}

\usepackage{float}
\usepackage{hyperref}
\hypersetup{breaklinks=true,colorlinks=true,urlcolor=blue, linkcolor=blue,  citecolor=blue, bookmarksopen=true}
\usepackage{placeins}
\usepackage{tikz}
\usetikzlibrary{positioning}
\usepackage{xcolor} 
\usepackage{siunitx}

\begin{document}

\title{MINDS survey of silicates in T Tauri disks: Correlation between dust and gas}
   \author{
	J.~Varga\inst{\ref{inst_K}} \and
	Th.~Henning\inst{\ref{inst_H}} \and
	L.~B.~F.~M.~Waters\inst{\ref{inst_R},\ref{inst_SR}} \and
	I.~Kamp\inst{\ref{inst_Gr}} \and
	Á.~Kóspál\inst{\ref{inst_K},\ref{inst_EL},\ref{inst_H}} \and
	P.~Ábrahám\inst{\ref{inst_K},\ref{inst_EL},\ref{inst_Vi}} \and
	O.~Absil\inst{\ref{inst_Liege}} \and
	A.~M.~Arabhavi\inst{\ref{inst_Gr}} \and
	D.~Gasman\inst{\ref{inst_Leu}} \and
	S.~L.~Grant\inst{\ref{inst_Ga},\ref{inst_Car}}  \and
	M.~Güdel\inst{\ref{inst_Vi},\ref{inst_Zur},\ref{inst_astron}} \and
	H.~Jang\inst{\ref{inst_R}} \and
	T.~Kaeufer\inst{\ref{inst_Ex}} \and
	J.~Kanwar\inst{\ref{inst_Mich}} \and
	N.~T.~Kurtovic\inst{\ref{inst_Ga}} \and
	P.-O.~Lagage\inst{\ref{inst_ParCEA}} \and
	G.~Perotti\inst{\ref{inst_Co},\ref{inst_H}} \and
	A.~Somigliana\inst{\ref{inst_H}} \and
	L.~M.~Stapper\inst{\ref{inst_H}} \and
	B.~Tabone\inst{\ref{inst_ParOrs}} \and
	M.~Temmink\inst{\ref{inst_Le}} \and
	E.~F.~van~Dishoeck\inst{\ref{inst_Le},\ref{inst_Ga}} \and
	M.~Vlasblom\inst{\ref{inst_Le}} 
      	} 
   \institute{
	Konkoly Observatory, HUN-REN Research Centre for Astronomy and Earth Sciences, MTA Centre of Excellence, Konkoly-Thege Miklós út 15-17, 1121 Budapest, Hungary\label{inst_K}\\ \email{varga.jozsef@csfk.org} \and      
	Max-Planck-Institut für Astronomie (MPIA), Königstuhl 17, 69117 Heidelberg, Germany\label{inst_H} \and
	Department of Astrophysics/IMAPP, Radboud University, PO Box 9010, 6500 GL Nijmegen, The Netherlands\label{inst_R} \and
	SRON Netherlands Institute for Space Research, Niels Bohrweg 4, 2333 CA Leiden, The Netherlands\label{inst_SR} \and
	Kapteyn Astronomical Institute, Rijksuniversiteit Groningen, Postbus 800, 9700AV Groningen, The Netherlands\label{inst_Gr} \and
	Institute of Physics and Astronomy, ELTE Eötvös Loránd University, Pázmány Péter sétány 1/A, 1117 Budapest, Hungary\label{inst_EL} \and
	Dept. of Astrophysics, University of Vienna, Türkenschanzstr. 17, 1180 Vienna, Austria\label{inst_Vi} \and
	STAR Institute, Université de Liège, Allée du Six Août 19C, 4000, Liège, Belgium\label{inst_Liege} \and
	Institute of Astronomy, KU Leuven, Celestijnenlaan 200D, 3001 Leuven, Belgium \label{inst_Leu} \and
	Max-Planck-Institut für Extraterrestrische Physik, Giessenbachstrasse 1, D-85748 Garching, Germany \label{inst_Ga} \and
	Earth and Planets Laboratory, Carnegie Institution for Science, 5241 Broad Branch Road, NW, Washington, DC 20015, USA\label{inst_Car} \and
	ETH Zürich, Institute for Particle Physics and Astrophysics, Wolfgang-Pauli-Str. 27, 8093 Zürich, Switzerland\label{inst_Zur} \and
	ASTRON, Netherlands Institute for Radio Astronomy, Oude Hoogeveensedijk 4, 7991 PD Dwingeloo, The Netherlands\label{inst_astron} \and
	Department of Physics and Astronomy, University of Exeter, Exeter EX4 4QL, UK\label{inst_Ex} \and
	Department of Astronomy, University of Michigan, 1085 South University Avenue, Ann Arbor, MI 48109, USA\label{inst_Mich} \and
	Université Paris-Saclay, Université Paris Cité, CEA, CNRS, AIM, F-91191 Gif-sur-Yvette, France \label{inst_ParCEA} \and
	Niels Bohr Institute, University of Copenhagen, NBB BA2, Jagtvej 155A, 2200 Copenhagen, Denmark\label{inst_Co} \and	
	Université Paris-Saclay, CNRS, Institut d'Astrophysique Spatiale, 91405 Orsay, France\label{inst_ParOrs} \and	
	Leiden Observatory, Leiden University, P.O. Box 9513, 2300 RA Leiden, the Netherlands\label{inst_Le}
 	}

   \date{Received September 15, 1996; accepted March 16, 1997}

\abstract
{Silicates are key constituents of planet-forming disks and are among the most important building blocks of rocky planets. Mid-infrared spectral features of micron-sized silicate grains are powerful tracers of grain growth, mineralogy, and disk chemistry.}
{We characterized the dust mineralogy in T Tauri disks using James Webb Space Telescope (JWST)/Mid-Infrared Instrument (MIRI) observations. A further aim of ours was to investigate the connections between the dust and molecular gas compositions.} 
{We analyzed JWST/MIRI spectra of 26 disks as part of the MIRI mid-Infrared Disk Survey (MINDS). We employed spectral decomposition with our new \texttt{DustComp} tool to derive the mass fractions of individual dust species. We included in our fits Mg$_2$SiO$_4$ (forsterite), MgSiO$_3$ (enstatite), and SiO$_2$ (silica) together with amorphous silicates of corresponding stoichiometry.}
{We find that Mg-rich (and Fe-poor) silicates represent our data well. Fit residuals are typically within $\pm 3\%$. Grain size distributions are skewed toward larger sizes ($>\!2\ \mu$m), indicating significant growth. Large ($\sim\!5\ \mu$m-sized) amorphous Mg-silicates were robustly detected, whereas the presence of large crystalline grains could not be firmly established. The average dust composition is dominated by grains of Mg$_2$SiO$_4$ stoichiometry ($\sim\!60\%$, including amorphous and crystalline state), followed by MgSiO$_3$ ($\sim\!30\%$) and SiO$_2$ ($\sim\!10\%$). The mass fractions of crystalline grains are typically in the $5$--$24\%$ range, with a mean of $14\%$. We robustly detected annealed silica in nine objects, with cristobalite as the main polymorph. We found a correlation between dust and molecular gas composition: disks with strong annealed silica features show relatively strong CO$_2$ emission, while forsterite-rich disks display stronger H$_2$O emission. Disks with annealed silica features may also have elevated gas-phase C/O ratios, suggesting a process, such as dust sublimation and recondensation, that establishes thermo-chemical equilibrium between solids and gas.}
{The correlation between dust and gas may provide the first indication that the molecular gas composition regulates the availability of dust species in the inner disk.}

   \keywords{protoplanetary disks -- stars: pre-main sequence --  stars: variables: T Tauri, Herbig Ae/Be -- planets and satellites: formation -- methods: data analysis -- Infrared: planetary systems}

   \maketitle
   \nolinenumbers

\section{Introduction}
Cosmic dust is produced in the outflows of asymptotic giant branch stars and supernovae \citep{Salpeter1977,Gail1999,Todini2001,Hofner2018} and subsequently enriches the interstellar medium (ISM). There are strong indications that dust is also formed in the denser phases of the ISM \citep[e.g.,][]{Draine2003,Zhukovska2018}. The ISM dust is rich in Mg-silicates, in the form of submicron-sized amorphous grains with olivine and pyroxene stoichiometry \citep{Kemper_silicate,Min2007,Henning2010}. These silicate grains can be readily detected due to their spectral resonances in the mid-infrared (mid-IR), which are seen as absorption \citep{Kemper_silicate} or emission features \citep{Tielens1998}. Silicates are not the only type of solids in astrophysical environments, though. There is direct and indirect evidence of other cosmic dust materials, including iron, various oxides, carbides, sulfides, carbonates, ices, and various forms of carbon \citep{DorschnerHenning1995,HenningSalama1998,Henning2011book,Tielens2022}.

During star formation, ISM dust is incorporated into the planet-forming disks encircling young stars. In these disks, as a first step of planet formation, dust grains start to grow by sticking together \citep{Brauer2008,Testi2014,Birnstiel2024}. Moreover, in high temperature disk regions ($T\!\gtrsim\!1000$~K), grains can be annealed, change their chemical composition, or even sublimate \citep[e.g.,][]{Fabian2000,DAlessio2005,Houge2025_decomp}. Dust can also form in situ in disks by condensation from the gas phase. In the hot inner regions, the sublimation of dust followed by recondensation can lead to chemical reprocessing of the original material. This process has been extensively studied in the context of equilibrium condensation, which predicts the sequence of mineral formation as a function of temperature and pressure \citep[e.g.,][]{Lodders2002,Lodders2003,Ebel2006}. The studies established a condensation sequence in which corundum (Al$_2$O$_3$), Ca-Al-oxides, and various Ca-Ti-minerals condense first (at $T\!\sim\!1500$--$1700$~K) from a cooling gas of solar composition gas. At lower temperatures ($T\!\sim\!1300$--$1400$~K) iron and Mg-silicates condense, followed by troilite (FeS) at $T\!\sim\!700$~K. 

Laboratory measurements of cosmic silicate analogs have shown that the positions and shapes of the mid-IR dust spectral features are influenced by the size, shape, lattice structure, and temperature of the grains, in addition to their chemical composition \citep{Henning1993,Jager1994,Fabian2001,Min2005,Min2007,Zeidler2015}. For example, crystalline grains can be identified by their sharp spectral features, such as the peaks at $10.0$, $11.3$, $16.3$, $19.5$, and $23.5\ \mu$m from forsterite and the $9.3$, $10.4$, and $19.4\ \mu$m peaks from enstatite \citep{Jager1998}. By comparing these laboratory data to observed disk spectra, it became possible to infer the physical and chemical properties of the grains emitting those features. 

Observations showed that there is a large variety in the spectral feature shapes of planet-forming disks, suggesting that their dust underwent processing \citep[e.g.,][]{Waelkens1996,Malfait1998,vanBoekel2005survey,Furlan2011}. Indeed, sample studies provided ample evidence of grain growth and crystallization in disks \citep{Bouwman2001,vanBoekel2003,Przygodda2003,Bouwman2008,Watson2009}. The mass fraction in crystalline silicates (relative to the total silicate mass) was found to be typically $\sim\!10\%$, although there is a large diversity among the individual sources (between $\sim\!1\%$ to $\sim\!30\%$ \citealp{Honda2006,Bouwman2008,Meeus2009,Juhasz2010,Olofsson2010}). For detailed reviews on cosmic silicates and dust processing, we refer to \citet{Henning2010} and \citet{Henning2011book}.

To date, the c2d survey with Spitzer/IRS has provided the most comprehensive spectroscopic study of silicates in T Tauri disks \citep{Kessler-Silacci2006}. The survey provided several important results. The shapes of the silicate spectral features indicate fast grain growth from $0.1\ \mu$m to $\sim\!1\ \mu$m in radius. The $\sim\!10\ \mu$m feature strength versus feature shape trend is not correlated with the ages of the systems, indicating that mixing and regeneration of small grains are occurring.
\cite{Olofsson2009} found that the silicates in their sample of T Tauri and Herbig Ae/Be disks are Mg-rich and Fe-poor, and the majority of the objects show at least some emission from crystalline grains. They noted that the $10\ \mu$m and $\lambda > 20\ \mu$m spectral features originate from different disk zones: the shorter wavelength is from a warm zone at $r\sim1$~au, and the longer wavelengths are from a colder zone extending to $r\sim10$~au. They also discovered a crystallinity paradox: Crystalline silicate features occur at longer wavelengths ($\lambda > 20\ \mu$m) more frequently than at $10\ \mu$m, suggesting a significant reservoir of cold crystals in the outer disk zones. This is contrary to the expectations, which predict an increased crystallinity fraction in the innermost disk zones due to thermal annealing at $T \!\gtrsim\!1000$~K \citep{Hallenbeck1998,Fabian2000}. 

\cite{Olofsson2010} subsequently suggested that the presence of cold crystals can be attributed to efficient radial mixing, which distributes crystalline grains at large distances from the central star. Alternatively, shock waves either by gravitational instability or by embedded planets may also be able to anneal the grains in the outer disk \citep{Harker2002}. 
The 2008 outburst of the young eruptive star EX Lup provided backing to the radial mixing scenario. \cite{Abraham2009} detected newly formed crystalline silicates in the disk after the outburst of EX Lup. Following a careful analysis of dust spectral features prior, during, and after the outburst, \cite{Juhasz2012} concluded that the crystals that formed in the inner disk within $\sim\!1$~au had been transported to larger radii, giving rise to the longer ($\lambda > 20\ \mu$m) wavelength crystalline silicate features a year after the outburst.

It has turned out to be difficult to find firm relations between the mineralogy and other system parameters (e.g., age, spectral type, disk mass) in planet-forming disks \citep[e.g.,][]{Honda2006,Watson2009,Oliveira2011}. \cite{Kessler-Silacci2006} and later \cite{Furlan2011} found that the $10\ \mu$m silicate feature tends to be weaker and flatter for disks around very low mass stars (VLMSs, spectral types M6--M9), and \cite{Furlan2011} attributed these characteristics to the presence of generally more processed dust in those objects. \cite{Glauser2009} found that high-energy particles are relevant for amorphizing crystals in young disks based on both observations of T Tauri stars and simulations. 

For a sample of Herbig Ae/Be disks observed with Spitzer/IRS, \citet{Juhasz2010} found that their dust mainly consists of amorphous silicates with olivine and pyroxene stoichiometry, crystalline forsterite, crystalline enstatite, and silica. This is in line with the results of the c2d survey for T Tauri disks. They also showed that enstatite is more concentrated toward the warm inner regions, compared to forsterite, which is in contrast to predictions of equilibrium condensation models. This result is in agreement with the findings of \cite{Bouwman2008}, who compared the relative abundances of forsterite and enstatite in the warm inner versus cold outer disk zones for a sample of seven planet-forming disks, mostly T Tauris. 
They found that more enstatite than forsterite occurs in the inner warm zone (at $\sim\!1$~au), while forsterite dominates the cold outer zone. They argued that this pattern is more consistent with a localized crystallization scenario rather than with radial mixing. To sum up these Spitzer/IRS results, pathways of dust processing are highly diverse in planet-forming disks: Crystals may form locally in the outer disk zones, but radial mixing can also occur, transporting crystals from the warm inner regions to the cold outer disk. 

While Spitzer/IRS spectra provide indirect spatial information, IR interferometric instruments such as VLTI/MIDI and VLTI/MATISSE can directly probe the radial distribution of dust in disks. These VLTI observations confirmed an increased crystallinity fraction in the innermost disk regions ($r\!\lesssim\!1$~au), in agreement with the thermal annealing scenario \citep{vanBoekel2004nature,vanBoekel2006SPIE,Varga2018,Varga2024}.
The spatial distribution of cold crystals emitting at $\lambda > 20\ \mu$m could not be studied with these instruments, as they do not cover that wavelength range.


\begin{table*} 
	\caption[]{\label{tab:obj_lst}Overview of the objects in our sample.}
	\centering
{\small
   	\begin{tabular}{lllllllc}
   \hline \hline
	
  Object & RA (J2000) &  Dec (J2000) &  SpT &  $d$ (pc) & $T_\mathrm{eff}$ (K) & $L_*$ ($L_\sun$) & Reference \\
   \hline
CX Tau  & 04:14:47.862 & 26:48:11.014 & M1.5        & 126.7 & 3487 & 0.34 & (1)\\ 
CY Tau  & 04:17:33.728 & 28:20:46.812 & M1.5        & 124.3 & 3516 & 0.37 & (1) \\ 
BP Tau  & 04:19:15.834 & 29:06:26.926 & K5/7        & 128.2 & 3777 & 0.83 & (1) \\ 
FT Tau  & 04:23:39.189 & 24:56:14.250 & M2.8        & 129.9 & 3415 & 0.44 & (1) \\ 
DF Tau  & 04:27:02.793 & 25:42:22.453 & M3          & 176.4 & 3900 & 3.89 & (1) \\ 
DL Tau  & 04:33:39.077 & 25:20:38.101 & K7          & 159.5 & 4276 & 1.47 & (1) \\
DM Tau  & 04:33:48.734 & 18:10:09.973 & M2          & 144.8 &  3715 & 0.16 & (2,3) \\
AA Tau & 04:34:55.420 & 24:28:53.033 & K5           & 137.7 & 3762 & 0.72 & (1) \\ 
DN Tau & 04:35:27.378 & 24:14:58.910 & M1:          & 127.2 & 3806 & 0.69 & (1) \\ 
LkCa 15 & 04:39:17.791 & 22:21:03.390 & K5          & 154.8 & 4276 & 1.12 & (1) \\ 
DR Tau & 04:47:06.215 & 16:58:42.813 & K5           & 186.9 & 4202 & 3.71 & (1) \\ 
RW Aur A & 05:07:49.565 & 30:24:05.131 & K0\tablefootmark{a}      &  150.0  & 4870 & 0.83 & (1,4) \\
RW Aur B & 05:07:49.456 & 30:24:04.775  & K6.5\tablefootmark{a}      &  150.0  & 4160 & 0.50 & (1,4) \\ 
SY Cha & 10:56:30.388 & -77:11:39.401 & K5          &  180.7 & 4060 & 0.43 & (1) \\
TW Hya & 11:01:51.905 & -34:42:17.033 & K6          &   59.9 & 4000 & 0.34 & (1) \\
VW Cha A & 11:08:01.482 & -77:42:28.567 &  K7    &  188.1 & 4060 & 2.21 & (1,5) \\ 
WX Cha A & 11:09:58.853 & -77:37:08.631 & M1      &  189.3  & 3710 & 1.13 & (1,5) \\ 
XX Cha & 11:11:39.673 & -76:20:15.032 & M3          &  194.6 & 3340 & 0.29 & (1) \\ 
Sz 50 & 13:00:55.382 & -77:10:22.247  & M3          &  148.9 & 3400 & 0.41 & (6) \\ 
PDS 70 & 14:08:10.155 & -41:23:52.573 & K7          &  112.3 & 4138 & 0.38\\ 
GW Lup & 15:46:44.729 & -34:30:35.677 & M1.5        & 155.2 & 3632 & 0.33 & (1) \\ 
IM Lup & 15:56:09.207 & -37:56:06.126 & M0          & 153.8 & 4350 & 2.57 & (1) \\ 
Sz 98 & 16:08:22.494 & -39:04:46.427 & M0.4         & 156.2 & 4060 & 1.51 & (1) \\ 
V1094 Sco & 16:08:36.177 & -39:23:02.464 & K6/K5    & 152.4 & 4205 & 1.15 & (1) \\ 
RNO 90 & 16:34:09.170 & -15:48:16.776 & G5          &  114.9 & 5662 & 5.7 & (7) \\
WA Oph 6 & 16:48:45.633 & -14:16:35.849 & K6        & 122.5 & 4169 & 2.88 & (8) \\ 
   \hline
   	\end{tabular}
}
    \tablefoot{This compilation of stellar parameters is based on \cite{Henning2024}. Coordinates (ICRS) and spectral types (SpT) are taken from SIMBAD, unless otherwise noted. Distances ($d$), effective temperatures ($T_\mathrm{eff}$), and stellar luminosities ($L_*$) are collected from Gaia DR3 (not noted), (1) \cite{Testi2022}, (2) \cite{Kenyon1995}, (3) \cite{Bertout2007}, (4) \cite{Herczeg2014}, (5) \cite{Daemgen2013}, (6) \cite{Sartori2003}, (7) \cite{Ghez1993}, and (8) \cite{Andrews2009}. 
    \tablefoottext{a}{\cite{Herczeg2014}}
	}
\end{table*}

Silica (SiO$_2$) is a minor constituent of dust in planet-forming disks, occurring mostly in amorphous form, but \cite{Sargent2009_silica} also detected crystalline SiO$_2$ in five T Tauri disks. They suggested that the formation of annealed silica requires high temperatures ($1200$--$1300$~K, \citealp{Fabian2000}) followed by rapid cooling, implying transient heating events similar to those hypothesized to have created Solar System chondrules. 

Compared to Spitzer/IRS, the Mid Infrared Instrument (MIRI) on board the James Webb Space Telescope  (JWST) \citep{Rigby2023} offers a much higher spectral resolution -- $R\!\sim\!3000$ with the medium-resolution spectrometer (MRS; \citealp{Argyriou2023}) -- and a more than hundredfold gain in sensitivity. This has opened up the exploration of mid-IR molecular lines emitted by the warm gas component in planet-forming disks with an unprecedented accuracy \citep[e.g.,][]{Perotti2023,Grant2023,Banzatti2023,Kospal2023,Tabone2023,Arabhavi2024,Arulanantham2025,Temmink2025,Gasman2025,Kospal2025}. 

As of now, the dust mineralogy aspect with JWST remains largely untapped. Early MIRI studies focusing on dust include \cite{Kospal2023,Perotti2023,Jang2024,Kaeufer2024_Sz28,Grant2025,Jang2025,Kospal2025,Liu2025}. Most of these studies focused on individual sources rather than on a larger sample. \cite{Grant2025} found an anticorrelation between the strength of the $10\ \mu$m silicate feature and the C$_2$H$_2$/H$_2$O line flux ratio, while \cite{Arabhavi2025} reported that objects with weaker silicate features tend to have higher gas optical depths, traced by the $^{13}$CCH$_2$/C$_2$H$_2$ flux ratio. \citet{Tabone2026} identified four C$_2$H$_2$-bright disks (V1094~Sco, DL~Tau, CY~Tau, and DoAr~33) that also exhibit a strong silica dust component, as revealed by their $9\ \mu$m spectral features.

There are two key advantages of MIRI/MRS over Spitzer/IRS for dust studies. First, the higher spectral resolution of MIRI allows for better separation of the solid state features from the molecular lines. Second, the excellent sensitivity of MIRI makes it possible to search for weaker dust components.
The Mid-INfrared Disk Survey (MINDS; \citealp{Henning2024}) is one of the large Guaranteed Time Observations (GTO) programs with JWST/MIRI, and it is aimed at exploring both the gas and dust content in planet-forming disks. 

In this work, we present a detailed mineralogy study of the T Tauri subset of the MINDS sample with the aims of (1) determining the dust composition for these disks with a precision surpassing what was possible with Spitzer, (2) searching for less common minerals, and (3) looking for correlations between the mineralogy and the molecular gas chemistry. For the data analysis, we applied spectral decomposition, similar to what was employed in \cite{Juhasz2010} for Herbig Ae/Be disks observed with Spitzer/IRS.

The paper is structured as follows. In Sect.~\ref{sec:obs} we describe the sample and the data. In Sect.~\ref{sec:model} we present our model to fit the data along with the  dust opacity templates. The results are shown in Sect.~\ref{sec:res} and followed by a discussion in Sect.~\ref{sec:disc}. Finally, Sect.~\ref{sec:concl} provides a summary.

\section{Sample and data}
\label{sec:obs}

The sample of the MINDS GTO program (ID: 1282, PI: Th. Henning, \citealp{Henning2024,Kamp2023}) consists of $52$ targets, mostly Class II T Tauri objects, but also Herbig Ae disks, and young debris disks. From these, we selected 26 T Tauri disks that are not edge-on and that all show silicate spectral features in emission.
Our sample covers spectral types G5--M3, stellar luminosities of $0.16$--$5.7\ L_\sun$, and distances of $60$--$195$~pc. It also includes the transitional disks LkCa 15, PDS 70, DM Tau, and SY Cha. Table~\ref{tab:obj_lst} shows an overview of the sample. 
The objects were observed with MIRI \citep{Rieke2015,Wright2015,Wright2023} in the MRS mode \citep{Wells2015,Argyriou2023}. The observations, which were taken with a four-point dither pattern using all three grating settings covering the whole $4.9$--$27.9\ \mu$m wavelength range, have been reduced following the same approach as in \cite{Temmink2025}. The data were reduced using the standard JWST pipeline reduction (version 1.16.1; \citealp{Bushouse2024}) with pmap 1315. The spectra were extracted using aperture photometry with the recommended aperture size of 2$\times$ the full width at half maximum (FWHM), and residual fringes were corrected using the pipeline's standard implementation.

There are a few known multiple stellar systems in our sample. RW Aur and WX Cha are binaries with separations of $1.5\arcsec$ and $0.75\arcsec$, respectively \citep{Daemgen2013}. DF Tau is an equal-mass binary with a semimajor axis of $97$~mas ($17$~au; \citealp{Allen2017,Kutra2025}). VW Cha is a quadruple system: Its component A is a spectroscopic binary with a tentative 10-day period, and the close pair BC is located $0.7\arcsec$ away from A \citep{Melo2003,Nguyen2012,Zsidi2022}. B and C are separated by $0.1\arcsec$ \citep{Brandeker2001}. \cite{Kurtovic2026} published MIRI spectra of RW Aur, VW Cha, and WX Cha, where they applied special pipeline processing to separate the components in the data. Here we use these component separated spectra of RW Aur A, RW Aur B, VW Cha A, and WX Cha A.
We do not study the spectra of VW Cha BC and WX Cha B, because they do not show silicate features. The components of the spectroscopic binary VW Cha A might be so close to each other that they do not possess individual disks. Thus, the IR continuum emission we see in the MIRI data likely originates from a circumbinary disk. In the case of DF Tau, both components have disks, but they are not resolved by JWST \citep{Grant2024}. Since the component spectra cannot be disentangled, the dust composition we derive reflects the average composition of the two disks.

The MRS spectra have a spectral resolution between 1000 and 4000, while the dust features, including the narrower and weaker subpeaks, can already be resolved at a resolving power of $R\!\sim\!30$--$50$.
Since our goal is to model the dust emission, we only require the full spectral resolution to separate the gas lines from the dust continuum. 
After removing the gas lines, the spectra can therefore be rebinned to a lower resolving power. To achieve this, we applied the following procedure: (1) We smoothed the spectrum using a Savitzky-Golay filter \citep{Savitzky1964}, with a polynomial degree of three and a variable window length in wavelength space, starting at $0.07\ \mu$m at the short-wavelength edge, increasing to $0.11\ \mu$m at $\lambda=10\ \mu$m, and reaching $0.91\ \mu$m at the long-wavelength edge\footnote{The window length in pixels was $85$ at $\lambda<22\ \mu$m and $151$ at $\lambda\gtrsim22\ \mu$m. For RW Aur A and DF Tau, we used slightly larger window lengths ($113$ pixels at $\lambda<22\ \mu$m and $201$ pixels at $\lambda\gtrsim22\ \mu$m), because their spectra are dominated by stronger molecular lines than those in the other objects. The corresponding window length in wavelength space ranges from $0.09$ to $1.2\ \mu$m, with a value of $0.15\ \mu$m at $\lambda=10\ \mu$m.}. (2) We calculated the residual by subtracting the smoothed spectrum from the original one. (3) We masked all the wavelength locations where the residual values were outside the $\left[0.62\%,93.3\%\right]$ percentile range. 
The percentiles were calculated using a $\sim\!1\ \mu$m wide rolling window. (4) We repeated the previous steps four more times, updating the mask each time to filter out the remaining spikes. (5) We rebinned the final smoothed spectrum to a lower spectral resolution of $R\!\approx\!100$--$200$. The advantage of this procedure is its model independence, as it does not require fitting the data with gas or dust spectral templates. This approach works well for our sample of T Tauri disks, where the gas emission is dominated by discrete lines. In contrast, VLMS disks tend to have a molecular pseudo-continuum that cannot be removed with such a procedure \citep[cf.][]{Jang2025}.
We also estimated the uncertainties of the data by taking the standard deviation of the non-masked residual values in each bin. However, at wavelengths where the uncertainties provided by the pipeline were larger than our estimates, we kept the original uncertainty values. In this way, we accounted for the inherent uncertainty in the continuum estimation caused by the spectral line residuals.

We did not use the spectral range beyond $27\ \mu$m, because of its low signal-to-noise ratio (S/N). There are specific wavelength regions where we noticed that the removal of gas lines were inadequate, for example, $13$--$15\ \mu$m (due to C$_2$H$_2$, HCN, and CO$_2$), $6$--$7\ \mu$m (due to H$_2$O), and around $26\ \mu$m (also due to H$_2$O). The majority of our spectra are affected by this to varying degrees. To mitigate this issue, we removed the affected spectral regions from our analysis (indicated, e.g., in Fig.~\ref{fig:fit_example}). Since silicates do not have significant features in these wavelength ranges, this exclusion does not impact our dust analysis.

\section{Modeling}
\label{sec:model}

\subsection{The art of spectral decomposition}

The aim of our modeling is to infer the composition, crystallinity, and typical grain sizes of the dust that produces the spectral features in our MIRI spectra. We employed spectral decomposition, which is a widely used approach for interpreting this kind of data. At its core, this technique involves fitting a set of spectral templates, each corresponding to a specific dust component, by linearly combining them with adjustable weights to find the best-fitting result. 

The fact that we see the silicate features in emission suggests optically thin radiation that arises from dust in a surface layer warmer than the underlying material, heated by the stellar illumination \citep[e.g.,][]{Chiang1997,Dullemond2001}. However, the combined contribution from the central star and the optically thin emission from micron-sized Mg-silicate grains alone is usually insufficient to account for all the observed thermal IR radiation\footnote{It is because Mg-silicates have very low opacities at $\lambda<8\ \mu$m wavelengths.}. The additional featureless IR continuum originates from dust components lacking spectral resonances (such as iron or carbon), or from optically thick dust emission arising in deeper disk layers, or both. We note that the disk's gas is not expected to provide significant continuum emission at these wavelengths. Additionally, polycyclic aromatic hydrocarbons (PAHs), which can be identified by their strong IR spectral bands, can also emit a spectral continuum \citep{Li2001}. However, we do not detect PAH bands in any of our objects. The directly irradiated sublimation front (hot inner rim of the dusty disk, at $T\!\sim\!1500$~K) can add a distinct continuum emission component which peaks in the near-IR \citep{Dullemond2001,Natta2001}. To summarize, spectral decomposition models need to include continuum emission sources, and the inference on the dust composition is therefore only partial \citep[cf.][]{Varga2024}. Constituents with featureless opacity curves could still be identified when they contribute to absorption, but this requires detailed radiative transfer calculations.

There are various implementations for spectral decomposition in the literature. The simpler ones \citep[e.g.,][]{vanBoekel2005survey} assumed a single characteristic temperature for the dust emission, and the flux of the dust components was calculated as
\begin{equation}
	F_{\nu,\mathrm{dust}} \propto \sum_{i} C_i \kappa_\mathrm{abs,i} B_\nu \left(T_\mathrm{c} \right),
\end{equation}
where $B_\nu \left(T_\mathrm{c} \right)$ is the Planck function at the characteristic temperature $T_\mathrm{c}$, $\kappa_\mathrm{abs,i}$ is the opacity curve of the dust component $i$, and $C_i$ is the weighting factor\footnote{The underlying assumption here is that the emission is optically thin. This model becomes unphysical if $\sum_{i} C_i \kappa_\mathrm{abs,i}$, i.e., the optical depth, becomes larger than $1$.} which is proportional to the dust mass in component $i$. The continuum emission components were represented by single-temperature blackbodies. As a next step in complexity, \cite{Olofsson2010} in their B2C model included two distinct dust populations, a warm and a cold one, responsible for the $10\ \mu$m and $20$--$30\ \mu$m spectral features, respectively. \cite{Juhasz2009} introduced the two-layer temperature distribution (TLTD) model, where the emission components, instead of having a single temperature, have power-law temperature distributions. This model can better represent the radially decreasing temperature profile of planet-forming disks. The TLTD model was developed into the TGMdust code in \cite{Varga2024}, who used it to fit mid-IR spectro-interferometric data from the VLTI/MATISSE instrument. MIRI spectra show dust spectral features and gas lines in an unprecedented detail, requiring novel analysis methods. \cite{Kaeufer2024} presented the Dust Continuum Kit with Line emission from Gas (DuCKLinG), which can model both the gas and dust emission at the same time. This unified approach is especially important for VLMS disks, where the molecular lines tend to overpower the dust features, producing a pseudo-continuum \citep{Tabone2023,Kanwar2024,Arabhavi2024,Jang2025,Kaeufer2024_Sz28}.

Circumstellar dust grains likely exist in the form of fractal aggregates composed of multiple different minerals \citep[e.g.,][]{Blum2008,Testi2014}. However, it is not yet feasible to take into account the full complexity of the shapes and materials of grains when analyzing IR dust emission, because of computational costs and too many free parameters. Therefore, spectral decomposition techniques require a simplified approach, by defining distinct dust components, each with its own chemical composition, mineralogy, size, and assuming simple grain shapes. Then for each component, a spectral template (opacity, scattering) to be fitted is computed.

\begin{table*}
 \caption[]{\label{tab:dustopac}Overview of the dust optical data used in this work.}
 {\small
 \centering
\begin{tabular}{llll}
 \hline \hline
  Species      &  DHS$\_$synth set & DHS$\_$nat set & GRF set \\
  \hline
Amorphous Mg$_2$SiO$_4$ & \cite{Jager2003} & \cite{Jager2003} & \cite{Henning1996} \\
Crystalline forsterite  & \cite{Suto2006}\tablefootmark{a} & \cite{Zeidler2015}\tablefootmark{b,\textdagger}& \cite{Servoin1973}\tablefootmark{a} \\
Amorphous MgSiO$_3$ & \cite{Dorschner1995}& \cite{Dorschner1995} & \cite{Dorschner1995}\\
Crystalline enstatite  & \cite{Jager1998}\tablefootmark{c} & \cite{Zeidler2015}\tablefootmark{d,\textdagger}& \cite{Jager1998}\tablefootmark{c}\\
Amorphous SiO$_2$ & \cite{Kitamura2007} & \cite{Kitamura2007} & \cite{Henning1997}\\
Annealed SiO$_2$ & \cite{Fabian2000}\tablefootmark{\textdaggerdbl} & \cite{Fabian2000}\tablefootmark{\textdaggerdbl} & \cite{Fabian2000}\tablefootmark{\textdaggerdbl}\\
  \hline 
\end{tabular}
\tablefoot{
Formulae:
\tablefoottext{a}{Mg$_{2}$SiO$_4$,}
\tablefoottext{b}{Mg$_{1.72}$Fe$_{0.2}$SiO$_4$,}
\tablefoottext{c}{MgSiO$_3$,}
\tablefoottext{d}{Mg$_{0.92}$Fe$_{0.09}$SiO$_3$.}
\tablefoottext{\textdagger}{Measured at 300~K.} 
\tablefoottext{\textdaggerdbl}{5-hour annealed sample.}
}}
\end{table*}

\begin{table}
 \caption[]{\label{tab:grain_size}Grain size sets.}
 \small
 \centering
\begin{tabular}{ll}
 \hline \hline
  Grain sizes ($\mu$m)  & Note \\ 
  \hline
$\left[0.1, 2\right]$ & \\
$\left[0.1, 1, 2\right]$ & \\
$\left[0.1, 2, 5\ \mathrm{am.}\right]$ & $5\ \mu$m-size for amorphous species only\\ 
$\left[0.1, 2, 5\right]$ & $5\ \mu$m-size for all species\\ 
$\left[0.1, 1, 2, 5\right]$ & $5\ \mu$m-size for all species\\ 
$\left[0.1, 1, 2, 3, 4, 5\right]$ & $5\ \mu$m-size for all species\\ 
  \hline
\end{tabular}

\end{table}

\subsection{Our \texttt{DustComp} model}

Our python modeling tool, named \texttt{DustComp}\footnote{The code, along with the opacity data, is publicly available at \url{https://github.com/jvarga42/DustComp}.} is based on \cite{Juhasz2009} and \cite{Kaeufer2024}. 
Our model consists of the following components: central star, inner rim, disk midplane, and disk surface. The central star is represented by a blackbody with an effective temperature ($T_\mathrm{eff}$) taken from the literature, and an emitting area on the sky calculated from the stellar radius ($R_*$) and the distance ($d$) of the source. The stellar radius is estimated from the luminosity ($L_*$), as reported in the literature. $T_\mathrm{eff}$, $d$, and $L_*$ are listed in Table~\ref{tab:obj_lst}. The flux density of the central star is given by 
\begin{equation}
F_{\nu,*} = \frac{R_*^2 \pi}{d^2} B_\nu \left(T_\mathrm{eff}\right).
\end{equation}
The flux of the central star is fixed in our model.
The rim component accounts for the inner rim of the dusty disk that gets direct illumination from the central star, and hence it is the hottest region where dust grains can still exist.  
We modeled the rim with a single-temperature blackbody ($T_\mathrm{rim}$) in the shape of an annulus:
\begin{equation}
F_{\nu,\mathrm{rim}} = \frac{\pi}{d^2} B_\nu \left(T_\mathrm{rim} \right) w_\textrm{rim}\left( w_\textrm{rim} + 2 r_\textrm{in} \right),
\end{equation}
where $r_\textrm{in}$ is the inner radius of the model disk and $w_\textrm{rim}$ is the radial width of the annulus. The corresponding surface area is thus $\pi w_\textrm{rim}\left( w_\textrm{rim} + 2 r_\textrm{in}\right)$. The disk midplane component has a power-law radial temperature profile:
\begin{equation}
	T_\mathrm{midplane}\left(r\right) = T_\mathrm{midplane,in} \left(\frac{r}{r_\mathrm{in}}\right)^{q_\mathrm{midplane}},
\label{eq:T_midplane}
\end{equation}
where $T_\mathrm{midplane,in}$ is the temperature of the midplane at $r_\textrm{in}$, and $q_\mathrm{midplane}$ is the power-law index.  
The resulting flux density of the midplane is given by
\begin{equation}
F_{\nu,\mathrm{midplane}} = \frac{1}{d^2} \int_{r_\textrm{in}}^{r_\textrm{out}} 2 \pi r B_\nu \left(T_\mathrm{midplane}\left(r\right) \right) dr.
\end{equation}
Here $r_\textrm{out}$ is the outer radius of the model disk, which we fixed to $20$~au, where the dust is too cold to contribute to the emission at MIRI wavelengths. The disk surface component also has a power-law temperature profile:
\begin{equation}
	T_\mathrm{surface}\left(r\right) = T_\mathrm{surface,in} \left(\frac{r}{r_\mathrm{in}}\right)^{q_\mathrm{surface}},
\end{equation}
where $T_\mathrm{surface,in}$ is the temperature of the surface at $r_\textrm{in}$, and $q_\mathrm{surface}$ is the power-law index. To obtain the surface brightness profile, the blackbody source function was multiplied with the weighted sum of the opacities of the dust components:
\begin{equation}
	I_{\nu,\mathrm{surface}} \left(r\right) = \sum_{i} C_i \kappa_\mathrm{abs,i} B_\nu \left(T_\mathrm{surface}\left(r\right) \right),
\end{equation}
where $C_i$ are the weights and $\kappa_\mathrm{abs,i}$ are the opacities. The index $i$ denotes the various dust components, which differ in stoichiometry, grain size, and crystallinity. Here we assume that all dust components share the same radial temperature profile ($T_\mathrm{surface}$). This is a simplification, as in radiative equilibrium the temperature of a grain depends on its size and optical properties, so different grains located at the same disk radius may have different temperatures.
The flux density of the surface component is the following:
\begin{equation}
F_{\nu,\mathrm{surface}} = \frac{1}{d^2} \int_{r_\textrm{in}}^{r_\textrm{out}} 2 \pi r I_{\nu,\mathrm{surface}}  \left(r\right) dr.
\end{equation}
Finally, the total model flux density was obtained by summing the flux densities of all the components:
\begin{equation}
	F_{\nu,\mathrm{model}} = F_{\nu,\mathrm{surface}} + F_{\nu,\mathrm{midplane}} + F_{\nu,\mathrm{rim}} + F_{\nu,*}
\end{equation}

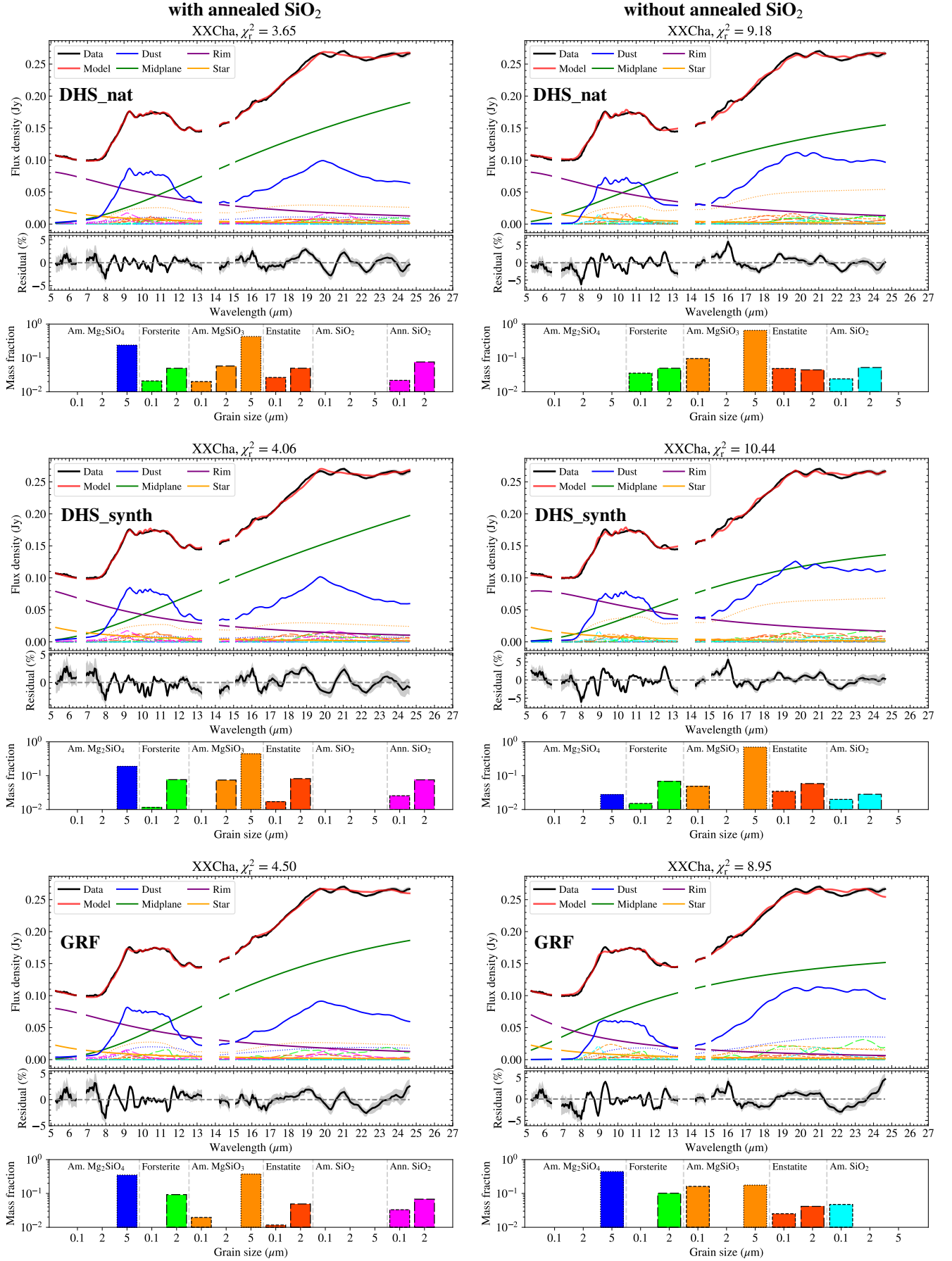
\begin{figure*}
	\centering
 
	\begin{tikzpicture}[
image/.style = {text width=0.48\textwidth, 
                 inner sep=0pt, outer sep=0pt},
node distance = 1mm and 1mm 
                        ] 
\node [image] (frame1)
    {\includegraphics[width=0.99\hsize]{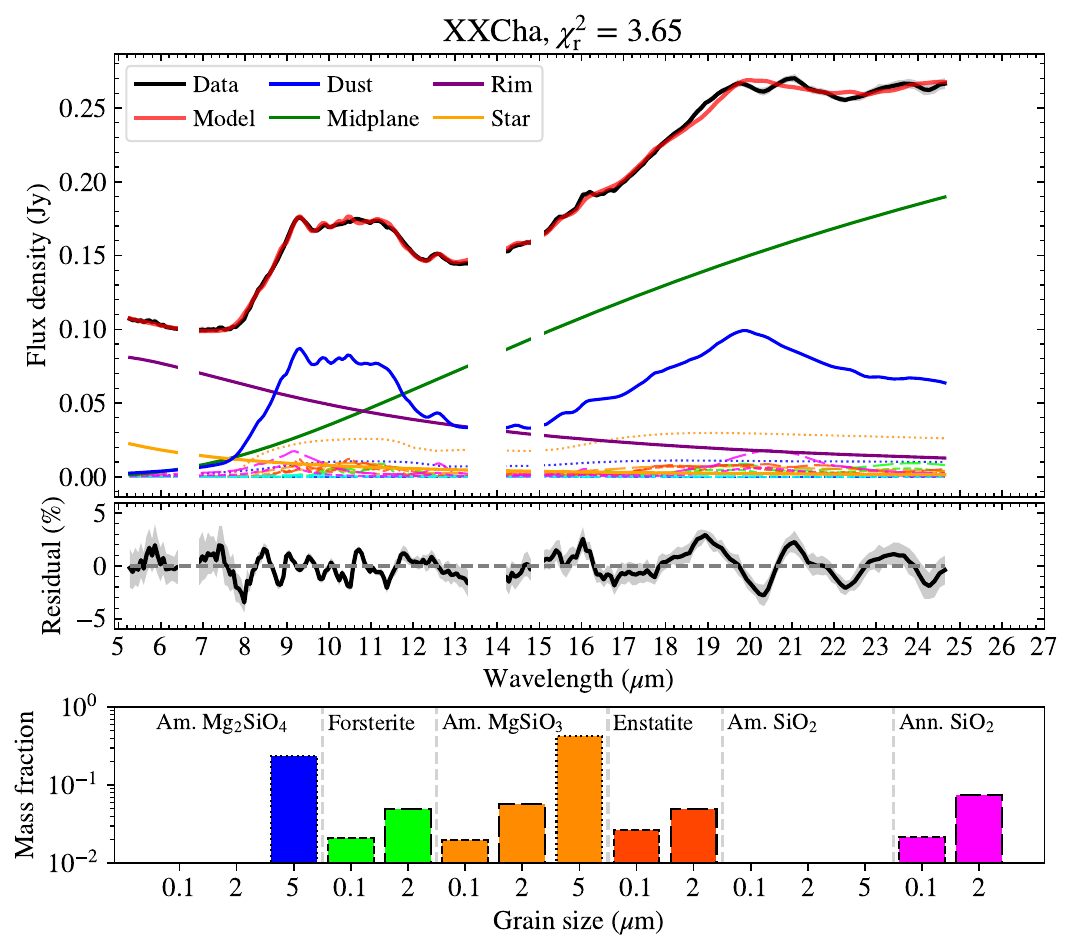 }};
\node at (-2.6,2.4) {\textbf{DHS$\_$nat}};
\node at (0.2,4.0) {\textbf{with annealed SiO$_2$}};
\node [image,right=of frame1] (frame2) 
    {\includegraphics[width=0.99\hsize]{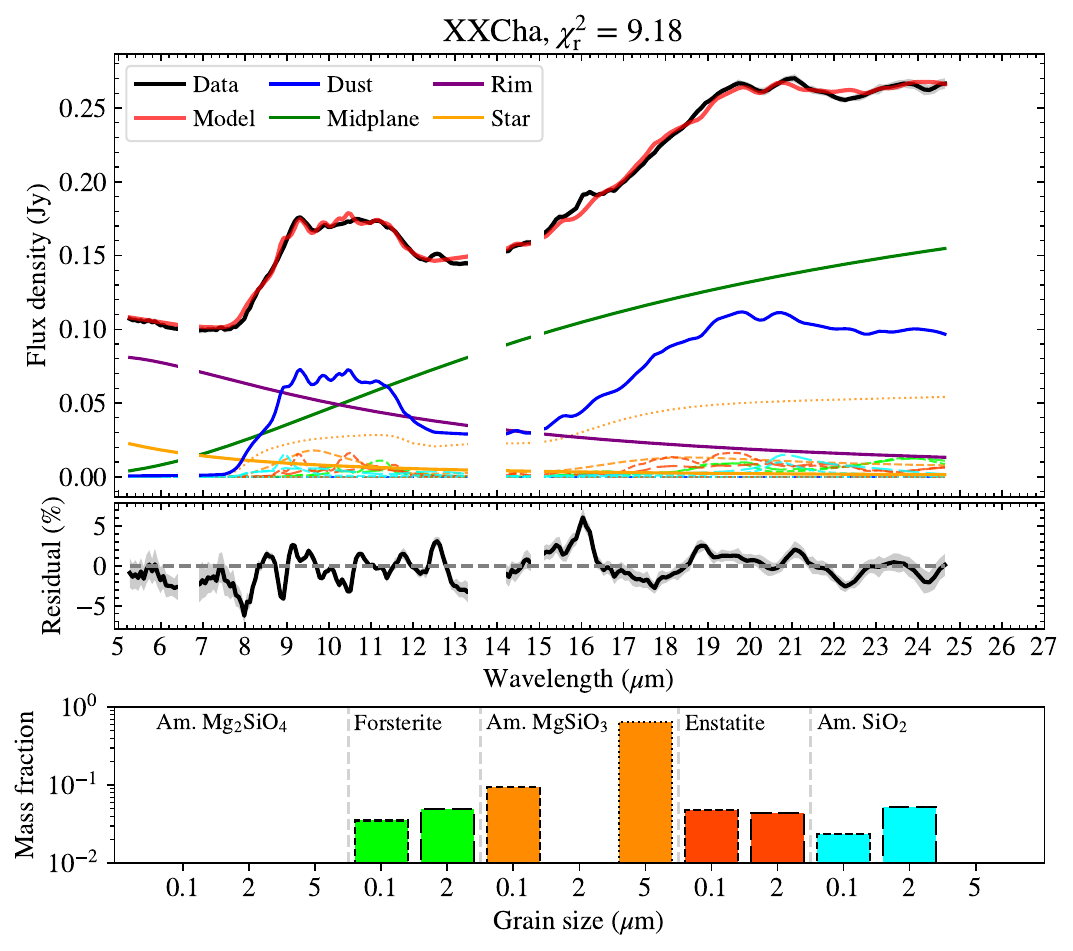 }};
\node at (9.0,4.0) {\textbf{without annealed SiO$_2$}};

\node at (6.3,2.4) {\textbf{DHS$\_$nat}};
\node[image,below=of frame1] (frame3)
    {\includegraphics[width=0.99\hsize]{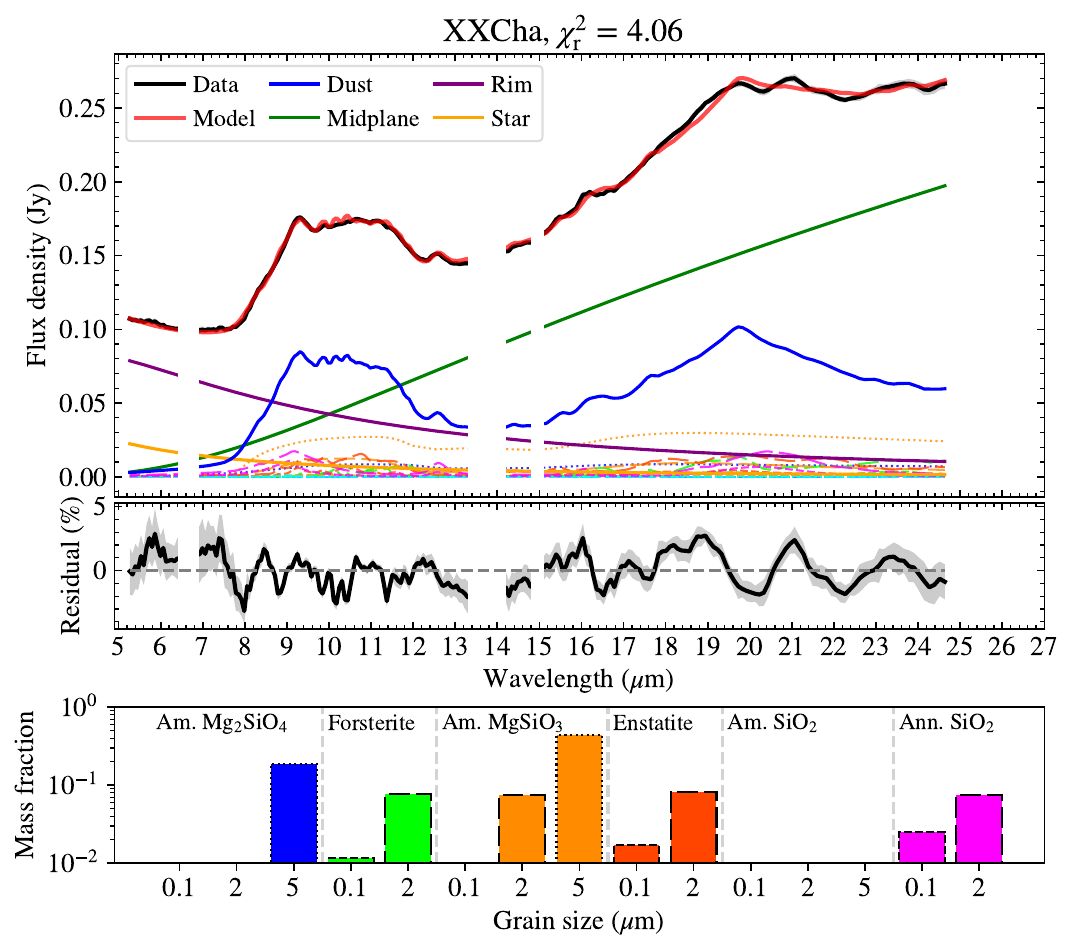 }};
\node at (-2.4,-5.5) {\textbf{DHS$\_$synth}};
\node[image,right=of frame3] (frame4)
    {\includegraphics[width=0.99\hsize]{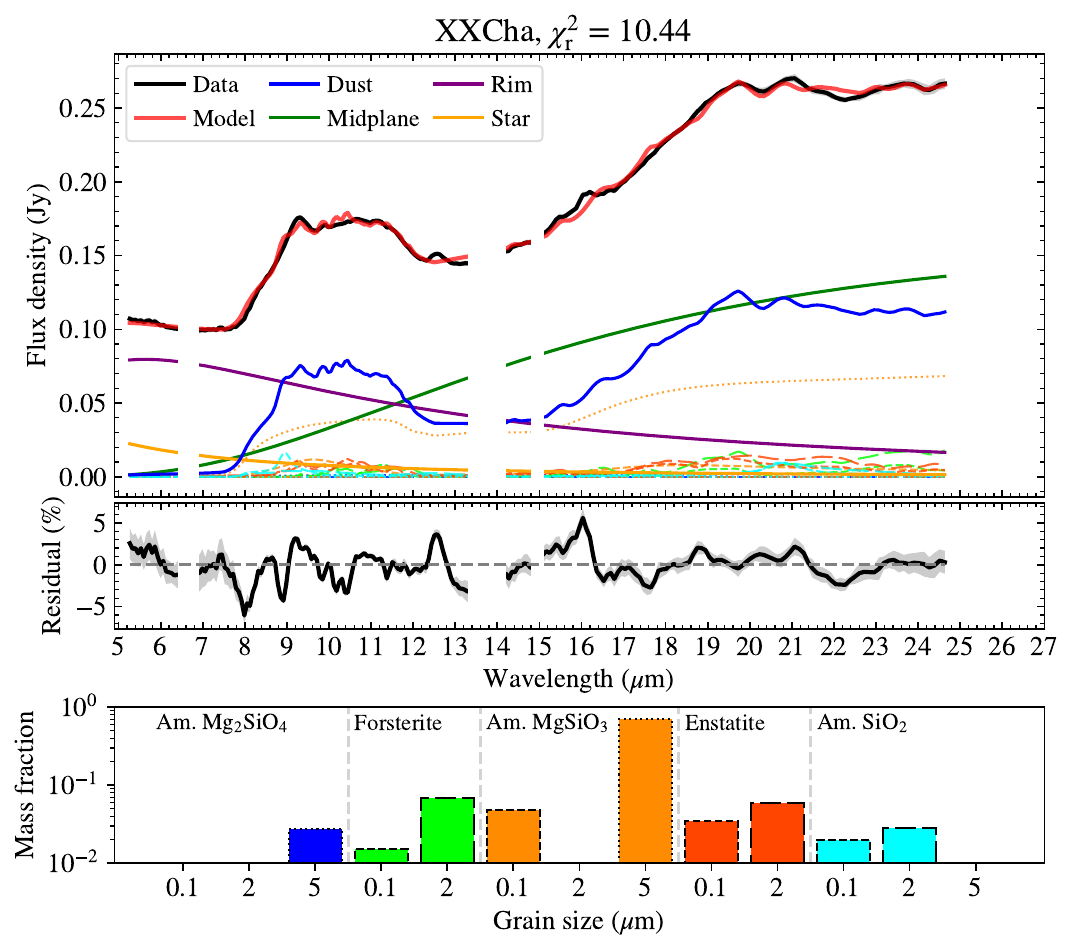 }};
\node at (6.5,-5.5) {\textbf{DHS$\_$synth}};

\node[image,below=of frame3] (frame5)
    {\includegraphics[width=0.99\hsize]{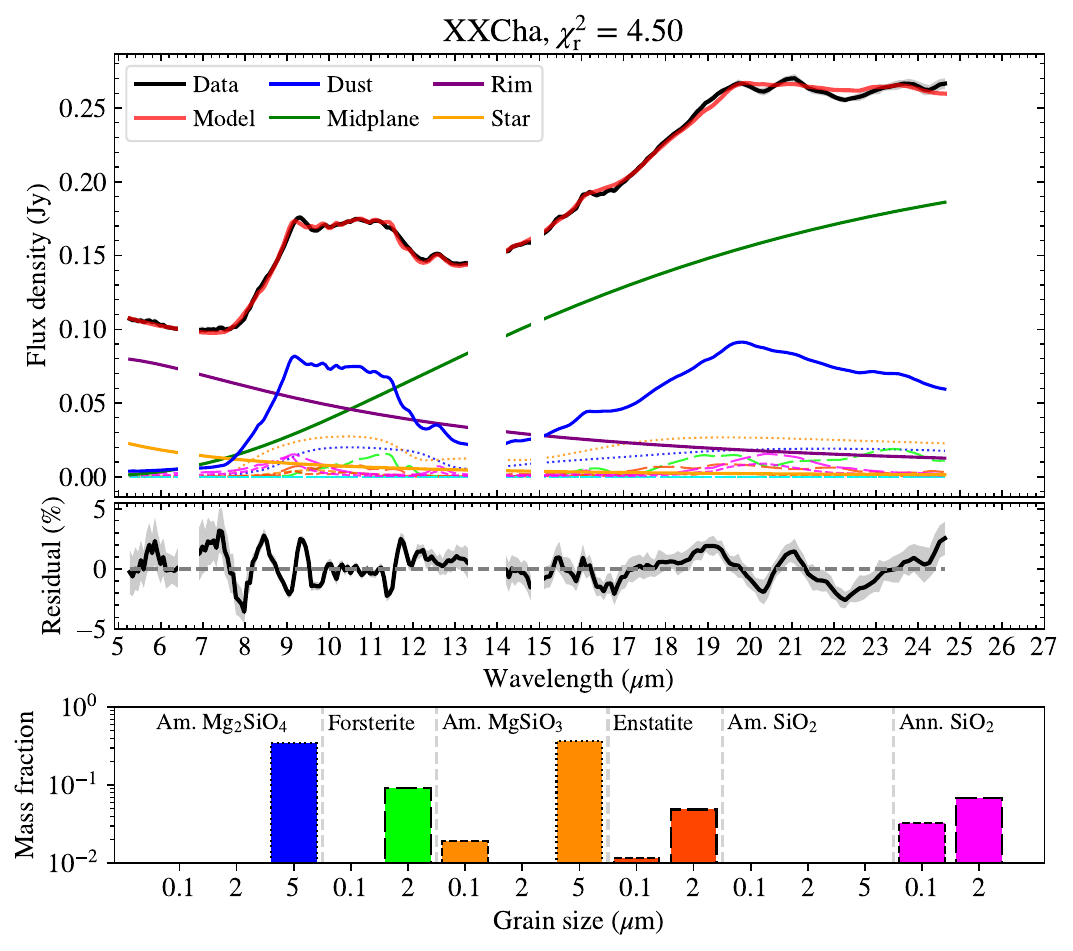  }};
\node at (-2.9,-13.5) {\textbf{GRF}};
\node[image,right=of frame5] (frame6)
    {\includegraphics[width=0.99\hsize]{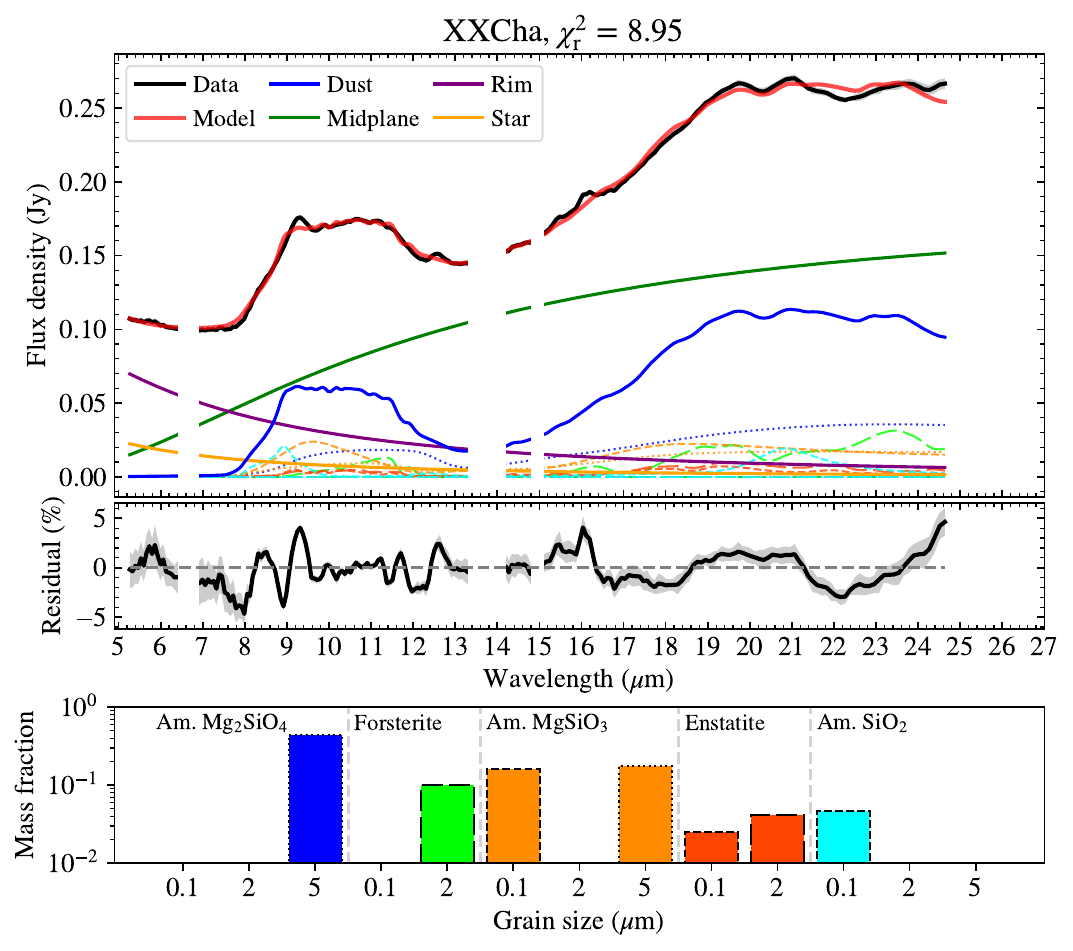  }};
\node at (6.0,-13.5) {\textbf{GRF}};

\end{tikzpicture}

	\caption{Fits to the spectrum of XX~Cha with the $\left[0.1, 2, 5\ \mathrm{am.}\right]$ grain size set. The panels in the top, middle, and bottom rows show the fits with the DHS$\_$nat, DHS$\_$synth, and GRF sets of opacity curves, respectively. In the left (right) column, fits with (without) annealed SiO$_2$ are shown. Spectral regions excluded from our fits are not shown.}
         	\label{fig:fit_example}
\end{figure*}

Our model assumes that the relative abundances of the dust components are the same everywhere in the disk. 
We know from previous studies that the dust composition can in fact change with radius \citep[e.g.,][]{vanBoekel2004nature,Olofsson2009,Varga2024}. However, we refrain from testing this here, and it will be addressed in another study.																	
\subsection{Dust opacities}

Laboratory measurements provide optical data of cosmic dust analogs, from which one can derive spectral templates for the compositional fits. Most of the laboratory optical data are available as complex refractive indices measured on bulk materials \citep[e.g.,][]{Jager1994,Dorschner1995,Henning1997,Zeidler2015}. Absorption and scattering efficiencies can be computed from these data, by taking into account the grain size and applying a grain shape model. Alternatively, laboratory-measured absorption spectra of micron-sized dust particles are also available \citep[e.g.,][]{Hallenbeck1998,Tamanai2006}. 

Despite the enormous variety of minerals on Earth and in the Solar System, only a handful have been detected in circumstellar environments \citep{Tielens2022}. Thus, a carefully selected small set of materials may be able to describe our data. 
Following earlier studies \citep[e.g.,][]{Juhasz2010,Olofsson2010,Jang2024}, we used in our modeling amorphous Mg-silicates of olivine (Mg$_2$SiO$_4$) and pyroxene (MgSiO$_3$) stoichiometry, as well as silica (SiO$_2$). In addition, we included the crystalline silicates forsterite and enstatite, the Mg-rich end members of the olivine and pyroxene mineral series, together with annealed silica.
We do not include materials that lack mid-IR spectral features, such as iron or amorphous carbon, because our data cannot constrain their abundances.
Table~\ref{tab:dustopac} lists the materials used in our study, with references to the original laboratory experiments that provided the complex refractive index data measured on bulk materials. The actual data were obtained from the Database of Optical Constants for Cosmic Dust\footnote{\url{https://www2.astro.uni-jena.de/Laboratory/OCDB/}} compiled by the Laboratory Astrophysics Group of the Astrophysical Institute and University Observatory (AIU) at Jena.

\begin{figure*}
	\centering
	\includegraphics[width=0.32\hsize]{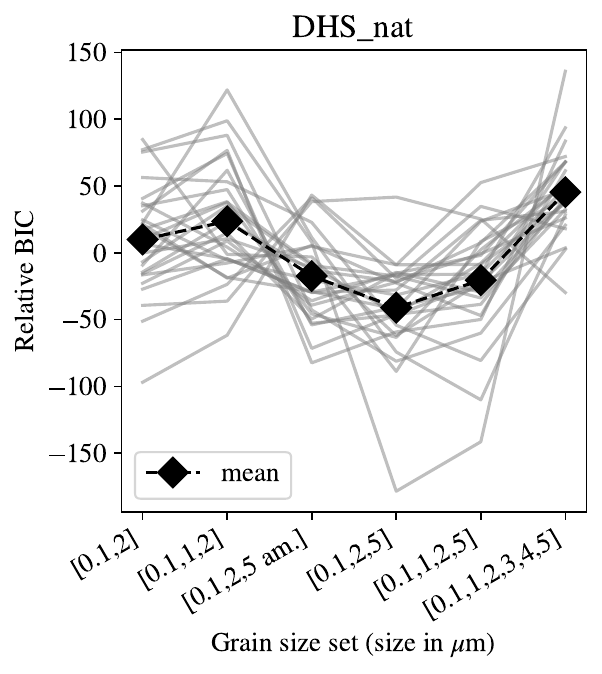} 
	\includegraphics[width=0.32\hsize]{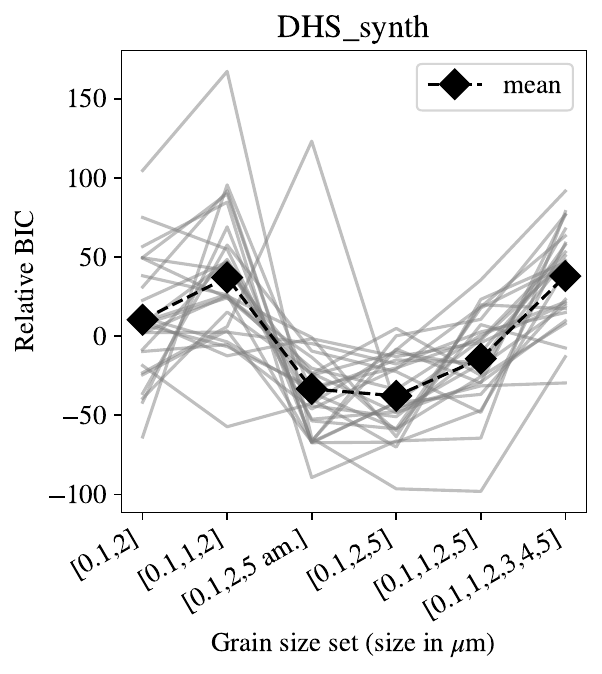} 
	\includegraphics[width=0.32\hsize]{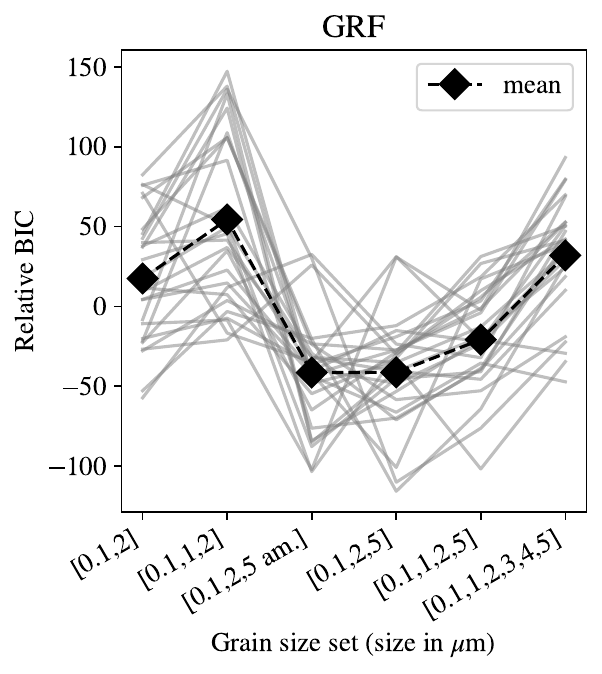} 
	\caption{Relative BIC values for our fits (with opacities including annealed silica) as a function of the grain size sets. The gray lines correspond to individual objects, and the dashed black line is the mean of them. The grain size set $\left[0.1, 2, 5\ \mathrm{am.}\right]$ does not contain $5\ \mu$m-sized crystalline grains.}
         	\label{fig:BIC_vs_Ngs}
\end{figure*}

\begin{figure}
	\centering
	\includegraphics[width=1.0\columnwidth]{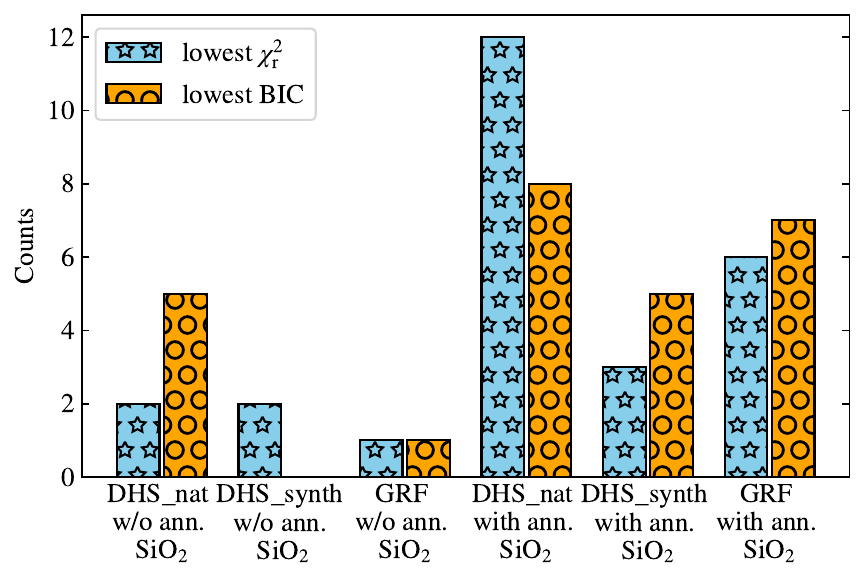} 
	\caption{Bar chart showing how often different opacity sets provide the best fit, as determined by the lowest $\chi^2_\mathrm{r}$ (orange) or the minimum BIC (blue).
	} \label{fig:BIC_vs_opac_types}
         	
\end{figure}

Grain shape has a significant impact on the emission properties of dust. Following \cite{Jang2024}, we considered two grain shape models, one is the distribution of hollow spheres \citep[DHS,][]{Min2005}, and the other is the Gaussian random field \citep[GRF,][]{Min2007}. We used \texttt{optool} \citep{Dominik2021optool} to calculate the DHS opacity curves. The maximum vacuum filling factor ($f_\mathrm{max}$) was set to $0.7$ for amorphous grains and $0.99$ for crystalline grains. 
This choice follows \citet{Juhasz2010}, who found empirically that $f_\mathrm{max} \approx 0.7$ provides the best agreement with the observed amorphous silicate features, while values close to unity are required for crystalline grains to match their characteristic band positions and shapes \citep[e.g.,][]{Min2005}. 

Grain size also plays an important role: as grain size increases, the spectral features become weaker, and for grains larger than $\sim\!10\ \mu$m, the features are so weak that it is difficult to distinguish their emission from a featureless continuum. For each material, we calculated opacities for six distinct grain sizes between $0.1$ and $5\ \mu$m. The resulting opacity curves are presented in Fig.~\ref{fig:opac}. The figure illustrates that crystalline silicate grains produce more pronounced and sharper spectral features compared to their amorphous counterparts. For more information on the grain shape models and on the calculation of opacities, we refer to \cite{Jang2024}. 
An important aim of our modeling is to determine which grain sizes are minimally required to reproduce the MIRI data. Accordingly, we defined six different sets of grain sizes, as shown in Table~\ref{tab:grain_size}. 
Motivated by previous studies that did not find evidence of large crystalline grains in Spitzer spectra of disks \citep[e.g.,][]{Bouwman2008}, we include a variant of the $\left[0.1, 2, 5\right]\ \mu$m grain-size set in which crystalline grains are restricted to sizes of $0.1$ and $2\ \mu$m only, and thus in the $5\ \mu$m size, we have only amorphous grains (denoted as the $\left[0.1, 2, 5\ \mathrm{am.}\right]$ set).

From the opacity curves, we constructed three sets, each of which was fitted to the data in separate runs. Each opacity set contains three amorphous (Mg$_2$SiO$_4$, MgSiO$_3$, and SiO$_2$), and three crystalline materials (forsterite, enstatite, annealed silica). Two sets contain opacities with the DHS grain shape model, while the third one (GRF set) includes GRF opacities\footnote{However, since no GRF opacity data were available for annealed silica, we adopted a DHS opacity curve for that instead.}. Table~\ref{tab:dustopac} provides an overview of the selection of optical data for each opacity set. The two DHS sets, named DHS$\_$synth and DHS$\_$nat, differ in the choices for forsterite and enstatite. While the DHS$\_$synth set includes measurements of high-purity synthetic crystals, for the DHS$\_$nat set, we used measurements of natural crystals that also contained a small amount of iron. The GRF set represents pure, iron-free silicates, as does the DHS$\_$synth set. We further note that the presence of irregular-shaped grains with some Fe content was not tested by these opacity sets and grain shape models.

As shown in Fig.~\ref{fig:opac}, the spectral features of the natural crystals \citep{Zeidler2015} appear somewhat smoother than those of the synthetic crystals \citep{Jager1998,Suto2006}. For example, synthetic forsterite emits quite strong subpeaks at $9.3$, $10.4$, and $11.9\ \mu$m, whereas these peaks in the spectrum of natural forsterite are much more attenuated. By comparison, observations of debris disk spectra with high forsterite content also show gentler features in the $8$--$13\ \mu$m range \citep{Olofsson2012}, more closely resembling the optical data of the natural forsterite sample. For each opacity set, we also defined a variant that excludes annealed SiO$_2$. Comparing fits with and without annealed SiO$_2$ allowed us to directly test for its presence.
In summary, we have six sets of dust materials (DHS$\_$synth, DHS$\_$nat, and GRF, each with or without annealed SiO$_2$), and six sets of grain sizes, resulting in a total of $36$ opacity sets.

\subsection{Fitting approach}

In total, we performed $36$ fitting runs per object, one for each opacity set. There are seven fitted parameters corresponding to the disk structure: inner radius ($r_\textrm{in}$), the inner temperatures of the disk components ($T_\mathrm{midplane,in}$, $T_\mathrm{surface,in}$, $T_\mathrm{rim}$), the width of the rim ($w_\textrm{rim}$), and the power-law gradients of the temperature profiles ($q_\mathrm{midplane}$, $q_\mathrm{surface}$). These parameters are optimized with a dynamic nested sampling algorithm \citep[\texttt{dynesty},][]{Higson2019,Speagle2020dynesty}. In each call of the sampler, when the model was evaluated, the weights of the dust components ($C_i$) were fitted with a nonnegative least squares algorithm \citep[\texttt{scipy.optimize.nnls,}][]{Lawson1995}. This approach is analogous to what was employed in \cite{Juhasz2010} and \cite{Jang2024}.

\subsection{Differences between \texttt{DustComp} and DuCKLinG}

\texttt{DustComp} is similar in functionality to DuCKLinG \citep[Dust Continuum Kit with Line emission from Gas,][]{Kaeufer2024}. DuCKLinG can be used to fit both dust spectral features and gas lines, while our tool is limited to fitting dust features.
Considering the dust fitting, minor differences between the two tools exist in the parameterization and optimization. In DuCKLinG, the grids are defined in temperature space, while \texttt{DustComp} employs radial grids, from which the temperature profile is calculated. Accordingly, \texttt{DustComp} uses the inner and outer radii as parameters, while DuCKLinG uses the inner and outer temperatures.

Both tools use nested sampling algorithms for global optimization and a linear least-squares algorithm for spectral decomposition. A further difference is that DuCKLinG performs a normalization prior to the least-squares step, while \texttt{DustComp} does not. Finally, DuCKLinG maximizes the full likelihood function, while \texttt{DustComp} uses a simplified likelihood that neglects the normalization term.

\section{Results} 
\label{sec:res}

In Fig.~\ref{fig:fit_example}, as an example, we show the fits for a single object, XX~Cha, with the $\left[0.1, 2, 5\ \mathrm{am.}\right]$ grain size set, using the six different opacity sets. Presenting all the $936$~fits ($6$ opacity sets $\times$ $6$ grain size sets $\times$ $26$ objects) in the paper would be superfluous. Thus, in the following subsection we perform a quantitative analysis of the fit performance to identify the opacity and grain size sets that provide the globally optimal fit to the data with the smallest number of parameters. 

\begin{figure*} 
	\centering
	\begin{tikzpicture}[
image/.style = {text width=0.48\textwidth, 
                 inner sep=0pt, outer sep=0pt},
node distance = 1mm and 1mm 
                        ] 
\node [image] (frame1)
    {\includegraphics[width=0.99\hsize]{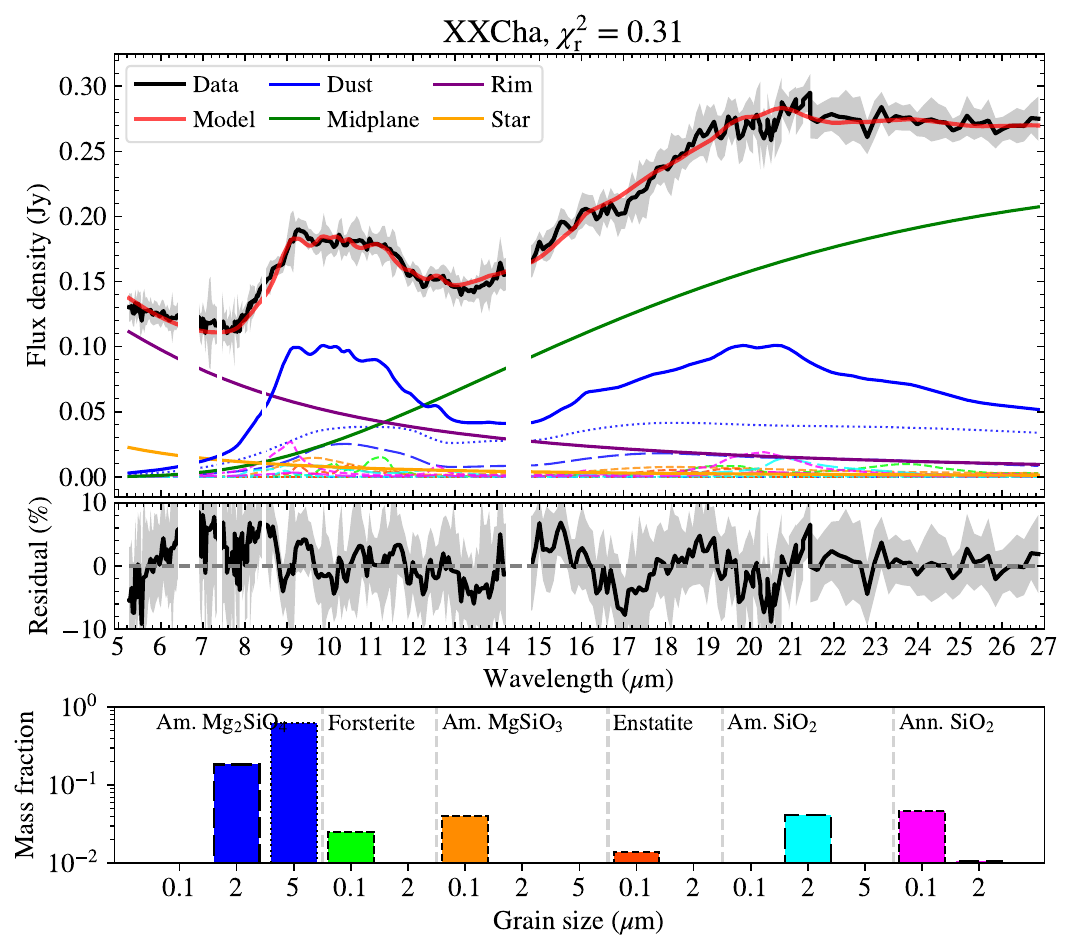}};
\node at (0.2,4.0) {\textbf{Fit to Spitzer IRS data}};
\node at (9.0,4.0) {\textbf{Fit to JWST/MIRI data}};
\node [image,right=of frame1] (frame2) 
    {\includegraphics[width=0.99\hsize]{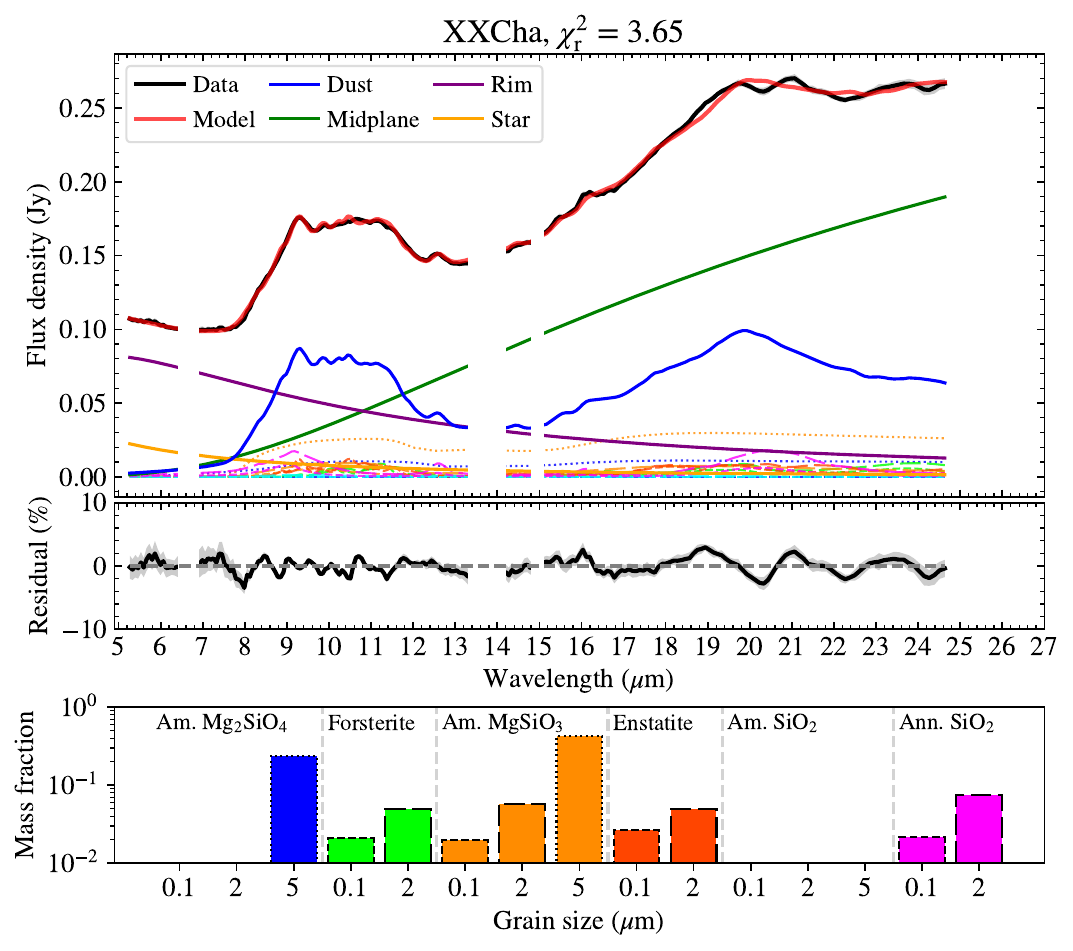}};
\end{tikzpicture}
	\caption{Comparison of spectral fits for XX~Cha using Spitzer (left) and MIRI (right) spectra. }
         	\label{fig:Spitzer_comparison}
\end{figure*}

\subsection{Selecting the best opacity set}
\label{sec:best_opac}

To compare the goodness of fit of the models (with differing number of free parameters), we calculated the Bayesian information criterion (BIC) as
\begin{equation}
 \mathrm{BIC} = N_\mathrm{data} \ln \left( \mathrm{RSS}/N_\mathrm{data} \right) + N_\mathrm{free} \ln \left( N_\mathrm{data} \right),
\end{equation} 
where $N_\mathrm{free}$ is the number of free parameters (including the weights of the dust components), $N_\mathrm{data}$ is the number of data points, and RSS is the residual sum of squares between the model and the data, $\mathrm{RSS} = \sum_{i}^{N_\mathrm{data}} \left( D_i - M_i \right)^2 $, $D_i$ being the data values, and $M_i$ being the model values. A lower BIC value indicates a fit improvement that is justified accounting for the increase in the number of free parameters. The BIC can be used to decide whether increasing the number of parameters (such as including more grain sizes) leads to significantly improved fits. 

First, we investigated the dependence of the BIC on the grain size sets. For a better comparison, we defined the relative BIC, which is $ \mathrm{BIC}_{j,k} - \sum_{k}^{N_\mathrm{gs}} \mathrm{BIC}_{j,k} / N_\mathrm{gs}$, where j, k are the indices of the objects and grain size sets, respectively, and $N_\mathrm{gs}$ is the number of grain size sets. The average is taken per object, and over the BIC values for the different grain size sets. With this, the inherent differences in the goodness of fit between the objects are removed, and only the impact of the grain size selection is emphasized. 
In Fig.~\ref{fig:BIC_vs_Ngs} we plot the relative BIC values for as a function of the grain size sets. For all three opacity sets, we see the same behavior: First, there is no improvement when the number of grains is increased from two to three by adding the $1\ \mu$m size. This is illustrated by Fig.~\ref{fig:opac}, which shows that the opacity curves of $1\ \mu$m-sized grains are very similar to those of $0.1\ \mu$m-sized grains for most species.
However, if we add $5\ \mu$m sized grains instead of the $1\ \mu$m sized ones, the decrease in the BIC values indicate significantly improved fits. 

The lowest average BIC values are obtained with the $\left[0.1, 2, 5\right]$ grain-size set for the runs using the DHS\_nat and DHS\_synth opacity sets. For the GRF opacity set, the $\left[0.1, 2, 5\ \mathrm{am.}\right]$ and $\left[0.1, 2, 5\right]$ sets yield nearly identical average BIC values. These two grain-size sets differ only in whether $5\ \mu$m-sized crystalline grains are included.
There is considerable source-to-source variation in how the BIC changes as a function of the grain-size set, reflecting the mineralogical diversity and thus varying complexity of the spectra in our sample. Some spectra may show signatures consistent with large crystalline grains, while others do not; consequently, including large crystalline grains does not lead to a uniform improvement in fit quality across all objects. However, we clearly find that including more than three grain sizes significantly increases the average BIC, indicating that such models are over-parametrized.
As a conservative choice, we therefore adopt the $\left[0.1, 2, 5\ \mathrm{am.}\right]$ grain-size set, which does not include large crystalline grains (see also Sect.~\ref{sec:disc:spitzer}).

Next, we compared the fit performance for different choices of opacity set. Fig.~\ref{fig:fit_example} presents, for a representative object (XX~Cha), the fits obtained with each of the six opacity sets. 
The figure shows that most of the retrieved mass fractions are consistent across the different opacity choices. One exception is the fraction of the $5\ \mu$m-sized amorphous Mg$_2$SiO$_4$: When using the DHS\_nat opacity set without annealed silica, the fit prefers no $5\ \mu$m-sized Mg$_2$SiO$_4$. Instead, that fit yields a significantly higher fraction of $5\ \mu$m-sized MgSiO$_3$. This behavior indicates a degeneracy between these components, arising from the similarity of their opacity templates. For this reason, the combined mass fraction of large Mg-silicate grains is often a more robust quantity. Further details on these degeneracies are given in Sect.~\ref{sec:res:degener}.
Fig.~\ref{fig:BIC_vs_opac_types}  shows how often different opacity sets provide the best fit, based on either the lowest reduced $\chi$ square ($\chi^2_\mathrm{r}$) or the lowest BIC. From this comparison, the DHS\_nat set with annealed silica emerges as a winner. 
We therefore adopted this opacity set as our preferred choice and present results based on this set in the following. 

\subsection{Fit results}
\label{sec:res:fit}

The fits to the spectra for all objects are shown in Fig.~\ref{fig:fit_all}. Overall, the models reproduce the data well, with residuals mostly within $\pm\!3\%$. The $\chi^2_\mathrm{r}$ values range from $0.5$ (DR~Tau) to $65$ (PDS~70), with a median value of $5.9$, indicating statistically significant deviations between the model and the data for most of our targets. The residuals generally show a wavy pattern, with the largest deviations typically occurring around $8$, $16$, $19$, and $21\ \mu$m (see also Fig.~\ref{fig:app:residual}). 	
The mismatch between the model and the data can be attributed to the following causes: a) the spectral shapes and peak positions of minor subpeaks in the crystalline silicate and silica opacities (Fig.~\ref{fig:opac}) do not exactly match those in the data, including mismatches in subpeak widths, strengths, and central wavelengths; b) the presence of materials that are not represented in our opacity data; and c) radial variations in dust composition, as our model assumes a homogeneous composition across the entire disk.

Furthermore, for several sources with very high silicate feature amplitudes (e.g., LkCa 15, PDS 70, Sz 98) it is difficult to match the model opacity curves with the data at the short wavelength edge ($\sim\!8\ \mu$m) of the silicate feature. \citet{Juhasz2010} showed that deviations of the synthetic spectra from the observations are most likely related to grain shape effects and uncertainties in the iron content of the dust grains.
Differences in the data uncertainties among the sample objects also affect the $\chi^2_\mathrm{r}$ values. For objects exhibiting weaker molecular line emission, the dust continuum is more tightly constrained (cf. Sect.~\ref{sec:obs}), which makes the fits more sensitive to small mismatches and can therefore lead to higher $\chi^2_\mathrm{r}$ values.
From our fits to the MIRI spectra, we derived the mass fraction of each dust component, and the results are listed in Table~\ref{tab:mass_fractions}.
The best-fit disk parameters are given in Table~\ref{tab:disk_params}.
The reported parameter values correspond to the best-fitting solution, while the uncertainties are derived from the 16th--84th percentiles of the equal-weight posterior distributions.
We consider a species detected if its mass fraction exceeds $1\%$. 

Our fits reproduce the dust spectra well using Mg-rich (and thus Fe-poor) silicates, in agreement with previous studies \citep[e.g.,][]{Bouwman2008,Olofsson2009,Juhasz2010}.  
However, we note that opacities of Fe-rich silicates were not included in our fits, and therefore their presence cannot be explicitly ruled out by our analysis. Herschel observations of the $69\ \mu$m forsterite feature generally indicate Fe-poor composition, typically containing less than a few percent Fe, although exceptions have also been reported \citep[e.g.,][]{Sturm2013}.
We also note that our adopted opacity set (DHS$\_$nat) contains some Fe and provides better fits than the other sets (DHS$\_$synth and GRF), which consist of pure synthetic Mg-silicates and silica.

\subsection{Comparison with Spitzer/IRS}

We performed \texttt{DustComp} fits to Spitzer IRS data for several targets in our sample, following the same procedure as for the MIRI data. A comparison for XX~Cha is shown in Fig.~\ref{fig:Spitzer_comparison}, where the left panel displays the fit to the Spitzer spectrum (AORkey: 12696576, \citealp{Lebouteiller2011}\footnote{The Spitzer spectrum was obtained from the LR7 version of the CASSIS spectral atlas.}), and the right panel shows the fit to the MIRI spectrum. 
Although the shapes of the major dust features is reproduced well in the Spitzer spectrum, the detection of minor dust components, such as silica, is made clear with the improved S/N of the MIRI data. 
For example, the smaller features that are clearly identifiable in the MIRI spectrum (at $12.6$, $16$, $19.8$, and $21\ \mu$m) are barely noticeable, if at all, in the Spitzer spectrum. In the Appendix~\ref{sec:app:MIRI_degraded_to_Spitzer}, we present a further test where we degrade the MIRI spectrum of XX~Cha to simulate Spitzer/IRS low-resolution spectroscopic data. 
The higher resolving power of MIRI makes it possible to separate the contributions of molecular line emission from those of the dust, which was not feasible with IRS low-resolution spectroscopy.
This, and in particular the improvement in S/N, are the main enabling factors for detecting weaker dust constituents in MIRI spectra.

\subsection{Degeneracies between dust components}
\label{sec:res:degener}

As mentioned in Sect.~\ref{sec:best_opac}, the mass fractions of amorphous species and/or large ($\gtrsim 2\ \mu$m) grains can be ambiguous because their spectral features are relatively similar, introducing degeneracies in the fitting process. In such cases, the posterior distributions of the corresponding mass fractions often become strongly correlated or multi-modal. We frequently observe that one mode of the posterior favors a solution in which one component of a degenerate pair has a substantial mass fraction while the other is absent, whereas another mode favors the opposite configuration, with intermediate solutions also present. These degeneracies are also reflected in the uncertainties, with the 16th--84th percentile ranges becoming very broad.
In these situations, the combined mass fractions of the degenerate components provides a more robust parameter. Typical degenerate pairs include $5\ \mu$m-sized Mg$_2$SiO$_4$ and $5\ \mu$m-sized MgSiO$_3$, as well as the $2\ \mu$m and $5\ \mu$m grain opacities within a given material. This is illustrated by Fig.~\ref{fig:opac}, where these opacity curves show similar spectral shapes.

The ambiguities caused by these degeneracies also appear when comparing spectra of the same object with different S/N, such as in the case of XX~Cha observed with Spitzer/IRS and MIRI (Fig.~\ref{fig:Spitzer_comparison}). In the Spitzer fit, the model favors substantial contributions from both $2\ \mu$m and $5\ \mu$m-sized Mg$_2$SiO$_4$. In contrast, in the MIRI fit only the $5\ \mu$m-sized Mg$_2$SiO$_4$ component is detected, while $2\ \mu$m and $5\ \mu$m-sized MgSiO$_3$ grains are present instead. Nevertheless, the combined mass fraction of the $2\ \mu$m and $5\ \mu$m Mg-silicate grains remains similar between the two fits ($82\%$ in the Spitzer fit and $73\%$ in the MIRI fit). This example illustrates that the improved S/N of MIRI does not necessarily resolve the degeneracies between large amorphous grains. Dust components with more distinct opacity curves, such as crystalline grains and $0.1\ \mu$m-sized amorphous grains, are much less affected by this degeneracy, and their mass fraction estimates are therefore more reliable. For a more detailed analysis of these degeneracies, we refer to \citet{vanBoekel2005survey} and \citet{Sargent2009_Taurus}.

\begin{figure}
	\centering
	\includegraphics[width=\hsize]{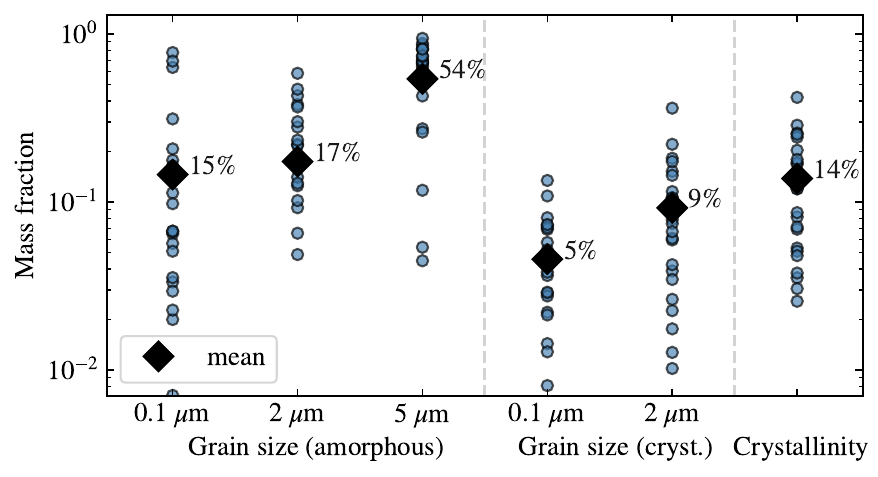}
	\includegraphics[width=\hsize]{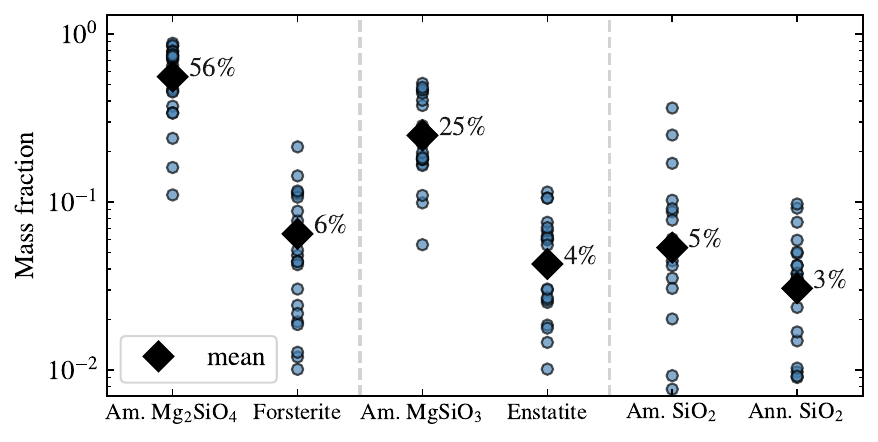}
	\caption{Top: Mass fractions with respect to grain size (separately for the amorphous and crystalline species) and crystallinity. Bottom: Mass fractions for the individual dust components. Colored circles represent individual objects, while black diamonds denote the mean values. The numerical percentages indicate these means.  } 
         	\label{fig:mf_statistics_plot}
\end{figure}

An additional factor contributing to these ambiguities is that the opacity decreases with increasing grain size. As a result, large ($>2\ \mu$m) grains may contribute only weakly to the observed flux despite representing a substantial fraction of the dust mass. For example, if the spectral features can be reproduced either by $2\ \mu$m-sized or $5\ \mu$m-sized grains of the same material; the latter case requires approximately twice as much mass. Integrated fluxes or luminosities of the dust components therefore may provide a more robust diagnostic since they are largely insensitive to the ambiguity in mass fractions arising from grain-size-dependent opacities.

\begin{figure*}
	\centering
	\includegraphics[width=\hsize,trim={0 0.6cm 0 0.4cm},clip]{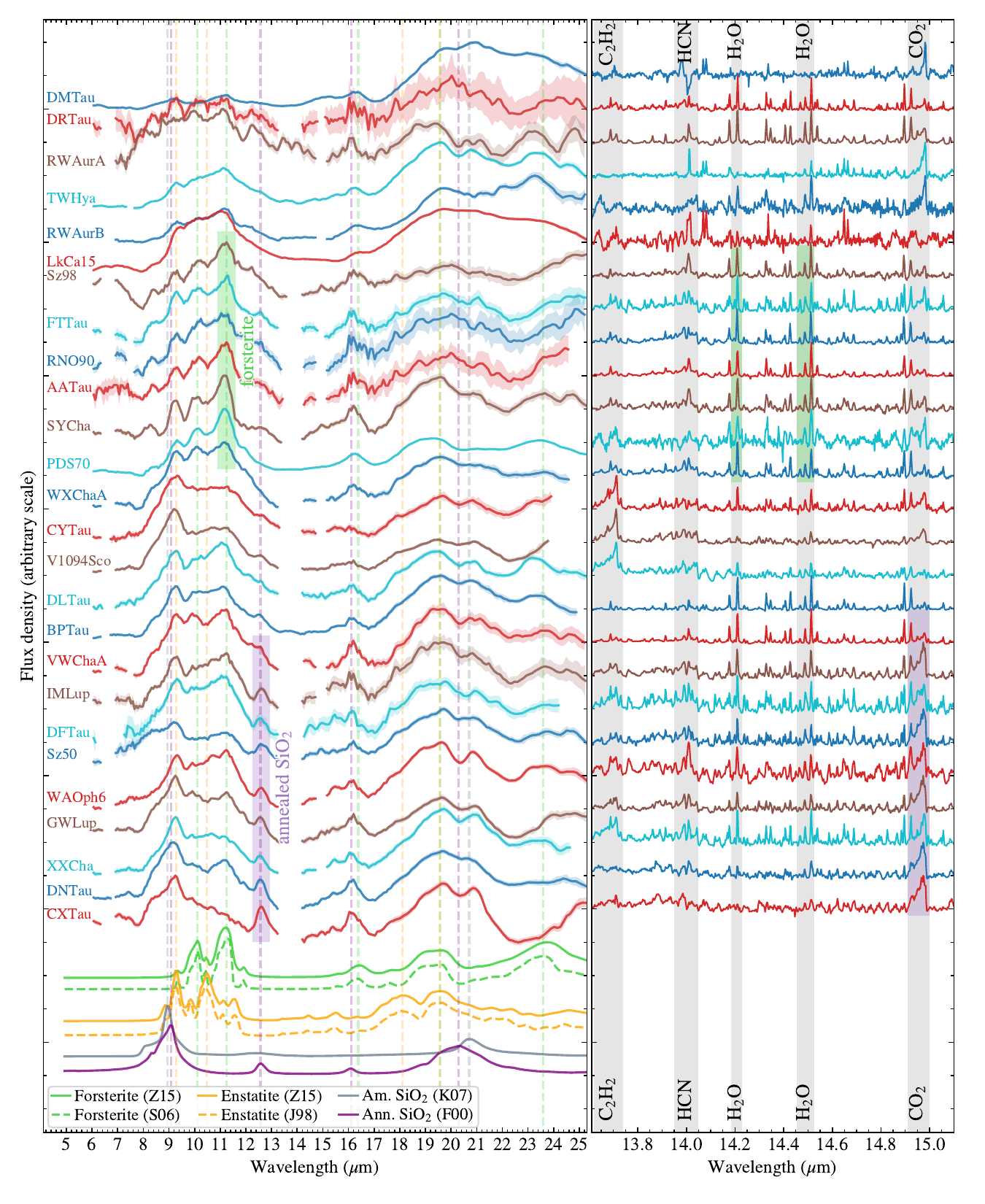}
	\caption{Left: Residual dust spectra after subtracting the modeled contributions of the amorphous Mg-silicates and featureless components. Each residual spectrum has been normalized to its maximum value. The residuals show features of crystalline Mg-silicates and of SiO$_2$ (both amorphous and annealed). For comparison, at the bottom of the plot, we have plotted the opacity curves of $0.1\ \mu$m-sized forsterite, enstatite, amorphous silica, and annealed silica. The positions of major feature peaks are indicated by vertical dashed lines. The green and purple rectangles highlight spectra with prominent features from forsterite and from annealed silica, respectively. Right: Continuum-subtracted spectra at full spectral resolution, showing molecular emission lines. The spectra have been normalized to their maximum value within the plotted wavelength range. Most objects showing strong annealed silica features in their dust spectra have relatively strong CO$_2$ lines, while those with prominent forsterite features tend to be water-rich. } 
         	\label{fig:dust_and_gas_lines} 
\end{figure*}

\begin{figure*}
	\centering 
	\includegraphics[width=0.45\hsize,trim={0.45cm 0.5cm 0.4cm 0.4cm},clip]{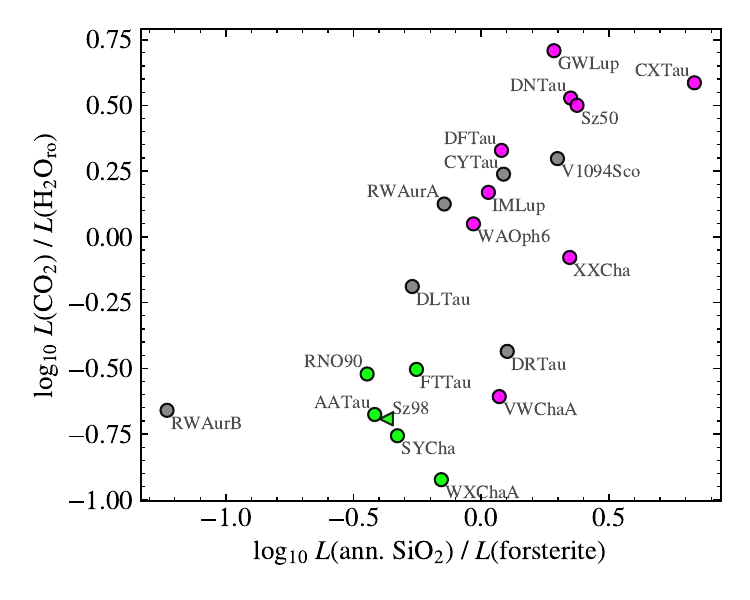}
	\includegraphics[width=0.45\hsize,trim={0.45cm 0.5cm 0.4cm 0.4cm},clip]{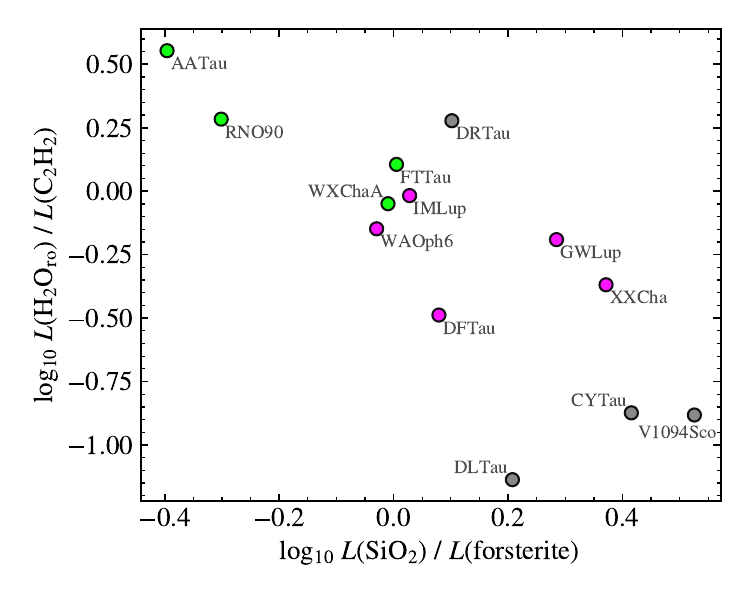}

	\caption{Left panel: Luminosity ratio of CO$_2$ to H$_2$O$_\mathrm{ro}$  versus the luminosity ratio of annealed silica to forsterite. 
	Right panel: Luminosity ratio of H$_2$O$_\mathrm{ro}$ to C$_2$H$_2$ versus the luminosity ratio of the total silica content to forsterite.
	The purple and green symbols denote objects belonging to the annealed silica-rich and forsterite-rich groups, respectively, identified in Fig.~\ref{fig:dust_and_gas_lines}. Triangles indicate limits.} 
         	\label{fig:gas_vs_dust}
\end{figure*}

\subsection{Sample statistics}

Fig.~\ref{fig:mf_statistics_plot} shows mass fractions with respect to grain size, stoichiometry, and crystallinity for the complete sample. Looking at the sample averages, we see an increasing trend towards larger grain sizes (both for amorphous and crystalline grains), clearly deviating from the Mathis--Rumpl--Nordsieck (MRN) grain size distribution \citep{Mathis1977_MRN} used to describe interstellar dust. Thus, our result on the increased fraction of $\gtrsim\!2\ \mu$m grains suggest that significant grain growth has happened in the disks of our sample. 
With regard to stoichiometry, we see that Mg$_2$SiO$_4$ (which includes crystalline forsterite and its amorphous counterpart) is the most abundant on average, followed by Mg$_2$SiO$_3$. The average value of the crystallinity is $14\%$, with typical values in the $5$--$24\%$ range, which is comparable to what was found in earlier studies based on Spitzer spectra \citep{Olofsson2010,Juhasz2010}.

There is growing evidence of significant variability in the IR spectra of planet forming disks \citep{Espaillat2011,kospal_atlas,Varga2017}, including variability of the silicate feature \citep{Abraham2009,Bary_spitzer,Kospal2023,Perotti2023,Jang2024,Sameshima2026}. Thus, it is plausible that the composition of the dust emitting the IR spectral features changes on observable timescales. \citet{Kospal2025} analyzed a four-epoch spectral monitoring dataset of the T Tauri source DQ Tau obtained with MIRI between January and March 2025. Their dust compositional fits to each epoch did not reveal significant changes in the dust mineralogy. The IR variability aspect of our MINDS sample will be addressed in a future study.

\subsection{Forsterite and annealed silica}
\label{sec:res:fors_silica}

Forsterite, which has its main spectral peak at $11.3\ \mu$m, is the most commonly found crystalline mineral in planet forming disks. In our sample, the average mass fraction of forsterite is $6.5\%$, with typical values in the $2$--$11\%$ range (corresponding to the 16th and 84th percentiles). To enhance the visibility of less abundant dust constituents, we show in the left panel of Fig.~\ref{fig:dust_and_gas_lines} the residual dust spectra obtained after subtracting the modeled contributions of the amorphous Mg-silicates and featureless components. These residual spectra therefore emphasize the spectral signatures of crystalline Mg-silicates and silica (both amorphous and annealed). We ordered the spectra to highlight two characteristic groups: one exhibiting strong forsterite features (indicated by a green rectangle), and another showing a well-defined feature at $12.6\ \mu$m (indicated by a purple rectangle).
We exclude that the $11.3\ \mu$m and $12.6\ \mu$m are PAH features, since no other PAH bands (e.g., at $6.2\ \mu$m or $7.7\ \mu$m) can be detected in the data.

Among silicates, the $12.6\ \mu$m peak is unique to the annealed silica template. When it is present in the object spectra, it is consistently accompanied by an additional peak at $16.1\ \mu$m, moreover, the $20\ \mu$m feature exhibits a double-peaked structure, with subpeaks at $19.7$ and $20.9\ \mu$m. In contrast, forsterite shows bands near $19.5$ and $16.3\ \mu$m, but does not exhibit prominent features near $12.6$ or $20.9\ \mu$m. Additionally, sources displaying strong forsterite emission at $11.3\ \mu$m show a minor peak at $16.2$--$16.3\ \mu$m, which is clearly offset from the $16.1\ \mu$m peak  observed in sources that exhibit the $12.6\ \mu$m feature. 
Amorphous (glassy) silica does have a band in the $12$--$13\ \mu$m range; however, this feature is broad and weak compared to the narrow and prominent peak observed here. The prominent $9\ \mu$m peak in the data is consistent with both amorphous and crystalline silica templates.
Taken together, these characteristics provide strong evidence that the features observed at $12.6$, $16.1$, and $20.9\ \mu$m originate from annealed silica. 

This mineral has already been found in a few T Tauri disks \cite{Sargent2009_silica,Sargent2009_Taurus}, but the improved S/N of MIRI enables to detect it in a larger fraction of objects (at least $35\%$ within our sample) than was previously possible with Spitzer. The average mass fraction of annealed silica in our sample is $3\%$, while the highest fraction, $10\%$, is found in XX~Cha.

Most of the residual dust spectra in Fig.~\ref{fig:dust_and_gas_lines} show a broad spectral feature between $17$ and $23\ \mu$m. Although all of our opacity templates show features in this wavelength range, they are not able to precisely reproduce the observed spectra, as the positions of the peaks do not coincide with those seen in the data.
A likely explanation for this mismatch is that spectral features in this region are particularly sensitive to the exact grain composition (e.g., iron content) and
morphology \citep{Juhasz2010,Jang2024}, and that our opacity set, while sufficient to reproduce the $10\ \mu$m feature, lacks the complexity required to do so around $20\ \mu$m. Radial variations in dust composition may also contribute to these deviations.

\subsection{Relation between the dust and gas content}
\label{sec:res:gas_dust}

Thanks to the excellent resolving power and S/N of MIRI, we can now investigate possible relations between the molecular gas content and the dust mineralogy of planet forming disks. In the right panel of Fig.~\ref{fig:dust_and_gas_lines}, we show the continuum-subtracted spectra in the $13.6$–$15.1\ \mu$m wavelength range. These can be compared to the left panel, which displays the flux contributions from crystalline Mg-silicates and silica (both amorphous and annealed). 
The figure reveals that most objects showing strong annealed silica features in their dust spectra (highlighted by the purple rectangle) also show relatively strong CO$_2$ emission lines (with VW Cha A being the only exception). In contrast, objects with prominent forsterite features tend to have relatively strong H$_2$O emission and weak CO$_2$ lines. Several sources do not fall into either group; for example, TW Hya and RW Aur B show relatively strong CO$_2$ emission but weak annealed silica features. 
Additionally, CY Tau, V1094 Sco, and DL Tau constitute a separate group characterized by strong C$_2$H$_2$ emission and similar dust residual spectra indicating a significant fraction of amorphous silica grains. This confirms the finding of \citet{Tabone2026}.

Following \citet{Gasman2025}, we fitted the continuum-subtracted spectra with slab models to measure line luminosities of CO$_2$, C$_2$H$_2$, and H$_2$O. For H$_2$O, we used the pure rotational lines at $23.789$--$23.912\ \mu$m (H$_2$O$_\mathrm{ro}$).
In the left panel of Fig.~\ref{fig:gas_vs_dust}, we present the luminosity ratio of CO$_2$ to H$_2$O$_\mathrm{ro}$ as a function of the luminosity ratio of annealed silica to forsterite\footnote{The dust feature luminosities are calculated in the fitted wavelength range of $5$--$27\ \mu$m.}. 
As noted in Sect.~\ref{sec:res:degener}, we prefer the luminosity ratios of the dust components over the ratios of their mass fractions, because degeneracies in the mass fractions arising from grain-size-dependent opacities can bias the inferred values. In contrast, feature luminosities provide a more robust diagnostic. We note, however, that luminosity ratios are intrinsically weighted toward smaller grains, which exhibit stronger emission features per unit mass.
The figure supports our interpretation that higher CO$_2$ luminosity (with respect to that of H$_2$O) is preferentially associated with a larger annealed silica contribution. We observe a very similar trend when annealed silica is replaced by the total silica luminosity.

In the right panel of Fig.~\ref{fig:gas_vs_dust}, we plotted the luminosity ratio of H$_2$O$_\mathrm{ro}$ to C$_2$H$_2$ as a function of the luminosity ratio of the silica to forsterite. Here we again see a correlation: H$_2$O-rich and forsterite-rich objects are located in the upper-left part of the diagram, whereas C$_2$H$_2$-rich and silica-rich objects lie in the lower-right part. In this case we consider the total silica luminosity, because the trend becomes less clear when only annealed silica is used. The main reason is that the most C$_2$H$_2$-rich objects (CY~Tau, V1094~Sco, and DL~Tau) are not particularly rich in annealed silica but instead show a higher contribution from its amorphous form.

To quantify these trends, we performed linear regression in logarithmic space. For the $L$(CO$_2$)/$L$(H$_2$O$_\mathrm{ro}$) diagram we obtained a Pearson correlation coefficient of $\rho = 0.71$ with a $p$-value of $0.00047$. For the $L$(H$_2$O$_\mathrm{ro}$)/$L$(C$_2$H$_2$) plot we got $\rho = -0.79$ with a $p$-value of $0.0014$.
Excluding the two leftmost points yields $\rho = -0.58$ with a $p$-value of $0.098$. We therefore conclude that the observed trends are statistically significant. 
In Sect.~\ref{sec:disc:gas_dust}, we discuss the correlation between the dust and molecular gas content in more detail.

\section{Discussion}
\label{sec:disc} 

\subsection{Grain growth}

\begin{figure}
	\centering
	\includegraphics[width=\hsize]{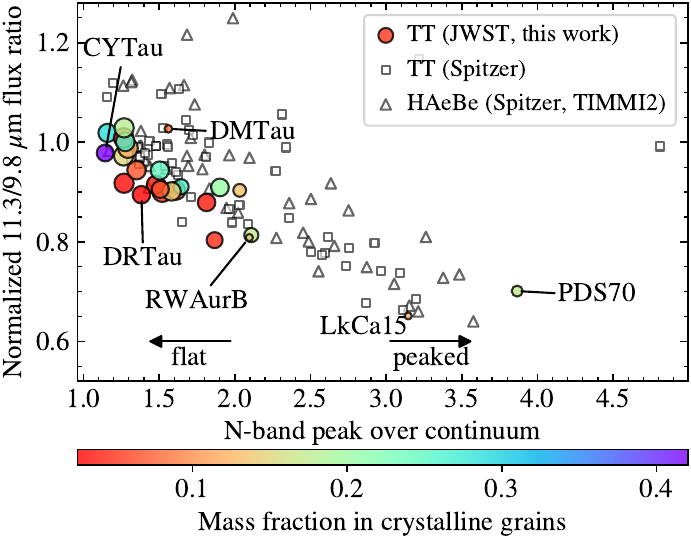}
	\caption{Normalized $11.3/9.8\ \mu$m flux ratio versus the amplitude of the $10\ \mu$m silicate feature. Colored circles indicate our new JWST observations. The size of the circles scales with the mass-averaged grain size, while the color indicates the crystallinity ratio. For comparison, corresponding points for samples of Spitzer and TIMMI2 spectra are also plotted \citep{vanBoekel2005survey,Kessler-Silacci2006,Juhasz2010}. } 
         	\label{fig:flux_ratio_plot}
\end{figure}

\begin{figure}
	\centering
	\includegraphics[width=\hsize]{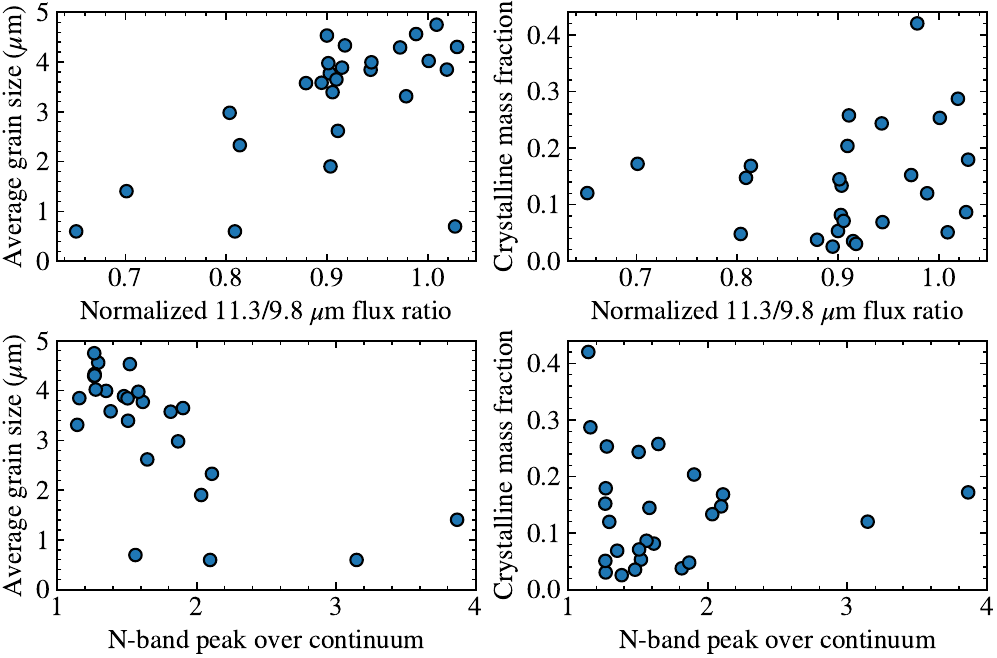}
	\caption{Scatter plots showing the relationships among the average grain size, crystalline mass fraction, normalized $11.3/9.8\ \mu$m flux ratio, and the amplitude of the $10\ \mu$m silicate feature.
	 } 
         	\label{fig:gs_cryst_vs_fluxratio_ampl}
\end{figure}

As can be seen in Fig.~\ref{fig:opac}, an increase in grain size causes the opacity peaks of silicate minerals to become weaker and flatter. Increased crystallinity can also flatten the shape of the $10\ \mu$m feature through the appearance of narrow crystalline subfeatures.
Based on these effects, a simple plot showing the $11.3/9.8\ \mu$m flux ratio versus the $10\ \mu$m feature strength has proven to be a useful diagnostic tool for tracing the degree of dust processing in a wide variety of planet forming disks \citep{Bouwman2001}.
In Fig.~\ref{fig:flux_ratio_plot}, we present this diagram for our sample\footnote{Following \citet{vanBoekel2003}, we normalized the spectra using $1 + (F_\nu-F_{\nu,\mathrm{cont}})/\langle F_{\nu,\mathrm{cont}}\rangle$, where $F_{\nu,\mathrm{cont}}$ is a linear continuum under the $10\ \mu$m silicate feature, with anchor points at $7.85$ and $13.15\ \mu$m.
The flux ratio and the feature amplitude are then calculated from this normalized spectrum.}. For comparison, we also include corresponding data points for samples of T Tauri and Herbig disks from earlier studies. 
The data points in the figure are distributed along a line from the upper left to the lower right, indicating that sources with a strong $10\ \mu$m silicate emission feature preferentially exhibit a triangular shape which peaking near $9.8\ \mu$m (lower right), while sources weaker emission tend to show a flatter silicate feature (upper left). This correlation was first reported by \citet{vanBoekel2003} and \citet{Przygodda2003} and has been interpreted as evidence of dust grain growth and increasing crystallinity fraction in disks. 

Strong silicate emission with low $11.3/9.8\ \mu$m flux ratios is characteristic of small, amorphous grains similar to ISM dust, while weaker and flatter features with higher ratios trace larger grains within the amorphous dust population. Increased crystallinity can also contribute to flattening of the $10\ \mu$m feature through narrow crystalline subfeatures superimposed on the broad amorphous silicate emission. In our sample, most objects have relatively flat silicate features, accordingly, their data points are concentrated in the upper left region of the diagram. Our data points overlap with the distributions of the Spitzer and TIMMI2 samples; however, we lack objects showing high amplitude dust emission peaks. The only two objects in our sample occupying this region are LkCa~15 and PDS~70. 
Since these two are transitional disks, they are considered more evolved than most of our sample. Yet, they occupy the locus where one would expect disks with pristine ISM-like dust. This demonstrates that older systems can still contain dust with “pristine” characteristics, implying that the trend in dust processing cannot be interpreted as a tracer of disk age. 
A possible explanation for the silicate features in transitional disks is the presence of mechanisms that maintain a population of small, amorphous grains, such as size sorting at an outer pressure bump or secondary generation of small grains in the inner disk.

In Fig.~\ref{fig:flux_ratio_plot}, the size of the plot symbols scales with the mass-averaged grain size derived from our spectral decomposition. We indeed find that the fraction of large grains tends to increase toward the upper left part of the figure. In addition, we indicate the crystallinity fraction by color coding. With that, no clear trend is apparent. The relationships between these parameters are further highlighted in Fig.~\ref{fig:gs_cryst_vs_fluxratio_ampl}. 
We therefore conclude that increasing grain size, rather than crystallinity, is the primary cause of the observed distribution of data points in the flux ratio versus feature amplitude diagram.
This result is in agreement with previous findings from Spitzer data \citep[e.g.,][]{Kessler-Silacci2006,Bouwman2008}.

\subsection{PDS 70}

PDS 70 has the strongest $10\ \mu$m silicate feature with respect to the continuum in our sample. \cite{Jang2024} presented a detailed mineralogical analysis of PDS 70, using a fitting process very similar to ours. Similar to us, they tested different opacity sets for the fits (GRF, DHS, and aerosol), and they found that the data are reproduced best with the GRF set. Our GRF fit to PDS 70 is presented in Fig.~\ref{fig:app:fit_PDS70}.
Comparing this to the fit with our adopted DHS\_nat opacity set, we find that GRF opacities can better reproduce the features between $16$ and $21\ \mu$m, while the DHS\_nat opacity set reproduces better the $10\ \mu$m feature.
Overall, we also find, in agreement with \cite{Jang2024}, that the GRF set provides a better fit to the MIRI spectrum of PDS 70 than the DHS sets.  
We also note that our DHS models prefer to have some enstatite, but the GRF models not, which was the same conclusion in \cite{Jang2024}. Otherwise, the mass fractions derived from our fits are generally consistent with the ones reported by \cite{Jang2024}, despite the different choice of grain sizes. For example, we find a forsterite fraction of $12\%$, compared to $\sim\!15\%$ in their study.
\cite{Liu2025} analyzed the Spitzer and MIRI spectra of PDS 70 using both spectral decomposition and radiative transfer modeling, and their best-fit mass fractions are in good agreement with those we obtained using our DHS\_nat opacity set.

\subsection{Comparison with Spitzer-based studies}
\label{sec:disc:spitzer}

Our findings are in good agreement with earlier mineralogical studies of disks based on Spitzer/IRS data. Consistent with previous work, we find that Mg-rich (and Fe-poor) silicates represent well our data \citep[e.g.,][]{Bouwman2008,Olofsson2009,Juhasz2010}. 
The crystallinity fractions derived in our analysis, ranging from $3$--$42\%$, are also comparable to those reported by \cite{Olofsson2010} for T Tauri disks (mean value $16\%$) and by \cite{Juhasz2010} for Herbig Ae/Be disks (range $1$--$30\%$).
One notable difference is that we do not require separate warm and cold dust populations to reproduce the observed spectra. This contrasts with the approaches of \cite{Bouwman2008} and \cite{Juhasz2010}, who modeled narrower wavelength regions by fitting the $10\ \mu$m and $20\ \mu$m silicate features separately. Similarly, \cite{Sargent2009_Taurus} and \cite{Olofsson2010} fitted most of the Spitzer spectral range ($5$--$37\ \mu$m) using two dust components, warm and cold, each with its own independent composition. Such analyses led to the so-called crystallinity paradox, suggesting the presence of a compositionally distinct reservoir of cold crystalline material responsible for the longer-wavelength ($\lambda > 20\ \mu$m) silicate bands \citep{Olofsson2009,Olofsson2010}.

In our modeling, we adopted a single dust composition across the entire radial and temperature range. Addressing the crystallinity paradox is beyond the scope of this study. Nevertheless, the fact that our model reproduces the spectra very well with a single dust population (with typical root mean square residuals of $1$--$2\%$) suggests that compositional differences between the warm ($T\!\sim\!300$--$600$~K) and cold ($T\!\sim\!100$--$200$~K) disk regions may be limited. That said, some of the residual features in our fits (Fig.~\ref{fig:app:residual}), particularly at wavelengths longer than $18\ \mu$m, could be compatible with the presence of cold crystalline material. We also note that our analysis does not cover the $27$--$35\ \mu$m wavelength range included in earlier Spitzer studies, which may provide a more sensitive probe of cold outer disk dust reservoirs.

Previous studies \citep[e.g.,][]{Sargent2009_Taurus,Olofsson2010} derived dust temperatures from their fits to Spitzer spectra. However, a direct comparison with our results is not straightforward because the models employ significantly different parameterizations. While these studies used separate warm and cold dust components, each characterized by a single temperature, our model features radial temperature gradients, such that no single temperature can be uniquely associated with a given emission component. To enable a comparison with previous work, we estimated characteristic temperatures for the dust emission using the following approach. First, we computed the half-flux radius, defined as the radius within which half of the flux at a given wavelength originates. This is a wavelength-dependent quantity. We then evaluated the temperature at the half-flux radius at two selected wavelengths ($10$ and $23\ \mu$m), which can be interpreted as characteristic temperatures of the emission at those wavelengths.
We found that at $10\ \mu$m the inferred dust temperatures range from 350 to 630~K in our sample (16th--84th percentile range), while at $23\ \mu$m they are in the range of 300--500~K. Our $10\ \mu$m temperatures are broadly consistent with the warm dust component reported by \citet{Sargent2009_Taurus} (490--720~K). In contrast, their cold dust component (110--150~K) is significantly cooler than our inferred temperatures at $23\ \mu$m. Both the warm (190--240~K) and cold (60--90~K) dust components reported by \citet{Olofsson2010} are also significantly cooler than the temperatures inferred in our study and by \citet{Sargent2009_Taurus}.

\begin{figure} 
	\centering 
	\includegraphics[width=\hsize,trim={0.45cm 0.6cm 0.6cm 0.4cm},clip]{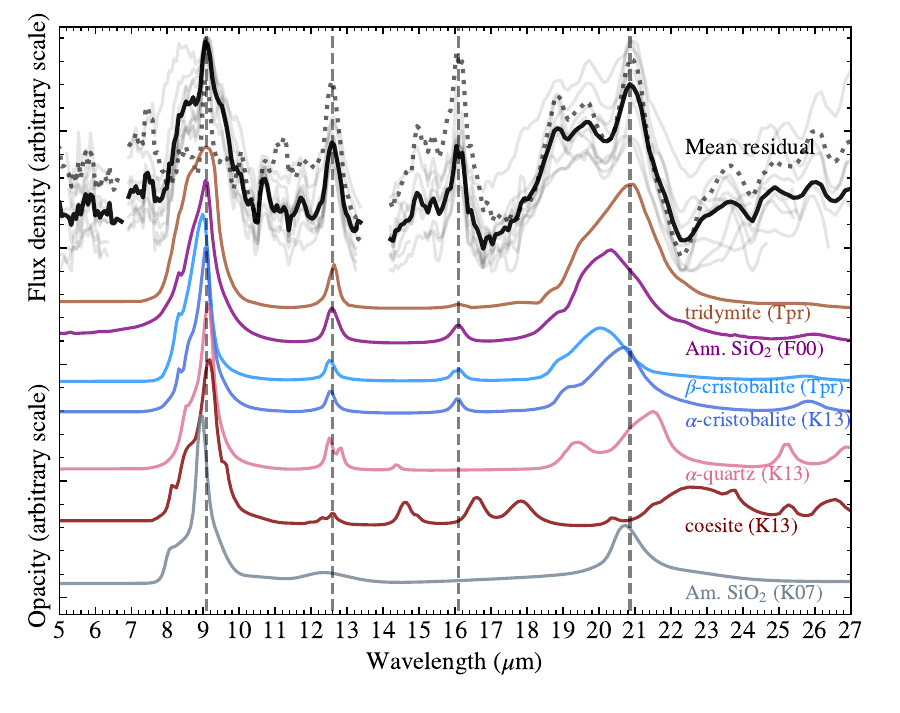}
	\caption{Upper part: Residual spectra obtained after subtracting the modeled contributions of all Mg-silicates and featureless components. Gray lines correspond to individual sources in the subset for which annealed silica is robustly detected. The solid black line shows the average residual spectrum of this subset, highlighting the spectral features of silica. For comparison, the dotted line shows the subset-averaged residuals from a fitting run performed without including any silica opacity templates.
	Dashed lines indicate the locations of the major spectral peaks in the residuals. Lower part: For comparison, laboratory-measured opacity curves of various silica polymorphs as well as that of amorphous silica are shown. Sources of the opacity data: Tpr: A. Tamanai (priv. comm.), F00: \citet{Fabian2000}, K13: \citet{Koike2013}, K07: \citet{Kitamura2007}.} 
         	\label{fig:silica_residuals}
\end{figure}

\begin{figure*}
	\centering 
	\includegraphics[width=0.45\hsize,trim={0.45cm 0.5cm 0.4cm 0.4cm},clip]{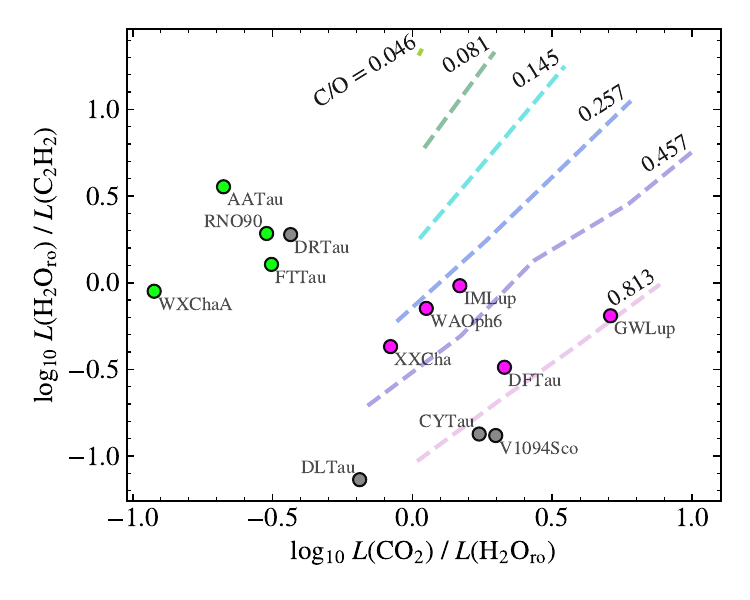}
	\includegraphics[width=0.45\hsize,trim={0.45cm 0.5cm 0.4cm 0.4cm},clip]{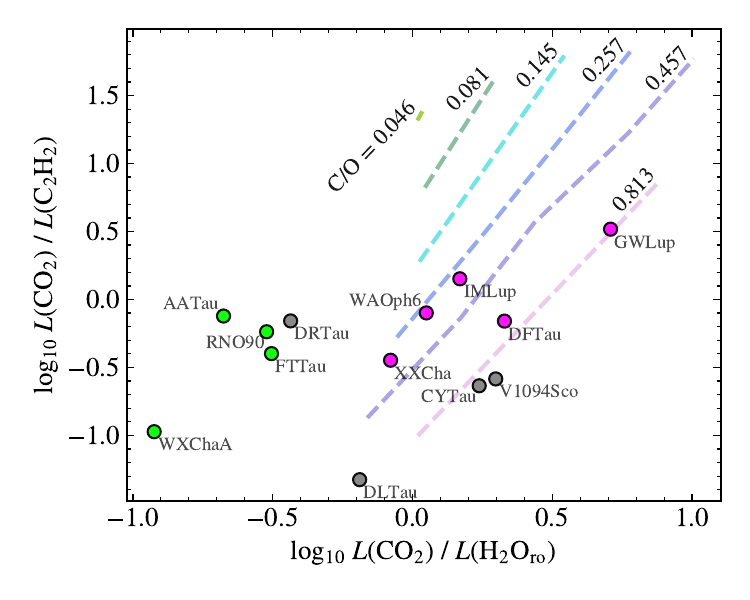}
	\caption{C/O diagnostic plots following \cite{Arabhavi2026}:
	luminosity ratios H$_2$O$_\mathrm{ro}$/C$_2$H$_2$ versus CO$_2$/ H$_2$O$_\mathrm{ro}$ (left panel) and CO$_2$/C$_2$H$_2$ versus CO$_2$/H$_2$O$_\mathrm{ro}$ (right panel). 
	The purple and green symbols denote objects belonging to the annealed silica-rich and forsterite-rich groups, respectively, identified in Fig.~\ref{fig:dust_and_gas_lines}. The dashed lines indicate the C/O ratio derived from the thermo-chemical modeling grid of \cite{Arabhavi2026}.} 
         	\label{fig:C_O_diag_Arabhavi}
\end{figure*}

Both \cite{Bouwman2008} for T Tauri disks and \cite{Juhasz2010} for Herbig Ae/Be disks found that the average size of crystalline grains is smaller than that of the amorphous ones. Other studies, such as \cite{Olofsson2010}, did not require large ($5$--$6\ \mu$m-sized) crystalline grains to reproduce their spectra. These findings are also supported by theoretical crystallization models \citep[e.g.,][]{Gail2004}. 
A further complication is that the opacity curves of large crystalline grains exhibit significant degeneracy with those of large amorphous grains, making them difficult to distinguish spectroscopically \citep{Olofsson2010}.
As described in Sect.~\ref{sec:best_opac}, there is substantial evidence of the presence of large, $\sim\!5\ \mu$m-sized amorphous grains in our sample. In contrast, the presence of large crystalline grains is not uniformly supported by our fits; therefore, we did not adopt them in our preferred opacity set.
To further examine this issue, we inspected fits obtained with the $\left[0.1, 2, 5\right]$ grain-size set, which includes large crystalline grains. This comparison shows that while large crystals are not required in most sources, there are four cases, AA Tau, CY Tau, RNO 90, and Sz 50, where their inclusion noticeably improves the fit.

\subsection{Silica mineralogy} 

Silica occurs in several polymorphic forms that show distinct spectral signatures in the mid-IR \citep[e.g.,][]{Koike2013}. The presence of specific silica polymorphs in circumstellar dust can provide insight into the dust's origin, as each polymorph forms under characteristic pressure and temperature conditions. In Fig.~\ref{fig:silica_residuals}, we plot the residual spectra after subtracting the modeled contributions of all Mg-silicates and featureless components. The black line is the mean of the residuals in the subset for which annealed silica is robustly detected (highlighted by the purple rectangle in Fig.~\ref{fig:dust_and_gas_lines}). Our specific choice of silica opacity templates in the modeling could introduce a bias, potentially causing the residual spectra to resemble the adopted opacity profiles. To assess whether this affects our analysis, we performed an additional modeling run in which silica was completely excluded from the opacity set. The resulting average residual spectrum from this run is shown as a dotted line in the figure. We note that the $9\ \mu$m peak is significantly weaker in this case, while the other peaks remain in good agreement with those seen in the original residual spectrum. We attribute the reduced strength of the $9\ \mu$m feature to the fitting routine attempting to reproduce the silica contribution at this wavelength using enstatite opacity, which also exhibits a prominent band in this region. We therefore conclude that the prominent residual peaks are not artifacts introduced by the adopted opacity set.

To compare the residual peaks with laboratory measurements of cosmic dust analogs, opacity curves of several silica polymorphs, as well as amorphous silica, are shown in Fig.~\ref{fig:silica_residuals}. The opacity curve by \citet{Fabian2000}, which we have included in our opacity set, is measured on samples which have been annealed at $T\!\sim\!1200$~K from amorphous silica nanoparticles. During the experiment, they observed the appearance of cristobalite and tridymite features in the IR spectra of the samples. Subsequent IR and X-ray diffraction analysis proved cristobalite to be the major component of the annealed silica end product. 

The $9.08\ \mu$m peak in our data matches all the presented opacity curves; however, the peaks at longer wavelengths allowed us to distinguish between them. The features at $12.58$ and $16.1\ \mu$m are best matched by cristobalite opacities, while the $20.86\ \mu$m feature is more consistent with tridymite. We note, however, that other species, such as enstatite and forsterite, also exhibit features around $20\ \mu$m, making the identification of this peak with a specific silica polymorph less certain. We therefore conclude that cristobalite is detected as a major component of the annealed silica content in our disk sample, with tridymite also potentially contributing. This is in agreement with the findings of \citet{Sargent2009_silica} who also found cristobalite as the dominant polymorph of silica in five T Tauri disks. Cristobalite is a high-temperature (forming at $T \gtrsim 1200$~K), low-pressure polymorph, and it was found in chondritic meteorites \citep{Dodd1981}. Furthermore, crystalline silica (predominantly cristobalite) has been found in five out of 25 cometary samples from the Stardust mission \citep{Roskosz2015}.
Silica, in particular annealed silica, is regarded as a tracer of dust processing because it is not observed in the ISM \citep{Kemper_silicate} and is therefore most likely to have formed within planet-forming disks. 

Further progress in modeling silica polymorphs will require laboratory measurements of their complex refractive indices, from which opacity curves for arbitrary grain sizes can be computed. Currently, for minerals such as cristobalite, only opacity measurements of powder samples are available, representing a single effective grain size.

\subsection{Origin of the dust-gas correlation}
\label{sec:disc:gas_dust}

\cite{Arabhavi2026} constructed a grid of thermo-chemical models to explore how variations in C and O abundances influence the mid-IR spectra of disks. 
They identified the molecular flux ratios CO$_2$/H$_2$O and H$_2$O/C$_2$H$_2$ as efficient tracers of the C/H and O/H ratios, respectively, and demonstrated that their combination can be used to infer the underlying C/O ratio.
In Fig.~\ref{fig:C_O_diag_Arabhavi}, we reproduce their diagnostic plots and place our objects on them. In these plots, the forsterite-rich and annealed silica-rich groups are clearly separated, with the latter located toward the region traced by models with higher C/O ratios. Furthermore, the C$_2$H$_2$- and amorphous silica-rich trio of CY~Tau, V1094~Sco, and DL~Tau is located in the region corresponding to the highest C/O.
We therefore suggest that the annealed silica-rich objects in our sample, along with CY~Tau, V1094~Sco, and DL~Tau, have elevated gas-phase C/O ratios. 

This correlation implies some form of coupling between the dust and molecular gas.
Thermo-chemical equilibrium calculations of condensates using GGchem \citep{Woitke2018,Woitke2021_GGchem} can provide the expected abundances of minerals under the pressure and temperature conditions typical of planet-forming disks. For Solar gas-phase elemental abundances, the equilibrium condensates in the warm and hot inner disk regions ($T>400$~K) are rich in forsterite and enstatite, while silica is confined to a narrow radial region where $T>1200$~K \citep[Fig.~9 in][]{Varga2024}. However, we noticed that increasing the C/O ratio makes silica formation more favorable over a broader temperature range. For example, at $\mathrm{C}/\mathrm{O} \approx 0.9$, silica is a major dust component at $T>900$~K, while the abundances of forsterite and enstatite are correspondingly reduced in these regions. 
The underlying reason is that increasing the carbon abundance alters the gas-phase chemistry by binding a larger fraction of oxygen into CO, thereby reducing the amount of free oxygen available for silicate formation \citep[e.g.,][]{Larimer1975}. This suppresses the formation of Mg-rich silicates, which require relatively large amounts of oxygen, while making silica formation more favorable.

This implies that an increasing gas-phase C/O ratio enhances the silica-to-forsterite and silica-to-enstatite abundance ratios, consistent with what we inferred from the C/O diagnostic plots.
Taken together, these results suggest that a process operating in the inner disk regions ($T \gtrsim 400$~K) establishes thermo-chemical equilibrium between the gas and solid phases.
One possible mechanism is dust sublimation and subsequent recondensation near the silicate sublimation front, which could bring the gas and dust in the inner disk close to chemical equilibrium. 
In support of this idea, \cite{Kanwar2025} investigated thermodynamic equilibrium in the inner disk using GGchem and thermo-chemical models and concluded that the very inner disk, close to the dust inner rim, can approach thermodynamic chemical equilibrium.

It remains unclear whether elevated gas-phase C/O ratios are primarily driven by carbon enrichment or by oxygen depletion.
Depletion of gas-phase oxygen may occur if water becomes locked in planetesimals that efficiently form just outside the water snowline \citep{Tabone2023}. 
We note that thermo-chemical models \citep[e.g.,][]{Woitke2018_mol,Arabhavi2026} show that CO$_2$ fluxes are particularly sensitive to the O/H abundance ratio because CO$_2$ contains two oxygen atoms. Consequently, oxygen depletion would reduce the CO$_2$ flux more strongly than the H$_2$O flux.
Alternatively, destruction of carbon dust grains provides another pathway for increasing the gas-phase C/O ratio \citep[e.g.,][]{Houge2025_decomp,Borderies2025,Raul2025}.
\citet{Tabone2026} also proposed that silica could form in a high gas-phase C/O environment, and they offered an explanation for the elevated C$_2$H$_2/\mathrm{H}_2$O flux ratios observed in the disks of V1094 Sco and DL Tau. In their scenario, a reduced delivery of water-rich icy pebbles from the outer disk, combined with the advection of chemically evolved gas with an elevated C/O ratio, leads to carbon enrichment in the inner disk regions of these objects. 

Pressure bumps associated with rings and gaps may slow or halt radial pebble drift, leading to local dust accumulation and altering the transport of solids to the inner disk \citep[e.g.,][]{Pinilla2012_trap}. Such substructures are commonly observed in ALMA images \citep[e.g.,][]{Huang2018}, yet their precise impact on the flow of molecular gas and volatile material to the inner disk remains actively debated \citep[e.g.,][]{Banzatti2023,Arulanantham2025,Gasman2025,Krijt2025}.
In a recent study, \cite{Mallaney2026} reported a dichotomy between molecule-rich and molecule-poor millimeter dust cavities, showing that hydrocarbon-rich chemistry is more prevalent in full disks than in disks with large cavities. Similarly, \cite{Gasman2025} analyzed a sample of disks dominated by gapped millimeter substructures and found that when a substructure is located within the CO and CH$_4$ snowlines, most disks show detectable mid-IR emission from carbon-bearing species. In contrast, disks with large cavities or very wide gaps tend to show stronger cold water emission and weaker warm water emission. Meanwhile, \cite{Volz2026} reported a transition disk with particularly rich hydrocarbon chemistry.
Although we do not explore these correlations in detail, we note that a significant fraction of our forsterite-rich objects (their spectra are highlighted by a green rectangle in Fig.~\ref{fig:dust_and_gas_lines}) possess millimeter cavities, including LkCa 15, SY Cha, and PDS 70. These sources also belong to the group we associate with relatively strong water emission. Conversely, the group characterized by prominent annealed silica and CO$_2$ emission shows no known millimeter cavities in most cases. 
On a related note, dust substructures in the inner few au disk region, co-located with the gas probed by MIRI, may also strongly influence the molecular chemistry, as demonstrated by \cite{Vlasblom2024} using thermo-chemical modeling.
Infrared interferometers, such as the VLTI, are capable of resolving these regions at sub-astronomical-unit spatial resolution \citep[e.g.,][]{Varga2024}, thereby directly probing inner-disk substructures.

\subsection{Comparison with VLMS disks}
\label{sec:VLMS}

As reported by \cite{Arabhavi2025_hidden,Arabhavi2025,Grant2025}, MIRI spectra of VLMS disks typically show weak water lines but strong C$_2$H$_2$ emission. \cite{Grant2025} suggested that this C$_2$H$_2$ enhancement is driven by an increase in the volatile C/O ratio, potentially caused by enrichment in carbon-rich gas rather than by oxygen depletion. In contrast, detailed thermo-chemical modeling of the VLMS disk J1605 indicated that oxygen depletion is the favored scenario to reproduce all observed molecular fluxes \citep{Kanwar2026}.
Considering both the MINDS VLMS and T Tauri samples, \cite{Grant2025} identified an anticorrelation between the C$_2$H$_2/\mathrm{H}_2$O line flux ratio and the strength of the $10\ \mu$m silicate feature: sources with weaker silicate features tend to show stronger C$_2$H$_2$ emission.

\cite{Jang2025} presented a detailed mineralogical analysis of VLMS disks from the MINDS survey. They confirmed that VLMS disks generally display weaker silicate features than T Tauri disks, although their overall silicate compositions are not fundamentally different. Interpreting the mid-IR spectral slopes, silicate feature strengths, and derived gas column densities within a framework of vertical dust settling, they found that less-settled disks show relatively stronger silicate emission, whereas more-settled disks exhibit higher gas column densities. 
They noted the presence of a feature near $21\ \mu$m in three objects (NC9, HK Cha, and J1605), which resembles silica emission at that wavelength. However, the absence of the accompanying feature at $9\ \mu$m makes the detection of silica uncertain.

\cite{Grant2025} and \cite{Jang2025} both suggest that dust in VLMS disks may undergo more rapid vertical settling, thereby exposing deeper disk layers that remain hidden in more massive systems. This would imply that mid-IR observations of VLMS disks probe layers closer to the disk midplane, which may also be present in higher-mass systems but are typically obscured in those disks.
This scenario might explain the enhanced presence of carbon-bearing species in VLMS mid-IR spectra. For example, 2D thermo-chemical models indicate that the mid-IR C$_2$H$_2$ emitting region lies deeper in the disk than the H$_2$O emitting region \citep[e.g.,][]{Woitke2018_mol}.
\cite{Kanwar2024} also showed this specifically for VLMS disks.
Nevertheless, further thermo-chemical modeling is required to robustly test this interpretation. Such modeling should cover a broad stellar mass range, including both VLMS and T Tauri disks, and account for the diversity in the gas-phase C/O ratio as well as the interplay between gas and dust.

\section{Summary}
\label{sec:concl}
 
We have studied JWST/MIRI spectra of 26 T Tauri disks from the MINDS survey. Compared to Spitzer/IRS, MIRI provides a significant improvement in both S/N ratio and spectral resolving power, enabling the detection of weaker dust features and more precise constraints on dust composition and grain size distribution.
For our analysis, we performed spectral decomposition using our new \texttt{DustComp} tool, which fits a set of dust opacity templates to the spectra within a flat-disk geometry while assuming emission components with power-law temperature profiles and derives the mass fractions of the individual dust constituents.
Following earlier studies, we included amorphous and crystalline Mg-silicates and silica in our opacity set. The opacities were calculated from laboratory measurements of cosmic dust analog materials, assuming the DHS grain-shape model and three representative grain sizes of $0.1$, $2$, and $5\ \mu$m. 
Our main results are as follows:
\begin{itemize}
	\item Our fits reproduce the data accurately, with residuals typically within $\pm 3\%$ and root mean square values around $1$--$2\ \%$.
	\item In agreement with previous studies, we find that Mg-rich (and Fe-poor) silicates represent our spectra very well. However, since Fe-rich silicates were not included in our fits, their presence cannot be explicitly ruled out by our analysis.
	\item We find strong evidence of the presence of large ($\sim\!5\ \mu$m) amorphous Mg-silicate grains in our sample objects. However, the presence of large crystalline grains cannot be firmly established. Only in four disks, AA Tau, CY Tau, RNO 90, and Sz 50, does the inclusion of $5\ \mu$m-sized crystals noticeably improve the fit quality.
	\item The dust mass distribution shows an enhanced fraction of $\gtrsim 2\ \mu$m grains, clearly deviating from the Mathis--Rumpl--Nordsieck distribution. This indicates that significant grain growth has occurred in the disks of our sample.
	\item In the $11.3/9.8\ \mu$m flux ratio versus $10\ \mu$m feature amplitude diagram, we find that the observed trend is primarily driven by increasing grain size rather than crystallinity.
	\item On average, the dust composition is dominated by grains of Mg$_2$SiO$_4$ stoichiometry ($\sim\!60\%$, including amorphous and crystalline state), followed by MgSiO$_3$ ($\sim\!30\%$) and SiO$_2$ ($\sim\!10\%$) stoichiometries. The average mass fraction of crystalline forsterite is $6.5\%$, with typical values in the $2$--$11\%$ range.
	\item The average crystalline mass fraction is $14\%$, with typical values between $5$ and $24\%$, in agreement with earlier studies based on Spitzer.
	\item We robustly detected annealed silica, identified by its characteristic spectral peaks at $12.6$, $16.1$, and $20.9\ \mu$m, in nine out of 26 objects ($35\%$). The average mass fraction of annealed silica is $3\%$, with the highest value ($10\%$) found in XX~Cha. Although this mineral has been previously reported in a few T Tauri disks, the improved S/N of MIRI allowed us to detect it in a larger fraction of objects.
	\item Having detected silica in our spectra, we further investigated its specific polymorphic forms and found that cristobalite is the dominant component and that tridymite is a potential minor contributor.
	\item In contrast to previous studies based on Spitzer, we did not require separate warm and cold dust populations to reproduce the observed spectra. This suggests that compositional differences between the warm ($T\!\sim\!300$--$600$~K) and cold ($T\!\sim\!100$--$200$~K) disk regions may be limited. Nevertheless, we note that our analysis does not cover the $27$--$35\ \mu$m wavelength range included in earlier Spitzer studies, which provided a more sensitive probe of cold outer disk dust reservoirs.
	\item We found a correlation between dust composition and molecular gas emission: Most objects with strong annealed silica features also exhibit relatively strong CO$_2$ emission, whereas objects with prominent forsterite features tend to show strong H$_2$O emission and weak CO$_2$ lines.
	\item Using the C/O diagnostics by \cite{Arabhavi2026}, we find that our objects with strong annealed silica features as well as CY~Tau, DL~Tau, and V1094~Sco may have elevated gas-phase C/O ratios.
	\item The correlation between silica content and C/O ratio suggests a coupling between the dust and molecular gas components. We propose a scenario in which gas and dust approach a thermo-chemical equilibrium such that an increased gas-phase C/O ratio in the warm and hot disk regions ($T > 400$~K) enhances the silica abundance while reducing that of forsterite. We further suggest that dust sublimation and subsequent recondensation near the silicate sublimation front may be the process driving the gas and dust in the inner disk toward chemical equilibrium. 
\end{itemize}

\begin{acknowledgements}
	We thank A. Tamanai for providing silica opacity data and C. Jäger for helpful insights on silica.

	This work is based on observations made with the NASA/ESA/CSA James Webb Space Telescope. The data were obtained from the Mikulski Archive for Space Telescopes at the Space Telescope Science Institute, which is operated by the Association of Universities for Research in Astronomy, Inc., under NASA contract NAS 5-03127 for JWST. These observations are associated with program \#1282. The data described here may be obtained from \url{https://dx.doi.org/10.17909/g10s-6e67}.
	The following National and International Funding Agencies funded and supported the MIRI development: NASA; ESA; Belgian Science Policy Office (BELSPO); Centre Nationale d'Etudes Spatiales (CNES); Danish National Space Centre; Deutsches Zentrum fur Luft- und Raumfahrt (DLR); Enterprise Ireland; Ministerio De Economía y Competividad; Netherlands Research School for Astronomy (NOVA); Netherlands Organisation for Scientific Research (NWO); Science and Technology Facilities Council; Swiss Space Office; Swedish National Space Agency; and UK Space Agency. 

	This work received funding from the Hungarian NKFIH OTKA project no. K-147380.
	This work was also supported by the NKFIH NKKP grant ADVANCED 149943 and the NKFIH excellence grant TKP2021-NKTA-64. Project no.149943 has been implemented with the support provided by the Ministry of Culture and Innovation of Hungary from the National Research, Development and Innovation Fund, financed under the NKKP ADVANCED funding scheme.
	This research was supported in part by grant NSF PHY-2309135 to the Kavli Institute for Theoretical Physics (KITP). 
	TH is grateful for the support by the Hungarian Academy of Sciences in the framework of the MTA Distinguished Guest Fellow Scientist Programme 2025 (VK-4\_2025).

	TH acknowledges support from the European Research Council under the Horizon 2020 Framework Program via the ERC Advanced Grant Origins 83 24 28. 
	IK and JK acknowledge funding from H2020-MSCA-ITN-2019, grant no. 860470 (CHAMELEON).
	IK, AMA, and EvD acknowledge support from grant TOP-1 614.001.751 from the Dutch Research Council (NWO). 
	EvD acknowledges support from the ERC grant 101019751 MOLDISK and the Danish National Research Foundation through the Center of Excellence `InterCat' (DNRF150). 
	DG would like to thank the Research Foundation Flanders for co-financing the present research (grant number V435622N) and the European Space Agency (ESA) and the Belgian Federal Science Policy Office (BELSPO) for their support in the framework of the PRODEX Programme. 
	OA thanks the European Space Agency (ESA) and the Belgian Federal Science Policy Office (BELSPO) for their support in the framework of the PRODEX Programme.
	OA is a Senior Research Associate of the Fonds de la Recherche Scientifique - FNRS
	TK acknowledges support from STFC Grant ST/Y002415/1.
	NK thanks the Deutsche Forschungsgemeinschaft (DFG) - grant 138 325594231, FOR 2634/2. 
	GP gratefully acknowledges support from the Carlsberg Foundation, grant CF23-0481 and from the Max Planck Society. 
	LMS has received funding from the European Research Council (ERC) under the European Union's Horizon 2020 research and innovation programme (PROTOPLANETS, grant agreement No. 101002188). 
	BT acknowledges support of the Programme National PCMI of CNRS/INSU with INC/INP cofunded by CEA and CNES. 
	MT and MV acknowledge support from the ERC grant 101019751 MOLDISK. 
	
	This work made use of the Database of Optical Constants for Cosmic Dust created by the Laboratory Astrophysics Group of the AIU Jena, Germany. 

	The Combined Atlas of Sources with Spitzer IRS Spectra (CASSIS) is a product of the IRS instrument team, supported by NASA and JPL. CASSIS is supported by the "Programme National de Physique Stellaire" (PNPS) of CNRS/INSU co-funded by CEA and CNES and through the "Programme National Physique et Chimie du Milieu Interstellaire" (PCMI) of CNRS/INSU with INC/INP co-funded by CEA and CNES.

	This research has made use of the SIMBAD database, CDS, Strasbourg Astronomical Observatory, France. This research has made use of the VizieR catalogue access tool, CDS, Strasbourg Astronomical Observatory, France (DOI : 10.26093/cds/vizier).
 	
	This work made use of the online tool WebPlotDigitizer v5.2 \citep{WebPlotDigitizer}.

\end{acknowledgements}

%
%

\bibliographystyle{aa}
\bibliography{ref_YSO_JV}

@ARTICLE{Bary_spitzer,
   author = {{Bary}, J.~S. and {Leisenring}, J.~M. and {Skrutskie}, M.~F.
	},
    title = "{Variations of the 10 {$\mu$}m Silicate Features in the Actively Accreting T Tauri Stars: DG Tau and XZ Tau}",
  journal = {\apjl},
archivePrefix = "arXiv",
   eprint = {0910.3454},
 primaryClass = "astro-ph.SR",
     year = 2009,
    month = nov,
   volume = 706,
    pages = {L168-L172},
      doi = {10.1088/0004-637X/706/1/L168},
   adsurl = {http://adsabs.harvard.edu/abs/2009ApJ...706L.168B}
}

@ARTICLE{Juhasz2009,
   author = {{Juh{\'a}sz}, A. and {Henning}, T. and {Bouwman}, J. and {Dullemond}, C.~P. and 
	{Pascucci}, I. and {Apai}, D.},
    title = "{Do We Really Know the Dust? Systematics and Uncertainties of the Mid-Infrared Spectral Analysis Methods}",
  journal = {\apj},
archivePrefix = "arXiv",
   eprint = {0902.0405},
 primaryClass = "astro-ph.EP",
     year = 2009,
    month = apr,
   volume = 695,
    pages = {1024-1041},
      doi = {10.1088/0004-637X/695/2/1024},
   adsurl = {http://adsabs.harvard.edu/abs/2009ApJ...695.1024J}
}

@ARTICLE{Juhasz2010,
       author = {{Juh{\'a}sz}, A. and {Bouwman}, J. and {Henning}, Th. and {Acke}, B. and {van den Ancker}, M.~E. and {Meeus}, G. and {Dominik}, C. and {Min}, M. and {Tielens}, A.~G.~G.~M. and {Waters}, L.~B.~F.~M.},
        title = "{Dust Evolution in Protoplanetary Disks Around Herbig Ae/Be Stars{\textemdash}the Spitzer View}",
      journal = {\apj},
         year = 2010,
        month = sep,
       volume = {721},
       number = {1},
        pages = {431-455},
          doi = {10.1088/0004-637X/721/1/431},
archivePrefix = {arXiv},
       eprint = {1008.0083},
 primaryClass = {astro-ph.SR},
       adsurl = {https://ui.adsabs.harvard.edu/abs/2010ApJ...721..431J}
}

@ARTICLE{vanBoekel2004nature,
   author = {{van Boekel}, R. and {Min}, M. and {Leinert}, C. and {Waters}, L.~B.~F.~M. and 
	{Richichi}, A. and {Chesneau}, O. and {Dominik}, C. and {Jaffe}, W. and 
	{Dutrey}, A. and {Graser}, U. and {Henning}, T. and {de Jong}, J. and 
	{K{\"o}hler}, R. and {de Koter}, A. and {Lopez}, B. and {Malbet}, F. and 
	{Morel}, S. and {Paresce}, F. and {Perrin}, G. and {Preibisch}, T. and 
	{Przygodda}, F. and {Sch{\"o}ller}, M. and {Wittkowski}, M.},
    title = "{The building blocks of planets within the `terrestrial' region of protoplanetary disks}",
  journal = {\nat},
     year = 2004,
    month = nov,
   volume = 432,
    pages = {479-482},
      doi = {10.1038/nature03088},
   adsurl = {http://adsabs.harvard.edu/abs/2004Natur.432..479V}
}

@ARTICLE{Kemper_silicate,
   author = {{Kemper}, F. and {Vriend}, W.~J. and {Tielens}, A.~G.~G.~M.},
    title = "{The Absence of Crystalline Silicates in the Diffuse Interstellar Medium}",
  journal = {\apj},
   eprint = {astro-ph/0403609},
     year = 2004,
    month = jul,
   volume = 609,
    pages = {826-837},
      doi = {10.1086/421339},
   adsurl = {http://adsabs.harvard.edu/abs/2004ApJ...609..826K}
}

@ARTICLE{kospal_atlas,
   author = {{K{\'o}sp{\'a}l}, {\'A}. and {{\'A}brah{\'a}m}, P. and {Acosta-Pulido}, J.~A. and 
	{Dullemond}, C.~P. and {Henning}, T. and {Kun}, M. and {Leinert}, C. and 
	{Mo{\'o}r}, A. and {Turner}, N.~J.},
    title = "{Mid-infrared Spectral Variability Atlas of Young Stellar Objects}",
  journal = {\apjs},
archivePrefix = "arXiv",
   eprint = {1204.3473},
 primaryClass = "astro-ph.SR",
     year = 2012,
    month = aug,
   volume = 201,
      eid = {11},
    pages = {11},
      doi = {10.1088/0067-0049/201/2/11},
   adsurl = {http://adsabs.harvard.edu/abs/2012ApJS..201...11K}
}

@ARTICLE{Abraham2009,
   author = {{{\'A}brah{\'a}m}, P. and {Juh{\'a}sz}, A. and {Dullemond}, C.~P. and 
	{K{\'o}sp{\'a}l}, {\'A}. and {van Boekel}, R. and {Bouwman}, J. and 
	{Henning}, T. and {Mo{\'o}r}, A. and {Mosoni}, L. and {Sicilia-Aguilar}, A. and 
	{Sipos}, N.},
    title = "{Episodic formation of cometary material in the outburst of a young Sun-like star}",
  journal = {\nat},
archivePrefix = "arXiv",
   eprint = {0906.3161},
 primaryClass = "astro-ph.SR",
     year = 2009,
    month = may,
   volume = 459,
    pages = {224-226},
      doi = {10.1038/nature08004},
   adsurl = {http://adsabs.harvard.edu/abs/2009Natur.459..224A}
}

@ARTICLE{Juhasz2012,
   author = {{Juh{\'a}sz}, A. and {Dullemond}, C.~P. and {van Boekel}, R. and 
	{Bouwman}, J. and {{\'A}brah{\'a}m}, P. and {Acosta-Pulido}, J.~A. and 
	{Henning}, T. and {K{\'o}sp{\'a}l}, A. and {Sicilia-Aguilar}, A. and 
	{Jones}, A. and {Mo{\'o}r}, A. and {Mosoni}, L. and {Reg{\'a}ly}, Z. and 
	{Szokoly}, G. and {Sipos}, N.},
    title = "{The 2008 Outburst of EX Lup{\mdash}Silicate Crystals in Motion}",
  journal = {\apj},
archivePrefix = "arXiv",
   eprint = {1110.3754},
 primaryClass = "astro-ph.SR",
     year = 2012,
    month = jan,
   volume = 744,
      eid = {118},
    pages = {118},
      doi = {10.1088/0004-637X/744/2/118},
   adsurl = {http://adsabs.harvard.edu/abs/2012ApJ...744..118J}
}

@ARTICLE{Varga2017,
   author = {{Varga}, J. and {Gab{\'a}nyi}, K.~{\'E}. and {{\'A}brah{\'a}m}, P. and 
	{Chen}, L. and {K{\'o}sp{\'a}l}, {\'A}. and {Menu}, J. and {Ratzka}, T. and 
	{van Boekel}, R. and {Dullemond}, C.~P. and {Henning}, T. and 
	{Jaffe}, W. and {Juh{\'a}sz}, A. and {Mo{\'o}r}, A. and {Mosoni}, L. and 
	{Sipos}, N.},
    title = "{Mid-infrared interferometric variability of <ASTROBJ>DG Tauri</ASTROBJ>: Implications for the inner-disk structure}",
  journal = {\aap},
archivePrefix = "arXiv",
   eprint = {1704.05675},
 primaryClass = "astro-ph.SR",
     year = 2017,
    month = aug,
   volume = 604,
      eid = {A84},
    pages = {A84},
      doi = {10.1051/0004-6361/201630287},
   adsurl = {http://adsabs.harvard.edu/abs/2017A%26A...604A..84V}
}

@ARTICLE{Przygodda2003,
   author = {{Przygodda}, F. and {van Boekel}, R. and {{\'A}brah{\'a}m}, P. and 
	{Melnikov}, S.~Y. and {Waters}, L.~B.~F.~M. and {Leinert}, C.
	},
    title = "{Evidence for grain growth in T Tauri disks}",
  journal = {\aap},
   eprint = {astro-ph/0311587},
     year = 2003,
    month = dec,
   volume = 412,
    pages = {L43-L46},
      doi = {10.1051/0004-6361:20034606},
   adsurl = {http://adsabs.harvard.edu/abs/2003A%26A...412L..43P}
}

@ARTICLE{vanBoekel2003,
   author = {{van Boekel}, R. and {Waters}, L.~B.~F.~M. and {Dominik}, C. and 
	{Bouwman}, J. and {de Koter}, A. and {Dullemond}, C.~P. and 
	{Paresce}, F.},
    title = "{Grain growth in the inner regions of Herbig Ae/Be star disks}",
  journal = {\aap},
     year = 2003,
    month = mar,
   volume = 400,
    pages = {L21-L24},
      doi = {10.1051/0004-6361:20030141},
   adsurl = {http://adsabs.harvard.edu/abs/2003A%26A...400L..21V}
}

@ARTICLE{Bouwman2001,
   author = {{Bouwman}, J. and {Meeus}, G. and {de Koter}, A. and {Hony}, S. and 
	{Dominik}, C. and {Waters}, L.~B.~F.~M.},
    title = "{Processing of silicate dust grains in Herbig Ae/Be systems}",
  journal = {\aap},
     year = 2001,
    month = sep,
   volume = 375,
    pages = {950-962},
      doi = {10.1051/0004-6361:20010878},
   adsurl = {http://adsabs.harvard.edu/abs/2001A%26A...375..950B}
}

@ARTICLE{vanBoekel2005survey,
   author = {{van Boekel}, R. and {Min}, M. and {Waters}, L.~B.~F.~M. and 
	{de Koter}, A. and {Dominik}, C. and {van den Ancker}, M.~E. and 
	{Bouwman}, J.},
    title = "{A 10 {$\mu$}m spectroscopic survey of Herbig Ae star disks: Grain growth and crystallization}",
  journal = {\aap},
   eprint = {astro-ph/0503507},
     year = 2005,
    month = jul,
   volume = 437,
    pages = {189-208},
      doi = {10.1051/0004-6361:20042339},
   adsurl = {http://adsabs.harvard.edu/abs/2005A%26A...437..189V}
}

@ARTICLE{Espaillat2011,
   author = {{Espaillat}, C. and {Furlan}, E. and {D'Alessio}, P. and {Sargent}, B. and 
	{Nagel}, E. and {Calvet}, N. and {Watson}, D.~M. and {Muzerolle}, J.
	},
    title = "{A Spitzer IRS Study of Infrared Variability in Transitional and Pre-transitional Disks Around T Tauri Stars}",
  journal = {\apj},
archivePrefix = "arXiv",
   eprint = {1012.3500},
 primaryClass = "astro-ph.SR",
     year = 2011,
    month = feb,
   volume = 728,
      eid = {49},
    pages = {49},
      doi = {10.1088/0004-637X/728/1/49},
   adsurl = {http://adsabs.harvard.edu/abs/2011ApJ...728...49E}
}

@ARTICLE{Lebouteiller2011,
   author = {{Lebouteiller}, V. and {Barry}, D.~J. and {Spoon}, H.~W.~W. and 
	{Bernard-Salas}, J. and {Sloan}, G.~C. and {Houck}, J.~R. and 
	{Weedman}, D.~W.},
    title = "{CASSIS: The Cornell Atlas of Spitzer/Infrared Spectrograph Sources}",
  journal = {\apjs},
archivePrefix = "arXiv",
   eprint = {1108.3507},
 primaryClass = "astro-ph.IM",
     year = 2011,
    month = sep,
   volume = 196,
      eid = {8},
    pages = {8},
      doi = {10.1088/0067-0049/196/1/8},
   adsurl = {http://adsabs.harvard.edu/abs/2011ApJS..196....8L}
}

@ARTICLE{Chiang1997,
   author = {{Chiang}, E.~I. and {Goldreich}, P.},
    title = "{Spectral Energy Distributions of T Tauri Stars with Passive Circumstellar Disks}",
  journal = {\apj},
   eprint = {astro-ph/9706042},
     year = 1997,
    month = nov,
   volume = 490,
    pages = {368-376},
      doi = {10.1086/304869},
   adsurl = {http://adsabs.harvard.edu/abs/1997ApJ...490..368C}
}

@ARTICLE{Dullemond2001,
   author = {{Dullemond}, C.~P. and {Dominik}, C. and {Natta}, A.},
    title = "{Passive Irradiated Circumstellar Disks with an Inner Hole}",
  journal = {\apj},
   eprint = {astro-ph/0106470},
     year = 2001,
    month = oct,
   volume = 560,
    pages = {957-969},
      doi = {10.1086/323057},
   adsurl = {http://adsabs.harvard.edu/abs/2001ApJ...560..957D}
}

@ARTICLE{Varga2018,
       author = {{Varga}, J. and {{\'A}brah{\'a}m}, P. and {Chen}, L. and {Ratzka}, Th. and
         {Gab{\'a}nyi}, K. {\'E}. and {K{\'o}sp{\'a}l}, {\'A}. and {Matter}, A. and
         {van Boekel}, R. and {Henning}, Th. and {Jaffe}, W. and
         {Juh{\'a}sz}, A. and {Lopez}, B. and {Menu}, J. and {Mo{\'o}r}, A. and
         {Mosoni}, L. and {Sipos}, N.},
        title = "{VLTI/MIDI atlas of disks around low- and intermediate-mass young stellar objects}",
      journal = {\aap},
         year = 2018,
        month = sep,
       volume = {617},
          eid = {A83},
        pages = {A83},
          doi = {10.1051/0004-6361/201832599},
archivePrefix = {arXiv},
       eprint = {1805.02939},
 primaryClass = {astro-ph.SR},
       adsurl = {https://ui.adsabs.harvard.edu/abs/2018A&A...617A..83V}
}

@ARTICLE{Huang2018,
       author = {{Huang}, Jane and {Andrews}, Sean M. and {Dullemond}, Cornelis P. and
         {Isella}, Andrea and {P{\'e}rez}, Laura M. and
         {Guzm{\'a}n}, Viviana V. and {{\"O}berg}, Karin I. and {Zhu}, Zhaohuan and
         {Zhang}, Shangjia and {Bai}, Xue-Ning and {Benisty}, Myriam and
         {Birnstiel}, Tilman and {Carpenter}, John M. and {Hughes}, A. Meredith and
         {Ricci}, Luca and {Weaver}, Erik and {Wilner}, David J.},
        title = "{The Disk Substructures at High Angular Resolution Project (DSHARP). II. Characteristics of Annular Substructures}",
      journal = {\apjl},
         year = 2018,
        month = dec,
       volume = {869},
       number = {2},
          eid = {L42},
        pages = {L42},
          doi = {10.3847/2041-8213/aaf740},
archivePrefix = {arXiv},
       eprint = {1812.04041},
 primaryClass = {astro-ph.EP},
       adsurl = {https://ui.adsabs.harvard.edu/abs/2018ApJ...869L..42H}
}

@INPROCEEDINGS{Testi2014,
       author = {{Testi}, L. and {Birnstiel}, T. and {Ricci}, L. and {Andrews}, S. and {Blum}, J. and {Carpenter}, J. and {Dominik}, C. and {Isella}, A. and {Natta}, A. and {Williams}, J.~P. and {Wilner}, D.~J.},
        title = "{Dust Evolution in Protoplanetary Disks}",
    booktitle = {Protostars and Planets VI},
         year = 2014,
       editor = {{Beuther}, Henrik and {Klessen}, Ralf S. and {Dullemond}, Cornelis P. and {Henning}, Thomas},
        month = jan,
        pages = {339},
          doi = {10.2458/azu\_uapress\_9780816531240-ch015},
archivePrefix = {arXiv},
       eprint = {1402.1354},
 primaryClass = {astro-ph.SR},
       adsurl = {https://ui.adsabs.harvard.edu/abs/2014prpl.conf..339T}
}

@ARTICLE{Harker2002,
       author = {{Harker}, David E. and {Desch}, Steven J.},
        title = "{Annealing of Silicate Dust by Nebular Shocks at 10 AU}",
      journal = {\apjl},
         year = 2002,
        month = feb,
       volume = {565},
       number = {2},
        pages = {L109-L112},
          doi = {10.1086/339363},
archivePrefix = {arXiv},
       eprint = {astro-ph/0112494},
 primaryClass = {astro-ph},
       adsurl = {https://ui.adsabs.harvard.edu/abs/2002ApJ...565L.109H}
}

@ARTICLE{Hallenbeck1998,
       author = {{Hallenbeck}, Susan L. and {Nuth}, Joseph A. and {Daukantas}, Patricia L.},
        title = "{Mid-Infrared Spectral Evolution of Amorphous Magnesium Silicate Smokes Annealed in Vacuum: Comparison to Cometary Spectra}",
      journal = {\icarus},
         year = 1998,
        month = jan,
       volume = {131},
       number = {1},
        pages = {198-209},
          doi = {10.1006/icar.1997.5854},
       adsurl = {https://ui.adsabs.harvard.edu/abs/1998Icar..131..198H}
}

@ARTICLE{Henning2010,
       author = {{Henning}, Thomas},
        title = "{Cosmic Silicates}",
      journal = {\araa},
         year = 2010,
        month = sep,
       volume = {48},
        pages = {21-46},
          doi = {10.1146/annurev-astro-081309-130815},
       adsurl = {https://ui.adsabs.harvard.edu/abs/2010ARA&A..48...21H}
}

@ARTICLE{Jager2003,
       author = {{J{\"a}ger}, C. and {Dorschner}, J. and {Mutschke}, H. and {Posch}, Th. and {Henning}, Th.},
        title = "{Steps toward interstellar silicate mineralogy. VII. Spectral properties and crystallization behaviour of magnesium silicates produced by the sol-gel method}",
      journal = {\aap},
         year = 2003,
        month = sep,
       volume = {408},
        pages = {193-204},
          doi = {10.1051/0004-6361:20030916},
       adsurl = {https://ui.adsabs.harvard.edu/abs/2003A&A...408..193J}
}

@ARTICLE{Jager1998,
       author = {{J{\"a}ger}, C. and {Molster}, F.~J. and {Dorschner}, J. and {Henning}, Th. and {Mutschke}, H. and {Waters}, L.~B.~F.~M.},
        title = "{Steps toward interstellar silicate mineralogy. IV. The crystalline revolution}",
      journal = {\aap},
         year = 1998,
        month = nov,
       volume = {339},
        pages = {904-916},
       adsurl = {https://ui.adsabs.harvard.edu/abs/1998A&A...339..904J}
}

@ARTICLE{Dorschner1995,
       author = {{Dorschner}, J. and {Begemann}, B. and {Henning}, T. and {J{\"a}ger}, C. and {Mutschke}, H.},
        title = "{Steps toward interstellar silicate mineralogy. II. Study of Mg-Fe-silicate glasses of variable composition.}",
      journal = {\aap},
         year = 1995,
        month = aug,
       volume = {300},
        pages = {503},
       adsurl = {https://ui.adsabs.harvard.edu/abs/1995A&A...300..503D}
}

@ARTICLE{Henning1996,
       author = {{Henning}, T. and {Stognienko}, R.},
        title = "{Dust opacities for protoplanetary accretion disks: influence of dust aggregates.}",
      journal = {\aap},
         year = 1996,
        month = jul,
       volume = {311},
        pages = {291-303},
       adsurl = {https://ui.adsabs.harvard.edu/abs/1996A&A...311..291H}
}

@MISC{Dominik2021optool,
       author = {{Dominik}, Carsten and {Min}, Michiel and {Tazaki}, Ryo},
        title = "{OpTool: Command-line driven tool for creating complex dust opacities}",
 howpublished = {Astrophysics Source Code Library, record ascl:2104.010},
         year = 2021,
        month = apr,
          eid = {ascl:2104.010},
        pages = {ascl:2104.010},
archivePrefix = {ascl},
       eprint = {2104.010},
       adsurl = {https://ui.adsabs.harvard.edu/abs/2021ascl.soft04010D}
}

@ARTICLE{Min2005,
       author = {{Min}, M. and {Hovenier}, J.~W. and {de Koter}, A.},
        title = "{Modeling optical properties of cosmic dust grains using a distribution of hollow spheres}",
      journal = {\aap},
         year = 2005,
        month = mar,
       volume = {432},
       number = {3},
        pages = {909-920},
          doi = {10.1051/0004-6361:20041920},
archivePrefix = {arXiv},
       eprint = {astro-ph/0503068},
 primaryClass = {astro-ph},
       adsurl = {https://ui.adsabs.harvard.edu/abs/2005A&A...432..909M}
}

@ARTICLE{Woitke2018,
       author = {{Woitke}, P. and {Helling}, Ch. and {Hunter}, G.~H. and {Millard}, J.~D. and {Turner}, G.~E. and {Worters}, M. and {Blecic}, J. and {Stock}, J.~W.},
        title = "{Equilibrium chemistry down to 100 K. Impact of silicates and phyllosilicates on the carbon to oxygen ratio}",
      journal = {\aap},
         year = 2018,
        month = jun,
       volume = {614},
          eid = {A1},
        pages = {A1},
          doi = {10.1051/0004-6361/201732193},
archivePrefix = {arXiv},
       eprint = {1712.01010},
 primaryClass = {astro-ph.EP},
       adsurl = {https://ui.adsabs.harvard.edu/abs/2018A&A...614A...1W}
}

@ARTICLE{Sargent2009_Taurus,
       author = {{Sargent}, B.~A. and {Forrest}, W.~J. and {Tayrien}, C. and {McClure}, M.~K. and {Watson}, Dan M. and {Sloan}, G.~C. and {Li}, A. and {Manoj}, P. and {Bohac}, C.~J. and {Furlan}, E. and {Kim}, K.~H. and {Green}, J.~D.},
        title = "{Dust Processing and Grain Growth in Protoplanetary Disks in the Taurus-Auriga Star-Forming Region}",
      journal = {\apjs},
         year = 2009,
        month = jun,
       volume = {182},
       number = {2},
        pages = {477-508},
          doi = {10.1088/0067-0049/182/2/477},
archivePrefix = {arXiv},
       eprint = {0811.3622},
 primaryClass = {astro-ph},
       adsurl = {https://ui.adsabs.harvard.edu/abs/2009ApJS..182..477S}
}

@ARTICLE{Sargent2009_silica,
       author = {{Sargent}, B.~A. and {Forrest}, W.~J. and {Tayrien}, C. and {McClure}, M.~K. and {Li}, A. and {Basu}, A.~R. and {Manoj}, P. and {Watson}, D.~M. and {Bohac}, C.~J. and {Furlan}, E. and {Kim}, K.~H. and {Green}, J.~D. and {Sloan}, G.~C.},
        title = "{Silica in Protoplanetary Disks}",
      journal = {\apj},
         year = 2009,
        month = jan,
       volume = {690},
       number = {2},
        pages = {1193-1207},
          doi = {10.1088/0004-637X/690/2/1193},
archivePrefix = {arXiv},
       eprint = {0811.3590},
 primaryClass = {astro-ph},
       adsurl = {https://ui.adsabs.harvard.edu/abs/2009ApJ...690.1193S}
}

@ARTICLE{Roskosz2015,
       author = {{Roskosz}, Mathieu and {Leroux}, Hugues},
        title = "{A Significant Amount of Crystalline Silica in Returned Cometary Samples: Bridging the Gap between Astrophysical and Meteoritical Observations}",
      journal = {\apjl},
         year = 2015,
        month = mar,
       volume = {801},
       number = {1},
          eid = {L7},
        pages = {L7},
          doi = {10.1088/2041-8205/801/1/L7},
       adsurl = {https://ui.adsabs.harvard.edu/abs/2015ApJ...801L...7R}
}

@ARTICLE{Tielens2022,
       author = {{Tielens}, A.~G.~G.~M.},
        title = "{Dust Formation in Astrophysical Environments: The Importance of Kinetics}",
      journal = {Frontiers in Astronomy and Space Sciences},
         year = 2022,
        month = may,
       volume = {9},
          eid = {908217},
        pages = {908217},
          doi = {10.3389/fspas.2022.908217},
archivePrefix = {arXiv},
       eprint = {2206.01548},
 primaryClass = {astro-ph.EP},
       adsurl = {https://ui.adsabs.harvard.edu/abs/2022FrASS...9.8217T}
}

@ARTICLE{Olofsson2010,
       author = {{Olofsson}, J. and {Augereau}, J. -C. and {van Dishoeck}, E.~F. and {Mer{\'\i}n}, B. and {Grosso}, N. and {M{\'e}nard}, F. and {Blake}, G.~A. and {Monin}, J. -L.},
        title = "{C2D Spitzer-IRS spectra of disks around T Tauri stars. V. Spectral decomposition}",
      journal = {\aap},
         year = 2010,
        month = sep,
       volume = {520},
          eid = {A39},
        pages = {A39},
          doi = {10.1051/0004-6361/200913909},
archivePrefix = {arXiv},
       eprint = {1007.0644},
 primaryClass = {astro-ph.SR},
       adsurl = {https://ui.adsabs.harvard.edu/abs/2010A&A...520A..39O}
}

@ARTICLE{Mathis1977_MRN,
       author = {{Mathis}, J.~S. and {Rumpl}, W. and {Nordsieck}, K.~H.},
        title = "{The size distribution of interstellar grains.}",
      journal = {\apj},
         year = 1977,
        month = oct,
       volume = {217},
        pages = {425-433},
          doi = {10.1086/155591},
       adsurl = {https://ui.adsabs.harvard.edu/abs/1977ApJ...217..425M}
}

@ARTICLE{Gail2004,
       author = {{Gail}, H. -P.},
        title = "{Radial mixing in protoplanetary accretion disks. IV. Metamorphosis of the silicate dust complex}",
      journal = {\aap},
         year = 2004,
        month = jan,
       volume = {413},
        pages = {571-591},
          doi = {10.1051/0004-6361:20031554},
       adsurl = {https://ui.adsabs.harvard.edu/abs/2004A&A...413..571G}
}

@ARTICLE{Varga2024,
       author = {{Varga}, J. and {Waters}, L.~B.~F.~M. and {Hogerheijde}, M. and {van Boekel}, R. and {Matter}, A. and {Lopez}, B. and {Perraut}, K. and {Chen}, L. and {Nadella}, D. and {Wolf}, S. and {Dominik}, C. and {K{\'o}sp{\'a}l}, {\'A}. and {{\'A}brah{\'a}m}, P. and {Augereau}, J. -C. and {Boley}, P. and {Bourdarot}, G. and {Caratti O Garatti}, A. and {Cruz-S{\'a}enz de Miera}, F. and {Danchi}, W.~C. and {G{\'a}mez Rosas}, V. and {Henning}, Th. and {Hofmann}, K. -H. and {Houll{\'e}}, M. and {Isbell}, J.~W. and {Jaffe}, W. and {Juh{\'a}sz}, T. and {Kecskem{\'e}thy}, V. and {Kobus}, J. and {Kokoulina}, E. and {Labadie}, L. and {Lykou}, F. and {Millour}, F. and {Mo{\'o}r}, A. and {Moruj{\~a}o}, N. and {Pantin}, E. and {Schertl}, D. and {Scheuck}, M. and {van Haastere}, L. and {Weigelt}, G. and {Woillez}, J. and {Woitke}, P. and {Matisse Collaboration} and {Gravity Collaboration}},
        title = "{Mid-infrared evidence for iron-rich dust in the multi-ringed inner disk of HD 144432}",
      journal = {\aap},
         year = 2024,
        month = jan,
       volume = {681},
          eid = {A47},
        pages = {A47},
          doi = {10.1051/0004-6361/202347535},
archivePrefix = {arXiv},
       eprint = {2401.03437},
 primaryClass = {astro-ph.SR},
       adsurl = {https://ui.adsabs.harvard.edu/abs/2024A&A...681A..47V}
}

@article{Kessler-Silacci2006,
	title = {c2d {Spitzer} {IRS} {Spectra} of {Disks} around {T} {Tauri} {Stars}. {I}. {Silicate} {Emission} and {Grain} {Growth}},
	volume = {639},
	url = {https://ui.adsabs.harvard.edu/#abs/2006ApJ...639..275K/abstract},
	doi = {10.1086/499330},
	number = {1},
	urldate = {2017-12-19},
	journal = {The Astrophysical Journal},
	author = {Kessler-Silacci, Jacqueline and Augereau, Jean-Charles and Dullemond, Cornelis P. and Geers, Vincent and Lahuis, Fred and Evans, Neal J. and Dishoeck, Van and F, Ewine and Blake, Geoffrey A. and Boogert, A. C. Adwin and Brown, Joanna and Jørgensen, Jes K. and Knez, Claudia and Pontoppidan, Klaus M.},
	month = mar,
	year = {2006},
	pages = {275},
}

@ARTICLE{Kaeufer2024,
       author = {{Kaeufer}, T. and {Min}, M. and {Woitke}, P. and {Kamp}, I. and {Arabhavi}, A.~M.},
        title = "{Bayesian analysis of the molecular emission and dust continuum of protoplanetary disks}",
      journal = {\aap},
         year = 2024,
        month = jul,
       volume = {687},
          eid = {A209},
        pages = {A209},
          doi = {10.1051/0004-6361/202449936},
archivePrefix = {arXiv},
       eprint = {2405.06486},
 primaryClass = {astro-ph.EP},
       adsurl = {https://ui.adsabs.harvard.edu/abs/2024A&A...687A.209K}
}

@ARTICLE{Tamanai2006,
       author = {{Tamanai}, A. and {Mutschke}, H. and {Blum}, J. and {Meeus}, G.},
        title = "{The 10 {\ensuremath{\mu}}m Infrared Band of Silicate Dust: A Laboratory Study Comparing the Aerosol and KBr Pellet Techniques}",
      journal = {\apjl},
         year = 2006,
        month = sep,
       volume = {648},
       number = {2},
        pages = {L147-L150},
          doi = {10.1086/508164},
archivePrefix = {arXiv},
       eprint = {astro-ph/0609231},
 primaryClass = {astro-ph},
       adsurl = {https://ui.adsabs.harvard.edu/abs/2006ApJ...648L.147T}
}

@ARTICLE{Birnstiel2024,
       author = {{Birnstiel}, Tilman},
        title = "{Dust Growth and Evolution in Protoplanetary Disks}",
      journal = {\araa},
         year = 2024,
        month = sep,
       volume = {62},
       number = {1},
        pages = {157-202},
          doi = {10.1146/annurev-astro-071221-052705},
archivePrefix = {arXiv},
       eprint = {2312.13287},
 primaryClass = {astro-ph.EP},
       adsurl = {https://ui.adsabs.harvard.edu/abs/2024ARA&A..62..157B}
}

@ARTICLE{Jang2024,
       author = {{Jang}, Hyerin and {Waters}, Rens and {Kaeufer}, Till and {Tamanai}, Akemi and {Perotti}, Giulia and {Christiaens}, Valentin and {Kamp}, Inga and {Henning}, Thomas and {Min}, Michiel and {Arabhavi}, Aditya M. and {Barrado}, David and {van Dishoeck}, Ewine F. and {Gasman}, Danny and {Grant}, Sierra L. and {G{\"u}del}, Manuel and {Lagage}, Pierre-Olivier and {Lahuis}, Fred and {Schwarz}, Kamber and {Tabone}, Beno{\^\i}t and {Temmink}, Milou},
        title = "{Dust mineralogy and variability of the inner PDS 70 disk: Insights from JWST/MIRI MRS and Spitzer IRS observations}",
      journal = {\aap},
         year = 2024,
        month = nov,
       volume = {691},
          eid = {A148},
        pages = {A148},
          doi = {10.1051/0004-6361/202451589},
archivePrefix = {arXiv},
       eprint = {2408.16367},
 primaryClass = {astro-ph.EP},
       adsurl = {https://ui.adsabs.harvard.edu/abs/2024A&A...691A.148J}
}

@ARTICLE{Bertout2007,
       author = {{Bertout}, C. and {Siess}, L. and {Cabrit}, S.},
        title = "{The evolution of stars in the Taurus-Auriga T association}",
      journal = {\aap},
         year = 2007,
        month = oct,
       volume = {473},
       number = {3},
        pages = {L21-L24},
          doi = {10.1051/0004-6361:20078276},
archivePrefix = {arXiv},
       eprint = {0708.3040},
 primaryClass = {astro-ph},
       adsurl = {https://ui.adsabs.harvard.edu/abs/2007A&A...473L..21B}
}

@ARTICLE{Kenyon1995,
       author = {{Kenyon}, Scott J. and {Hartmann}, Lee},
        title = "{Pre-Main-Sequence Evolution in the Taurus-Auriga Molecular Cloud}",
      journal = {\apjs},
         year = 1995,
        month = nov,
       volume = {101},
        pages = {117},
          doi = {10.1086/192235},
       adsurl = {https://ui.adsabs.harvard.edu/abs/1995ApJS..101..117K}
}

@ARTICLE{Testi2022,
       author = {{Testi}, L. and {Natta}, A. and {Manara}, C.~F. and {de Gregorio Monsalvo}, I. and {Lodato}, G. and {Lopez}, C. and {Muzic}, K. and {Pascucci}, I. and {Sanchis}, E. and {Miranda}, A. Santamaria and {Scholz}, A. and {De Simone}, M. and {Williams}, J.~P.},
        title = "{The protoplanetary disk population in the {\ensuremath{\rho}}-Ophiuchi region L1688 and the time evolution of Class II YSOs}",
      journal = {\aap},
         year = 2022,
        month = jul,
       volume = {663},
          eid = {A98},
        pages = {A98},
          doi = {10.1051/0004-6361/202141380},
archivePrefix = {arXiv},
       eprint = {2201.04079},
 primaryClass = {astro-ph.SR},
       adsurl = {https://ui.adsabs.harvard.edu/abs/2022A&A...663A..98T}
}

@ARTICLE{Sartori2003,
       author = {{Sartori}, M.~J. and {L{\'e}pine}, J.~R.~D. and {Dias}, W.~S.},
        title = "{Formation scenarios for the young stellar associations between galactic longitudes l = 280degr - 360degr}",
      journal = {\aap},
         year = 2003,
        month = jun,
       volume = {404},
        pages = {913-926},
          doi = {10.1051/0004-6361:20030581},
archivePrefix = {arXiv},
       eprint = {astro-ph/0304426},
 primaryClass = {astro-ph},
       adsurl = {https://ui.adsabs.harvard.edu/abs/2003A&A...404..913S}
}

@ARTICLE{Ghez1993,
       author = {{Ghez}, A.~M. and {Neugebauer}, G. and {Matthews}, K.},
        title = "{The Multiplicity of T Tauri Stars in the Star Forming Regions Taurus-Auriga and Ophiuchus-Scorpius: A 2.2 Micron Speckle Imaging Survey}",
      journal = {\aj},
         year = 1993,
        month = nov,
       volume = {106},
        pages = {2005},
          doi = {10.1086/116782},
       adsurl = {https://ui.adsabs.harvard.edu/abs/1993AJ....106.2005G}
}

@ARTICLE{Andrews2009,
       author = {{Andrews}, Sean M. and {Wilner}, D.~J. and {Hughes}, A.~M. and {Qi}, Chunhua and {Dullemond}, C.~P.},
        title = "{Protoplanetary Disk Structures in Ophiuchus}",
      journal = {\apj},
         year = 2009,
        month = aug,
       volume = {700},
       number = {2},
        pages = {1502-1523},
          doi = {10.1088/0004-637X/700/2/1502},
archivePrefix = {arXiv},
       eprint = {0906.0730},
 primaryClass = {astro-ph.EP},
       adsurl = {https://ui.adsabs.harvard.edu/abs/2009ApJ...700.1502A}
}

@ARTICLE{Henning2024,
       author = {{Henning}, Thomas and {Kamp}, Inga and {Samland}, Matthias and {Arabhavi}, Aditya M. and {Kanwar}, Jayatee and {van Dishoeck}, Ewine F. and {G{\"u}del}, Manuel and {Lagage}, Pierre-Olivier and {Waelkens}, Christoffel and {Abergel}, Alain and {Absil}, Olivier and {Barrado}, David and {Boccaletti}, Anthony and {Bouwman}, Jeroen and {Caratti o Garatti}, Alessio and {Geers}, Vincent and {Glauser}, Adrian M. and {Lahuis}, Fred and {Mueller}, Michael and {Nehm{\'e}}, Cyrine and {Olofsson}, G{\"o}ran and {Pantin}, Eric and {Ray}, Tom P. and {Scheithauer}, Silvia and {Vandenbussche}, Bart and {Waters}, L.~B.~F.~M. and {Wright}, Gillian and {Argyriou}, Ioannis and {Christiaens}, Valentin and {Franceschi}, Riccardo and {Gasman}, Danny and {Grant}, Sierra L. and {Guadarrama}, Rodrigo and {Jang}, Hyerin and {Morales-Calder{\'o}n}, Maria and {Pawellek}, Nicole and {Perotti}, Giulia and {Rodgers-Lee}, Donna and {Schreiber}, J{\"u}rgen and {Schwarz}, Kamber and {Tabone}, Beno{\^\i}t and {Temmink}, Milou and {Vlasblom}, Marissa and {Colina}, Luis and {Greve}, Thomas R. and {{\"O}stlin}, G{\"o}ran},
        title = "{MINDS: The JWST MIRI Mid-INfrared Disk Survey}",
      journal = {\pasp},
         year = 2024,
        month = may,
       volume = {136},
       number = {5},
          eid = {054302},
        pages = {054302},
          doi = {10.1088/1538-3873/ad3455},
archivePrefix = {arXiv},
       eprint = {2403.09210},
 primaryClass = {astro-ph.EP},
       adsurl = {https://ui.adsabs.harvard.edu/abs/2024PASP..136e4302H}
}

@ARTICLE{Zeidler2015,
       author = {{Zeidler}, S. and {Mutschke}, H. and {Posch}, Th.},
        title = "{Temperature-dependent Infrared Optical Constants of Olivine and Enstatite}",
      journal = {\apj},
         year = 2015,
        month = jan,
       volume = {798},
       number = {2},
          eid = {125},
        pages = {125},
          doi = {10.1088/0004-637X/798/2/125},
       adsurl = {https://ui.adsabs.harvard.edu/abs/2015ApJ...798..125Z}
}

@ARTICLE{Kitamura2007,
       author = {{Kitamura}, Rei and {Pilon}, Laurent and {Jonasz}, Miroslaw},
        title = "{Optical constants of silica glass from extreme ultraviolet to far infrared at near room temperature}",
      journal = {\ao},
         year = 2007,
        month = nov,
       volume = {46},
       number = {33},
        pages = {8118-8133},
          doi = {10.1364/AO.46.008118},
       adsurl = {https://ui.adsabs.harvard.edu/abs/2007ApOpt..46.8118K}
}

@ARTICLE{Fabian2000,
       author = {{Fabian}, D. and {J{\"a}ger}, C. and {Henning}, Th. and {Dorschner}, J. and {Mutschke}, H.},
        title = "{Steps toward interstellar silicate mineralogy. V. Thermal Evolution of Amorphous Magnesium Silicates and Silica}",
      journal = {\aap},
         year = 2000,
        month = dec,
       volume = {364},
        pages = {282-292},
       adsurl = {https://ui.adsabs.harvard.edu/abs/2000A&A...364..282F}
}

@ARTICLE{Servoin1973,
       author = {{Servoin}, J.~L. and {Piriou}, B.},
        title = "{Infrared Reflectivity and Raman Scattering of Mg2SiO4 Single Crystal}",
      journal = {Physica Status Solidi B Basic Research},
         year = 1973,
        month = feb,
       volume = {55},
       number = {2},
        pages = {677-686},
          doi = {10.1002/pssb.2220550224},
       adsurl = {https://ui.adsabs.harvard.edu/abs/1973PSSBR..55..677S}
}

@ARTICLE{Henning1997,
       author = {{Henning}, T. and {Mutschke}, H.},
        title = "{Low-temperature infrared properties of cosmic dust analogues.}",
      journal = {\aap},
         year = 1997,
        month = nov,
       volume = {327},
        pages = {743-754},
       adsurl = {https://ui.adsabs.harvard.edu/abs/1997A&A...327..743H}
}

@ARTICLE{Suto2006,
       author = {{Suto}, H. and {Sogawa}, H. and {Tachibana}, S. and {Koike}, C. and {Karoji}, H. and {Tsuchiyama}, A. and {Chihara}, H. and {Mizutani}, K. and {Akedo}, J. and {Ogiso}, K. and {Fukui}, T. and {Ohara}, S.},
        title = "{Low-temperature single crystal reflection spectra of forsterite}",
      journal = {\mnras},
         year = 2006,
        month = aug,
       volume = {370},
       number = {4},
        pages = {1599-1606},
          doi = {10.1111/j.1365-2966.2006.10594.x},
       adsurl = {https://ui.adsabs.harvard.edu/abs/2006MNRAS.370.1599S}
}

@ARTICLE{Min2007,
       author = {{Min}, M. and {Waters}, L.~B.~F.~M. and {de Koter}, A. and {Hovenier}, J.~W. and {Keller}, L.~P. and {Markwick-Kemper}, F.},
        title = "{The shape and composition of interstellar silicate grains}",
      journal = {\aap},
         year = 2007,
        month = feb,
       volume = {462},
       number = {2},
        pages = {667-676},
          doi = {10.1051/0004-6361:20065436},
archivePrefix = {arXiv},
       eprint = {astro-ph/0611329},
 primaryClass = {astro-ph},
       adsurl = {https://ui.adsabs.harvard.edu/abs/2007A&A...462..667M}
}

@ARTICLE{Speagle2020dynesty,
       author = {{Speagle}, Joshua S.},
        title = "{DYNESTY: a dynamic nested sampling package for estimating Bayesian posteriors and evidences}",
      journal = {\mnras},
         year = 2020,
        month = apr,
       volume = {493},
       number = {3},
        pages = {3132-3158},
          doi = {10.1093/mnras/staa278},
archivePrefix = {arXiv},
       eprint = {1904.02180},
 primaryClass = {astro-ph.IM},
       adsurl = {https://ui.adsabs.harvard.edu/abs/2020MNRAS.493.3132S}
}

@article{Higson2019,
	title = {Dynamic nested sampling: an improved algorithm for parameter estimation and evidence calculation},
	volume = {29},
	issn = {1573-1375},
	url = {https://doi.org/10.1007/s11222-018-9844-0},
	doi = {10.1007/s11222-018-9844-0},
	number = {5},
	journal = {Statistics and Computing},
	author = {Higson, Edward and Handley, Will and Hobson, Michael and Lasenby, Anthony},
	month = sep,
	year = {2019},
	pages = {891--913},
}

@book{Lawson1995,
	title = {Solving {Least} {Squares} {Problems}},
	url = {https://epubs.siam.org/doi/abs/10.1137/1.9781611971217},
	publisher = {Society for Industrial and Applied Mathematics},
	author = {Lawson, Charles L. and Hanson, Richard J.},
	year = {1995},
	doi = {10.1137/1.9781611971217},
	note = {\_eprint: https://epubs.siam.org/doi/pdf/10.1137/1.9781611971217},
}

@ARTICLE{Hofner2018,
       author = {{H{\"o}fner}, Susanne and {Olofsson}, Hans},
        title = "{Mass loss of stars on the asymptotic giant branch. Mechanisms, models and measurements}",
      journal = {\aapr},
         year = 2018,
        month = jan,
       volume = {26},
       number = {1},
          eid = {1},
        pages = {1},
          doi = {10.1007/s00159-017-0106-5},
       adsurl = {https://ui.adsabs.harvard.edu/abs/2018A&ARv..26....1H}
}

@ARTICLE{Salpeter1977,
       author = {{Salpeter}, E.~E.},
        title = "{Formation and destruction of dust grains.}",
      journal = {\araa},
         year = 1977,
        month = jan,
       volume = {15},
        pages = {267-293},
          doi = {10.1146/annurev.aa.15.090177.001411},
       adsurl = {https://ui.adsabs.harvard.edu/abs/1977ARA&A..15..267S}
}

@ARTICLE{Todini2001,
       author = {{Todini}, Paolo and {Ferrara}, Andrea},
        title = "{Dust formation in primordial Type II supernovae}",
      journal = {\mnras},
         year = 2001,
        month = aug,
       volume = {325},
       number = {2},
        pages = {726-736},
          doi = {10.1046/j.1365-8711.2001.04486.x},
archivePrefix = {arXiv},
       eprint = {astro-ph/0009176},
 primaryClass = {astro-ph},
       adsurl = {https://ui.adsabs.harvard.edu/abs/2001MNRAS.325..726T}
}

@ARTICLE{Tielens1998,
       author = {{Tielens}, A.~G.~G.~M. and {Waters}, L.~B.~F.~M. and {Molster}, F.~J. and {Justtanont}, K.},
        title = "{Circumstellar Silicate Mineralogy}",
      journal = {\apss},
         year = 1998,
        month = jan,
       volume = {255},
        pages = {415-426},
          doi = {10.1023/A:1001585120472},
       adsurl = {https://ui.adsabs.harvard.edu/abs/1998Ap&SS.255..415T}
}

@ARTICLE{Olofsson2009,
       author = {{Olofsson}, J. and {Augereau}, J.-C. and {van Dishoeck}, E.~F. and {Mer{\'\i}n}, B. and {Lahuis}, F. and {Kessler-Silacci}, J. and {Dullemond}, C.~P. and {Oliveira}, I. and {Blake}, G.~A. and {Boogert}, A.~C.~A. and {Brown}, J.~M. and {Evans}, II, N.~J. and {Geers}, V. and {Knez}, C. and {Monin}, J.-L. and {Pontoppidan}, K.},
        title = "{C2D Spitzer-IRS spectra of disks around T Tauri stars. IV. Crystalline silicates}",
      journal = {\aap},
         year = 2009,
        month = nov,
       volume = {507},
       number = {1},
        pages = {327-345},
          doi = {10.1051/0004-6361/200912062},
archivePrefix = {arXiv},
       eprint = {0908.4153},
 primaryClass = {astro-ph.SR},
       adsurl = {https://ui.adsabs.harvard.edu/abs/2009A&A...507..327O}
}

@ARTICLE{Gail1999,
       author = {{Gail}, H.-P. and {Sedlmayr}, E.},
        title = "{Mineral formation in stellar winds. I. Condensation sequence of silicate and iron grains in stationary oxygen rich outflows}",
      journal = {\aap},
         year = 1999,
        month = jul,
       volume = {347},
        pages = {594-616},
       adsurl = {https://ui.adsabs.harvard.edu/abs/1999A&A...347..594G}
}

@ARTICLE{Jager1994,
       author = {{J{\"a}ger}, C. and {Mutschke}, H. and {Begemann}, B. and {Dorschner}, J. and {Henning}, Th.},
        title = "{Steps toward interstellar silicate mineralogy. I. Laboratory results of a silicate glass of mean cosmic composition.}",
      journal = {\aap},
         year = 1994,
        month = dec,
       volume = {292},
        pages = {641-655},
       adsurl = {https://ui.adsabs.harvard.edu/abs/1994A&A...292..641J}
}

@ARTICLE{Fabian2001,
       author = {{Fabian}, D. and {Henning}, T. and {J{\"a}ger}, C. and {Mutschke}, H. and {Dorschner}, J. and {Wehrhan}, O.},
        title = "{Steps toward interstellar silicate mineralogy. VI. Dependence of crystalline olivine IR spectra on iron content and particle shape}",
      journal = {\aap},
         year = 2001,
        month = oct,
       volume = {378},
        pages = {228-238},
          doi = {10.1051/0004-6361:20011196},
       adsurl = {https://ui.adsabs.harvard.edu/abs/2001A&A...378..228F}
}

@INCOLLECTION{Henning2011book,
       author = {{Henning}, Thomas and {Meeus}, Gwendolyn},
        title = "{Dust Processing and Mineralogy in Protoplanetary Accretion Disks}",
    booktitle = {Physical Processes in Circumstellar Disks around Young Stars},
         year = 2011,
       editor = {{Garcia}, Paulo J.~V.},
        pages = {114-148},
       adsurl = {https://ui.adsabs.harvard.edu/abs/2011ppcd.book..114H}
}

@ARTICLE{Furlan2011,
       author = {{Furlan}, E. and {Luhman}, K.~L. and {Espaillat}, C. and {D'Alessio}, P. and {Adame}, L. and {Manoj}, P. and {Kim}, K.~H. and {Watson}, Dan M. and {Forrest}, W.~J. and {McClure}, M.~K. and {Calvet}, N. and {Sargent}, B.~A. and {Green}, J.~D. and {Fischer}, W.~J.},
        title = "{The Spitzer Infrared Spectrograph Survey of T Tauri Stars in Taurus}",
      journal = {\apjs},
         year = 2011,
        month = jul,
       volume = {195},
       number = {1},
          eid = {3},
        pages = {3},
          doi = {10.1088/0067-0049/195/1/3},
       adsurl = {https://ui.adsabs.harvard.edu/abs/2011ApJS..195....3F}
}

@INPROCEEDINGS{DAlessio2005,
       author = {{D'Alessio}, P. and {Calvet}, N. and {Woolum}, D.~S.},
        title = "{Thermal Structure of Protoplanetary Disks}",
    booktitle = {Chondrites and the Protoplanetary Disk},
         year = 2005,
       editor = {{Krot}, A.~N. and {Scott}, E.~R.~D. and {Reipurth}, B.},
       series = {Astronomical Society of the Pacific Conference Series},
       volume = {341},
        month = dec,
        pages = {353},
       adsurl = {https://ui.adsabs.harvard.edu/abs/2005ASPC..341..353D}
}

@ARTICLE{Henning1993,
       author = {{Henning}, Th. and {Stognienko}, R.},
        title = "{Porous grains and polarization of light : the silicate features.}",
      journal = {\aap},
         year = 1993,
        month = dec,
       volume = {280},
        pages = {609-616},
       adsurl = {https://ui.adsabs.harvard.edu/abs/1993A&A...280..609H}
}

@ARTICLE{Bouwman2008,
       author = {{Bouwman}, J. and {Henning}, Th. and {Hillenbrand}, L.~A. and {Meyer}, M.~R. and {Pascucci}, I. and {Carpenter}, J. and {Hines}, D. and {Kim}, J.~S. and {Silverstone}, M.~D. and {Hollenbach}, D. and {Wolf}, S.},
        title = "{The Formation and Evolution of Planetary Systems: Grain Growth and Chemical Processing of Dust in T Tauri Systems}",
      journal = {\apj},
         year = 2008,
        month = aug,
       volume = {683},
       number = {1},
        pages = {479-498},
          doi = {10.1086/587793},
archivePrefix = {arXiv},
       eprint = {0802.3033},
 primaryClass = {astro-ph},
       adsurl = {https://ui.adsabs.harvard.edu/abs/2008ApJ...683..479B}
}

@ARTICLE{Watson2009,
       author = {{Watson}, Dan M. and {Leisenring}, Jarron M. and {Furlan}, Elise and {Bohac}, C.~J. and {Sargent}, B. and {Forrest}, W.~J. and {Calvet}, Nuria and {Hartmann}, Lee and {Nordhaus}, Jason T. and {Green}, Joel D. and {Kim}, K.~H. and {Sloan}, G.~C. and {Chen}, C.~H. and {Keller}, L.~D. and {d'Alessio}, Paola and {Najita}, J. and {Uchida}, Keven I. and {Houck}, J.~R.},
        title = "{Crystalline Silicates and Dust Processing in the Protoplanetary Disks of the Taurus Young Cluster}",
      journal = {\apjs},
         year = 2009,
        month = jan,
       volume = {180},
       number = {1},
        pages = {84-101},
          doi = {10.1088/0067-0049/180/1/84},
archivePrefix = {arXiv},
       eprint = {0704.1518},
 primaryClass = {astro-ph},
       adsurl = {https://ui.adsabs.harvard.edu/abs/2009ApJS..180...84W}
}

@ARTICLE{Meeus2009,
       author = {{Meeus}, G. and {Juh{\'a}sz}, A. and {Henning}, Th. and {Bouwman}, J. and {Chen}, C. and {Lawson}, W. and {Apai}, D. and {Pascucci}, I. and {Sicilia-Aguilar}, A.},
        title = "{MBM 12: young protoplanetary discs at high galactic latitude}",
      journal = {\aap},
         year = 2009,
        month = apr,
       volume = {497},
       number = {2},
        pages = {379-392},
          doi = {10.1051/0004-6361/200811490},
archivePrefix = {arXiv},
       eprint = {0901.1668},
 primaryClass = {astro-ph.SR},
       adsurl = {https://ui.adsabs.harvard.edu/abs/2009A&A...497..379M}
}

@ARTICLE{Honda2006,
       author = {{Honda}, Mitsuhiko and {Kataza}, Hirokazu and {Okamoto}, Yoshiko K. and {Yamashita}, Takuya and {Min}, Michiel and {Miyata}, Takashi and {Sako}, Shigeyuki and {Fujiyoshi}, Takuya and {Sakon}, Itsuki and {Onaka}, Takashi},
        title = "{Subaru/COMICS Study on Silicate Dust Processing around Young Low-Mass Stars}",
      journal = {\apj},
         year = 2006,
        month = aug,
       volume = {646},
       number = {2},
        pages = {1024-1037},
          doi = {10.1086/505035},
       adsurl = {https://ui.adsabs.harvard.edu/abs/2006ApJ...646.1024H}
}

@ARTICLE{Rigby2023,
       author = {{Rigby}, Jane and {Perrin}, Marshall and {McElwain}, Michael and {Kimble}, Randy and {Friedman}, Scott and {Lallo}, Matt and {Doyon}, Ren{\'e} and {Feinberg}, Lee and {Ferruit}, Pierre and {Glasse}, Alistair and {Rieke}, Marcia and {Rieke}, George and {Wright}, Gillian and {Willott}, Chris and {Colon}, Knicole and {Milam}, Stefanie and {Neff}, Susan and {Stark}, Christopher and {Valenti}, Jeff and {Abell}, Jim and {Abney}, Faith and {Abul-Huda}, Yasin and {Acton}, D. Scott and {Adams}, Evan and {Adler}, David and {Aguilar}, Jonathan and {Ahmed}, Nasif and {Albert}, Lo{\"\i}c and {Alberts}, Stacey and {Aldridge}, David and {Allen}, Marsha and {Altenburg}, Martin and {{\'A}lvarez-M{\'a}rquez}, Javier and {Alves de Oliveira}, Catarina and {Andersen}, Greg and {Anderson}, Harry and {Anderson}, Sara and {Argyriou}, Ioannis and {Armstrong}, Amber and {Arribas}, Santiago and {Artigau}, Etienne and {Arvai}, Amanda and {Atkinson}, Charles and {Bacon}, Gregory and {Bair}, Thomas and {Banks}, Kimberly and {Barrientes}, Jaclyn and {Barringer}, Bruce and {Bartosik}, Peter and {Bast}, William and {Baudoz}, Pierre and {Beatty}, Thomas and {Bechtold}, Katie and {Beck}, Tracy and {Bergeron}, Eddie and {Bergkoetter}, Matthew and {Bhatawdekar}, Rachana and {Birkmann}, Stephan and {Blazek}, Ronald and {Blome}, Claire and {Boccaletti}, Anthony and {B{\"o}ker}, Torsten and {Boia}, John and {Bonaventura}, Nina and {Bond}, Nicholas and {Bosley}, Kari and {Boucarut}, Ray and {Bourque}, Matthew and {Bouwman}, Jeroen and {Bower}, Gary and {Bowers}, Charles and {Boyer}, Martha and {Bradley}, Larry and {Brady}, Greg and {Braun}, Hannah and {Breda}, David and {Bresnahan}, Pamela and {Bright}, Stacey and {Britt}, Christopher and {Bromenschenkel}, Asa and {Brooks}, Brian and {Brooks}, Keira and {Brown}, Bob and {Brown}, Matthew and {Brown}, Patricia and {Bunker}, Andy and {Burger}, Matthew and {Bushouse}, Howard and {Cale}, Steven and {Cameron}, Alex and {Cameron}, Peter and {Canipe}, Alicia and {Caplinger}, James and {Caputo}, Francis and {Cara}, Mihai and {Carey}, Larkin and {Carniani}, Stefano and {Carrasquilla}, Maria and {Carruthers}, Margaret and {Case}, Michael and {Catherine}, Riggs and {Chance}, Don and {Chapman}, George and {Charlot}, St{\'e}phane and {Charlow}, Brian and {Chayer}, Pierre and {Chen}, Bin and {Cherinka}, Brian and {Chichester}, Sarah and {Chilton}, Zack and {Chonis}, Taylor and {Clampin}, Mark and {Clark}, Charles and {Clark}, Kerry and {Coe}, Dan and {Coleman}, Benee and {Comber}, Brian and {Comeau}, Tom and {Connolly}, Dennis and {Cooper}, James and {Cooper}, Rachel and {Coppock}, Eric and {Correnti}, Matteo and {Cossou}, Christophe and {Coulais}, Alain and {Coyle}, Laura and {Cracraft}, Misty and {Curti}, Mirko and {Cuturic}, Steven and {Davis}, Katherine and {Davis}, Michael and {Dean}, Bruce and {DeLisa}, Amy and {deMeester}, Wim and {Dencheva}, Nadia and {Dencheva}, Nadezhda and {DePasquale}, Joseph and {Deschenes}, Jeremy and {Hunor Detre}, {\"O}rs and {Diaz}, Rosa and {Dicken}, Dan and {DiFelice}, Audrey and {Dillman}, Matthew and {Dixon}, William and {Doggett}, Jesse and {Donaldson}, Tom and {Douglas}, Rob and {DuPrie}, Kimberly and {Dupuis}, Jean and {Durning}, John and {Easmin}, Nilufar and {Eck}, Weston and {Edeani}, Chinwe and {Egami}, Eiichi and {Ehrenwinkler}, Ralf and {Eisenhamer}, Jonathan and {Eisenhower}, Michael and {Elie}, Michelle and {Elliott}, James and {Elliott}, Kyle and {Ellis}, Tracy and {Engesser}, Michael and {Espinoza}, Nestor and {Etienne}, Odessa and {Etxaluze}, Mireya and {Falini}, Patrick and {Feeney}, Matthew and {Ferry}, Malcolm and {Filippazzo}, Joseph and {Fincham}, Brian and {Fix}, Mees and {Flagey}, Nicolas and {Florian}, Michael and {Flynn}, Jim and {Fontanella}, Erin and {Ford}, Terrance and {Forshay}, Peter and {Fox}, Ori and {Franz}, David and {Fu}, Henry and {Fullerton}, Alexander and {Galkin}, Sergey and {Galyer}, Anthony and {Garc{\'\i}a Mar{\'\i}n}, Macarena and {Gardner}, Jonathan P. and {Gardner}, Lisa and {Garland}, Dennis and {Garrett}, Bruce and {Gasman}, Danny and {Gaspar}, Andras and {Gaudreau}, Daniel and {Gauthier}, Peter and {Geers}, Vincent and {Geithner}, Paul and {Gennaro}, Mario and {Giardino}, Giovanna and {Girard}, Julien and {Giuliano}, Mark and {Glassmire}, Kirk and {Glauser}, Adrian},
        title = "{The Science Performance of JWST as Characterized in Commissioning}",
      journal = {\pasp},
         year = 2023,
        month = apr,
       volume = {135},
       number = {1046},
          eid = {048001},
        pages = {048001},
          doi = {10.1088/1538-3873/acb293},
archivePrefix = {arXiv},
       eprint = {2207.05632},
 primaryClass = {astro-ph.IM},
       adsurl = {https://ui.adsabs.harvard.edu/abs/2023PASP..135d8001R}
}

@ARTICLE{Argyriou2023,
       author = {{Argyriou}, Ioannis and {Glasse}, Alistair and {Law}, David R. and {Labiano}, Alvaro and {{\'A}lvarez-M{\'a}rquez}, Javier and {Patapis}, Polychronis and {Kavanagh}, Patrick J. and {Gasman}, Danny and {Mueller}, Michael and {Larson}, Kirsten and {Vandenbussche}, Bart and {Glauser}, Adrian M. and {Royer}, Pierre and {Dicken}, Daniel and {Harkett}, Jake and {Sargent}, Beth A. and {Engesser}, Michael and {Jones}, Olivia C. and {Kendrew}, Sarah and {Noriega-Crespo}, Alberto and {Brandl}, Bernhard and {Rieke}, George H. and {Wright}, Gillian S. and {Lee}, David and {Wells}, Martyn},
        title = "{JWST MIRI flight performance: The Medium-Resolution Spectrometer}",
      journal = {\aap},
         year = 2023,
        month = jul,
       volume = {675},
          eid = {A111},
        pages = {A111},
          doi = {10.1051/0004-6361/202346489},
archivePrefix = {arXiv},
       eprint = {2303.13469},
 primaryClass = {astro-ph.IM},
       adsurl = {https://ui.adsabs.harvard.edu/abs/2023A&A...675A.111A}
}

@ARTICLE{Banzatti2023,
       author = {{Banzatti}, Andrea and {Pontoppidan}, Klaus M. and {Carr}, John S. and {Jellison}, Evan and {Pascucci}, Ilaria and {Najita}, Joan R. and {Romero-Mirza}, Carlos E. and {{\"O}berg}, Karin I. and {Kalyaan}, Anusha and {Pinilla}, Paola and {Krijt}, Sebastiaan and {Long}, Feng and {Lambrechts}, Michiel and {Rosotti}, Giovanni and {Herczeg}, Gregory J. and {Salyk}, Colette and {Zhang}, Ke and {Bergin}, Edwin A. and {Ballering}, Nicholas P. and {Meyer}, Michael R. and {Bruderer}, Simon and {Jdiscs Collaboration}},
        title = "{JWST Reveals Excess Cool Water near the Snow Line in Compact Disks, Consistent with Pebble Drift}",
      journal = {\apjl},
         year = 2023,
        month = nov,
       volume = {957},
       number = {2},
          eid = {L22},
        pages = {L22},
          doi = {10.3847/2041-8213/acf5ec},
archivePrefix = {arXiv},
       eprint = {2307.03846},
 primaryClass = {astro-ph.EP},
       adsurl = {https://ui.adsabs.harvard.edu/abs/2023ApJ...957L..22B}
}

@ARTICLE{Arulanantham2025,
       author = {{Arulanantham}, Nicole and {Salyk}, Colette and {Pontoppidan}, Klaus and {Banzatti}, Andrea and {Zhang}, Ke and {{\"O}berg}, Karin and {Long}, Feng and {Carr}, John and {Najita}, Joan and {Pascucci}, Ilaria and {Colmenares}, Mar{\'\i}a Jos{\'e} and {Xie}, Chengyan and {Huang}, Jane and {Green}, Joel and {Andrews}, Sean M. and {Blake}, Geoffrey A. and {Bergin}, Edwin A. and {Pinilla}, Paola and {Vioque}, Miguel and {Dahl}, Emma and {Raul}, Eshan and {Krijt}, Sebastiaan and {The Jdiscs Collaboration}},
        title = "{The JDISC Survey: Linking the Physics and Chemistry of Inner and Outer Protoplanetary Disk Zones}",
      journal = {\aj},
         year = 2025,
        month = aug,
       volume = {170},
       number = {2},
          eid = {67},
        pages = {67},
          doi = {10.3847/1538-3881/addd01},
archivePrefix = {arXiv},
       eprint = {2505.07562},
 primaryClass = {astro-ph.SR},
       adsurl = {https://ui.adsabs.harvard.edu/abs/2025AJ....170...67A}
}

@ARTICLE{Perotti2023,
       author = {{Perotti}, G. and {Christiaens}, V. and {Henning}, Th. and {Tabone}, B. and {Waters}, L.~B.~F.~M. and {Kamp}, I. and {Olofsson}, G. and {Grant}, S.~L. and {Gasman}, D. and {Bouwman}, J. and {Samland}, M. and {Franceschi}, R. and {van Dishoeck}, E.~F. and {Schwarz}, K. and {G{\"u}del}, M. and {Lagage}, P. -O. and {Ray}, T.~P. and {Vandenbussche}, B. and {Abergel}, A. and {Absil}, O. and {Arabhavi}, A.~M. and {Argyriou}, I. and {Barrado}, D. and {Boccaletti}, A. and {Caratti o Garatti}, A. and {Geers}, V. and {Glauser}, A.~M. and {Justannont}, K. and {Lahuis}, F. and {Mueller}, M. and {Nehm{\'e}}, C. and {Pantin}, E. and {Scheithauer}, S. and {Waelkens}, C. and {Guadarrama}, R. and {Jang}, H. and {Kanwar}, J. and {Morales-Calder{\'o}n}, M. and {Pawellek}, N. and {Rodgers-Lee}, D. and {Schreiber}, J. and {Colina}, L. and {Greve}, T.~R. and {{\"O}stlin}, G. and {Wright}, G.},
        title = "{Water in the terrestrial planet-forming zone of the PDS 70 disk}",
      journal = {\nat},
         year = 2023,
        month = aug,
       volume = {620},
       number = {7974},
        pages = {516-520},
          doi = {10.1038/s41586-023-06317-9},
archivePrefix = {arXiv},
       eprint = {2307.12040},
 primaryClass = {astro-ph.EP},
       adsurl = {https://ui.adsabs.harvard.edu/abs/2023Natur.620..516P}
}

@ARTICLE{Grant2023,
       author = {{Grant}, Sierra L. and {van Dishoeck}, Ewine F. and {Tabone}, Beno{\^\i}t and {Gasman}, Danny and {Henning}, Thomas and {Kamp}, Inga and {G{\"u}del}, Manuel and {Lagage}, Pierre-Olivier and {Bettoni}, Giulio and {Perotti}, Giulia and {Christiaens}, Valentin and {Samland}, Matthias and {Arabhavi}, Aditya M. and {Argyriou}, Ioannis and {Abergel}, Alain and {Absil}, Olivier and {Barrado}, David and {Boccaletti}, Anthony and {Bouwman}, Jeroen and {o Garatti}, Alessio Caratti and {Geers}, Vincent and {Glauser}, Adrian M. and {Guadarrama}, Rodrigo and {Jang}, Hyerin and {Kanwar}, Jayatee and {Lahuis}, Fred and {Morales-Calder{\'o}n}, Maria and {Mueller}, Michael and {Nehm{\'e}}, Cyrine and {Olofsson}, G{\"o}ran and {Pantin}, Eric and {Pawellek}, Nicole and {Ray}, Tom P. and {Rodgers-Lee}, Donna and {Scheithauer}, Silvia and {Schreiber}, J{\"u}rgen and {Schwarz}, Kamber and {Temmink}, Milou and {Vandenbussche}, Bart and {Vlasblom}, Marissa and {Waters}, L.~B.~F.~M. and {Wright}, Gillian and {Colina}, Luis and {Greve}, Thomas R. and {Justannont}, Kay and {{\"O}stlin}, G{\"o}ran},
        title = "{MINDS. The Detection of $^{13}$CO$_{2}$ with JWST-MIRI Indicates Abundant CO$_{2}$ in a Protoplanetary Disk}",
      journal = {\apjl},
         year = 2023,
        month = apr,
       volume = {947},
       number = {1},
          eid = {L6},
        pages = {L6},
          doi = {10.3847/2041-8213/acc44b},
archivePrefix = {arXiv},
       eprint = {2212.08047},
 primaryClass = {astro-ph.SR},
       adsurl = {https://ui.adsabs.harvard.edu/abs/2023ApJ...947L...6G}
}

@ARTICLE{Temmink2025,
       author = {{Temmink}, Milou and {Sellek}, Andrew D. and {Gasman}, Danny and {van Dishoeck}, Ewine F. and {Vlasblom}, Marissa and {Pranger}, Ang{\`e}l and {G{\"u}del}, Manuel and {Henning}, Thomas and {Lagage}, Pierre-Olivier and {Caratti o Garatti}, Alessio and {Kamp}, Inga and {Olofsson}, G{\"o}ran and {Arabhavi}, Aditya M. and {Grant}, Sierra L. and {Kaeufer}, Till and {Kurtovic}, Nicolas T. and {Perotti}, Giulia and {Samland}, Matthias and {Schwarz}, Kamber and {Tabone}, Beno{\^\i}t},
        title = "{MINDS: Water reservoirs of compact planet-forming dust discs: A diversity of H$_{2}$O distributions}",
      journal = {\aap},
         year = 2025,
        month = jul,
       volume = {699},
          eid = {A134},
        pages = {A134},
          doi = {10.1051/0004-6361/202554213},
archivePrefix = {arXiv},
       eprint = {2505.15237},
 primaryClass = {astro-ph.EP},
       adsurl = {https://ui.adsabs.harvard.edu/abs/2025A&A...699A.134T}
}

@ARTICLE{Gasman2025,
       author = {{Gasman}, Danny and {Temmink}, Milou and {van Dishoeck}, Ewine F. and {Kurtovic}, Nicolas T. and {Grant}, Sierra L. and {Sellek}, Andrew and {Tabone}, Beno{\^\i}t and {Henning}, Thomas and {Kamp}, Inga and {G{\"u}del}, Manuel and {Barrado}, David and {Caratti o Garatti}, Alessio and {Glauser}, Adrian M. and {Waters}, Laurens B.~F.~M. and {Arabhavi}, Aditya M. and {Jang}, Hyerin and {Kanwar}, Jayatee and {Lienert}, Julia L. and {Perotti}, Giulia and {Schwarz}, Kamber and {Vlasblom}, Marissa},
        title = "{MINDS: The influence of outer dust disc structure on the volatile delivery to the inner disc}",
      journal = {\aap},
         year = 2025,
        month = feb,
       volume = {694},
          eid = {A147},
        pages = {A147},
          doi = {10.1051/0004-6361/202452152},
archivePrefix = {arXiv},
       eprint = {2501.04587},
 primaryClass = {astro-ph.EP},
       adsurl = {https://ui.adsabs.harvard.edu/abs/2025A&A...694A.147G}
}

@ARTICLE{Kospal2023,
       author = {{K{\'o}sp{\'a}l}, {\'A}gnes and {{\'A}brah{\'a}m}, P{\'e}ter and {Diehl}, Lindsey and {Banzatti}, Andrea and {Bouwman}, Jeroen and {Chen}, Lei and {Cruz-S{\'a}enz de Miera}, Fernando and {Green}, Joel D. and {Henning}, Thomas and {Rab}, Christian},
        title = "{JWST/MIRI Spectroscopy of the Disk of the Young Eruptive Star EX Lup in Quiescence}",
      journal = {\apjl},
         year = 2023,
        month = mar,
       volume = {945},
       number = {1},
          eid = {L7},
        pages = {L7},
          doi = {10.3847/2041-8213/acb58a},
archivePrefix = {arXiv},
       eprint = {2301.08770},
 primaryClass = {astro-ph.SR},
       adsurl = {https://ui.adsabs.harvard.edu/abs/2023ApJ...945L...7K}
}

@ARTICLE{Kospal2025,
       author = {{K{\'o}sp{\'a}l}, {\'A}. and {{\'A}brah{\'a}m}, P. and {Akimkin}, V.~V. and {Chen}, L. and {Forbrich}, J. and {Getman}, K.~V. and {Portilla-Revelo}, B. and {Semenov}, D. and {van Terwisga}, S.~E. and {Varga}, J. and {Zwicky}, L. and {Bal{\'a}zs}, G.~G. and {Bora}, Zs. and {Horti-D{\'a}vid}, {\'A}. and {Jo{\'o}}, A.~P. and {Og{\l}oza}, W. and {Seli}, B. and {Siwak}, M. and {S{\'o}dor}, {\'A}. and {Tak{\'a}cs}, N.},
        title = "{Time-resolved protoplanetary disk physics in DQ Tau with JWST}",
      journal = {\aap},
    volume    =  703,
    pages     = "A20",
    month     =  nov,
    year      =  2025,     
          eid = {arXiv:2508.19701},
        pages = {arXiv:2508.19701},
          doi = {10.1051/0004-6361/202556016},
archivePrefix = {arXiv},
       eprint = {2508.19701},
 primaryClass = {astro-ph.SR},
       adsurl = {https://ui.adsabs.harvard.edu/abs/2025arXiv250819701K}
}

@ARTICLE{Grant2025,
       author = {{Grant}, S.~L. and {Temmink}, M. and {van Dishoeck}, E.~F. and {Gasman}, D. and {Arabhavi}, A.~M. and {Tabone}, B. and {Henning}, T. and {Kamp}, I. and {Caratti o Garatti}, A. and {Christiaens}, V. and {Esteve}, P. and {G{\"u}del}, M. and {Jang}, H. and {Kaeufer}, T. and {Kurtovic}, N.~T. and {Morales-Calder{\'o}n}, M. and {Perotti}, G. and {Schwarz}, K. and {Sellek}, A.~D. and {Stapper}, L.~M. and {Vlasblom}, M. and {Waters}, L.~B.~F.~M.},
        title = "{MINDS: A transition from H$_{2}$O to C$_{2}$H$_{2}$ dominated disk spectra with decreasing stellar luminosity}",
      journal = {\aap},
         year = 2025,
        month = oct,
       volume = {702},
          eid = {A126},
        pages = {A126},
          doi = {10.1051/0004-6361/202555862},
archivePrefix = {arXiv},
       eprint = {2508.04692},
 primaryClass = {astro-ph.EP},
       adsurl = {https://ui.adsabs.harvard.edu/abs/2025A&A...702A.126G}
}

@ARTICLE{Liu2025,
       author = {{Liu}, Yao and {Li}, Dafa and {Wang}, Hongchi and {Feng}, Haoran and {Fang}, Min and {Du}, Fujun and {Henning}, Thomas and {Perotti}, Giulia},
        title = "{Dust processing in the terrestrial planet-forming region of the PDS 70 disk}",
      journal = {Science China Physics, Mechanics, and Astronomy},
         year = 2025,
        month = may,
       volume = {68},
       number = {5},
          eid = {259511},
        pages = {259511},
          doi = {10.1007/s11433-024-2597-5},
archivePrefix = {arXiv},
       eprint = {2501.05913},
 primaryClass = {astro-ph.EP},
       adsurl = {https://ui.adsabs.harvard.edu/abs/2025SCPMA..6859511L}
}

@ARTICLE{Kurtovic2026,
       author = {{Kurtovic}, Nicol{\'a}s T. and {Grant}, Sierra L. and {Temmink}, Milou and {Sellek}, Andrew D. and {van Dishoeck}, Ewine F. and {Henning}, Thomas and {Kamp}, Inga and {Christiaens}, Valentin and {Banzatti}, Andrea and {Gasman}, Danny and {Kaeufer}, Till and {Stapper}, Lucas M. and {Franceschi}, Riccardo and {G{\"u}del}, Manuel and {Lagage}, Pierre-Olivier and {Vlasblom}, Marissa and {Perotti}, Giulia and {Schwarz}, Kamber and {Somigliana}, Alice},
        title = "{MINDS: Young binary systems with JWST/MIRI: Variable water-rich primaries and extended emission}",
      journal = {\aap},
         year = 2026,
        month = jan,
       volume = {705},
          eid = {A97},
        pages = {A97},
          doi = {10.1051/0004-6361/202554927},
archivePrefix = {arXiv},
       eprint = {2508.02576},
 primaryClass = {astro-ph.EP},
       adsurl = {https://ui.adsabs.harvard.edu/abs/2026A&A...705A..97K}
}

@ARTICLE{Jang2025,
       author = {{Jang}, Hyerin and {Arabhavi}, Aditya M. and {Kaeufer}, Till and {Waters}, Rens and {Kamp}, Inga and {Henning}, Thomas and {Caratti o Garatti}, Alessio and {van Dishoeck}, Ewine F. and {Perotti}, Giulia and {Kanwar}, Jayatee and {G{\"u}del}, Manuel and {Morales-Calder{\'o}n}, Maria and {Grant}, Sierra L. and {Christiaens}, Valentin},
        title = "{MINDS: The very low-mass star and brown dwarf sample: II. Probing disk settling, dust properties, and dust-gas interplay with JWST/MIRI}",
      journal = {\aap},
         year = 2025,
        month = nov,
       volume = {703},
          eid = {A53},
        pages = {A53},
          doi = {10.1051/0004-6361/202556193},
archivePrefix = {arXiv},
       eprint = {2509.16004},
 primaryClass = {astro-ph.EP},
       adsurl = {https://ui.adsabs.harvard.edu/abs/2025A&A...703A..53J}
}

@ARTICLE{Kamp2023,
       author = {{Kamp}, Inga and {Henning}, Thomas and {Arabhavi}, Aditya M. and {Bettoni}, Giulio and {Christiaens}, Valentin and {Gasman}, Danny and {Grant}, Sierra L. and {Morales-Calder{\'o}n}, Maria and {Tabone}, Beno{\^\i}t and {Abergel}, Alain and {Absil}, Olivier and {Argyriou}, Ioannis and {Barrado}, David and {Boccaletti}, Anthony and {Bouwman}, Jeroen and {Caratti o Garatti}, Alessio and {van Dishoeck}, Ewine F. and {Geers}, Vincent and {Glauser}, Adrian M. and {G{\"u}del}, Manuel and {Guadarrama}, Rodrigo and {Jang}, Hyerin and {Kanwar}, Jayatee and {Lagage}, Pierre-Olivier and {Lahuis}, Fred and {Mueller}, Michael and {Nehm{\'e}}, Cyrine and {Olofsson}, G{\"o}ran and {Pantin}, Eric and {Pawellek}, Nicole and {Perotti}, Giulia and {Ray}, Tom P. and {Rodgers-Lee}, Donna and {Samland}, Matthias and {Scheithauer}, Silvia and {Schreiber}, J{\"u}rgen and {Schwarz}, Kamber and {Temmink}, Milou and {Vandenbussche}, Bart and {Vlasblom}, Marissa and {Waelkens}, Christoffel and {Waters}, L.~B.~F.~M. and {Wright}, Gillian},
        title = "{The chemical inventory of the inner regions of planet-forming disks {\textendash} the JWST/MINDS program}",
      journal = {Faraday Discussions},
         year = 2023,
        month = sep,
       volume = {245},
        pages = {112-137},
          doi = {10.1039/D3FD00013C},
archivePrefix = {arXiv},
       eprint = {2307.16729},
 primaryClass = {astro-ph.EP},
       adsurl = {https://ui.adsabs.harvard.edu/abs/2023FaDi..245..112K}
}

@ARTICLE{Rieke2015,
       author = {{Rieke}, G.~H. and {Ressler}, M.~E. and {Morrison}, Jane E. and {Bergeron}, L. and {Bouchet}, Patrice and {Garc{\'\i}a-Mar{\'\i}n}, Macarena and {Greene}, T.~P. and {Regan}, M.~W. and {Sukhatme}, K.~G. and {Walker}, Helen},
        title = "{The Mid-Infrared Instrument for the James Webb Space Telescope, VII: The MIRI Detectors}",
      journal = {\pasp},
         year = 2015,
        month = jul,
       volume = {127},
       number = {953},
        pages = {665},
          doi = {10.1086/682257},
archivePrefix = {arXiv},
       eprint = {1508.02362},
 primaryClass = {astro-ph.IM},
       adsurl = {https://ui.adsabs.harvard.edu/abs/2015PASP..127..665R}
}

@ARTICLE{Wright2015,
       author = {{Wright}, G.~S. and {Wright}, David and {Goodson}, G.~B. and {Rieke}, G.~H. and {Aitink-Kroes}, Gabby and {Amiaux}, J. and {Aricha-Yanguas}, Ana and {Azzollini}, Ruym{\'a}n and {Banks}, Kimberly and {Barrado-Navascues}, D. and {Belenguer-Davila}, T. and {Blommaert}, J.~A.~D.~L. and {Bouchet}, Patrice and {Brandl}, B.~R. and {Colina}, L. and {Detre}, {\"O}rs and {Diaz-Catala}, Eva and {Eccleston}, Paul and {Friedman}, Scott D. and {Garc{\'\i}a-Mar{\'\i}n}, Macarena and {G{\"u}del}, Manuel and {Glasse}, Alistair and {Glauser}, Adrian M. and {Greene}, T.~P. and {Groezinger}, Uli and {Grundy}, Tim and {Hastings}, Peter and {Henning}, Th. and {Hofferbert}, Ralph and {Hunter}, Faye and {Jessen}, N.~C. and {Justtanont}, K. and {Karnik}, Avinash R. and {Khorrami}, Mori A. and {Krause}, Oliver and {Labiano}, Alvaro and {Lagage}, P.-O. and {Langer}, Ulrich and {Lemke}, Dietrich and {Lim}, Tanya and {Lorenzo-Alvarez}, Jose and {Mazy}, Emmanuel and {McGowan}, Norman and {Meixner}, M.~E. and {Morris}, Nigel and {Morrison}, Jane E. and {M{\"u}ller}, Friedrich and {rgaard-Nielson}, H.-U. N{\o} and {Olofsson}, G{\"o}ran and {O'Sullivan}, Brian and {Pel}, J.-W. and {Penanen}, Konstantin and {Petach}, M.~B. and {Pye}, J.~P. and {Ray}, T.~P. and {Renotte}, Etienne and {Renouf}, Ian and {Ressler}, M.~E. and {Samara-Ratna}, Piyal and {Scheithauer}, Silvia and {Schneider}, Analyn and {Shaughnessy}, Bryan and {Stevenson}, Tim and {Sukhatme}, Kalyani and {Swinyard}, Bruce and {Sykes}, Jon and {Thatcher}, John and {Tikkanen}, Tuomo and {van Dishoeck}, E.~F. and {Waelkens}, C. and {Walker}, Helen and {Wells}, Martyn and {Zhender}, Alex},
        title = "{The Mid-Infrared Instrument for the James Webb Space Telescope, II: Design and Build}",
      journal = {\pasp},
         year = 2015,
        month = jul,
       volume = {127},
       number = {953},
        pages = {595},
          doi = {10.1086/682253},
archivePrefix = {arXiv},
       eprint = {1508.02333},
 primaryClass = {astro-ph.IM},
       adsurl = {https://ui.adsabs.harvard.edu/abs/2015PASP..127..595W}
}

@ARTICLE{Wright2023,
       author = {{Wright}, Gillian S. and {Rieke}, George H. and {Glasse}, Alistair and {Ressler}, Michael and {Garc{\'\i}a Mar{\'\i}n}, Macarena and {Aguilar}, Jonathan and {Alberts}, Stacey and {{\'A}lvarez-M{\'a}rquez}, Javier and {Argyriou}, Ioannis and {Banks}, Kimberly and {Baudoz}, Pierre and {Boccaletti}, Anthony and {Bouchet}, Patrice and {Bouwman}, Jeroen and {Brandl}, Bernard R. and {Breda}, David and {Bright}, Stacey and {Cale}, Steven and {Colina}, Luis and {Cossou}, Christophe and {Coulais}, Alain and {Cracraft}, Misty and {De Meester}, Wim and {Dicken}, Daniel and {Engesser}, Michael and {Etxaluze}, Mireya and {Fox}, Ori D. and {Friedman}, Scott and {Fu}, Henry and {Gasman}, Danny and {G{\'a}sp{\'a}r}, Andr{\'a}s and {Gastaud}, Ren{\'e} and {Geers}, Vincent and {Glauser}, Adrian Michael and {Gordon}, Karl D. and {Greene}, Thomas and {Greve}, Thomas R. and {Grundy}, Timothy and {G{\"u}del}, Manuel and {Guillard}, Pierre and {Haderlein}, Peter and {Hashimoto}, Ryan and {Henning}, Thomas and {Hines}, Dean and {Holler}, Bryan and {Detre}, {\"O}rs Hunor and {Jahromi}, Amir and {James}, Bryan and {Jones}, Olivia C. and {Justtanont}, Kay and {Kavanagh}, Patrick and {Kendrew}, Sarah and {Klaassen}, Pamela and {Krause}, Oliver and {Labiano}, Alvaro and {Lagage}, Pierre-Olivier and {Lambros}, Scott and {Larson}, Kirsten and {Law}, David and {Lee}, David and {Libralato}, Mattia and {Lorenzo Alverez}, Jose and {Meixner}, Margaret and {Morrison}, Jane and {Mueller}, Migo and {Murray}, Katherine and {Mycroft}, Matthew and {Myers}, Richard and {Nayak}, Omnarayani and {Naylor}, Bret and {Nickson}, Bryony and {Noriega-Crespo}, Alberto and {{\"O}stlin}, G{\"o}ran and {O'Sullivan}, Brian and {Ottens}, Richard and {Patapis}, Polychronis and {Penanen}, Konstantin and {Pietraszkiewicz}, Martin and {Ray}, Tom and {Regan}, Michael and {Roteliuk}, Anthony and {Royer}, Pierre and {Samara-Ratna}, Piyal and {Samuelson}, Bridget and {Sargent}, Beth A. and {Scheithauer}, Silvia and {Schneider}, Analyn and {Schreiber}, J{\"u}rgen and {Shaughnessy}, Bryan and {Sheehan}, Evan and {Shivaei}, Irene and {Sloan}, G.~C. and {Tamas}, Laszlo and {Teague}, Kelly and {Temim}, Tea and {Tikkanen}, Tuomo and {Tustain}, Samuel and {van Dishoeck}, Ewine F. and {Vandenbussche}, Bart and {Weilert}, Mark and {Whitehouse}, Paul and {Wolff}, Schuyler},
        title = "{The Mid-infrared Instrument for JWST and Its In-flight Performance}",
      journal = {\pasp},
         year = 2023,
        month = apr,
       volume = {135},
       number = {1046},
          eid = {048003},
        pages = {048003},
          doi = {10.1088/1538-3873/acbe66},
       adsurl = {https://ui.adsabs.harvard.edu/abs/2023PASP..135d8003W}
}

@ARTICLE{Wells2015,
       author = {{Wells}, Martyn and {Pel}, J.-W. and {Glasse}, Alistair and {Wright}, G.~S. and {Aitink-Kroes}, Gabby and {Azzollini}, Ruym{\'a}n and {Beard}, Steven and {Brandl}, B.~R. and {Gallie}, Angus and {Geers}, V.~C. and {Glauser}, A.~M. and {Hastings}, Peter and {Henning}, Th. and {Jager}, Rieks and {Justtanont}, K. and {Kruizinga}, Bob and {Lahuis}, Fred and {Lee}, David and {Martinez-Delgado}, I. and {Mart{\'\i}nez-Galarza}, J.~R. and {Meijers}, M. and {Morrison}, Jane E. and {M{\"u}ller}, Friedrich and {Nakos}, Thodori and {O'Sullivan}, Brian and {Oudenhuysen}, Ad and {Parr-Burman}, P. and {Pauwels}, Evert and {Rohloff}, R.-R. and {Schmalzl}, Eva and {Sykes}, Jon and {Thelen}, M.~P. and {van Dishoeck}, E.~F. and {Vandenbussche}, Bart and {Venema}, Lars B. and {Visser}, Huib and {Waters}, L.~B.~F.~M. and {Wright}, David},
        title = "{The Mid-Infrared Instrument for the James Webb Space Telescope, VI: The Medium Resolution Spectrometer}",
      journal = {\pasp},
         year = 2015,
        month = jul,
       volume = {127},
       number = {953},
        pages = {646},
          doi = {10.1086/682281},
archivePrefix = {arXiv},
       eprint = {1508.03070},
 primaryClass = {astro-ph.IM},
       adsurl = {https://ui.adsabs.harvard.edu/abs/2015PASP..127..646W}
}

@ARTICLE{Draine2003,
       author = {{Draine}, B.~T.},
        title = "{Interstellar Dust Grains}",
      journal = {\araa},
         year = 2003,
        month = jan,
       volume = {41},
        pages = {241-289},
          doi = {10.1146/annurev.astro.41.011802.094840},
archivePrefix = {arXiv},
       eprint = {astro-ph/0304489},
 primaryClass = {astro-ph},
       adsurl = {https://ui.adsabs.harvard.edu/abs/2003ARA&A..41..241D}
}

@ARTICLE{Zhukovska2018,
       author = {{Zhukovska}, Svitlana and {Henning}, Thomas and {Dobbs}, Clare},
        title = "{Iron and Silicate Dust Growth in the Galactic Interstellar Medium: Clues from Element Depletions}",
      journal = {\apj},
         year = 2018,
        month = apr,
       volume = {857},
       number = {2},
          eid = {94},
        pages = {94},
          doi = {10.3847/1538-4357/aab438},
archivePrefix = {arXiv},
       eprint = {1803.01929},
 primaryClass = {astro-ph.GA},
       adsurl = {https://ui.adsabs.harvard.edu/abs/2018ApJ...857...94Z}
}

@ARTICLE{DorschnerHenning1995,
       author = {{Dorschner}, J. and {Henning}, T.},
        title = "{Dust metamorphosis in the galaxy}",
      journal = {\aapr},
         year = 1995,
        month = jan,
       volume = {6},
       number = {4},
        pages = {271-333},
          doi = {10.1007/BF00873686},
       adsurl = {https://ui.adsabs.harvard.edu/abs/1995A&ARv...6..271D}
}

@ARTICLE{HenningSalama1998,
       author = {{Henning}, T. and {Salama}, F.},
        title = "{Carbon in the Universe}",
      journal = {Science},
         year = 1998,
        month = dec,
       volume = {282},
        pages = {2204},
          doi = {10.1126/science.282.5397.2204},
       adsurl = {https://ui.adsabs.harvard.edu/abs/1998Sci...282.2204H}
}

@ARTICLE{Brauer2008,
       author = {{Brauer}, F. and {Dullemond}, C.~P. and {Henning}, Th.},
        title = "{Coagulation, fragmentation and radial motion of solid particles in protoplanetary disks}",
      journal = {\aap},
         year = 2008,
        month = mar,
       volume = {480},
       number = {3},
        pages = {859-877},
          doi = {10.1051/0004-6361:20077759},
archivePrefix = {arXiv},
       eprint = {0711.2192},
 primaryClass = {astro-ph},
       adsurl = {https://ui.adsabs.harvard.edu/abs/2008A&A...480..859B}
}

@ARTICLE{Natta2001,
       author = {{Natta}, A. and {Prusti}, T. and {Neri}, R. and {Wooden}, D. and {Grinin}, V.~P. and {Mannings}, V.},
        title = "{A reconsideration of disk properties in Herbig Ae stars}",
      journal = {\aap},
         year = 2001,
        month = may,
       volume = {371},
        pages = {186-197},
          doi = {10.1051/0004-6361:20010334},
       adsurl = {https://ui.adsabs.harvard.edu/abs/2001A&A...371..186N}
}

@ARTICLE{Li2001,
       author = {{Li}, Aigen and {Draine}, B.~T.},
        title = "{Infrared Emission from Interstellar Dust. II. The Diffuse Interstellar Medium}",
      journal = {\apj},
         year = 2001,
        month = jun,
       volume = {554},
       number = {2},
        pages = {778-802},
          doi = {10.1086/323147},
archivePrefix = {arXiv},
       eprint = {astro-ph/0011319},
 primaryClass = {astro-ph},
       adsurl = {https://ui.adsabs.harvard.edu/abs/2001ApJ...554..778L}
}

@ARTICLE{Kaeufer2024_Sz28,
       author = {{Kaeufer}, T. and {Woitke}, P. and {Kamp}, I. and {Kanwar}, J. and {Min}, M.},
        title = "{Disentangling the dust and gas contributions of the JWST/MIRI spectrum of Sz 28}",
      journal = {\aap},
         year = 2024,
        month = oct,
       volume = {690},
          eid = {A100},
        pages = {A100},
          doi = {10.1051/0004-6361/202450891},
archivePrefix = {arXiv},
       eprint = {2408.06077},
 primaryClass = {astro-ph.EP},
       adsurl = {https://ui.adsabs.harvard.edu/abs/2024A&A...690A.100K}
}

@ARTICLE{Blum2008,
       author = {{Blum}, J. and {Wurm}, G.},
        title = "{The growth mechanisms of macroscopic bodies in protoplanetary disks.}",
      journal = {\araa},
         year = 2008,
        month = sep,
       volume = {46},
        pages = {21-56},
          doi = {10.1146/annurev.astro.46.060407.145152},
       adsurl = {https://ui.adsabs.harvard.edu/abs/2008ARA&A..46...21B}
}

@software{Bushouse2024,
       author = {{Bushouse}, Howard and {Eisenhamer}, Jonathan and {Dencheva}, Nadia and {Davies}, James and {Greenfield}, Perry and {Morrison}, Jane and {Hodge}, Phil and {Simon}, Bernie and {Grumm}, David and {Droettboom}, Michael and {Slavich}, Edward and {Sosey}, Megan and {Pauly}, Tyler and {Miller}, Todd and {Jedrzejewski}, Robert and {Hack}, Warren and {Davis}, David and {Crawford}, Steven and {Law}, David and {Gordon}, Karl and {Regan}, Michael and {Cara}, Mihai and {MacDonald}, Ken and {Bradley}, Larry and {Shanahan}, Clare and {Jamieson}, William and {Teodoro}, Mairan and {Williams}, Thomas and {Pena-Guerrero}, Maria and {Graham}, Brett and {Molter}, Edward and {Brandt}, Timothy and {Hayes}, Christian and {Cooper}, Rachel and {Clarke}, Melanie},
        title = "{JWST Calibration Pipeline}",
         year = 2024,
        month = nov,
          eid = {10.5281/zenodo.14153298},
          doi = {10.5281/zenodo.14153298},
      version = {1.16.1},
    publisher = {Zenodo},
       adsurl = {https://ui.adsabs.harvard.edu/abs/2024zndo..14153298B}
}

@ARTICLE{Grant2024,
       author = {{Grant}, Sierra L. and {Kurtovic}, Nicolas T. and {van Dishoeck}, Ewine F. and {Henning}, Thomas and {Kamp}, Inga and {Nowacki}, Hugo and {Perraut}, Karine and {Banzatti}, Andrea and {Temmink}, Milou and {Christiaens}, Valentin and {Samland}, Matthias and {Gasman}, Danny and {Tabone}, Beno{\^\i}t and {G{\"u}del}, Manuel and {Lagage}, Pierre-Olivier and {Arabhavi}, Aditya M. and {Barrado}, David and {Caratti o Garatti}, Alessio and {Glauser}, Adrian M. and {Jang}, Hyerin and {Kanwar}, Jayatee and {Lahuis}, Fred and {Morales-Calder{\'o}n}, Maria and {Olofsson}, G{\"o}ran and {Perotti}, Giulia and {Schwarz}, Kamber and {Vlasblom}, Marissa and {Garcia Lopez}, Rebeca and {Long}, Feng},
        title = "{MINDS: A multi-instrument investigation into the molecule-rich JWST-MIRI spectrum of the DF Tau binary system}",
      journal = {\aap},
         year = 2024,
        month = sep,
       volume = {689},
          eid = {A85},
        pages = {A85},
          doi = {10.1051/0004-6361/202450768},
archivePrefix = {arXiv},
       eprint = {2406.10217},
 primaryClass = {astro-ph.EP},
       adsurl = {https://ui.adsabs.harvard.edu/abs/2024A&A...689A..85G}
}

@ARTICLE{Daemgen2013,
       author = {{Daemgen}, S. and {Petr-Gotzens}, M.~G. and {Correia}, S. and {Teixeira}, P.~S. and {Brandner}, W. and {Kley}, W. and {Zinnecker}, H.},
        title = "{Protoplanetary disk evolution and stellar parameters of T Tauri binaries in Chamaeleon I}",
      journal = {\aap},
         year = 2013,
        month = jun,
       volume = {554},
          eid = {A43},
        pages = {A43},
          doi = {10.1051/0004-6361/201321220},
archivePrefix = {arXiv},
       eprint = {1304.1150},
 primaryClass = {astro-ph.SR},
       adsurl = {https://ui.adsabs.harvard.edu/abs/2013A&A...554A..43D}
}

@ARTICLE{Allen2017,
       author = {{Allen}, T.~S. and {Prato}, L. and {Wright-Garba}, N. and {Schaefer}, G. and {Biddle}, L.~I. and {Skiff}, B. and {Avilez}, I. and {Muzzio}, R. and {Simon}, M.},
        title = "{Properties of the Closest Young Binaries. I. DF Tau{\textquoteright}s Unequal Circumstellar Disk Evolution}",
      journal = {\apj},
         year = 2017,
        month = aug,
       volume = {845},
       number = {2},
          eid = {161},
        pages = {161},
          doi = {10.3847/1538-4357/aa8094},
       adsurl = {https://ui.adsabs.harvard.edu/abs/2017ApJ...845..161A}
}

@ARTICLE{Kutra2025,
       author = {{Kutra}, Taylor and {Prato}, Lisa and {Tofflemire}, Benjamin M. and {Akeson}, Rachel and {Schaefer}, G.~H. and {Tang}, Shih-Yun and {Segura-Cox}, Dominique and {Johns-Krull}, Christopher M. and {Kraus}, Adam and {Andrews}, Sean and {Jensen}, Eric L.~N.},
        title = "{Sites of Planet Formation in Binary Systems. II. Double the Disks in DF Tau}",
      journal = {\aj},
         year = 2025,
        month = jan,
       volume = {169},
       number = {1},
          eid = {20},
        pages = {20},
          doi = {10.3847/1538-3881/ad900a},
archivePrefix = {arXiv},
       eprint = {2411.05203},
 primaryClass = {astro-ph.SR},
       adsurl = {https://ui.adsabs.harvard.edu/abs/2025AJ....169...20K}
}

@ARTICLE{Melo2003,
       author = {{Melo}, C.~H.~F.},
        title = "{The short period multiplicity among T Tauri stars}",
      journal = {\aap},
         year = 2003,
        month = oct,
       volume = {410},
        pages = {269-282},
          doi = {10.1051/0004-6361:20031242},
       adsurl = {https://ui.adsabs.harvard.edu/abs/2003A&A...410..269M}
}

@ARTICLE{Nguyen2012,
       author = {{Nguyen}, Duy Cuong and {Brandeker}, Alexis and {van Kerkwijk}, Marten H. and {Jayawardhana}, Ray},
        title = "{Close Companions to Young Stars. I. A Large Spectroscopic Survey in Chamaeleon I and Taurus-Auriga}",
      journal = {\apj},
         year = 2012,
        month = feb,
       volume = {745},
       number = {2},
          eid = {119},
        pages = {119},
          doi = {10.1088/0004-637X/745/2/119},
archivePrefix = {arXiv},
       eprint = {1112.0002},
 primaryClass = {astro-ph.SR},
       adsurl = {https://ui.adsabs.harvard.edu/abs/2012ApJ...745..119N}
}

@ARTICLE{Zsidi2022,
       author = {{Zsidi}, Gabriella and {Fiorellino}, Eleonora and {K{\'o}sp{\'a}l}, {\'A}gnes and {{\'A}brah{\'a}m}, P{\'e}ter and {B{\'o}di}, Attila and {Hussain}, Gaitee and {Manara}, Carlo F. and {P{\'a}l}, Andr{\'a}s},
        title = "{Accretion Variability of the Multiple T Tauri System VW Cha}",
      journal = {\apj},
         year = 2022,
        month = dec,
       volume = {941},
       number = {2},
          eid = {177},
        pages = {177},
          doi = {10.3847/1538-4357/ac7229},
archivePrefix = {arXiv},
       eprint = {2205.11435},
 primaryClass = {astro-ph.SR},
       adsurl = {https://ui.adsabs.harvard.edu/abs/2022ApJ...941..177Z}
}

@ARTICLE{Brandeker2001,
       author = {{Brandeker}, Alexis and {Liseau}, Ren{\'e} and {Artymowicz}, Pawel and {Jayawardhana}, Ray},
        title = "{Discovery of a New Companion and Evidence of a Circumprimary Disk: Adaptive Optics Imaging of the Young Multiple System VW Chamaeleon}",
      journal = {\apjl},
         year = 2001,
        month = nov,
       volume = {561},
       number = {2},
        pages = {L199-L202},
          doi = {10.1086/324676},
archivePrefix = {arXiv},
       eprint = {astro-ph/0110047},
 primaryClass = {astro-ph},
       adsurl = {https://ui.adsabs.harvard.edu/abs/2001ApJ...561L.199B}
}

@ARTICLE{Olofsson2012,
       author = {{Olofsson}, J. and {Juh{\'a}sz}, A. and {Henning}, Th. and {Mutschke}, H. and {Tamanai}, A. and {Mo{\'o}r}, A. and {{\'A}brah{\'a}m}, P.},
        title = "{Transient dust in warm debris disks. Detection of Fe-rich olivine grains}",
      journal = {\aap},
         year = 2012,
        month = jun,
       volume = {542},
          eid = {A90},
        pages = {A90},
          doi = {10.1051/0004-6361/201118735},
archivePrefix = {arXiv},
       eprint = {1204.2374},
 primaryClass = {astro-ph.SR},
       adsurl = {https://ui.adsabs.harvard.edu/abs/2012A&A...542A..90O}
}

@ARTICLE{Herczeg2014,
       author = {{Herczeg}, Gregory J. and {Hillenbrand}, Lynne A.},
        title = "{An Optical Spectroscopic Study of T Tauri Stars. I. Photospheric Properties}",
      journal = {\apj},
         year = 2014,
        month = may,
       volume = {786},
       number = {2},
          eid = {97},
        pages = {97},
          doi = {10.1088/0004-637X/786/2/97},
archivePrefix = {arXiv},
       eprint = {1403.1675},
 primaryClass = {astro-ph.SR},
       adsurl = {https://ui.adsabs.harvard.edu/abs/2014ApJ...786...97H}
}

@ARTICLE{Koike2013,
       author = {{Koike}, C. and {Noguchi}, R. and {Chihara}, H. and {Suto}, H. and {Ohtaka}, O. and {Imai}, Y. and {Matsumoto}, T. and {Tsuchiyama}, A.},
        title = "{Infrared Spectra of Silica Polymorphs and the Conditions of Their Formation}",
      journal = {\apj},
         year = 2013,
        month = nov,
       volume = {778},
       number = {1},
          eid = {60},
        pages = {60},
          doi = {10.1088/0004-637X/778/1/60},
       adsurl = {https://ui.adsabs.harvard.edu/abs/2013ApJ...778...60K}
}

@software{WebPlotDigitizer,
    author = {Ankit Rohatgi},
    title = {WebPlotDigitizer (\url{https://automeris.io})},
    url = {https://automeris.io},
    version = {5.2},
    year = 2024,
}

@BOOK{Dodd1981,
       author = {{Dodd}, Robert T.},
        title = "{Meteorites, a petrologic-chemical synthesis}",
         year = 1981,
       adsurl = {https://ui.adsabs.harvard.edu/abs/1981mpcs.book.....D}
}

@ARTICLE{Waelkens1996,
       author = {{Waelkens}, C. and {Waters}, L.~B.~F.~M. and {de Graauw}, M.~S. and {Huygen}, E. and {Malfait}, K. and {Plets}, H. and {Vandenbussche}, B. and {Beintema}, D.~A. and {Boxhoorn}, D.~R. and {Habing}, H.~J. and {Heras}, A.~M. and {Kester}, D.~J.~M. and {Lahuis}, F. and {Morris}, P.~W. and {Roelfsema}, P.~R. and {Salama}, A. and {Siebenmorgen}, R. and {Trams}, N.~R. and {van der Bliek}, N.~R. and {Valentijn}, E.~A. and {Wesselius}, P.~R.},
        title = "{SWS observations of young main-sequence stars with dusty circumstellar disks.}",
      journal = {\aap},
         year = 1996,
        month = nov,
       volume = {315},
        pages = {L245-L248},
       adsurl = {https://ui.adsabs.harvard.edu/abs/1996A&A...315L.245W}
}

@ARTICLE{Malfait1998,
       author = {{Malfait}, K. and {Waelkens}, C. and {Waters}, L.~B.~F.~M. and {Vandenbussche}, B. and {Huygen}, E. and {de Graauw}, M.~S.},
        title = "{The spectrum of the young star HD 100546 observed with the Infrared Space Observatory}",
      journal = {\aap},
         year = 1998,
        month = apr,
       volume = {332},
        pages = {L25-L28},
       adsurl = {https://ui.adsabs.harvard.edu/abs/1998A&A...332L..25M}
}

@ARTICLE{Oliveira2011,
       author = {{Oliveira}, Isa and {Olofsson}, Johan and {Pontoppidan}, Klaus M. and {van Dishoeck}, Ewine F. and {Augereau}, Jean-Charles and {Mer{\'\i}n}, Bruno},
        title = "{On the Evolution of Dust Mineralogy, from Protoplanetary Disks to Planetary Systems}",
      journal = {\apj},
         year = 2011,
        month = jun,
       volume = {734},
       number = {1},
          eid = {51},
        pages = {51},
          doi = {10.1088/0004-637X/734/1/51},
archivePrefix = {arXiv},
       eprint = {1104.3574},
 primaryClass = {astro-ph.EP},
       adsurl = {https://ui.adsabs.harvard.edu/abs/2011ApJ...734...51O}
}

@software{Woitke2021_GGchem,
       author = {{Woitke}, Peter and {Helling}, Christiane},
        title = "{GGchem: Fast thermo-chemical equilibrium code}",
 howpublished = {Astrophysics Source Code Library, record ascl:2104.018},
         year = 2021,
        month = apr,
          eid = {ascl:2104.018},
archivePrefix = {ascl},
       eprint = {2104.018},
       adsurl = {https://ui.adsabs.harvard.edu/abs/2021ascl.soft04018W}
}

@ARTICLE{Woitke2018_mol,
       author = {{Woitke}, P. and {Min}, M. and {Thi}, W.-F. and {Roberts}, C. and {Carmona}, A. and {Kamp}, I. and {M{\'e}nard}, F. and {Pinte}, C.},
        title = "{Modelling mid-infrared molecular emission lines from T Tauri stars}",
      journal = {\aap},
         year = 2018,
        month = oct,
       volume = {618},
          eid = {A57},
        pages = {A57},
          doi = {10.1051/0004-6361/201731460},
archivePrefix = {arXiv},
       eprint = {1807.05784},
 primaryClass = {astro-ph.SR},
       adsurl = {https://ui.adsabs.harvard.edu/abs/2018A&A...618A..57W}
}

@ARTICLE{Tabone2023,
       author = {{Tabone}, B. and {Bettoni}, G. and {van Dishoeck}, E.~F. and {Arabhavi}, A.~M. and {Grant}, S. and {Gasman}, D. and {Henning}, Th. and {Kamp}, I. and {G{\"u}del}, M. and {Lagage}, P.~O. and {Ray}, T. and {Vandenbussche}, B. and {Abergel}, A. and {Absil}, O. and {Argyriou}, I. and {Barrado}, D. and {Boccaletti}, A. and {Bouwman}, J. and {Caratti o Garatti}, A. and {Geers}, V. and {Glauser}, A.~M. and {Justannont}, K. and {Lahuis}, F. and {Mueller}, M. and {Nehm{\'e}}, C. and {Olofsson}, G. and {Pantin}, E. and {Scheithauer}, S. and {Waelkens}, C. and {Waters}, L.~B.~F.~M. and {Black}, J.~H. and {Christiaens}, V. and {Guadarrama}, R. and {Morales-Calder{\'o}n}, M. and {Jang}, H. and {Kanwar}, J. and {Pawellek}, N. and {Perotti}, G. and {Perrin}, A. and {Rodgers-Lee}, D. and {Samland}, M. and {Schreiber}, J. and {Schwarz}, K. and {Colina}, L. and {{\"O}stlin}, G. and {Wright}, G.},
        title = "{A rich hydrocarbon chemistry and high C to O ratio in the inner disk around a very low-mass star}",
      journal = {Nature Astronomy},
         year = 2023,
        month = jul,
       volume = {7},
        pages = {805-814},
          doi = {10.1038/s41550-023-01965-3},
archivePrefix = {arXiv},
       eprint = {2304.05954},
 primaryClass = {astro-ph.EP},
       adsurl = {https://ui.adsabs.harvard.edu/abs/2023NatAs...7..805T}
}

@ARTICLE{Houge2025_decomp,
       author = {{Houge}, Adrien and {Johansen}, Anders and {Bergin}, Edwin and {Ciesla}, Fred J. and {Bitsch}, Bertram and {Lambrechts}, Michiel and {Henning}, Thomas and {Perotti}, Giulia},
        title = "{Burned to ashes: How the thermal decomposition of refractory organics in the inner protoplanetary disc impacts the gas-phase C/O ratio}",
      journal = {\aap},
         year = 2025,
        month = jul,
       volume = {699},
          eid = {A227},
        pages = {A227},
          doi = {10.1051/0004-6361/202555164},
archivePrefix = {arXiv},
       eprint = {2505.20427},
 primaryClass = {astro-ph.EP},
       adsurl = {https://ui.adsabs.harvard.edu/abs/2025A&A...699A.227H}
}

@ARTICLE{Borderies2025,
       author = {{Borderies}, Antonin and {Commer{\c{c}}on}, Beno{\^\i}t and {Bourdon}, Bernard},
        title = "{Dust evolution by chemisputtering during protostellar formation}",
      journal = {\aap},
         year = 2025,
        month = feb,
       volume = {694},
          eid = {A89},
        pages = {A89},
          doi = {10.1051/0004-6361/202452228},
archivePrefix = {arXiv},
       eprint = {2501.17937},
 primaryClass = {astro-ph.GA},
       adsurl = {https://ui.adsabs.harvard.edu/abs/2025A&A...694A..89B}
}

@INPROCEEDINGS{vanBoekel2006SPIE,
       author = {{van Boekel}, Roy and {{\'A}brah{\'a}m}, P{\'e}ter and {Correia}, Serge and {de Koter}, Alex and {Dominik}, Carsten and {Dutrey}, Anne and {Henning}, Thomas and {K{\'o}sp{\'a}l}, {\'A}gnes and {Lachaume}, R{\'e}gis and {Leinert}, Christoph and {Linz}, Hendrik and {Min}, Michiel and {Mosoni}, L{\'a}szl{\'o} and {Preibisch}, Thomas and {Quanz}, Sascha and {Ratzka}, Thorsten and {Schegerer}, Alexander and {Waters}, Rens and {Wolf}, Sebastian and {Zinnecker}, Hans},
        title = "{Disks around young stars with VLTI/MIDI}",
    booktitle = {Advances in Stellar Interferometry},
         year = 2006,
       editor = {{Monnier}, John D. and {Sch{\"o}ller}, Markus and {Danchi}, William C.},
       series = {Society of Photo-Optical Instrumentation Engineers (SPIE) Conference Series},
       volume = {6268},
        month = jun,
          eid = {62680D},
        pages = {62680D},
          doi = {10.1117/12.673777},
archivePrefix = {arXiv},
       eprint = {astro-ph/0607387},
 primaryClass = {astro-ph},
       adsurl = {https://ui.adsabs.harvard.edu/abs/2006SPIE.6268E..0DV}
}

@ARTICLE{Savitzky1964,
       author = {{Savitzky}, A. and {Golay}, M.~J.~E.},
        title = "{Smoothing and differentiation of data by simplified least squares procedures}",
      journal = {Analytical Chemistry},
         year = 1964,
        month = jan,
       volume = {36},
        pages = {1627-1639},
          doi = {10.1021/ac60214a047},
       adsurl = {https://ui.adsabs.harvard.edu/abs/1964AnaCh..36.1627S}
}

@ARTICLE{Mallaney2026,
       author = {{Mallaney}, Patrick and {Banzatti}, Andrea and {Salyk}, Colette and {Pascucci}, Ilaria and {Pinilla}, Paola and {Najita}, Joan and {Pontoppidan}, Klaus M. and {Krijt}, Sebastiaan and {Blake}, Geoffrey A. and {Tabone}, Beno{\^\i}t and {Kaeufer}, Till and {Zhang}, Ke and {Long}, Feng and {Huang}, Jane and {Rosotti}, Giovanni and {{\"O}berg}, Karin I. and {Colmenares}, Mar{\'\i}a Jos{\'e} and {Lay}, Andrew and {Cieza}, Lucas A. and {Cleeves}, L. Ilsedore and {Williams}, Joe and {Xie}, Chengyan and {Vioque}, Miguel and {Narang}, Mayank and {Ballering}, Nicholas P. and {Kim}, Minjae and {Jdiscs Collaboration}},
        title = "{Protoplanetary Disk Cavities with JWST-MIRI: A Dichotomy in Molecular Emission}",
      journal = {\apj},
         year = 2026,
        month = feb,
       volume = {998},
       number = {2},
          eid = {255},
        pages = {255},
          doi = {10.3847/1538-4357/ae32f1},
archivePrefix = {arXiv},
       eprint = {2601.02344},
 primaryClass = {astro-ph.EP},
       adsurl = {https://ui.adsabs.harvard.edu/abs/2026ApJ...998..255M}
}

@ARTICLE{Pinilla2012_trap,
       author = {{Pinilla}, P. and {Birnstiel}, T. and {Ricci}, L. and {Dullemond}, C.~P. and {Uribe}, A.~L. and {Testi}, L. and {Natta}, A.},
        title = "{Trapping dust particles in the outer regions of protoplanetary disks}",
      journal = {\aap},
         year = 2012,
        month = feb,
       volume = {538},
          eid = {A114},
        pages = {A114},
          doi = {10.1051/0004-6361/201118204},
archivePrefix = {arXiv},
       eprint = {1112.2349},
 primaryClass = {astro-ph.EP},
       adsurl = {https://ui.adsabs.harvard.edu/abs/2012A&A...538A.114P}
}

@ARTICLE{Krijt2025,
       author = {{Krijt}, Sebastiaan and {Banzatti}, Andrea and {Zhang}, Ke and {Pinilla}, Paola and {Kaeufer}, Till and {Bergin}, Edwin A. and {Salyk}, Colette and {Pontoppidan}, Klaus and {Blake}, Geoffrey A. and {Long}, Feng and {Huang}, Jane and {Colmenares}, Mar{\'\i}a Jos{\'e} and {Williams}, Joe and {Houge}, Adrien and {Narang}, Mayank and {Vioque}, Miguel and {Lambrechts}, Michiel and {Cleeves}, L. Ilsedore and {{\"O}berg}, Karin and {The Jdiscs Collaboration}},
        title = "{Cosmic Cascades: How Disk Substructure Regulates the Flow of Water to Inner Planetary Systems}",
      journal = {\apjl},
         year = 2025,
        month = sep,
       volume = {990},
       number = {2},
          eid = {L72},
        pages = {L72},
          doi = {10.3847/2041-8213/adfbe3},
archivePrefix = {arXiv},
       eprint = {2508.10402},
 primaryClass = {astro-ph.EP},
       adsurl = {https://ui.adsabs.harvard.edu/abs/2025ApJ...990L..72K}
}

@ARTICLE{Glauser2009,
       author = {{Glauser}, A.~M. and {G{\"u}del}, M. and {Watson}, D.~M. and {Henning}, T. and {Schegerer}, A.~A. and {Wolf}, S. and {Audard}, M. and {Baldovin-Saavedra}, C.},
        title = "{Dust amorphization in protoplanetary disks}",
      journal = {\aap},
         year = 2009,
        month = dec,
       volume = {508},
       number = {1},
        pages = {247-257},
          doi = {10.1051/0004-6361/200912087},
archivePrefix = {arXiv},
       eprint = {0909.3183},
 primaryClass = {astro-ph.SR},
       adsurl = {https://ui.adsabs.harvard.edu/abs/2009A&A...508..247G}
}

@ARTICLE{Arabhavi2025,
       author = {{Arabhavi}, A.~M. and {Kamp}, I. and {Henning}, Th. and {van Dishoeck}, E.~F. and {Jang}, H. and {Waters}, L.~B.~F.~M. and {Christiaens}, V. and {Gasman}, D. and {Pascucci}, I. and {Perotti}, G. and {Grant}, S.~L. and {G{\"u}del}, M. and {Lagage}, P.-O. and {Barrado}, D. and {Caratti o Garatti}, A. and {Lahuis}, F. and {Kaeufer}, T. and {Kanwar}, J. and {Morales-Calder{\'o}n}, M. and {Schwarz}, K. and {Sellek}, A.~D. and {Tabone}, B. and {Temmink}, M. and {Vlasblom}, M. and {Patapis}, P.},
        title = "{MINDS: The very low-mass star and brown dwarf sample: Detections and trends in the inner disk gas}",
      journal = {\aap},
         year = 2025,
        month = jul,
       volume = {699},
          eid = {A194},
        pages = {A194},
          doi = {10.1051/0004-6361/202554109},
archivePrefix = {arXiv},
       eprint = {2506.02748},
 primaryClass = {astro-ph.EP},
       adsurl = {https://ui.adsabs.harvard.edu/abs/2025A&A...699A.194A}
}

@ARTICLE{Arabhavi2024,
       author = {{Arabhavi}, A.~M. and {Kamp}, I. and {Henning}, Th. and {van Dishoeck}, E.~F. and {Christiaens}, V. and {Gasman}, D. and {Perrin}, A. and {G{\"u}del}, M. and {Tabone}, B. and {Kanwar}, J. and {Waters}, L.~B.~F.~M. and {Pascucci}, I. and {Samland}, M. and {Perotti}, G. and {Bettoni}, G. and {Grant}, S.~L. and {Lagage}, P.~O. and {Ray}, T.~P. and {Vandenbussche}, B. and {Absil}, O. and {Argyriou}, I. and {Barrado}, D. and {Boccaletti}, A. and {Bouwman}, J. and {Caratti o Garatti}, A. and {Glauser}, A.~M. and {Lahuis}, F. and {Mueller}, M. and {Olofsson}, G. and {Pantin}, E. and {Scheithauer}, S. and {Morales-Calder{\'o}n}, M. and {Franceschi}, R. and {Jang}, H. and {Pawellek}, N. and {Rodgers-Lee}, D. and {Schreiber}, J. and {Schwarz}, K. and {Temmink}, M. and {Vlasblom}, M. and {Wright}, G. and {Colina}, L. and {{\"O}stlin}, G.},
        title = "{Abundant hydrocarbons in the disk around a very-low-mass star}",
      journal = {Science},
         year = 2024,
        month = jun,
       volume = {384},
       number = {6700},
        pages = {1086-1090},
          doi = {10.1126/science.adi8147},
archivePrefix = {arXiv},
       eprint = {2406.14293},
 primaryClass = {astro-ph.EP},
       adsurl = {https://ui.adsabs.harvard.edu/abs/2024Sci...384.1086A}
}

@ARTICLE{Vlasblom2024,
       author = {{Vlasblom}, Marissa and {van Dishoeck}, Ewine F. and {Tabone}, Beno{\^\i}t and {Bruderer}, Simon},
        title = "{Mid-infrared spectra of T Tauri disks: Modeling the effects of a small inner cavity on CO$_{2}$ and H$_{2}$O emission}",
      journal = {\aap},
         year = 2024,
        month = feb,
       volume = {682},
          eid = {A91},
        pages = {A91},
          doi = {10.1051/0004-6361/202348224},
archivePrefix = {arXiv},
       eprint = {2311.12445},
 primaryClass = {astro-ph.EP},
       adsurl = {https://ui.adsabs.harvard.edu/abs/2024A&A...682A..91V}
}

@ARTICLE{Raul2025,
       author = {{Raul}, Eshan and {Alarc{\'o}n}, Felipe and {Bergin}, Edwin A.},
        title = "{Tracking the Chemical Evolution of Hydrocarbons Through Carbon Grain Supply in Protoplanetary Disks}",
      journal = {\apj},
         year = 2025,
        month = apr,
       volume = {982},
       number = {2},
          eid = {155},
        pages = {155},
          doi = {10.3847/1538-4357/adaeae},
archivePrefix = {arXiv},
       eprint = {2502.01765},
 primaryClass = {astro-ph.EP},
       adsurl = {https://ui.adsabs.harvard.edu/abs/2025ApJ...982..155R}
}

@ARTICLE{Volz2026,
       author = {{Volz}, M. and {Espaillat}, C.~C. and {Pittman}, C.~V. and {Grant}, S.~L. and {Thanathibodee}, T. and {McClure}, M. and {Tabone}, B. and {Calvet}, N. and {Walter}, F.~M.},
        title = "{JWST Reveals Carbon-rich Chemistry in a Transitional Disk}",
      journal = {\aj},
         year = 2026,
        month = jan,
       volume = {171},
       number = {1},
          eid = {39},
        pages = {39},
          doi = {10.3847/1538-3881/ae1f16},
archivePrefix = {arXiv},
       eprint = {2511.08816},
 primaryClass = {astro-ph.SR},
       adsurl = {https://ui.adsabs.harvard.edu/abs/2026AJ....171...39V}
}

@ARTICLE{Kanwar2024,
       author = {{Kanwar}, Jayatee and {Kamp}, Inga and {Jang}, Hyerin and {Waters}, Laurens B.~F.~M. and {van Dishoeck}, Ewine F. and {Christiaens}, Valentin and {Arabhavi}, Aditya M. and {Henning}, Thomas and {G{\"u}del}, Manuel and {Woitke}, Peter and {Absil}, Olivier and {Barrado}, David and {Caratti o Garatti}, Alessio and {Glauser}, Adrian M. and {Lahuis}, Fred and {Scheithauer}, Silvia and {Vandenbussche}, Bart and {Gasman}, Danny and {Grant}, Sierra L. and {Kurtovic}, Nicolas T. and {Perotti}, Giulia and {Tabone}, Beno{\^\i}t and {Temmink}, Milou},
        title = "{MINDS. Hydrocarbons detected by JWST/MIRI in the inner disk of Sz28 consistent with a high C/O gas-phase chemistry}",
      journal = {\aap},
         year = 2024,
        month = sep,
       volume = {689},
          eid = {A231},
        pages = {A231},
          doi = {10.1051/0004-6361/202450078},
archivePrefix = {arXiv},
       eprint = {2407.14362},
 primaryClass = {astro-ph.EP},
       adsurl = {https://ui.adsabs.harvard.edu/abs/2024A&A...689A.231K}
}

@ARTICLE{Kanwar2025,
       author = {{Kanwar}, Jayatee and {Woitke}, Peter and {Kamp}, Inga and {Rimmer}, Paul and {Helling}, Christiane},
        title = "{Can thermodynamic equilibrium be established in planet-forming disks?}",
      journal = {\aap},
         year = 2025,
        month = jun,
       volume = {698},
          eid = {A294},
        pages = {A294},
          doi = {10.1051/0004-6361/202452249},
archivePrefix = {arXiv},
       eprint = {2505.13705},
 primaryClass = {astro-ph.EP},
       adsurl = {https://ui.adsabs.harvard.edu/abs/2025A&A...698A.294K}
}

@ARTICLE{Kanwar2026,
       author = {{Kanwar}, Jayatee and {Kamp}, Inga and {Woitke}, Peter and {van Dishoeck}, Ewine F. and {Henning}, Thomas and {Liu}, Yao and {Kaeufer}, Till and {Tabone}, Beno{\^\i}t and {G{\"u}del}, Manuel and {Barrado}, David and {Arabhavi}, Aditya M. and {Franceschi}, Riccardo and {Vlasblom}, Marissa},
        title = "{MINDS: Strong oxygen depletion in the inner regions of a very low-mass star disk?}",
      journal = {\aap},
         year = 2026,
        month = jan,
       volume = {705},
          eid = {A222},
        pages = {A222},
          doi = {10.1051/0004-6361/202451844},
archivePrefix = {arXiv},
       eprint = {2508.11761},
 primaryClass = {astro-ph.EP},
       adsurl = {https://ui.adsabs.harvard.edu/abs/2026A&A...705A.222K}
}

@ARTICLE{Arabhavi2025_hidden,
       author = {{Arabhavi}, Aditya M. and {Kamp}, Inga and {van Dishoeck}, Ewine F. and {Henning}, Thomas and {Jang}, Hyerin and {Christiaens}, Valentin and {Gasman}, Danny and {Pascucci}, Ilaria and {Perotti}, Giulia and {Grant}, Sierra L. and {Barrado}, David and {G{\"u}del}, Manuel and {Lagage}, Pierre-Olivier and {Caratti o Garatti}, Alessio and {Lahuis}, Fred and {Waters}, L.~B.~F.~M. and {Kaeufer}, Till and {Kanwar}, Jayatee and {Morales-Calder{\'o}n}, Maria and {Schwarz}, Kamber and {Sellek}, Andrew D. and {Tabone}, Beno{\^\i}t and {Temmink}, Milou and {Vlasblom}, Marissa},
        title = "{MINDS: The Very Low-mass Star and Brown Dwarf Sample Hidden Water in Carbon-dominated Protoplanetary Disks}",
      journal = {\apjl},
         year = 2025,
        month = may,
       volume = {984},
       number = {2},
          eid = {L62},
        pages = {L62},
          doi = {10.3847/2041-8213/adc692},
archivePrefix = {arXiv},
       eprint = {2504.11425},
 primaryClass = {astro-ph.EP},
       adsurl = {https://ui.adsabs.harvard.edu/abs/2025ApJ...984L..62A}
}

@ARTICLE{Sameshima2026,
       author = {{Sameshima}, N. and {Miyata}, T. and {Kamizuka}, T. and {Aikawa}, Y. and {Honda}, M. and {Cleeves}, L.~I. and {Ballering}, N.~P. and {Colmenares}, M.~J. and {Gonz{\'a}lez-Ruilova}, C. and {Guzman}, V.~V. and {Haworth}, T.~J. and {Law}, C.~J. and {Williams}, J.~P.},
        title = "{JWST─DECO: temporal variations in the mid-IR silicate features of two T Tauri discs based on Spitzer and JWST observations}",
      journal = {\mnras},
         year = 2026,
        month = may,
       volume = {548},
       number = {2},
          eid = {stag273},
        pages = {stag273},
          doi = {10.1093/mnras/stag273},
archivePrefix = {arXiv},
       eprint = {2602.22615},
 primaryClass = {astro-ph.SR},
       adsurl = {https://ui.adsabs.harvard.edu/abs/2026MNRAS.548ag273S}
}

@ARTICLE{Lodders2003,
       author = {{Lodders}, Katharina},
        title = "{Solar System Abundances and Condensation Temperatures of the Elements}",
      journal = {\apj},
         year = 2003,
        month = jul,
       volume = {591},
       number = {2},
        pages = {1220-1247},
          doi = {10.1086/375492},
       adsurl = {https://ui.adsabs.harvard.edu/abs/2003ApJ...591.1220L}
}

@ARTICLE{Lodders2002,
       author = {{Lodders}, Katharina and {Fegley}, Bruce},
        title = "{Atmospheric Chemistry in Giant Planets, Brown Dwarfs, and Low-Mass Dwarf Stars. I. Carbon, Nitrogen, and Oxygen}",
      journal = {\icarus},
         year = 2002,
        month = feb,
       volume = {155},
       number = {2},
        pages = {393-424},
          doi = {10.1006/icar.2001.6740},
       adsurl = {https://ui.adsabs.harvard.edu/abs/2002Icar..155..393L}
}

@INCOLLECTION{Ebel2006,
       author = {{Ebel}, D.~S.},
        title = "{Condensation of Rocky Material in Astrophysical Environments}",
    booktitle = {Meteorites and the Early Solar System II},
         year = 2006,
       editor = {{Lauretta}, Dante S. and {McSween}, Harry Y.},
        pages = {253},
       adsurl = {https://ui.adsabs.harvard.edu/abs/2006mess.book..253E}
}

@ARTICLE{Larimer1975,
       author = {{Larimer}, J.~W.},
        title = "{The effect of C / O ratio on the condensation of planetary material}",
      journal = {\gca},
         year = 1975,
        month = mar,
       volume = {39},
       number = {3},
        pages = {389-392},
          doi = {10.1016/0016-7037(75)90204-5},
       adsurl = {https://ui.adsabs.harvard.edu/abs/1975GeCoA..39..389L}
}

@ARTICLE{Sturm2013,
       author = {{Sturm}, B. and {Bouwman}, J. and {Henning}, Th. and {Evans}, N.~J. and {Waters}, L.~B.~F.~M. and {van Dishoeck}, E.~F. and {Green}, J.~D. and {Olofsson}, J. and {Meeus}, G. and {Maaskant}, K. and {Dominik}, C. and {Augereau}, J.~C. and {Mulders}, G.~D. and {Acke}, B. and {Merin}, B. and {Herczeg}, G.~J.},
        title = "{The 69 {\ensuremath{\mu}}m forsterite band in spectra of protoplanetary disks. Results from the Herschel DIGIT programme}",
      journal = {\aap},
         year = 2013,
        month = may,
       volume = {553},
          eid = {A5},
        pages = {A5},
          doi = {10.1051/0004-6361/201220243},
archivePrefix = {arXiv},
       eprint = {1303.3744},
 primaryClass = {astro-ph.SR},
       adsurl = {https://ui.adsabs.harvard.edu/abs/2013A&A...553A...5S}
}

@ARTICLE{Tabone2026,
       author = {{Tabone}, Beno{\^\i}t and {Temmink}, Milou and {Waters}, Laurens B.~F.~M. and {van Dishoeck}, Ewine F. and {Sellek}, Andrew and {Est{\`e}ve}, Pac{\^o}me and {Kurtovic}, Nicolas T. and {Kamp}, Inga and {Henning}, Thomas and {Gasman}, Danny and {Grant}, Sierra L. and {Varga}, J{\'o}zsef and {Guerras}, Alice and {Semenov}, Dmitry and {Arabhavi}, Aditya M. and {Garatti}, Alessio Caratti o and {Dutrey}, Anne and {Chapillon}, Edwige and {Guilloteau}, St{\'e}phane and {G{\"u}del}, Manuel and {Jang}, Hyerin and {Kaeufer}, Till and {Kanwar}, Jayatee and {Olofsson}, G{\"o}ran and {Perotti}, Giulia and {Pi{\'e}tu}, Vincent and {Ray}, Thomas P. and {Vlasblom}, Marissa},
        title = "{MINDS: Intertwined evolution of dust and gas in large planet-forming disks. A diversity driven by halted pebble drift?}",
      journal = {submitted to A\&A, arXiv e-prints},
         year = 2026,
        month = apr,
          eid = {arXiv:2604.21803},
        pages = {arXiv:2604.21803},
          doi = {10.48550/arXiv.2604.21803},
archivePrefix = {arXiv},
       eprint = {2604.21803},
 primaryClass = {astro-ph.EP},
       adsurl = {https://ui.adsabs.harvard.edu/abs/2026arXiv260421803T}
}

@ARTICLE{Arabhavi2026,
       author = {{Arabhavi}, Aditya M. and {Kamp}, Inga and {van Dishoeck}, Ewine F. and {Woitke}, Peter and {Rab}, Christian and {Thi}, Wing-Fai and {Kaeufer}, Till and {Kanwar}, Jayatee and {Tabone}, Beno{\^\i}t and {Esteve}, Pac{\^o}me and {Vlasblom}, Marissa},
        title = "{Molecular diagnostics for the mid-infrared emission of planet-forming disks: Carbon and oxygen elemental abundances}",
      journal = {\aap},
         year = 2026,
        month = mar,
       volume = {708},
          eid = {A82},
        pages = {A82},
          doi = {10.1051/0004-6361/202556335},
archivePrefix = {arXiv},
       eprint = {2602.16030},
 primaryClass = {astro-ph.EP},
       adsurl = {https://ui.adsabs.harvard.edu/abs/2026A&A...708A..82A}
}

\begin{appendix}

\onecolumn
\section{Opacity curves}
\label{sec:app_opac}

\begin{figure*}[h!]
	\centering
	
\includegraphics[width=0.48\hsize]{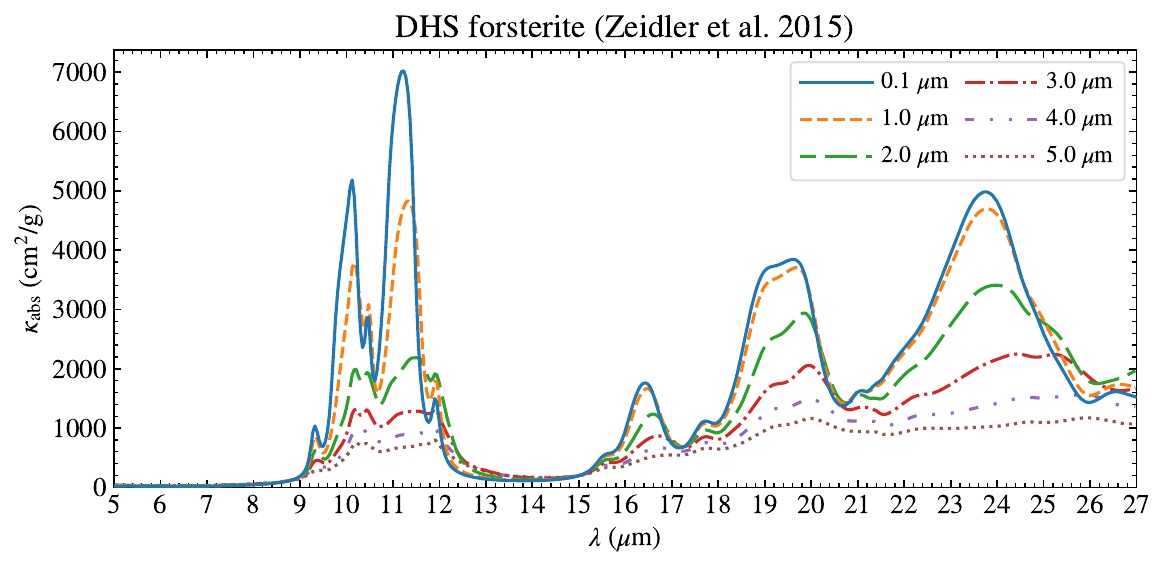 } 
\includegraphics[width=0.48\hsize]{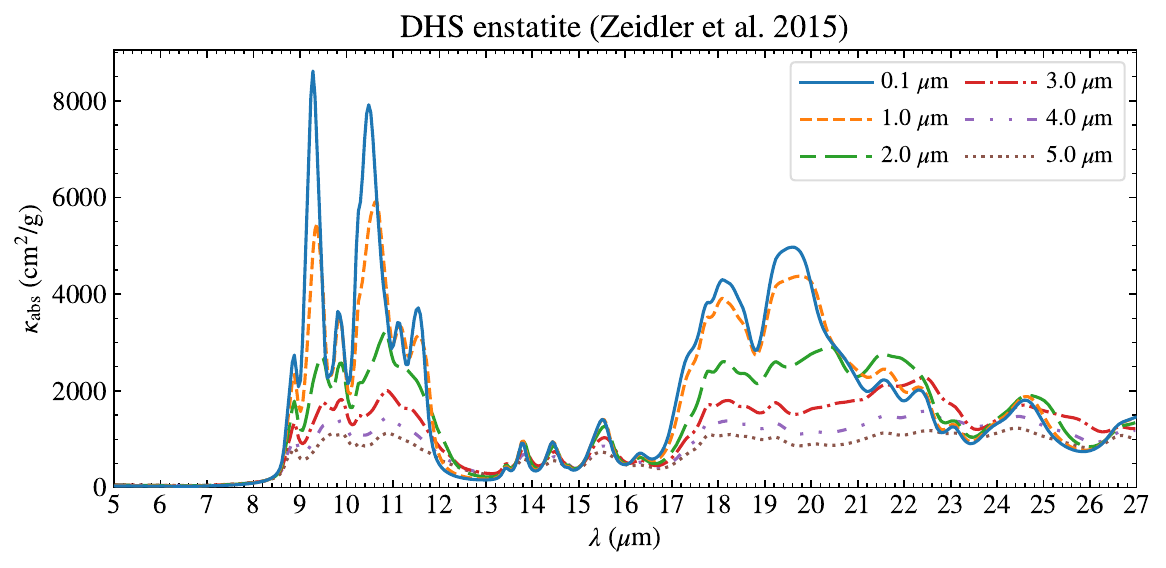} 

\includegraphics[width=0.48\hsize]{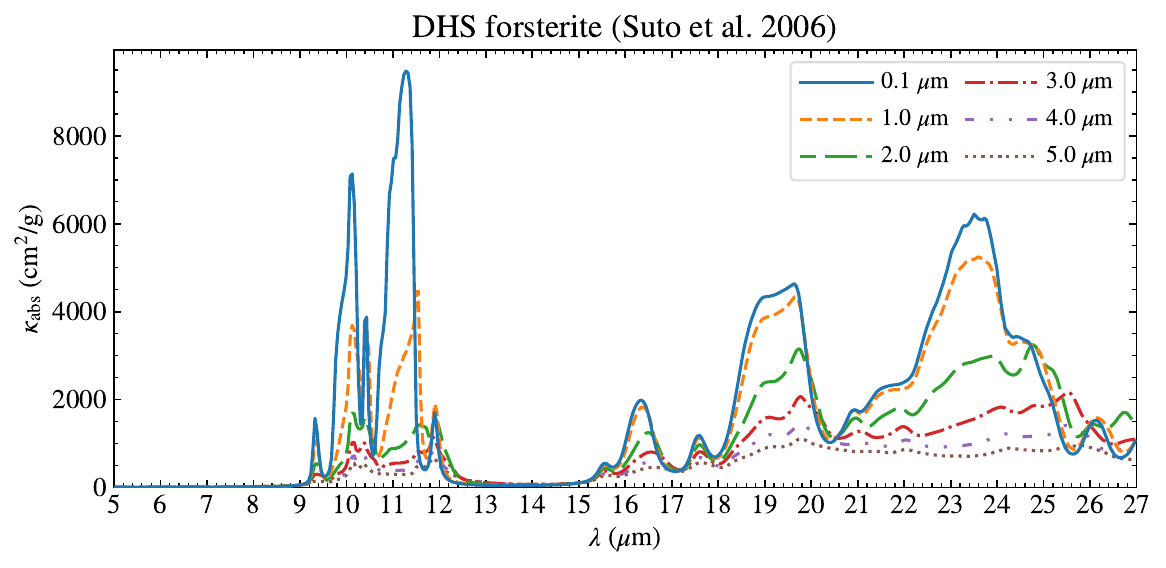 } 
\includegraphics[width=0.48\hsize]{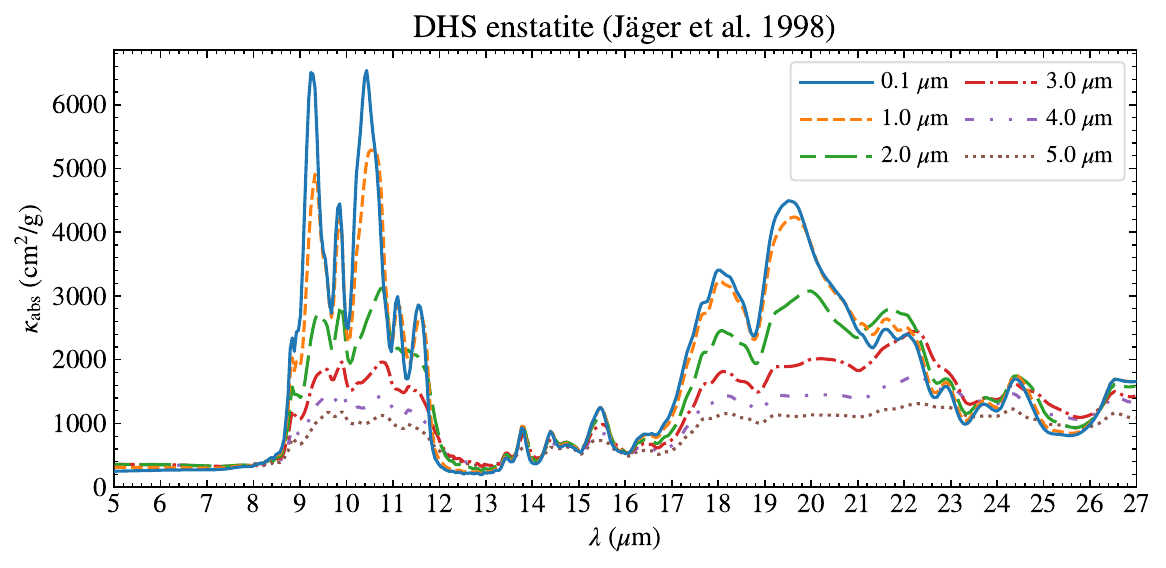 } 

\includegraphics[width=0.48\hsize]{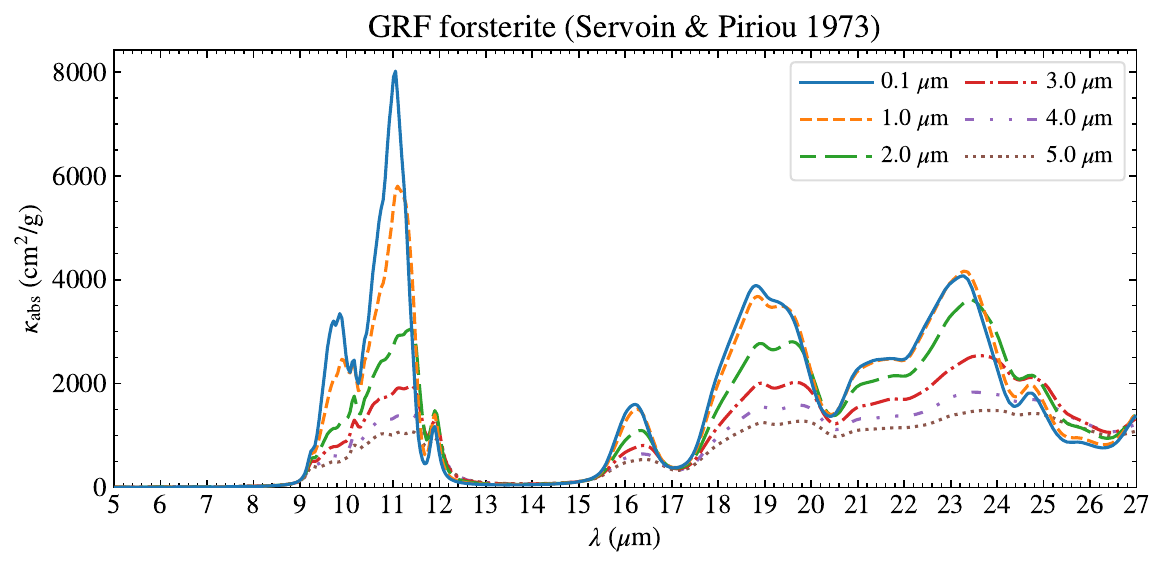} 
\includegraphics[width=0.48\hsize]{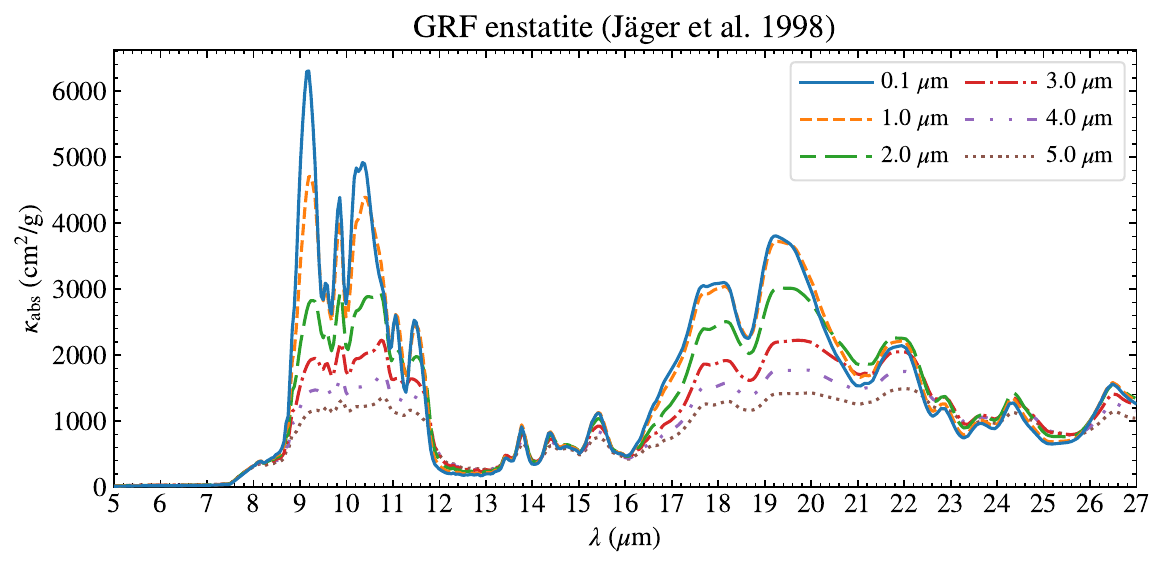} 

\includegraphics[width=0.48\hsize]{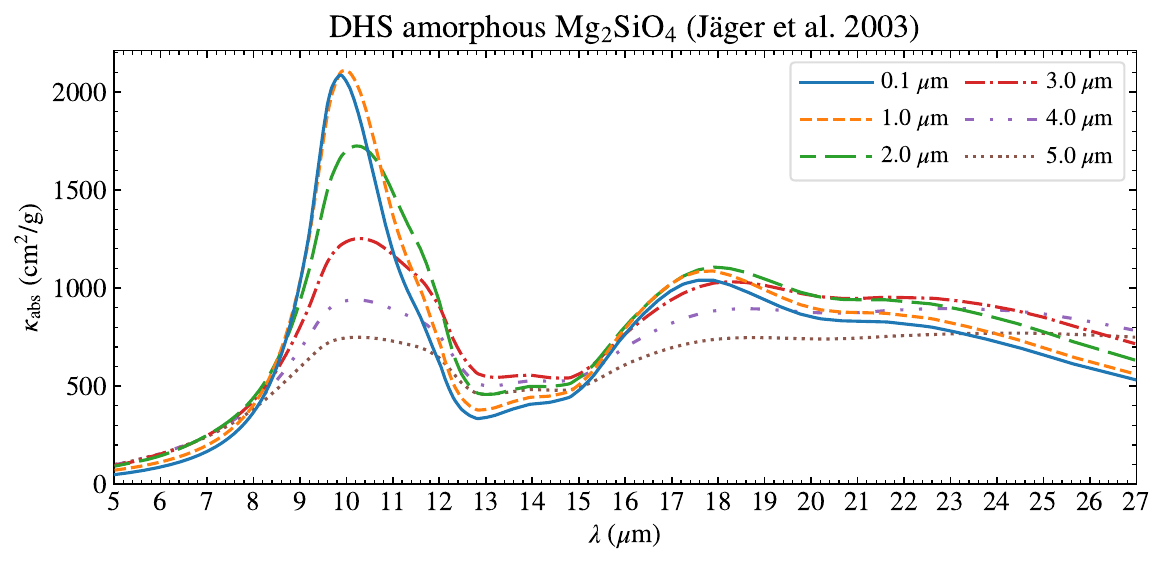} 
\includegraphics[width=0.48\hsize]{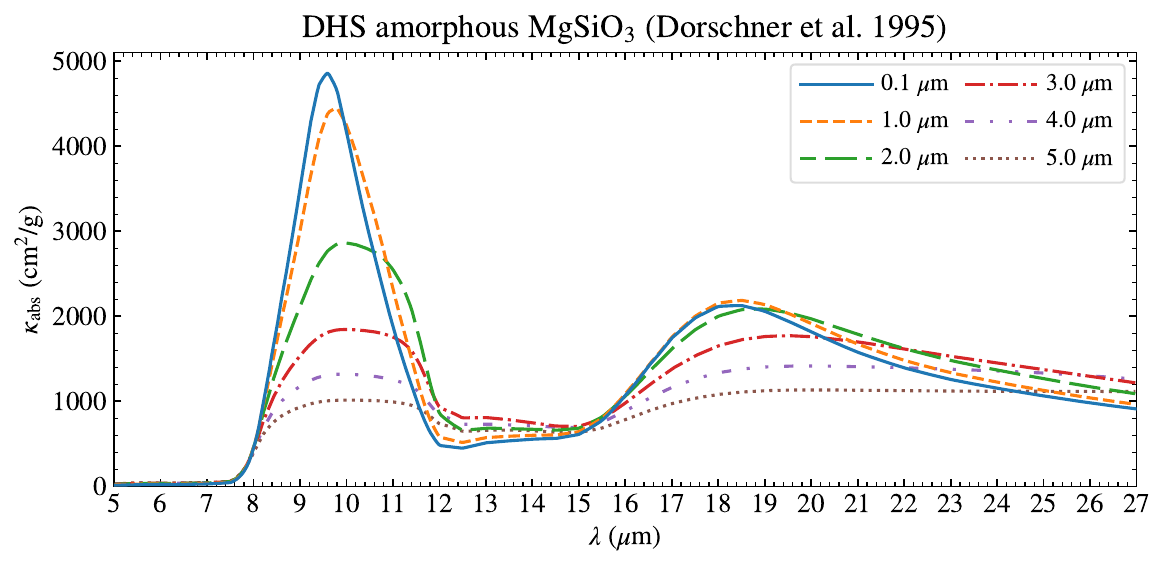 } 

\includegraphics[width=0.48\hsize]{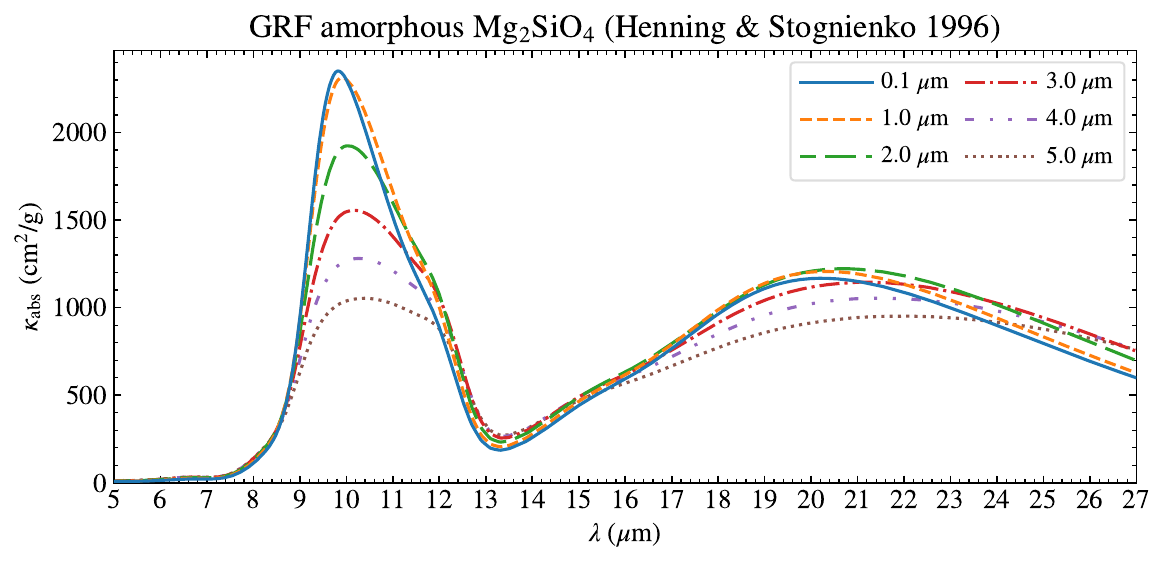} 
\includegraphics[width=0.48\hsize]{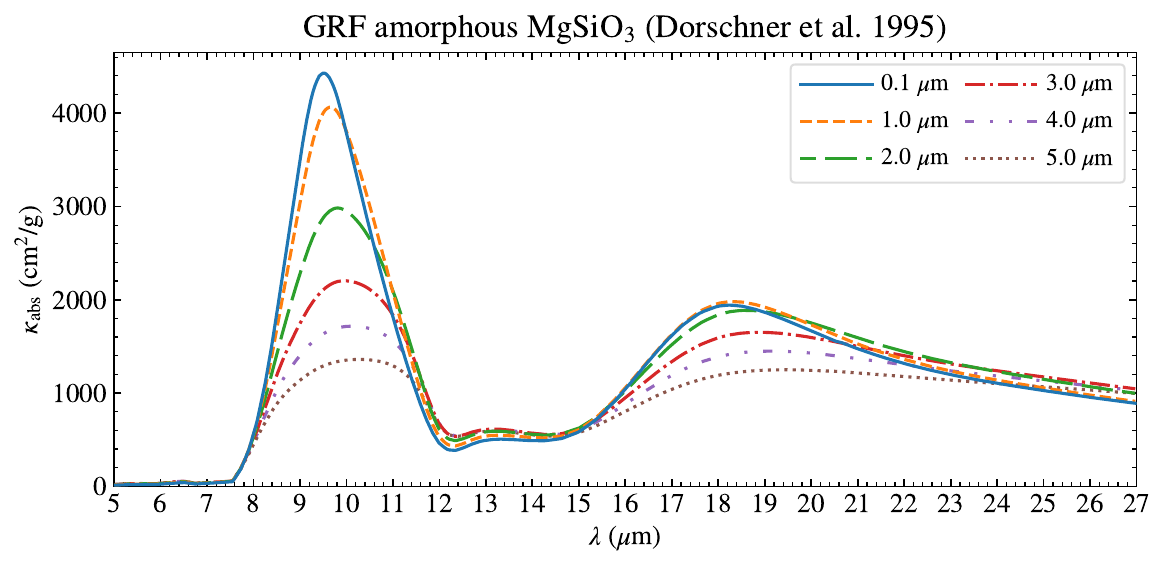} 

	\caption{Opacity curves used in our modeling.}
         	\label{fig:opac}	
\end{figure*}

\FloatBarrier
\twocolumn

\setcounter{figure}{0}
\onecolumn
\begin{figure}[h!]
\includegraphics[width=0.48\hsize]{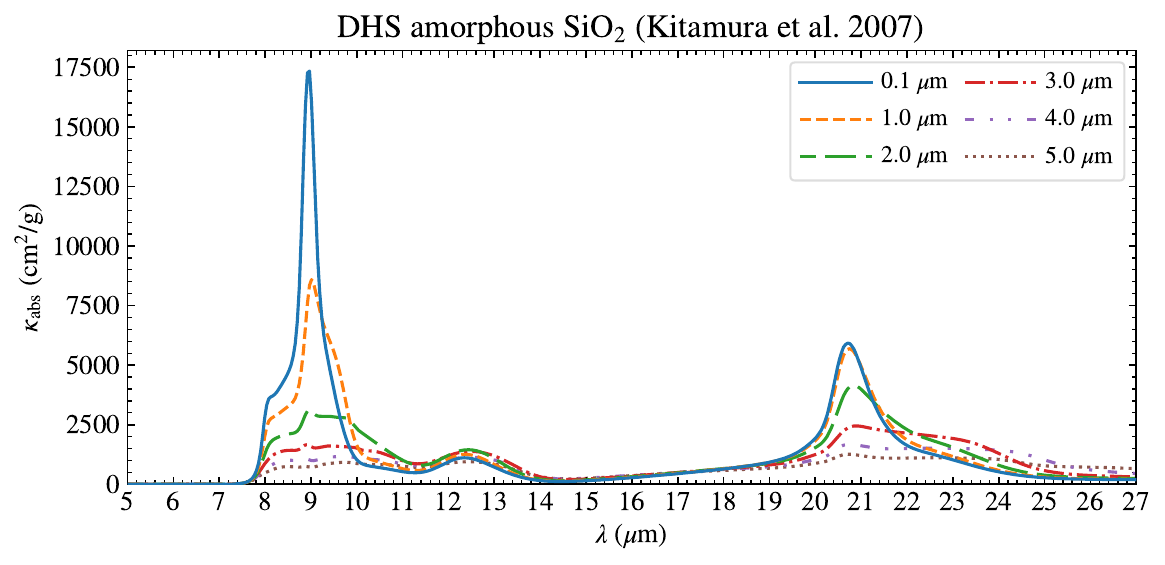 }
\includegraphics[width=0.48\hsize]{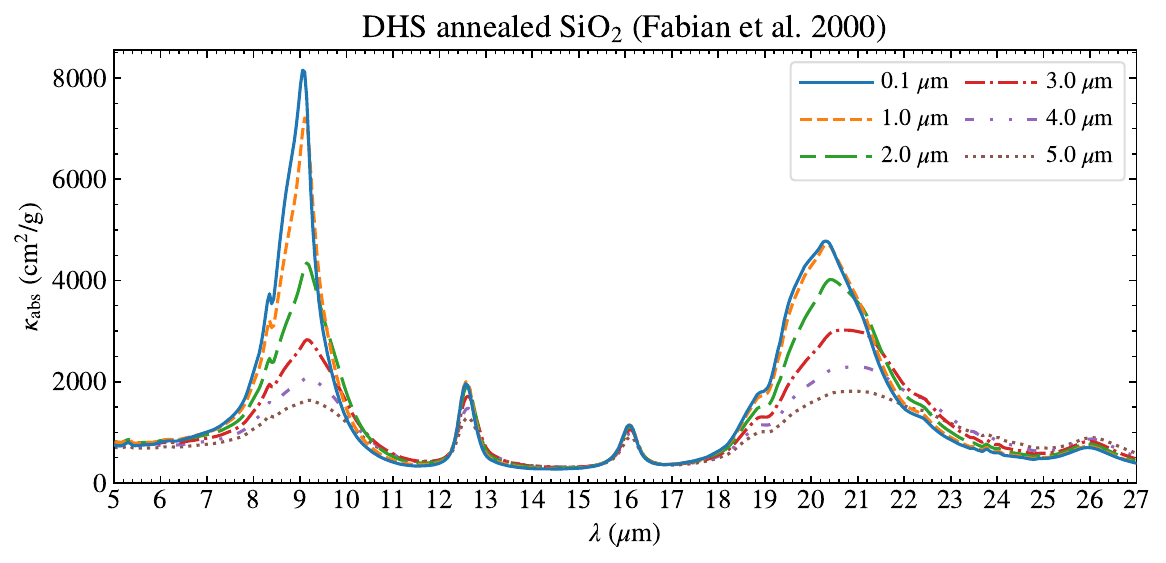 }
\includegraphics[width=0.48\hsize]{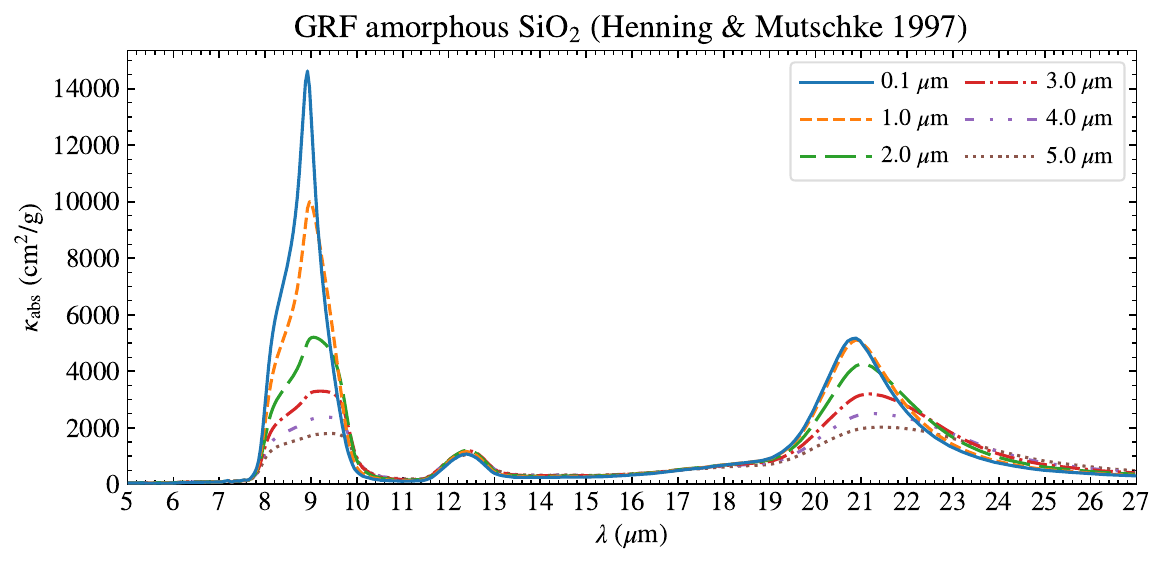}
	\caption{continued.}
         	\label{fig:opac2}	
\end{figure}

\begingroup
\let\clearpage\relax{
\onecolumn	
\section{Fit plots}
\label{sec:app_fitplots}

\begin{figure*}[h!]
	\vspace{-0.5cm}
	\centering
\includegraphics[width=0.48\hsize]{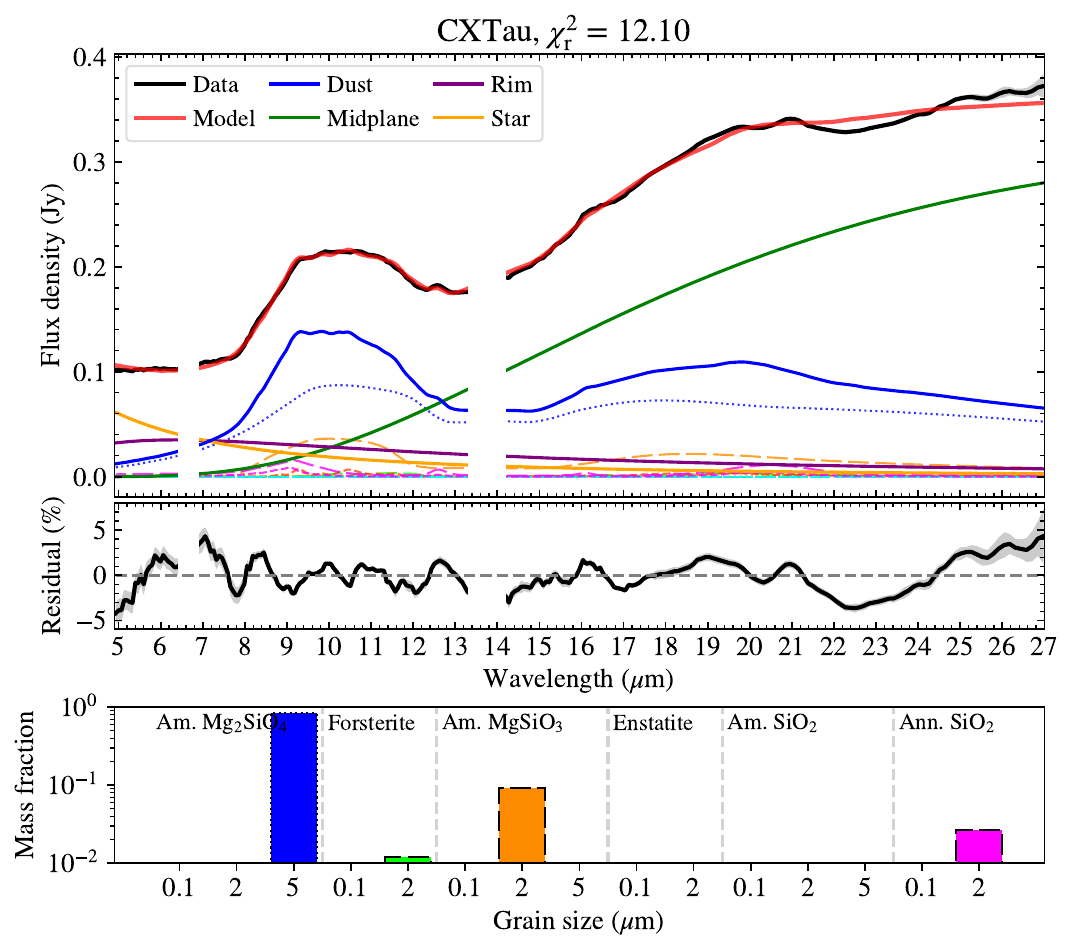}
\includegraphics[width=0.48\hsize]{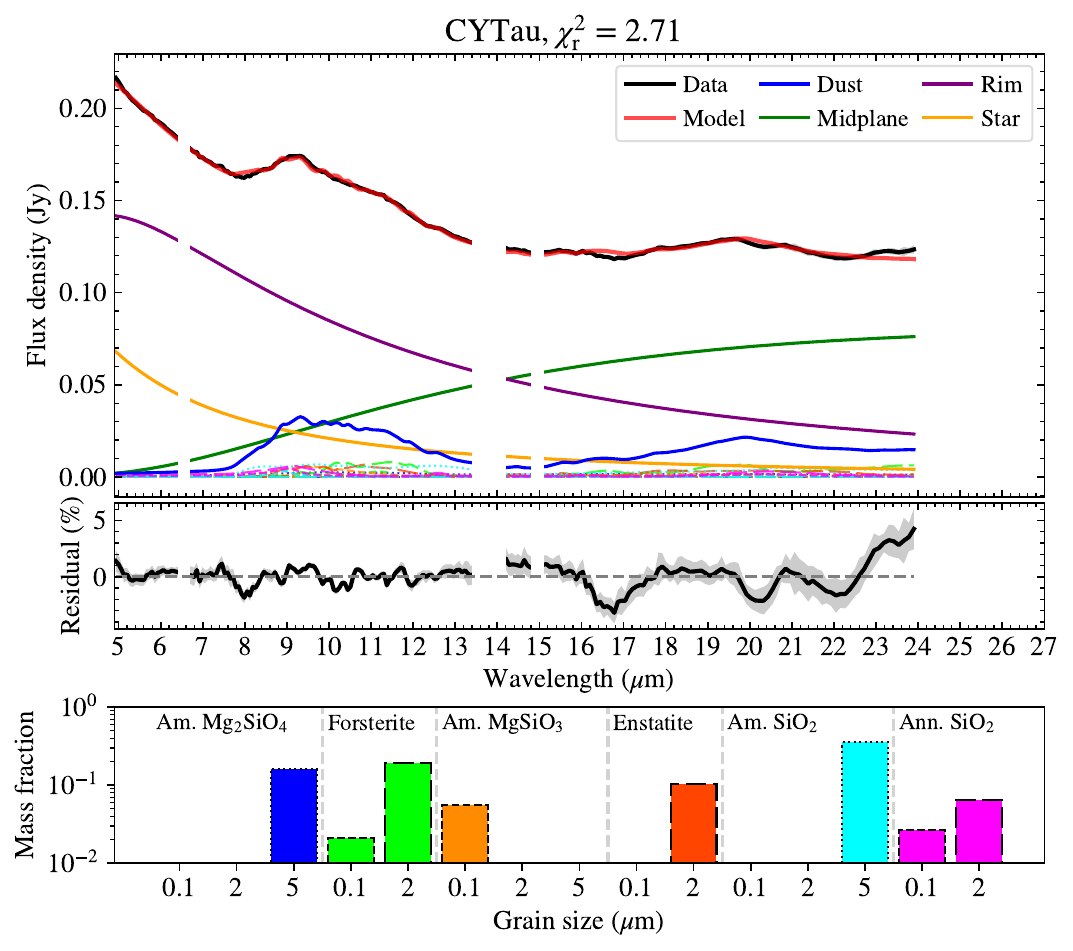}
\caption{Fits to all the spectra, with the DHS$\_$nat set of opacity curves, including annealed silica, using the $\left[0.1, 2, 5\ \mathrm{am.}\right]$ grain size set. 
}
         	\label{fig:fit_all}
\end{figure*}}
\twocolumn
\endgroup
\FloatBarrier

\setcounter{figure}{0}
\clearpage

\begin{figure*}[h!]
	\centering
\includegraphics[width=0.48\hsize]{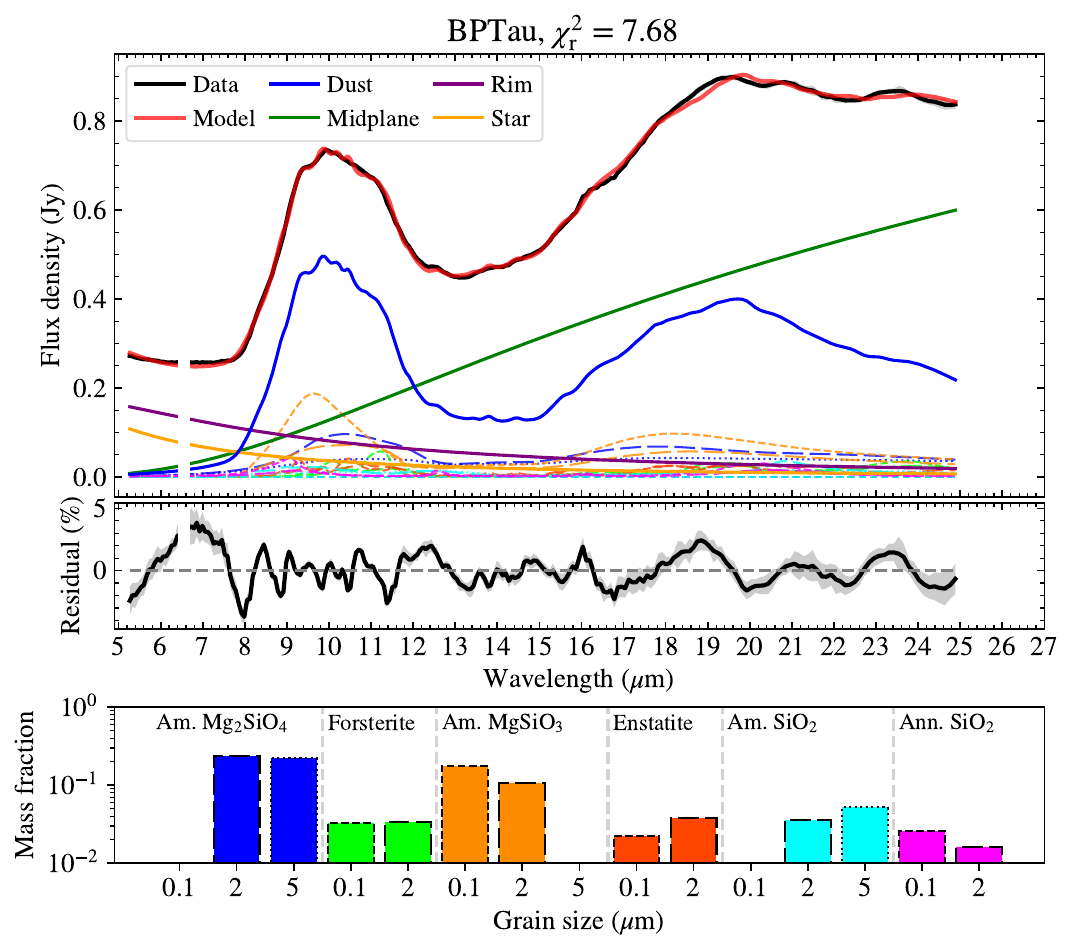}
\includegraphics[width=0.48\hsize]{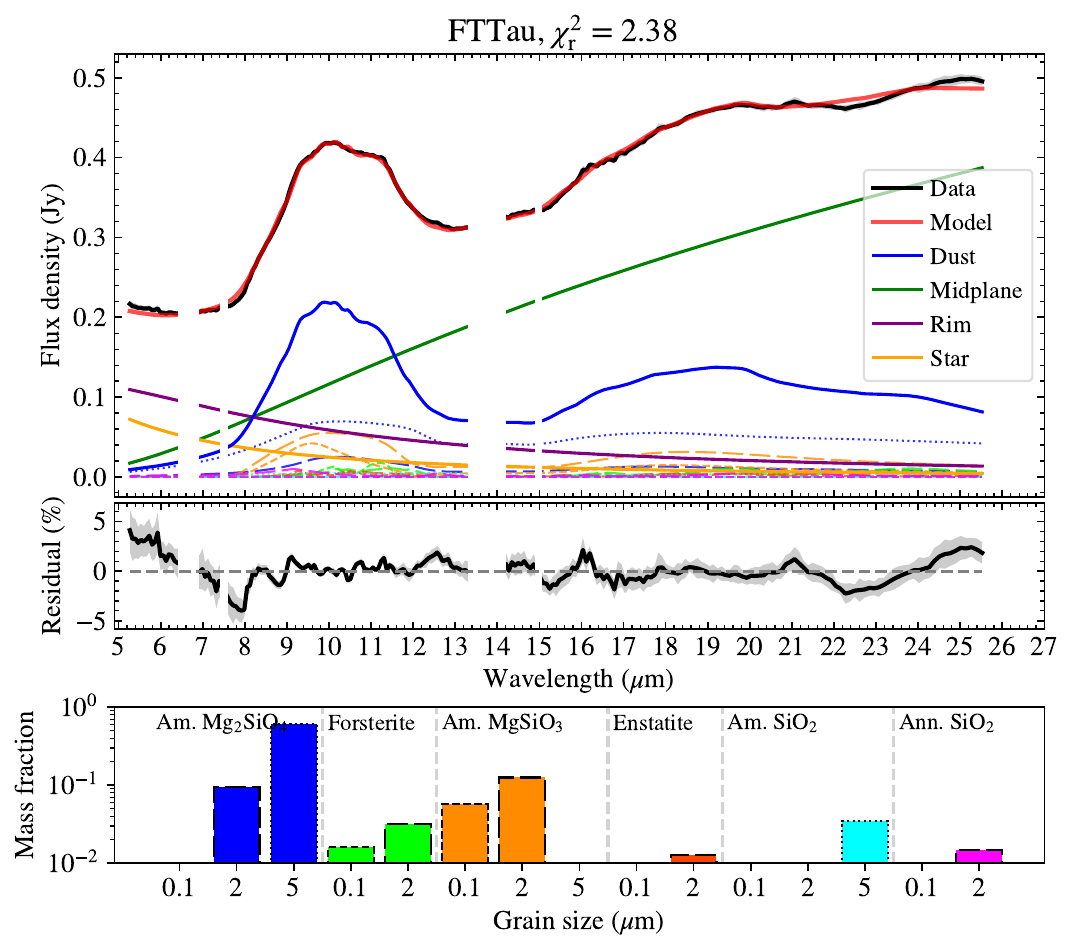}
\includegraphics[width=0.48\hsize]{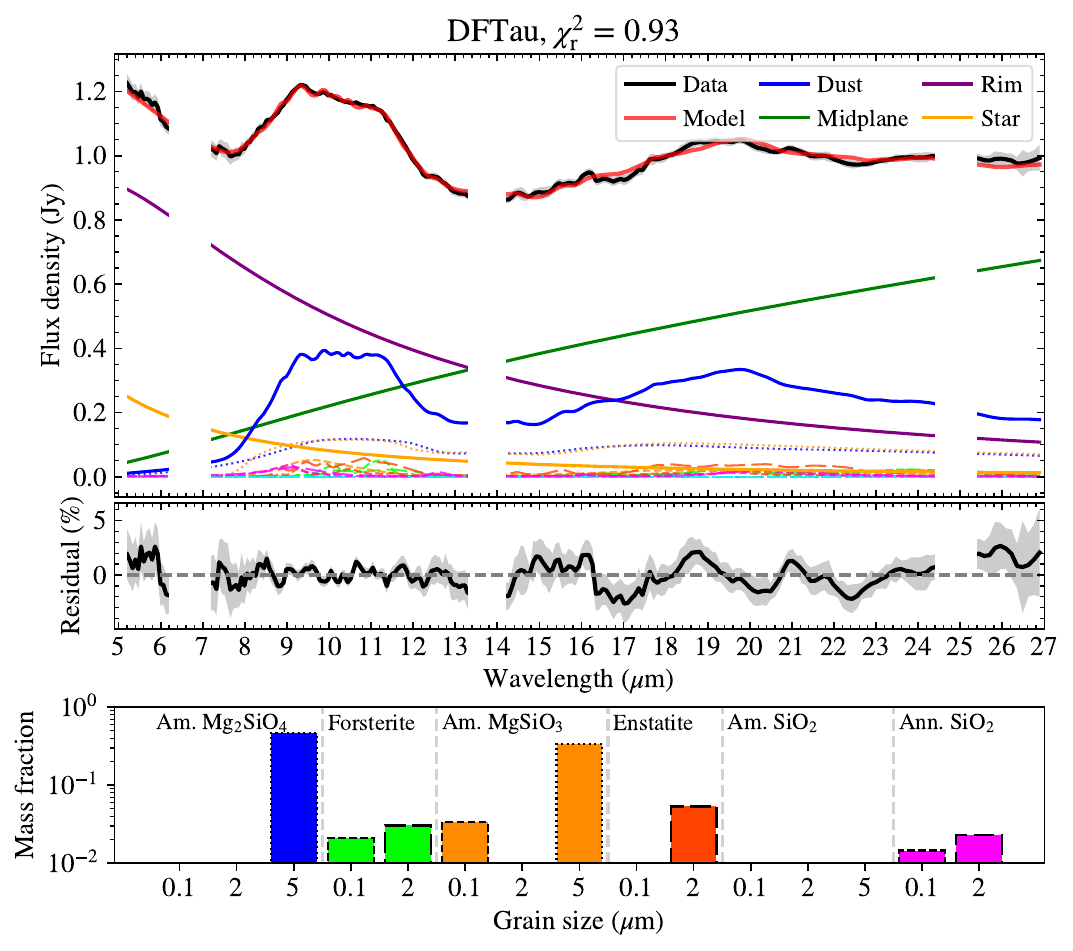}
\includegraphics[width=0.48\hsize]{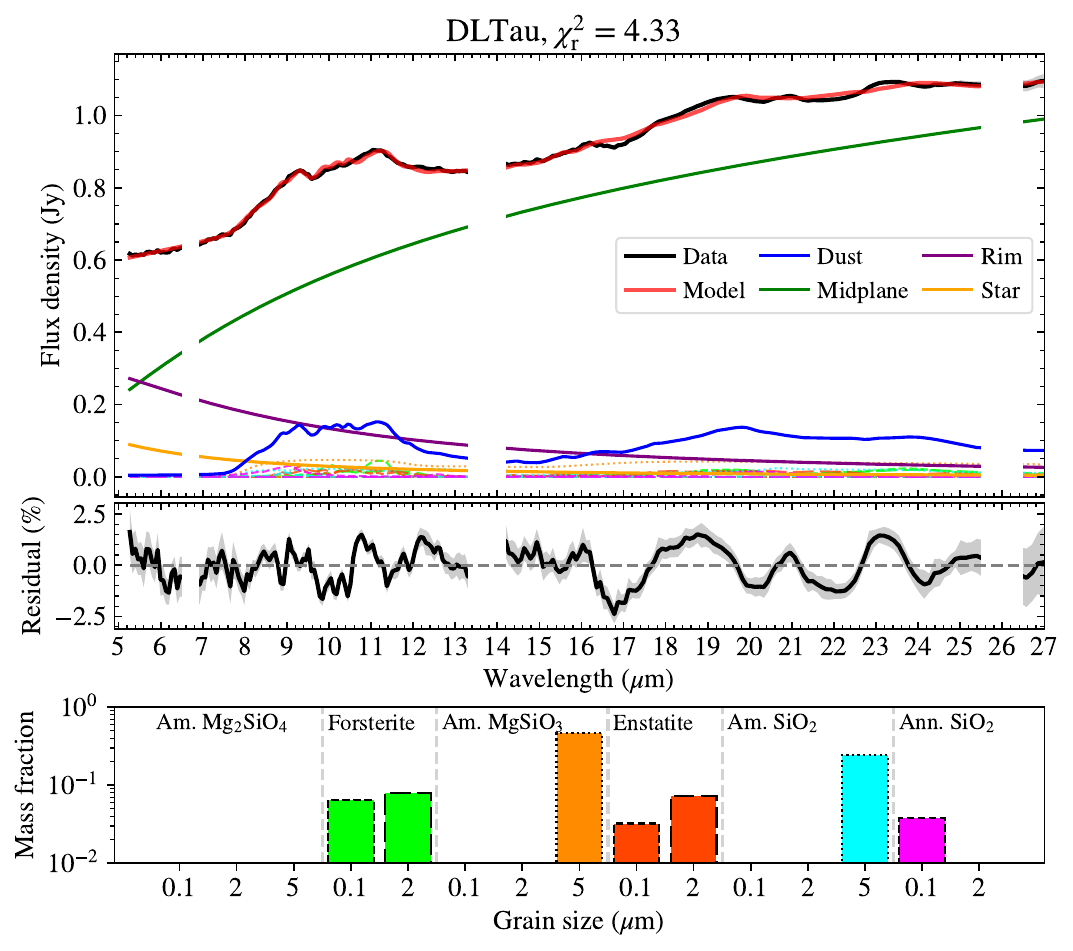}
\includegraphics[width=0.48\hsize]{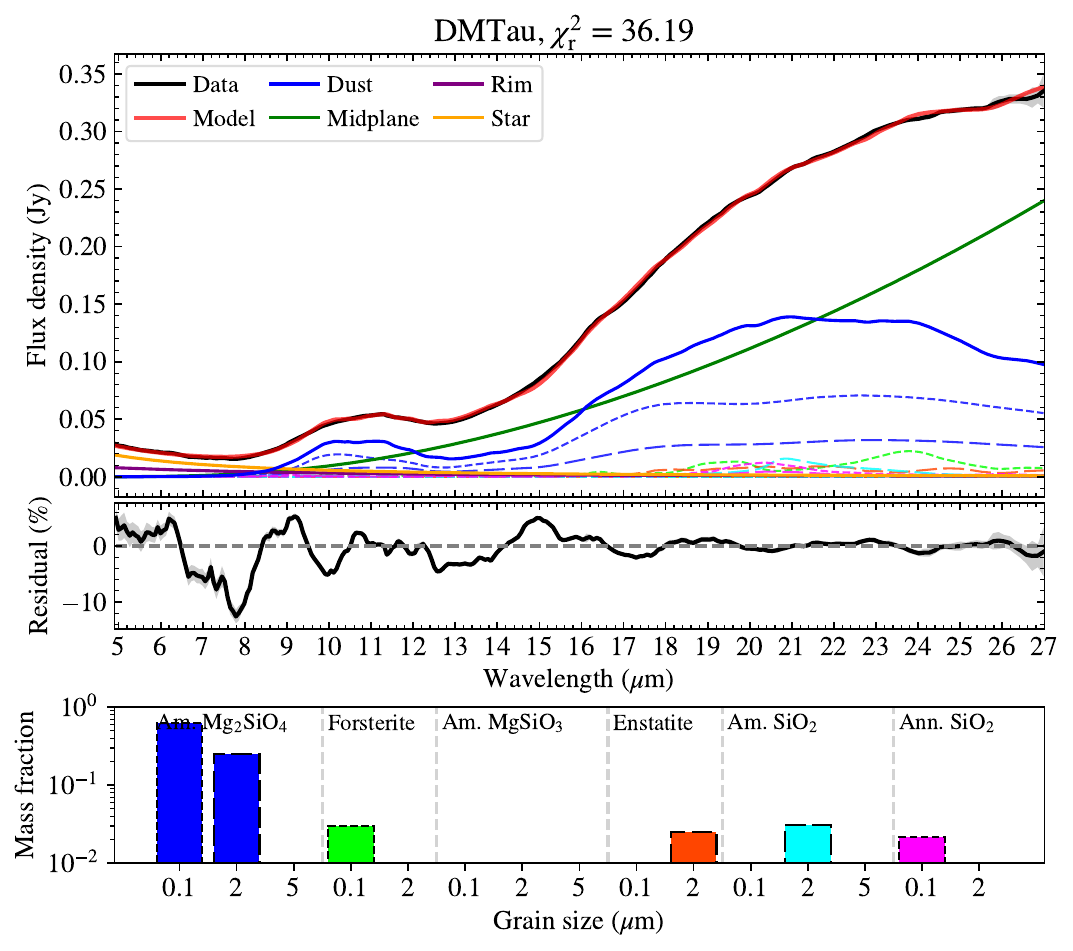}
\includegraphics[width=0.48\hsize]{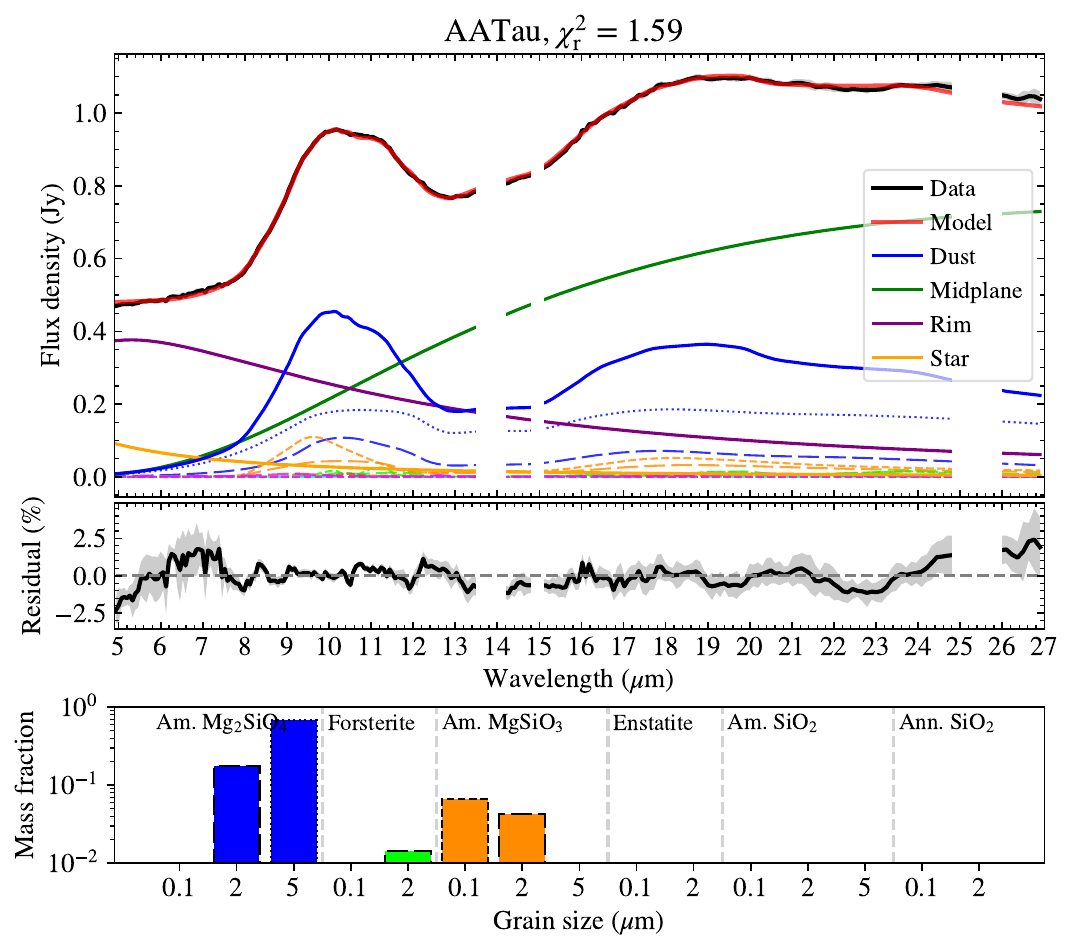}
\caption{continued.}
         	\label{fig:fit_all1}
\end{figure*}
\FloatBarrier

\setcounter{figure}{0}

\begin{figure*}
	\centering
\includegraphics[width=0.48\hsize]{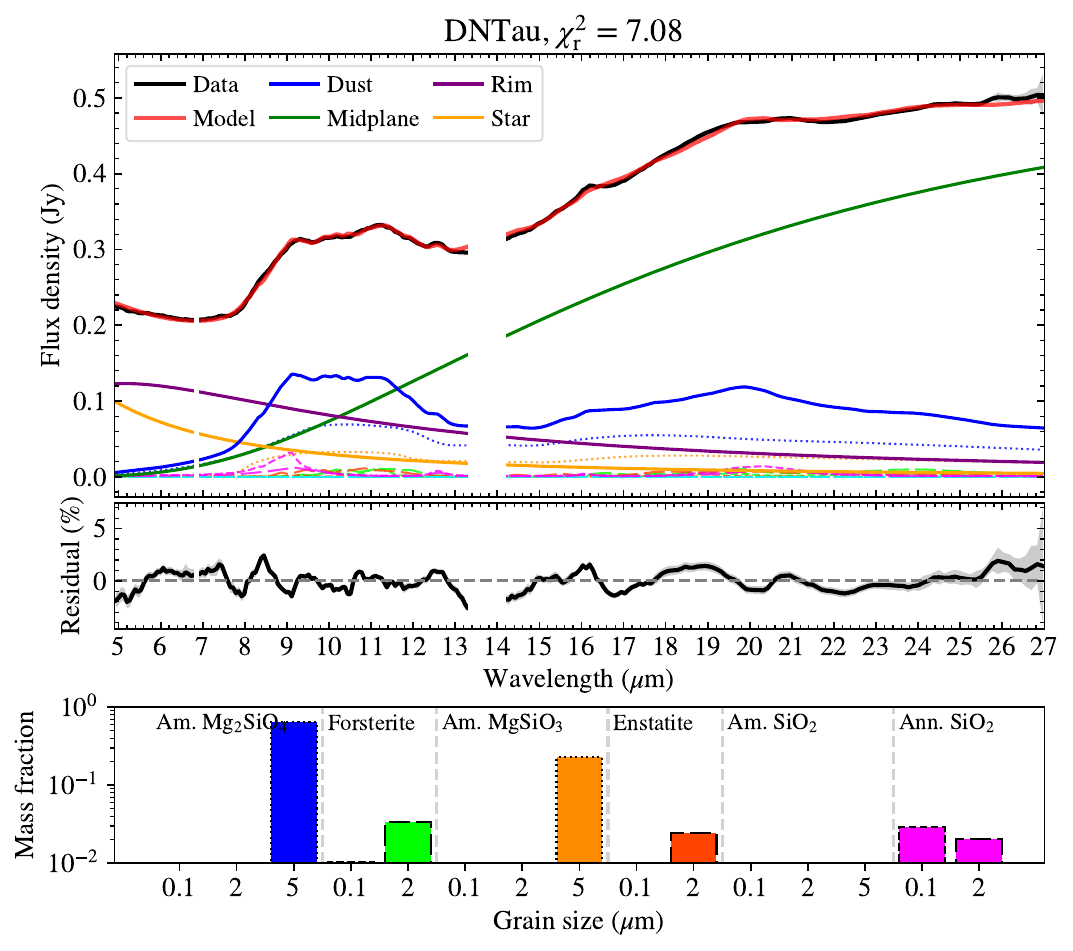}
\includegraphics[width=0.48\hsize]{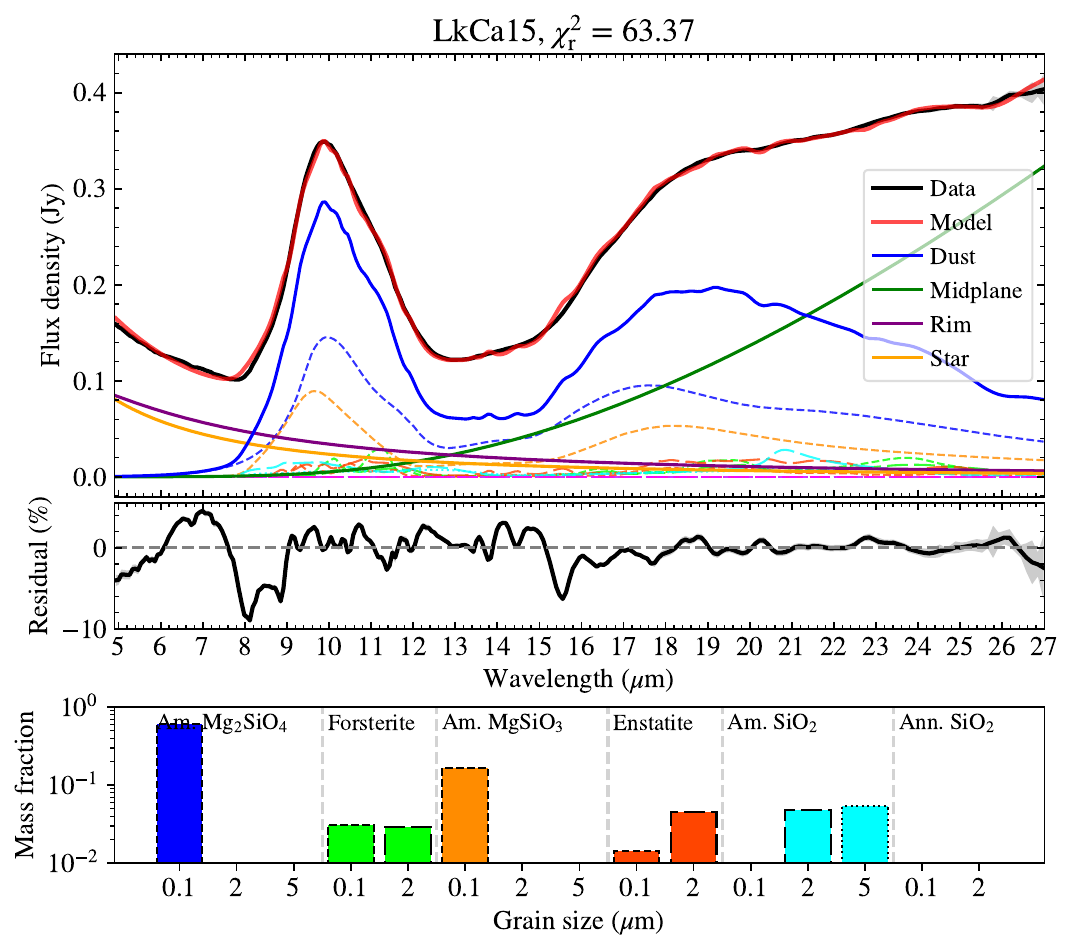}
\includegraphics[width=0.48\hsize]{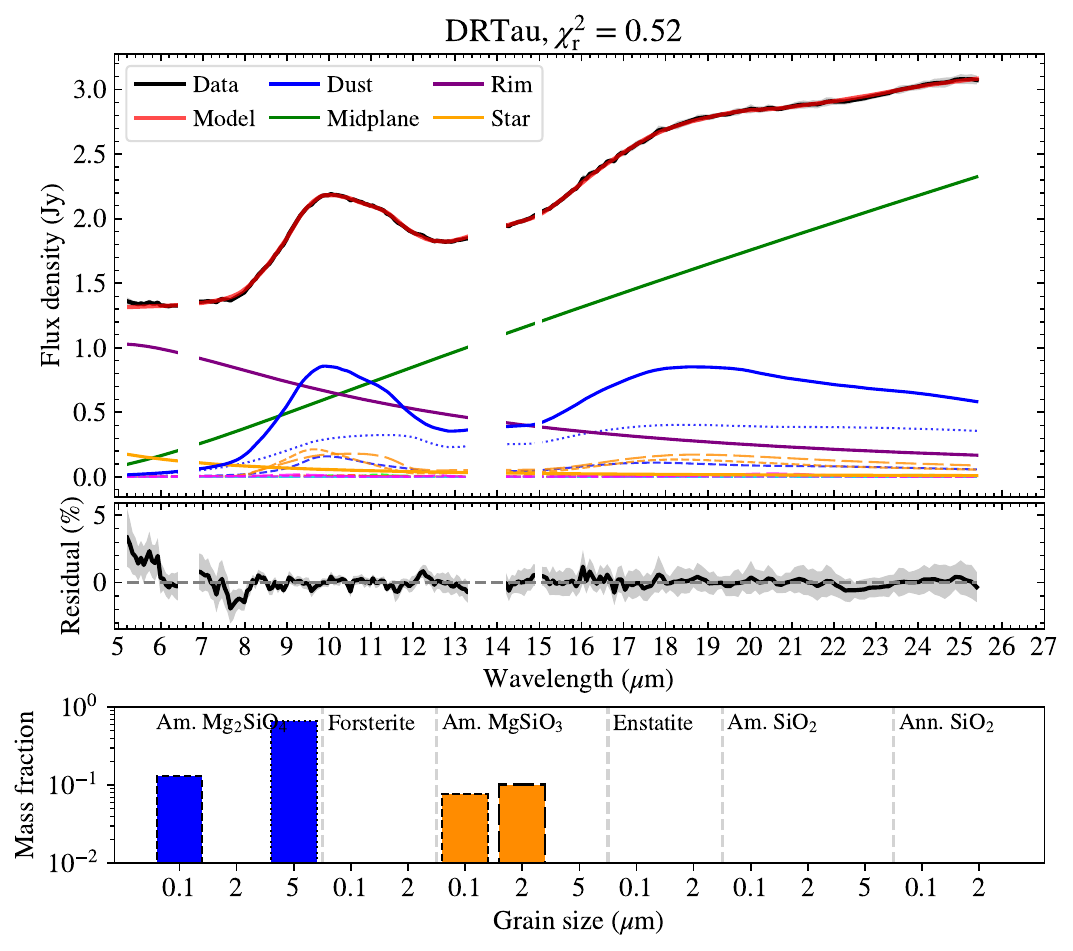}
\includegraphics[width=0.48\hsize]{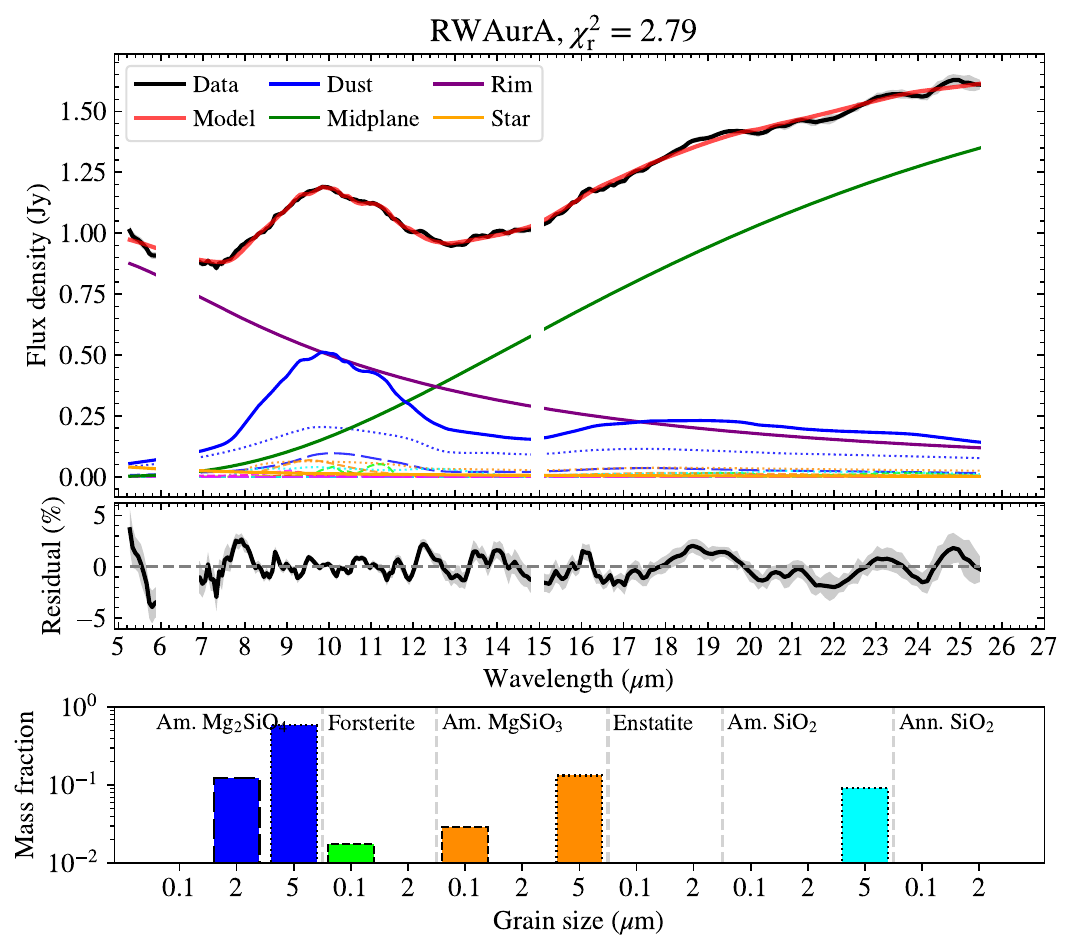}
\includegraphics[width=0.48\hsize]{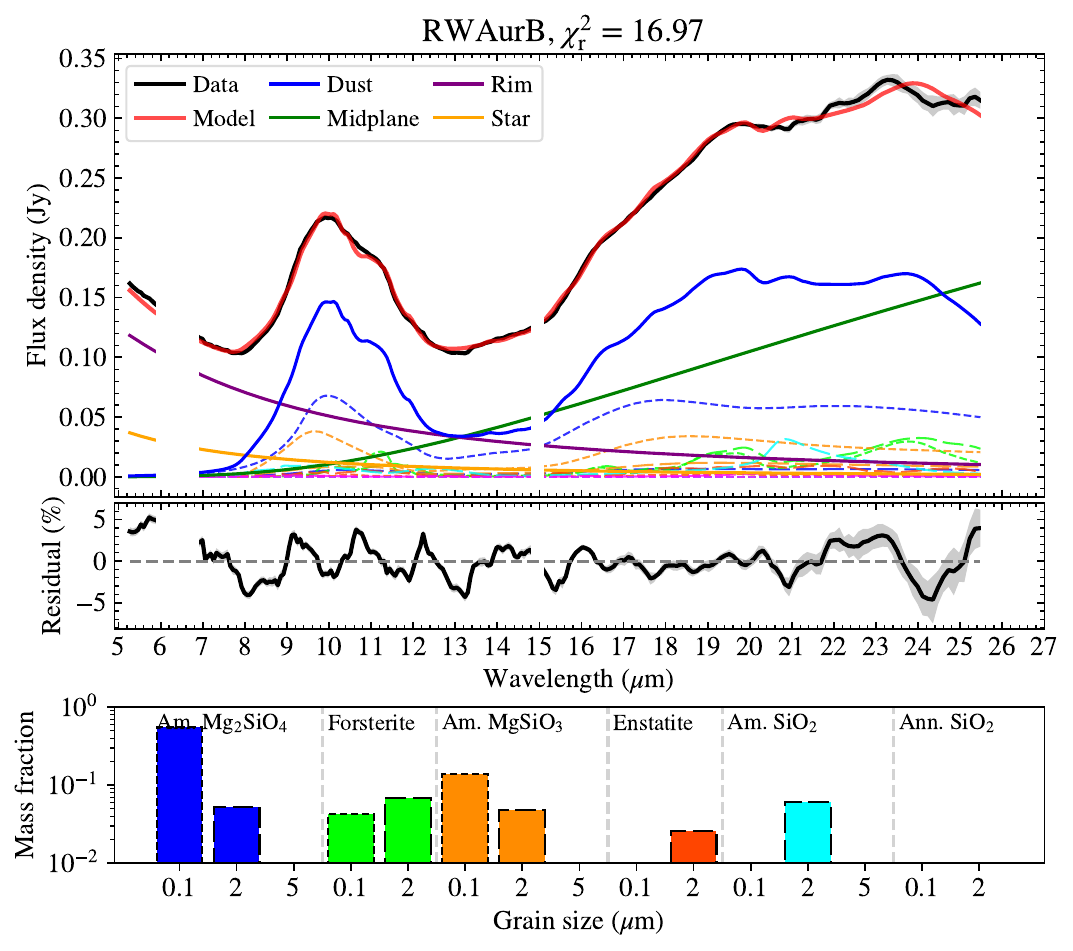}
\includegraphics[width=0.48\hsize]{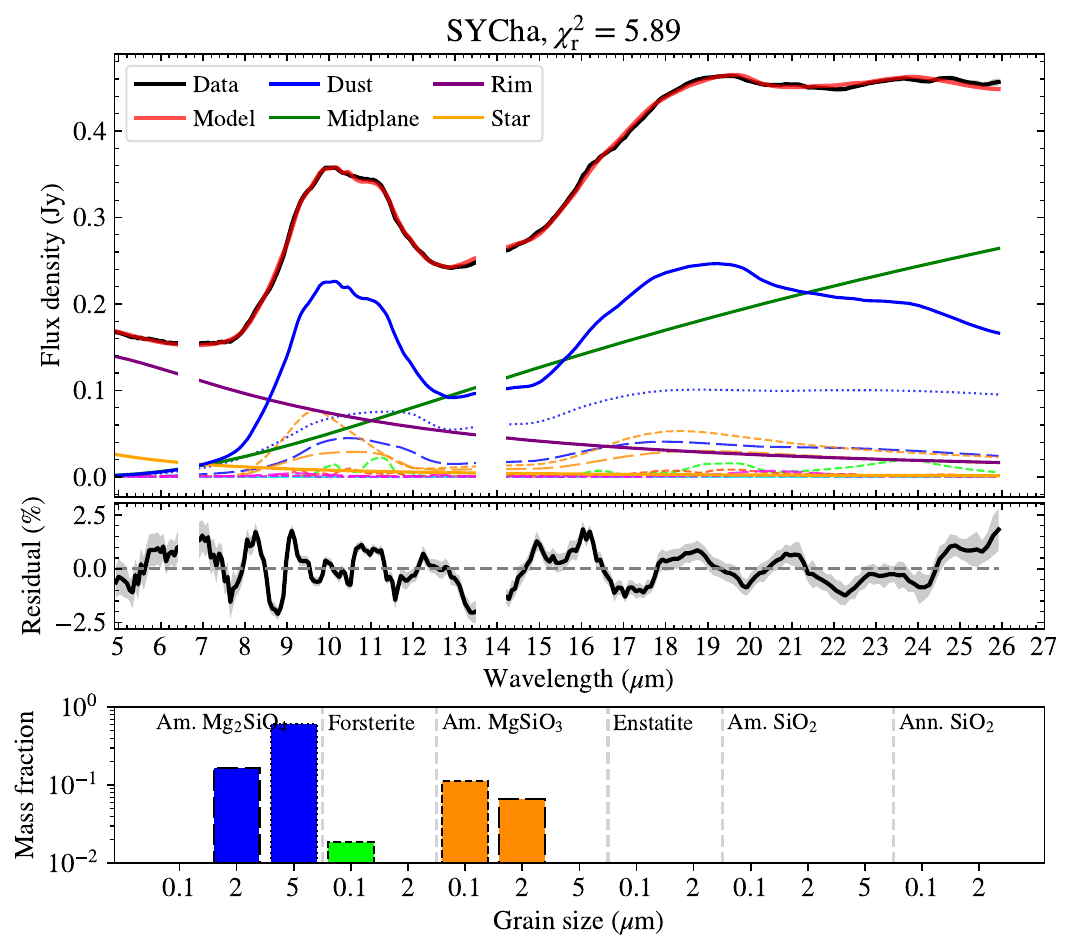}
	\caption{continued.}
         	\label{fig:fit_all2}
\end{figure*}

\FloatBarrier
\setcounter{figure}{0}

\begin{figure*}
	\centering
\includegraphics[width=0.48\hsize]{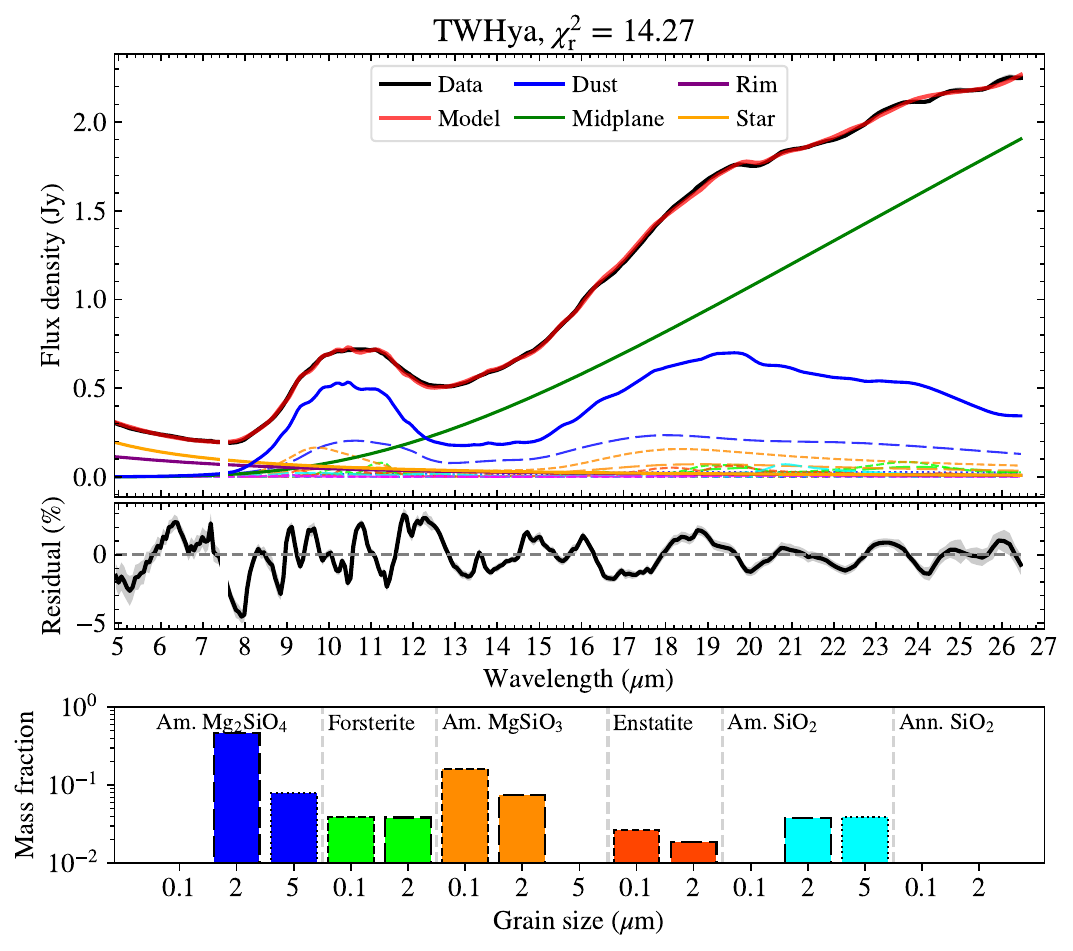}
\includegraphics[width=0.48\hsize]{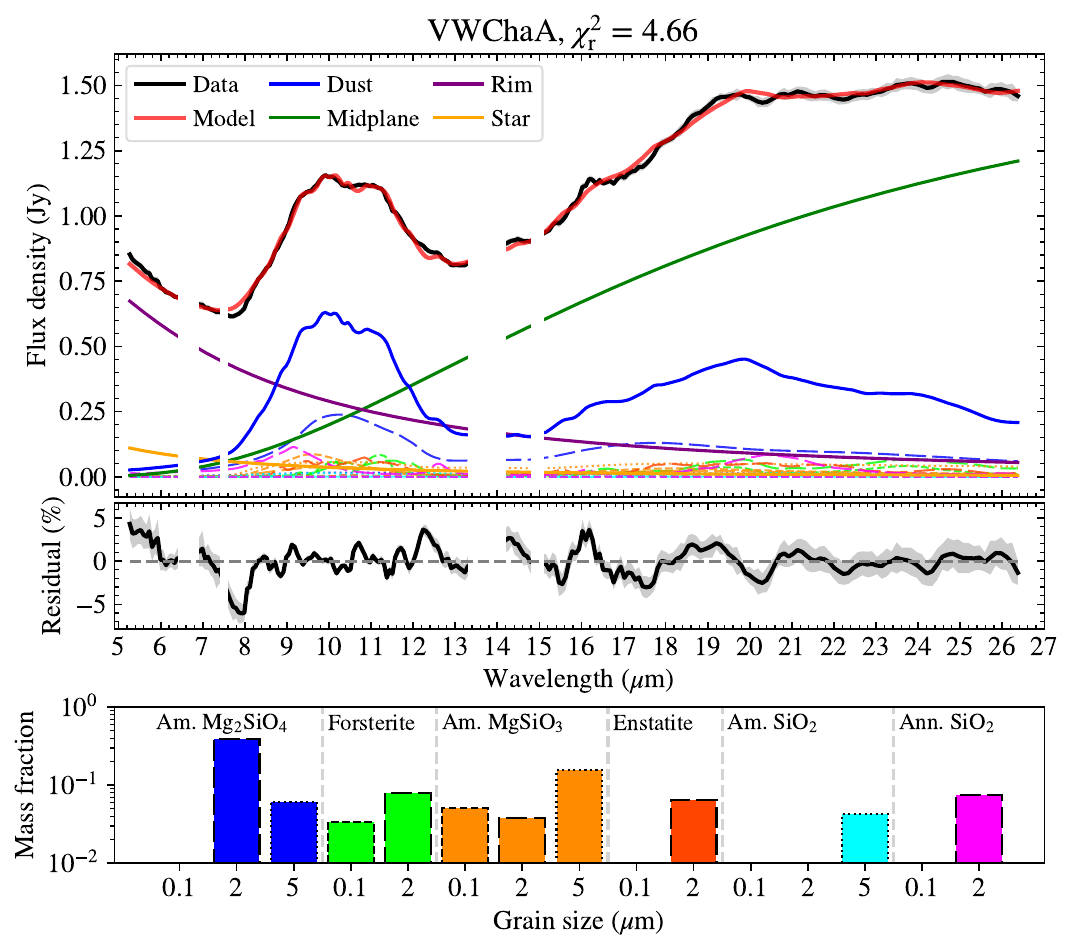}
\includegraphics[width=0.48\hsize]{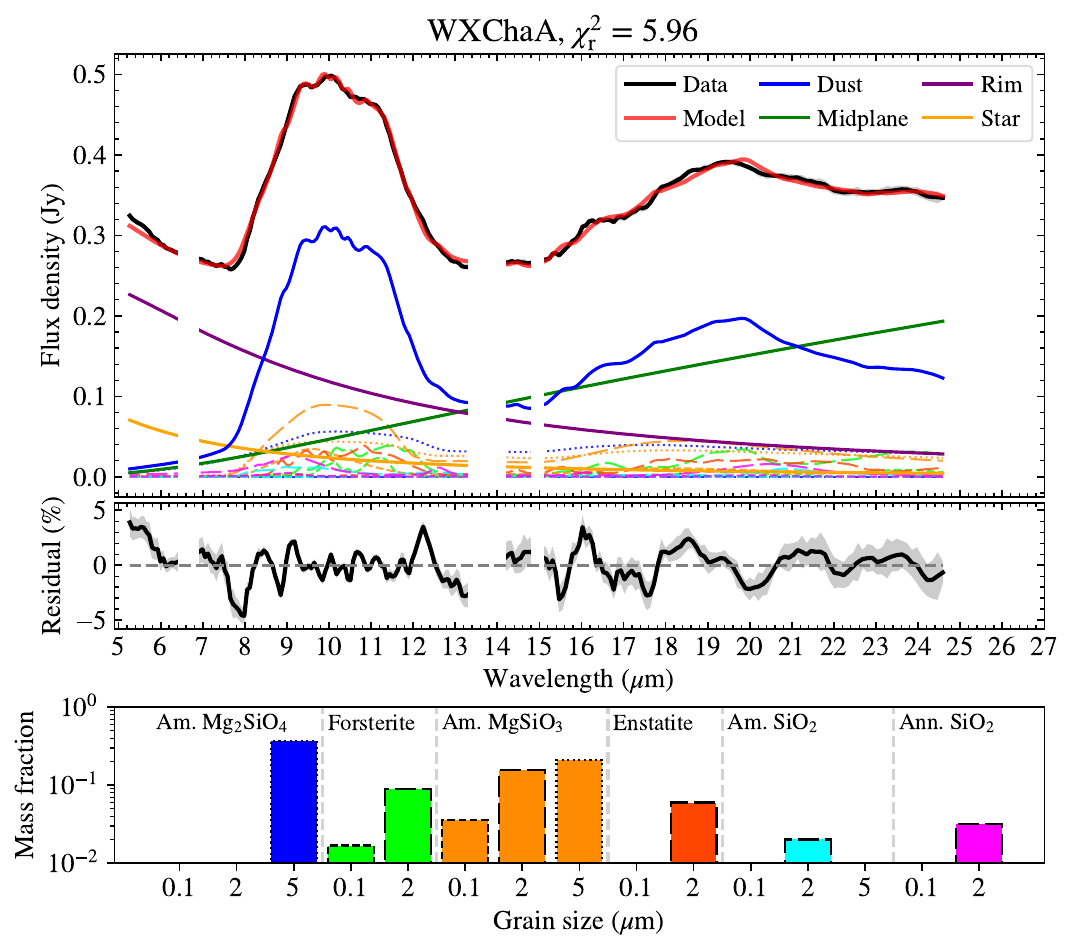}
\includegraphics[width=0.48\hsize]{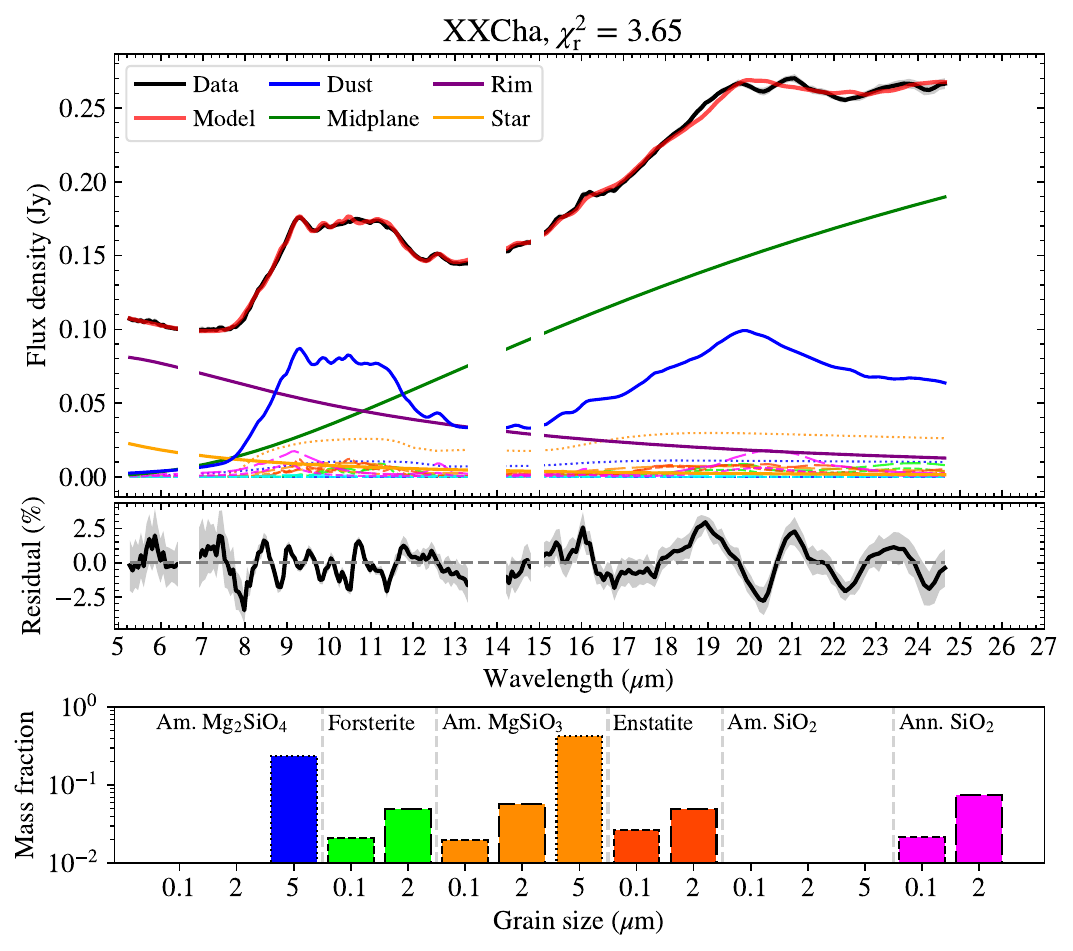}
\includegraphics[width=0.48\hsize]{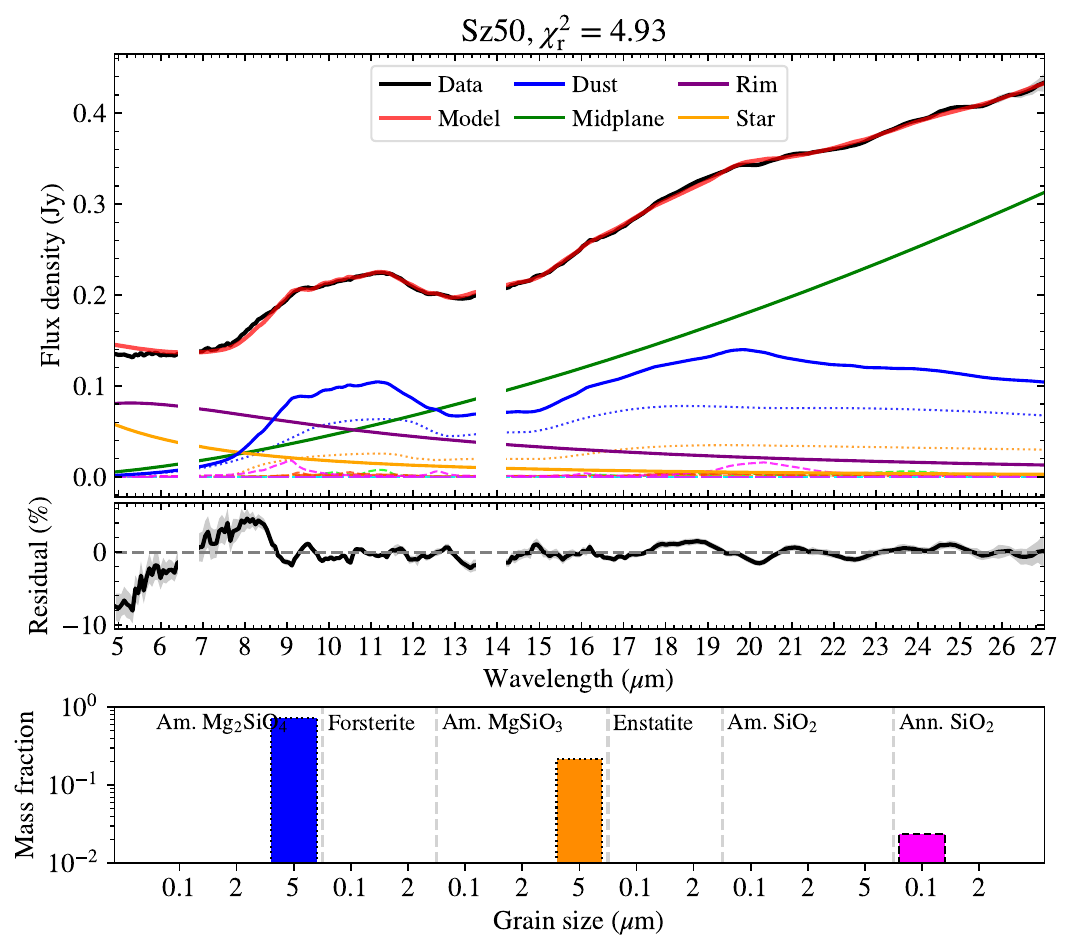}
\includegraphics[width=0.48\hsize]{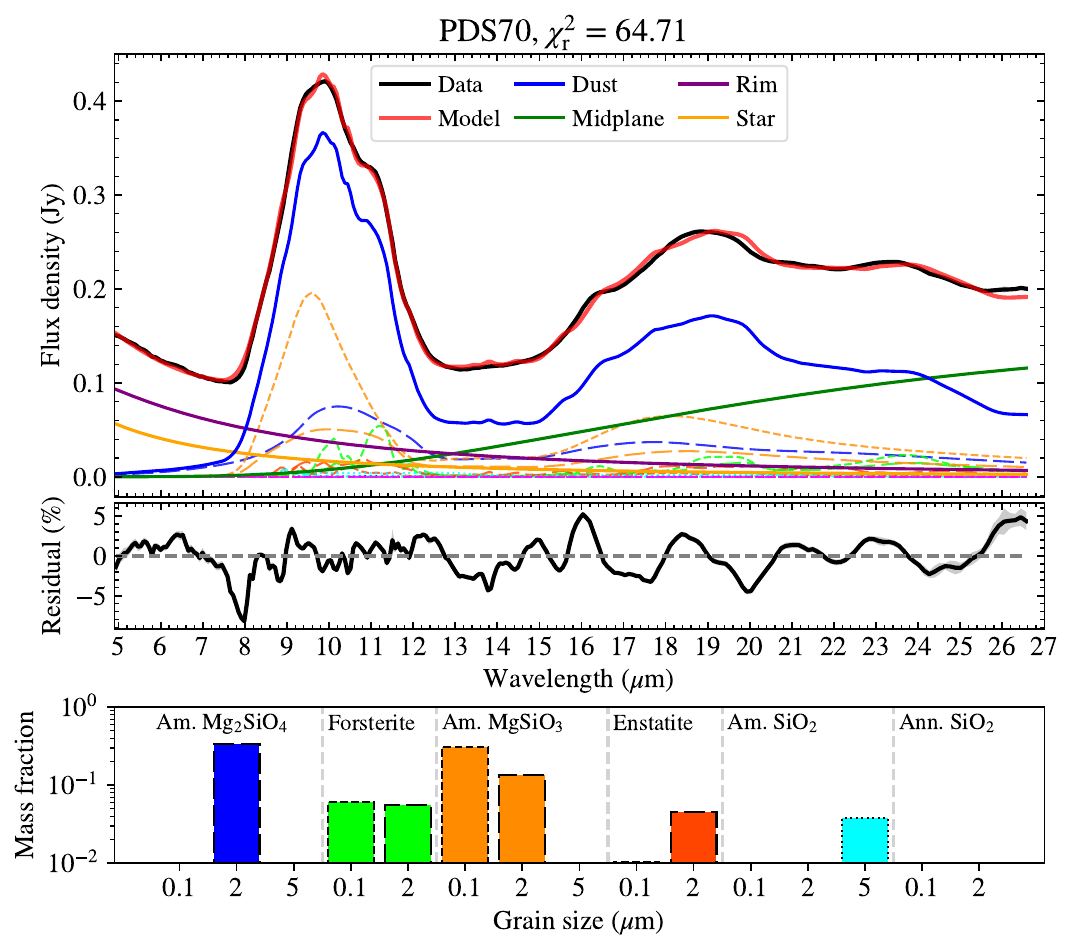}
	\caption{continued.}
         	\label{fig:fit_all3}
\end{figure*}

\FloatBarrier
\setcounter{figure}{0}

\begin{figure*}
	\centering
\includegraphics[width=0.48\hsize]{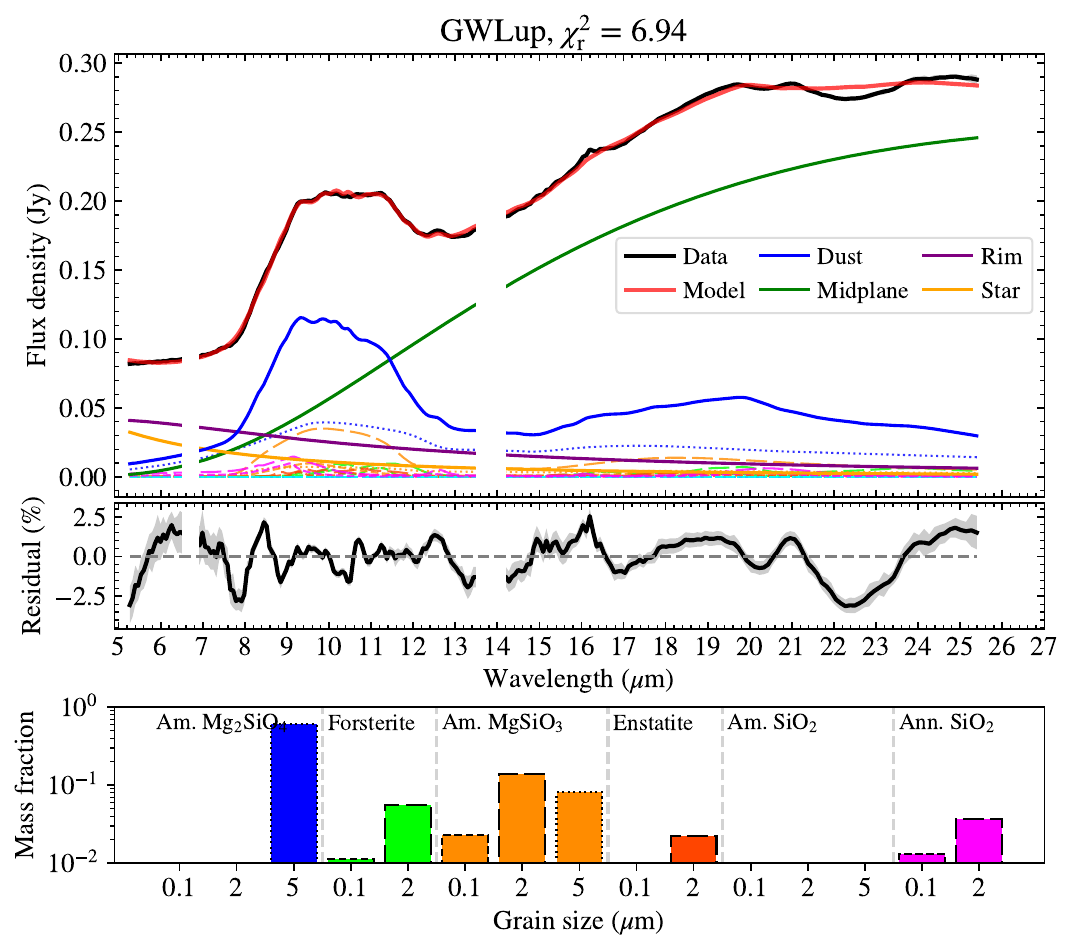}
\includegraphics[width=0.48\hsize]{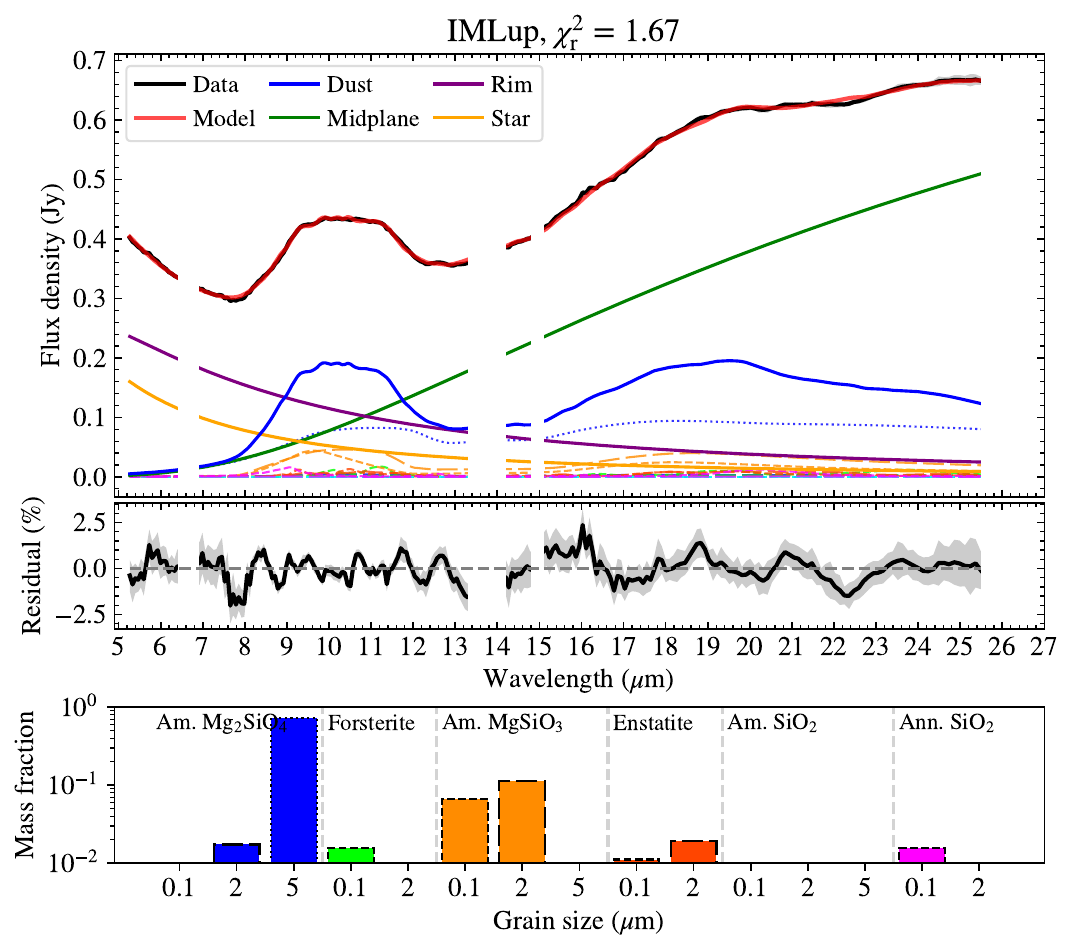}
\includegraphics[width=0.48\hsize]{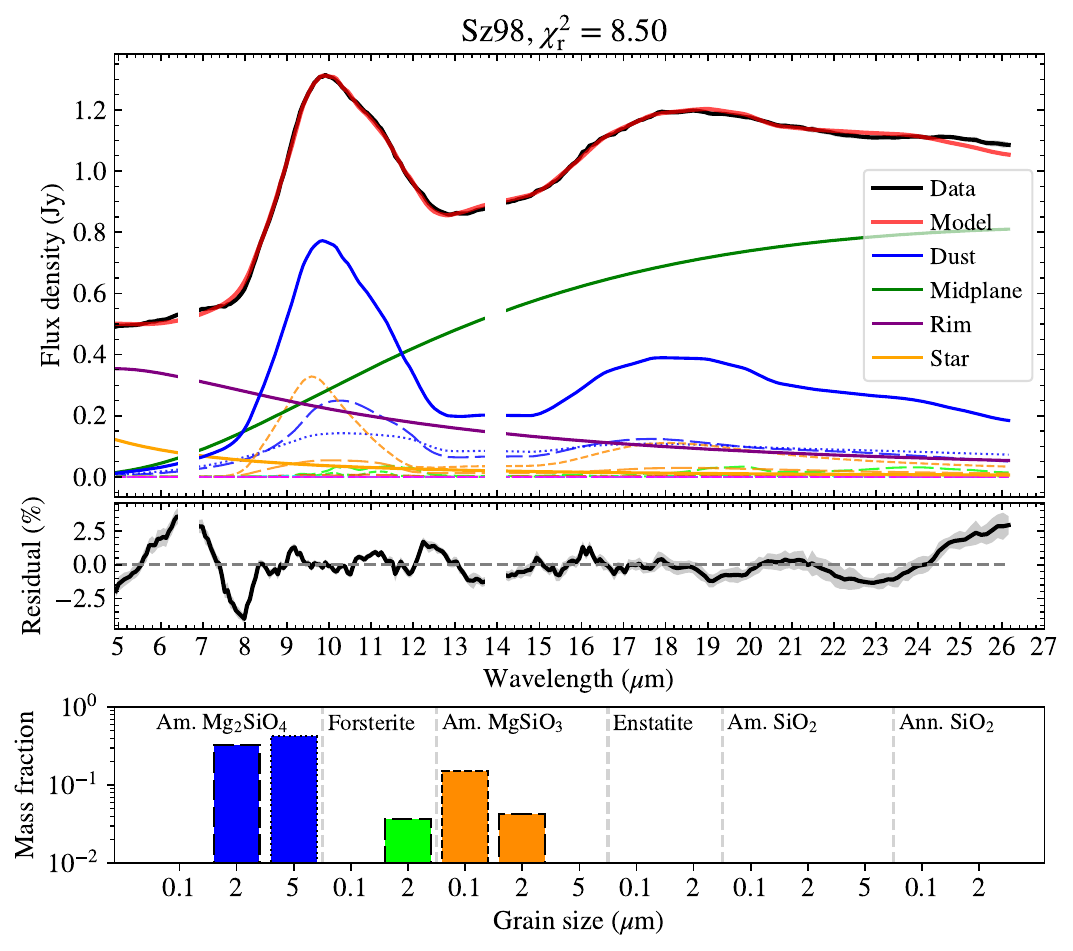}
\includegraphics[width=0.48\hsize]{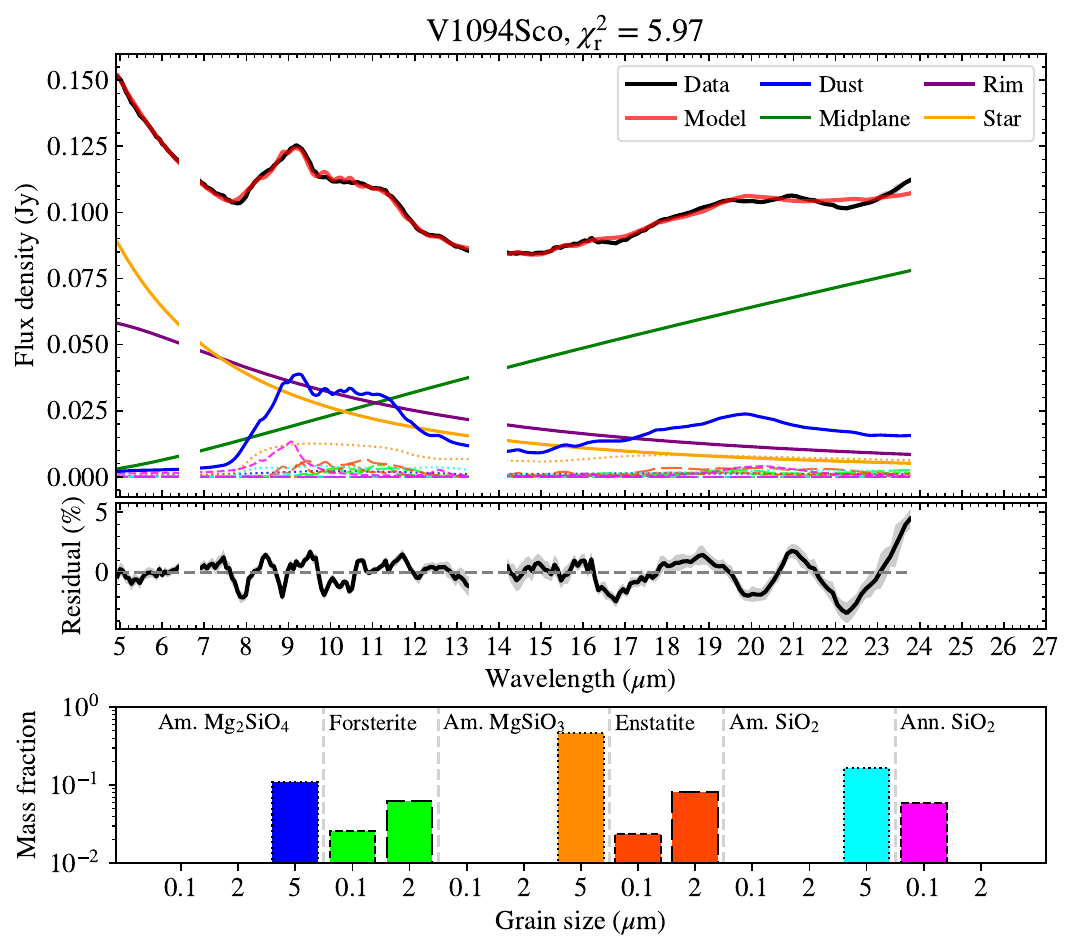}
\includegraphics[width=0.48\hsize]{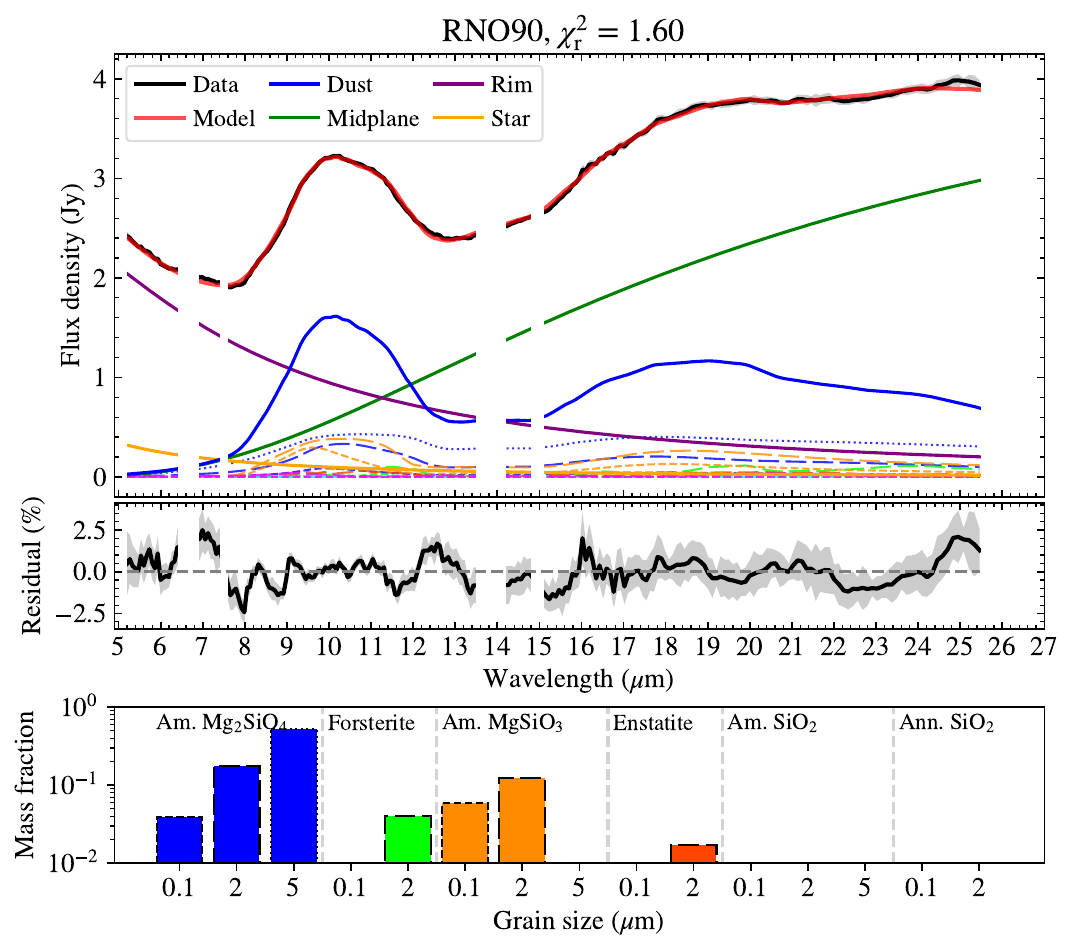}
\includegraphics[width=0.48\hsize]{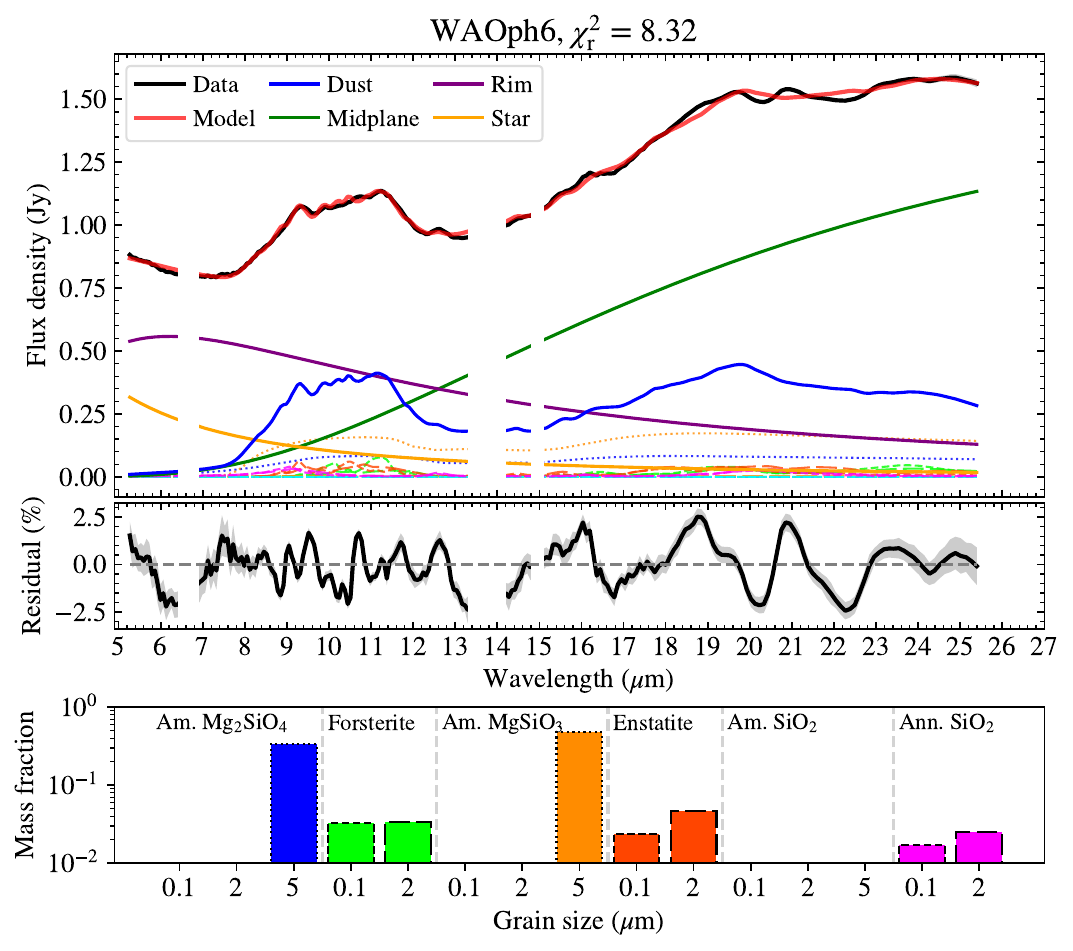}

	\caption{continued.}
         	\label{fig:fit_all4}
\end{figure*}

\FloatBarrier

\begingroup
\let\clearpage\relax{
\onecolumn	

\section{Derived mass fractions.}
\label{sec:app:mf}	

\begin{table*}[h!]
        \caption[]{\label{tab:mass_fractions}Mass fractions ($\%$) and $\chi^2_\mathrm{r}$ values from our fits with the DHS$\_$nat set of opacities. Grain sizes are indicated in $\mu$m. Dashes indicate dust species not detected in the fit.}
        \centering
    {\small
\renewcommand{\arraystretch}{1.1}
\begin{tabular}{llccccccccc}
\hline
\hline
Dust species & Grain \\ 
 & size &CX Tau & CY Tau & BP Tau & FT Tau & DF Tau & DL Tau & DM Tau & AA Tau & DN Tau \\ 
\hline
$\chi^2_\mathrm{r}$ &  &12.1 & 2.7 & 7.7 & 2.4 & 0.9 & 4.3 & 36.2 & 1.6 & 7.1 \\ 
\hline
Am. $\mathrm{Mg}_{2}\mathrm{SiO}_{4}$ & 0.1 &$-$ & $-$ & $-$ & $0.00^{+3.15}_{-0.00} $ & $-$ & $-$ & $63.4^{+-8.5}_{-26.6} $ & $0.00^{+9.84}_{-0.00} $ & $-$ \\
Am. $\mathrm{Mg}_{2}\mathrm{SiO}_{4}$ & 2  &$0.00^{+5.14}_{-0.00} $ & $-$ & $23.8^{+13.3}_{-23.3} $ & $9.5^{+24.2}_{-9.5} $ & $-$ & $-$ & $24.9^{+-4.6}_{-24.9} $ & $17.5^{+17.9}_{-17.5} $ & $<0.1$ \\
Am. $\mathrm{Mg}_{2}\mathrm{SiO}_{4}$ & 5  &$84.8^{+0.2}_{-23.0} $ & $16.1^{+58.6}_{-16.1} $ & $22.1^{+30.2}_{-20.3} $ & $60.6^{+8.9}_{-60.6} $ & $47.2^{+41.4}_{-47.2} $ & $0.00^{+61.22}_{-0.00} $ & $0.00^{+54.39}_{-+15.12} $ & $68.0^{+16.0}_{-30.0} $ & $65.1^{+13.0}_{-21.3} $ \\
Forsterite & 0.1 &$0.00^{+0.26}_{-0.00} $ & $2.1^{+1.7}_{-1.8} $ & $3.3^{+0.7}_{-0.9} $ & $1.6^{+1.1}_{-0.1} $ & $2.1^{+-0.4}_{-2.1} $ & $6.4^{+2.6}_{-4.5} $ & $3.0^{+-0.5}_{-1.5} $ & $0.98^{+0.29}_{-0.56} $ & $1.0^{+0.2}_{-0.5} $ \\
Forsterite & 2  &$1.2^{+3.0}_{-1.2} $ & $19.2^{+3.5}_{-18.8} $ & $3.4^{+4.5}_{-3.4} $ & $3.2^{+9.4}_{-2.2} $ & $3.0^{+4.4}_{-3.0} $ & $7.9^{+7.5}_{-7.9} $ & $-$ & $1.5^{+3.0}_{-1.5} $ & $3.4^{+4.3}_{-3.4} $ \\
Am. MgSiO$_3$ & 0.1 &$0.05^{+2.38}_{-0.05} $ & $5.6^{+3.8}_{-5.6} $ & $17.8^{+2.5}_{-6.1} $ & $5.7^{+3.8}_{-3.6} $ & $3.4^{+0.2}_{-3.4} $ & $-$ & $-$ & $6.7^{+0.9}_{-4.6} $ & $0.00^{+1.08}_{-0.00} $ \\
Am. MgSiO$_3$ & 2  &$9.3^{+3.8}_{-5.6} $ & $-$ & $10.8^{+4.2}_{-10.8} $ & $12.6^{+-1.6}_{-12.6} $ & $-$ & $-$ & $0.00^{+0.37}_{-0.00} $ & $4.3^{+1.3}_{-4.3} $ & $0.00^{+2.69}_{-0.00} $ \\
Am. MgSiO$_3$ & 5  &$0.59^{+14.49}_{-0.59} $ & $0.00^{+32.31}_{-0.00} $ & $0.00^{+3.28}_{-0.00} $ & $0.00^{+24.08}_{-0.00} $ & $34.2^{+41.4}_{-34.2} $ & $46.2^{+17.2}_{-46.2} $ & $-$ & $0.00^{+10.34}_{-0.00} $ & $22.9^{+15.8}_{-16.4} $ \\
Enstatite & 0.1 &$0.64^{+0.27}_{-0.07} $ & $0.98^{+0.81}_{-0.98} $ & $2.2^{+0.2}_{-0.5} $ & $0.59^{+0.43}_{-0.41} $ & $0.97^{+-0.04}_{-0.97} $ & $3.2^{+1.6}_{-3.1} $ & $<0.1$ & $0.31^{+0.29}_{-0.31} $ & $0.17^{+0.30}_{-0.17} $ \\
Enstatite & 2  &$0.00^{+1.17}_{-0.00} $ & $10.5^{+-3.1}_{-10.5} $ & $3.8^{+1.4}_{-2.0} $ & $1.3^{+3.3}_{-1.1} $ & $5.3^{+0.4}_{-5.3} $ & $7.4^{+3.9}_{-7.4} $ & $2.5^{+-0.6}_{-1.5} $ & $0.16^{+1.60}_{-0.16} $ & $2.4^{+1.6}_{-2.2} $ \\
Am. SiO$_2$ & 0.1 &$-$ & $0.90^{+2.42}_{-0.81} $ & $0.00^{+0.45}_{-0.00} $ & $0.00^{+0.18}_{-0.00} $ & $0.00^{+0.38}_{-0.00} $ & $0.71^{+2.30}_{-0.71} $ & $-$ & $<0.1$ & $<0.1$ \\
Am. SiO$_2$ & 2  &$-$ & $<0.1$ & $3.6^{+1.2}_{-3.1} $ & $-$ & $-$ & $-$ & $3.1^{+-1.3}_{-2.3} $ & $-$ & $-$ \\
Am. SiO$_2$ & 5  &$-$ & $35.4^{+-4.8}_{-29.9} $ & $5.2^{+7.1}_{-3.9} $ & $3.5^{+12.7}_{-3.5} $ & $-$ & $24.4^{+4.3}_{-20.3} $ & $-$ & $0.00^{+0.32}_{-0.00} $ & $0.00^{+4.30}_{-0.00} $ \\
Ann. SiO$_2$ & 0.1 &$0.79^{+1.22}_{-0.65} $ & $2.7^{+-0.8}_{-2.7} $ & $2.6^{+-0.2}_{-2.6} $ & $0.00^{+0.29}_{-0.00} $ & $1.5^{+-0.5}_{-1.5} $ & $3.8^{+1.3}_{-3.8} $ & $2.2^{+-2.2}_{-2.2} $ & $0.00^{+0.51}_{-0.00} $ & $2.9^{+1.0}_{-0.8} $ \\
Ann. SiO$_2$ & 2  &$2.7^{+1.0}_{-1.3} $ & $6.5^{+2.9}_{-6.5} $ & $1.6^{+1.9}_{-1.6} $ & $1.5^{+1.1}_{-1.5} $ & $2.3^{+-0.1}_{-2.3} $ & $-$ & $0.94^{+2.58}_{-+1.42} $ & $0.65^{+0.25}_{-0.65} $ & $2.1^{+0.7}_{-2.1} $ \\
\hline
        \end{tabular}
\vspace{\baselineskip}

\begin{tabular}{llccccccccc}
\hline
\hline 
Dust species & Grain \\ 
 & size &LkCa 15 & DR Tau & RW Aur A & RW Aur B & SY Cha & TW Hya & VW Cha A & WX Cha A & XX Cha \\ 
\hline
$\chi^2_\mathrm{r}$ &  &63.4 & 0.5 & 2.8 & 17.0 & 5.9 & 14.3 & 4.7 & 6.0 & 3.6 \\ 
\hline
Am. $\mathrm{Mg}_{2}\mathrm{SiO}_{4}$ & 0.1 &$60.9^{+2.9}_{-1.9} $ & $13.0^{+15.0}_{-12.3} $ & $0.00^{+7.78}_{-0.00} $ & $55.2^{+4.6}_{-5.4} $ & $0.00^{+8.83}_{-0.00} $ & $-$ & $0.00^{+9.52}_{-0.00} $ & $0.00^{+0.37}_{-0.00} $ & $-$ \\
Am. $\mathrm{Mg}_{2}\mathrm{SiO}_{4}$ & 2  &$-$ & $0.00^{+40.50}_{-0.00} $ & $12.3^{+16.8}_{-12.3} $ & $5.3^{+1.1}_{-5.3} $ & $16.8^{+17.1}_{-16.8} $ & $47.1^{+11.2}_{-16.3} $ & $39.1^{+3.4}_{-39.1} $ & $0.00^{+15.83}_{-0.00} $ & $0.00^{+7.77}_{-0.00} $ \\
Am. $\mathrm{Mg}_{2}\mathrm{SiO}_{4}$ & 5  &$-$ & $66.5^{+4.3}_{-66.5} $ & $59.0^{+20.5}_{-59.0} $ & $0.00^{+6.67}_{-0.00} $ & $61.5^{+12.3}_{-24.7} $ & $7.9^{+23.0}_{-7.9} $ & $6.2^{+48.7}_{-6.2} $ & $37.2^{+21.8}_{-37.2} $ & $24.0^{+40.1}_{-24.0} $ \\
Forsterite & 0.1 &$3.1^{+0.2}_{-0.6} $ & $0.39^{+0.39}_{-0.39} $ & $1.8^{+0.6}_{-1.5} $ & $4.3^{+1.0}_{-1.0} $ & $1.9^{+0.2}_{-0.6} $ & $3.9^{+0.8}_{-0.6} $ & $3.4^{+1.2}_{-0.9} $ & $1.7^{+1.2}_{-+0.2} $ & $2.1^{+1.3}_{-0.7} $ \\
Forsterite & 2  &$2.9^{+1.3}_{-1.6} $ & $0.62^{+6.38}_{-0.62} $ & $0.19^{+7.99}_{-0.19} $ & $6.9^{+3.7}_{-4.8} $ & $0.00^{+1.78}_{-0.00} $ & $3.9^{+1.7}_{-3.2} $ & $8.0^{+9.8}_{-6.9} $ & $9.0^{+3.1}_{-7.4} $ & $4.9^{+6.2}_{-4.9} $ \\
Am. MgSiO$_3$ & 0.1 &$16.8^{+0.4}_{-2.4} $ & $7.8^{+9.9}_{-4.2} $ & $2.9^{+10.3}_{-2.9} $ & $13.9^{+3.5}_{-2.9} $ & $11.3^{+2.0}_{-4.9} $ & $16.3^{+2.4}_{-3.0} $ & $5.1^{+9.7}_{-3.8} $ & $3.6^{+3.9}_{-0.7} $ & $2.0^{+3.7}_{-2.0} $ \\
Am. MgSiO$_3$ & 2  &$-$ & $10.2^{+2.4}_{-10.2} $ & $0.24^{+6.52}_{-0.24} $ & $4.8^{+-3.6}_{-4.8} $ & $6.6^{+4.6}_{-3.2} $ & $7.5^{+4.0}_{-7.5} $ & $3.8^{+7.0}_{-3.8} $ & $15.5^{+-2.1}_{-15.5} $ & $5.8^{+2.2}_{-5.8} $ \\
Am. MgSiO$_3$ & 5  &$-$ & $0.00^{+29.77}_{-0.00} $ & $13.3^{+37.2}_{-13.3} $ & $0.00^{+9.66}_{-0.00} $ & $0.00^{+4.25}_{-0.00} $ & $-$ & $15.7^{+15.1}_{-15.7} $ & $21.3^{+24.7}_{-8.2} $ & $43.2^{+11.9}_{-43.2} $ \\
Enstatite & 0.1 &$1.4^{+0.4}_{-0.5} $ & $0.30^{+0.62}_{-0.30} $ & $0.49^{+-0.25}_{-0.49} $ & $0.39^{+0.81}_{-0.39} $ & $0.78^{+0.19}_{-0.26} $ & $2.7^{+0.3}_{-0.8} $ & $0.30^{+0.59}_{-0.30} $ & $0.14^{+0.59}_{-0.08} $ & $2.6^{+0.7}_{-1.0} $ \\
Enstatite & 2  &$4.5^{+1.0}_{-0.5} $ & $0.29^{+3.01}_{-0.29} $ & $0.00^{+1.39}_{-0.00} $ & $2.6^{+0.6}_{-1.3} $ & $0.24^{+0.80}_{-0.24} $ & $1.9^{+0.6}_{-0.7} $ & $6.5^{+4.5}_{-3.9} $ & $6.0^{+2.1}_{-4.2} $ & $4.9^{+2.5}_{-4.5} $ \\
Am. SiO$_2$ & 0.1 &$-$ & $0.00^{+0.36}_{-0.00} $ & $0.00^{+0.69}_{-0.00} $ & $-$ & $0.00^{+0.15}_{-0.00} $ & $0.14^{+0.14}_{-0.14} $ & $0.00^{+0.14}_{-0.00} $ & $0.00^{+0.48}_{-0.00} $ & $0.00^{+0.80}_{-0.00} $ \\
Am. SiO$_2$ & 2  &$4.9^{+-0.3}_{-1.8} $ & $-$ & $<0.1$ & $6.0^{+0.7}_{-1.9} $ & $-$ & $3.8^{+1.2}_{-0.8} $ & $0.19^{+5.12}_{-0.19} $ & $2.0^{+1.9}_{-1.5} $ & $0.77^{+3.82}_{-0.77} $ \\
Am. SiO$_2$ & 5  &$5.4^{+3.2}_{-+0.2} $ & $-$ & $9.1^{+-1.0}_{-9.1} $ & $0.00^{+4.33}_{-0.00} $ & $-$ & $3.9^{+-0.7}_{-3.9} $ & $4.3^{+-1.6}_{-4.3} $ & $0.00^{+4.43}_{-0.00} $ & $0.00^{+8.01}_{-0.00} $ \\
Ann. SiO$_2$ & 0.1 &$-$ & $0.12^{+0.82}_{-0.12} $ & $0.63^{+1.19}_{-0.63} $ & $-$ & $0.12^{+0.43}_{-0.12} $ & $0.67^{+-0.13}_{-0.67} $ & $-$ & $0.30^{+0.23}_{-0.30} $ & $2.2^{+3.0}_{-2.2} $ \\
Ann. SiO$_2$ & 2  &$-$ & $0.86^{+-0.05}_{-0.86} $ & $0.00^{+2.49}_{-0.00} $ & $0.58^{+1.49}_{-0.58} $ & $0.79^{+-0.26}_{-0.79} $ & $0.36^{+1.15}_{-0.36} $ & $7.6^{+0.3}_{-5.5} $ & $3.2^{+-0.7}_{-3.2} $ & $7.6^{+0.6}_{-6.5} $ \\
\hline
        \end{tabular}
\vspace{\baselineskip}

\begin{tabular}{llcccccccc}
\hline
\hline
Dust species & Grain \\ 
 & size &Sz 50 & PDS 70 & GW Lup & IM Lup & Sz 98 & V1094 Sco & RNO 90 & WA Oph 6 \\ 
\hline
$\chi^2_\mathrm{r}$ &  &4.9 & 64.7 & 6.9 & 1.7 & 8.5 & 6.0 & 1.6 & 8.3 \\ 
\hline
Am. $\mathrm{Mg}_{2}\mathrm{SiO}_{4}$ & 0.1 &$-$ & $-$ & $-$ & $0.00^{+2.48}_{-0.00} $ & $0.00^{+9.40}_{-0.00} $ & $-$ & $3.9^{+18.1}_{-3.9} $ & $-$ \\
Am. $\mathrm{Mg}_{2}\mathrm{SiO}_{4}$ & 2  &$<0.1$ & $33.4^{+2.8}_{-8.4} $ & $0.00^{+10.02}_{-0.00} $ & $1.7^{+27.1}_{-1.7} $ & $32.6^{+25.5}_{-14.0} $ & $-$ & $17.7^{+36.9}_{-17.7} $ & $0.00^{+3.20}_{-0.00} $ \\
Am. $\mathrm{Mg}_{2}\mathrm{SiO}_{4}$ & 5  &$72.8^{+13.1}_{-31.4} $ & $0.63^{+6.30}_{-0.63} $ & $61.1^{+9.2}_{-33.3} $ & $73.4^{+2.2}_{-73.4} $ & $42.8^{+9.2}_{-42.6} $ & $11.0^{+41.3}_{-11.0} $ & $52.0^{+12.8}_{-52.0} $ & $33.9^{+44.7}_{-18.0} $ \\
Forsterite & 0.1 &$0.90^{+-0.23}_{-0.90} $ & $6.0^{+0.2}_{-0.3} $ & $1.1^{+0.6}_{-+0.1} $ & $1.6^{+0.7}_{-0.8} $ & $0.56^{+0.47}_{-0.09} $ & $2.6^{+0.6}_{-1.7} $ & $0.39^{+0.93}_{-0.24} $ & $3.3^{+0.7}_{-1.5} $ \\
Forsterite & 2  &$0.38^{+3.53}_{-0.38} $ & $5.6^{+2.5}_{-1.0} $ & $5.6^{+3.8}_{-4.8} $ & $0.60^{+9.17}_{-0.60} $ & $3.7^{+5.0}_{-2.6} $ & $6.2^{+4.0}_{-5.2} $ & $4.0^{+5.9}_{-4.0} $ & $3.4^{+4.4}_{-3.4} $ \\
Am. MgSiO$_3$ & 0.1 &$0.49^{+0.54}_{-0.49} $ & $31.0^{+1.4}_{-1.4} $ & $2.3^{+5.0}_{-0.3} $ & $6.7^{+6.7}_{-3.1} $ & $15.4^{+5.6}_{-4.6} $ & $0.00^{+0.55}_{-0.00} $ & $5.9^{+4.3}_{-4.1} $ & $0.09^{+1.82}_{-0.09} $ \\
Am. MgSiO$_3$ & 2  &$0.00^{+0.63}_{-0.00} $ & $13.6^{+3.3}_{-2.9} $ & $14.1^{+0.9}_{-14.1} $ & $11.2^{+9.9}_{-11.2} $ & $4.3^{+1.4}_{-4.3} $ & $-$ & $12.4^{+0.9}_{-12.4} $ & $0.48^{+7.22}_{-0.48} $ \\
Am. MgSiO$_3$ & 5  &$21.7^{+29.1}_{-16.9} $ & $0.00^{+3.16}_{-0.00} $ & $8.1^{+27.5}_{-8.1} $ & $0.00^{+32.63}_{-0.00} $ & $0.00^{+8.14}_{-0.00} $ & $46.6^{+18.6}_{-32.2} $ & $0.00^{+23.34}_{-0.00} $ & $47.6^{+1.5}_{-47.6} $ \\
Enstatite & 0.1 &$0.57^{+0.09}_{-0.57} $ & $1.0^{+0.1}_{-0.3} $ & $0.46^{+0.42}_{-0.17} $ & $1.1^{+0.7}_{-0.5} $ & $0.00^{+0.32}_{-0.00} $ & $2.3^{+0.3}_{-1.8} $ & $0.06^{+0.63}_{-0.06} $ & $2.4^{+0.3}_{-1.1} $ \\
Enstatite & 2  &$0.90^{+0.78}_{-0.90} $ & $4.5^{+0.9}_{-0.3} $ & $2.2^{+2.1}_{-1.6} $ & $1.9^{+4.9}_{-1.5} $ & $0.56^{+1.44}_{-0.56} $ & $8.2^{+-2.2}_{-7.8} $ & $1.7^{+2.9}_{-1.7} $ & $4.7^{+1.7}_{-4.0} $ \\
Am. SiO$_2$ & 0.1 &$0.00^{+0.21}_{-0.00} $ & $0.34^{+0.03}_{-0.34} $ & $-$ & $0.00^{+0.73}_{-0.00} $ & $0.05^{+0.24}_{-0.05} $ & $0.25^{+1.86}_{-+0.01} $ & $0.00^{+0.23}_{-0.00} $ & $0.01^{+0.95}_{-0.01} $ \\
Am. SiO$_2$ & 2  &$-$ & $0.00^{+0.71}_{-0.00} $ & $-$ & $-$ & $-$ & $0.00^{+0.31}_{-0.00} $ & $-$ & $-$ \\
Am. SiO$_2$ & 5  &$0.00^{+1.89}_{-0.00} $ & $3.8^{+-0.7}_{-3.8} $ & $-$ & $-$ & $0.00^{+1.32}_{-0.00} $ & $16.8^{+8.3}_{-12.3} $ & $0.93^{+3.92}_{-0.93} $ & $0.00^{+1.08}_{-0.00} $ \\
Ann. SiO$_2$ & 0.1 &$2.4^{+-0.2}_{-2.0} $ & $-$ & $1.3^{+1.5}_{-0.5} $ & $1.6^{+0.3}_{-1.6} $ & $<0.1$ & $5.9^{+-2.1}_{-5.0} $ & $<0.1$ & $1.7^{+1.6}_{-0.9} $ \\
Ann. SiO$_2$ & 2  &$-$ & $0.00^{+0.21}_{-0.00} $ & $3.7^{+0.5}_{-3.1} $ & $0.12^{+0.88}_{-0.12} $ & $0.00^{+0.15}_{-0.00} $ & $0.00^{+2.99}_{-0.00} $ & $0.92^{+0.10}_{-0.92} $ & $2.5^{+0.2}_{-2.5} $ \\
\hline
        \end{tabular}
		\renewcommand{\arraystretch}{1.0}
    }
    \end{table*}}
\twocolumn
\endgroup
\FloatBarrier
\clearpage

\section{Best-fit disk parameters.}
\label{sec:app:pm}	

\begingroup
\let\clearpage\relax{
\onecolumn

\begin{table*}[h!]
        \caption[]{\label{tab:disk_params}Disk parameters from our fit with the DHS\_nat set of opacities.}
        \centering
    {\small
\renewcommand{\arraystretch}{1.1}
\begin{tabular}{lccccccc}
\hline
\hline
 &$r_\mathrm{in}$ (au) & $T_\mathrm{surface,in}$ (K) & $q_\mathrm{surface}$ & $T_\mathrm{midplane,in}$ (K) & $q_\mathrm{midplane}$ & $w_\mathrm{rim}$ (au) & $T_\mathrm{rim}$ (K) \\ 
\hline
CX Tau & $0.92^{+0.06}_{-0.45} $ & $1700^{+-28}_{-421} $ & $-0.69^{+0.08}_{-0.37} $ & $212^{+22}_{-+2} $ & $-1.91^{+0.98}_{-0.35} $ & $0.0004^{+0.0002}_{-0.0002} $ & $806^{+385}_{-+18} $ \\
CY Tau & $0.09^{+0.16}_{-0.06} $ & $1022^{+460}_{-453} $ & $-2.79^{+2.19}_{-+0.53} $ & $371^{+209}_{-90} $ & $-0.85^{+0.06}_{-1.57} $ & $0.0064^{+0.0010}_{-0.0044} $ & $1079^{+455}_{-94} $ \\
BP Tau & $0.17^{+0.71}_{-+0.04} $ & $1084^{+-28}_{-586} $ & $-0.28^{+-0.32}_{-2.20} $ & $343^{+-20}_{-98} $ & $-0.57^{+-0.03}_{-1.04} $ & $0.0025^{+0.0000}_{-0.0017} $ & $1319^{+219}_{-405} $ \\
FT Tau & $0.07^{+0.72}_{-+0.03} $ & $730^{+859}_{-+211} $ & $-2.68^{+1.74}_{-+0.12} $ & $468^{+-52}_{-217} $ & $-0.55^{+-0.04}_{-1.26} $ & $0.0051^{+-0.0010}_{-0.0046} $ & $1243^{+208}_{-287} $ \\
DF Tau & $0.08^{+0.29}_{-0.02} $ & $1092^{+243}_{-526} $ & $-0.19^{+-0.35}_{-1.05} $ & $562^{+596}_{-242} $ & $-0.54^{+-0.07}_{-1.99} $ & $0.0582^{+0.0162}_{-0.0480} $ & $1167^{+276}_{-280} $ \\
DL Tau & $0.06^{+0.03}_{-0.02} $ & $1139^{+361}_{-580} $ & $-0.95^{+0.44}_{-1.56} $ & $859^{+239}_{-197} $ & $-0.60^{+0.03}_{-0.05} $ & $0.0158^{+0.0249}_{-0.0065} $ & $1401^{+168}_{-530} $ \\
DM Tau & $0.11^{+-0.05}_{-0.08} $ & $472^{+169}_{-+62} $ & $-0.26^{+0.02}_{-0.02} $ & $246^{+102}_{-+34} $ & $-0.39^{+0.03}_{-+0.01} $ & $0.0001^{+0.0008}_{-+0.0001} $ & $1712^{+-68}_{-717} $ \\
AA Tau & $0.51^{+0.40}_{-0.16} $ & $539^{+549}_{-+23} $ & $-2.96^{+2.24}_{-+1.03} $ & $314^{+53}_{-28} $ & $-1.02^{+0.11}_{-0.91} $ & $0.0055^{+-0.0002}_{-0.0036} $ & $963^{+301}_{-37} $ \\
DN Tau & $0.30^{+0.59}_{-0.23} $ & $1162^{+137}_{-579} $ & $-0.26^{+-0.02}_{-1.59} $ & $284^{+193}_{-46} $ & $-0.75^{+0.20}_{-1.15} $ & $0.0025^{+0.0024}_{-0.0014} $ & $987^{+378}_{-106} $ \\
LkCa 15 & $0.40^{+0.38}_{-+0.11} $ & $447^{+170}_{-6} $ & $-0.09^{+-0.01}_{-0.13} $ & $175^{+-8}_{-20} $ & $-0.42^{+-0.02}_{-0.12} $ & $0.0004^{+-0.0000}_{-0.0002} $ & $1760^{+-9}_{-237} $ \\
DR Tau & $0.16^{+0.27}_{-0.10} $ & $1354^{+-417}_{-898} $ & $-0.37^{+-0.15}_{-1.22} $ & $513^{+422}_{-150} $ & $-0.50^{+-0.01}_{-0.10} $ & $0.0630^{+-0.0023}_{-0.0427} $ & $1018^{+536}_{-135} $ \\
RW Aur A & $0.96^{+-0.13}_{-0.89} $ & $1495^{+-60}_{-961} $ & $-2.05^{+1.23}_{-0.61} $ & $241^{+307}_{-3} $ & $-0.80^{+0.34}_{-+0.06} $ & $0.0050^{+0.0484}_{-+0.0009} $ & $1149^{+356}_{-192} $ \\
RW Aur B & $0.27^{+0.44}_{-0.21} $ & $591^{+847}_{-101} $ & $-0.57^{+0.21}_{-0.06} $ & $220^{+58}_{-25} $ & $-0.56^{+0.21}_{-0.77} $ & $0.0009^{+0.0043}_{-0.0006} $ & $1758^{+-14}_{-245} $ \\
SY Cha & $0.16^{+0.42}_{-0.01} $ & $425^{+365}_{-+11} $ & $-1.86^{+1.24}_{-0.13} $ & $335^{+22}_{-90} $ & $-0.56^{+-0.03}_{-0.62} $ & $0.0057^{+-0.0008}_{-0.0043} $ & $1216^{+245}_{-163} $ \\
TW Hya & $0.40^{+-0.02}_{-0.25} $ & $447^{+340}_{-+47} $ & $-0.16^{+-0.03}_{-0.16} $ & $201^{+49}_{-1} $ & $-0.53^{+0.11}_{-+0.01} $ & $0.0001^{+0.0002}_{-+0.0000} $ & $1755^{+-102}_{-570} $ \\
VW Cha A & $0.82^{+0.00}_{-0.57} $ & $763^{+158}_{-165} $ & $-1.57^{+0.70}_{-0.46} $ & $274^{+101}_{-10} $ & $-0.75^{+0.20}_{-0.05} $ & $0.0027^{+0.0092}_{-+0.0001} $ & $1758^{+8}_{-308} $ \\
WX Cha A & $0.08^{+0.61}_{-0.05} $ & $1144^{+192}_{-433} $ & $-0.16^{+-0.62}_{-1.91} $ & $422^{+312}_{-177} $ & $-0.51^{+0.00}_{-1.67} $ & $0.0172^{+0.0149}_{-0.0134} $ & $1285^{+178}_{-216} $ \\
XX Cha & $0.21^{+0.74}_{-+0.00} $ & $568^{+237}_{-79} $ & $-1.51^{+0.67}_{-0.87} $ & $305^{+2}_{-83} $ & $-0.61^{+-0.03}_{-1.17} $ & $0.0041^{+0.0003}_{-0.0033} $ & $1079^{+202}_{-192} $ \\
Sz 50 & $0.02^{+0.18}_{-0.01} $ & $445^{+619}_{-+34} $ & $-2.39^{+1.92}_{-0.22} $ & $560^{+899}_{-282} $ & $-0.42^{+-0.03}_{-0.26} $ & $0.0222^{+-0.0036}_{-0.0191} $ & $967^{+135}_{-136} $ \\
PDS 70 & $0.52^{+0.10}_{-0.24} $ & $773^{+94}_{-8} $ & $-2.89^{+1.08}_{-+0.05} $ & $195^{+17}_{-4} $ & $-1.49^{+0.62}_{-0.46} $ & $0.0002^{+0.0002}_{-0.0000} $ & $1793^{+-16}_{-187} $ \\
GW Lup & $0.58^{+0.12}_{-0.28} $ & $1261^{+247}_{-456} $ & $-0.22^{+-0.46}_{-2.31} $ & $272^{+36}_{-8} $ & $-1.49^{+0.57}_{-0.37} $ & $0.0005^{+0.0006}_{-0.0002} $ & $1061^{+280}_{-184} $ \\
IM Lup & $0.30^{+0.59}_{-0.01} $ & $1274^{+43}_{-553} $ & $-0.40^{+-0.02}_{-0.59} $ & $287^{+18}_{-56} $ & $-0.60^{+-0.03}_{-0.59} $ & $0.0027^{+0.0006}_{-0.0019} $ & $1424^{+165}_{-335} $ \\
Sz 98 & $0.51^{+0.37}_{-0.12} $ & $524^{+611}_{-+117} $ & $-0.03^{+-0.10}_{-2.38} $ & $342^{+50}_{-29} $ & $-0.98^{+0.07}_{-0.64} $ & $0.0054^{+0.0004}_{-0.0036} $ & $1033^{+505}_{-131} $ \\
V1094 Sco & $0.03^{+0.65}_{-0.01} $ & $1232^{+118}_{-558} $ & $-1.87^{+0.25}_{-0.95} $ & $514^{+24}_{-305} $ & $-0.51^{+0.02}_{-2.08} $ & $0.0098^{+0.0053}_{-0.0091} $ & $1171^{+-61}_{-308} $ \\
RNO 90 & $0.61^{+0.29}_{-0.15} $ & $765^{+285}_{-278} $ & $-0.23^{+-0.29}_{-2.23} $ & $294^{+29}_{-24} $ & $-0.67^{+0.05}_{-0.23} $ & $0.0056^{+0.0040}_{-0.0017} $ & $1510^{+130}_{-304} $ \\
WA Oph 6 & $0.64^{+0.14}_{-0.51} $ & $540^{+600}_{-32} $ & $-3.00^{+2.41}_{-+0.22} $ & $254^{+233}_{-3} $ & $-0.79^{+0.22}_{-0.17} $ & $0.0084^{+0.0069}_{-0.0036} $ & $823^{+440}_{-+25} $ \\
\hline
        \end{tabular}
    }
    \end{table*}}
\twocolumn
\endgroup
\FloatBarrier
\clearpage

\section{Fit to PDS 70 with GRF opacities.}
\label{sec:app:fit_PDS70}	
\vspace{-0.4cm}
\begin{figure}[H]
\includegraphics[width=\hsize,trim={0.0cm 0.3cm 0.0cm 0.25cm},clip]{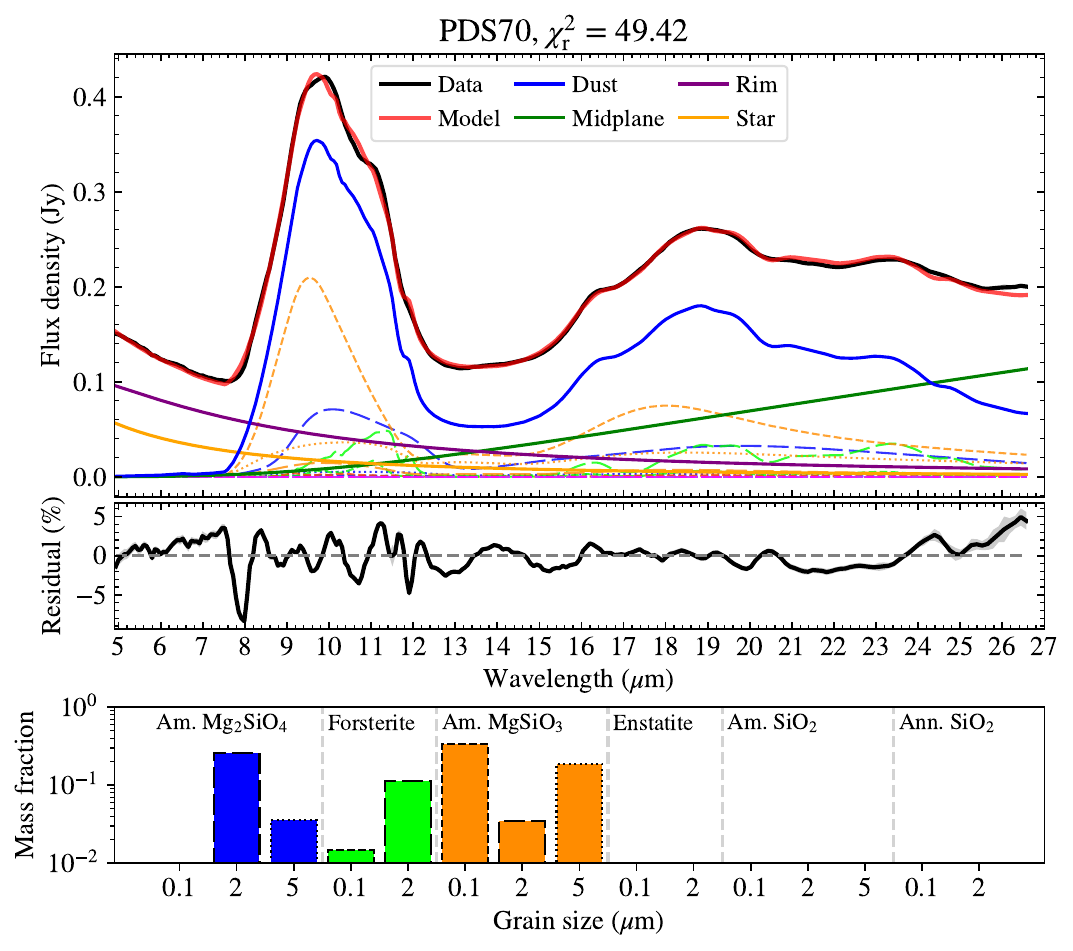}
	\caption{Fits to the spectrum of PDS~70 with GRS opacities (including annealed silica), using the $\left[0.1, 2, 5\ \mathrm{am.}\right]$ grain size set.}
         	\label{fig:app:fit_PDS70}
\end{figure}

\FloatBarrier

\section{Fit residuals}
\label{sec:app:residuals}	

\vspace{-0.5cm}

\begin{figure}[H] 
	\centering
	\includegraphics[width=1.0\hsize,trim={0.0cm 0.25cm 0.0cm 0.2cm},clip]{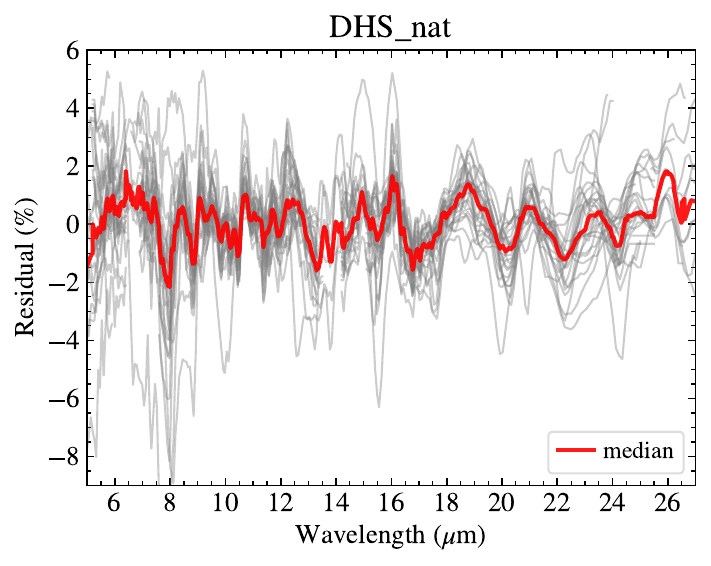 } 
	\caption{Fit residuals (gray lines) plotted on top of each other with the grain size set $\left[0.1, 2, 5\ \mathrm{am.}\right]$, with annealed silica. The red line shows the median of the residuals. } 
         	\label{fig:app:residual}
\end{figure}

\FloatBarrier

\section{Simulating Spitzer/IRS low-resolution spectroscopy}
\label{sec:app:MIRI_degraded_to_Spitzer}	

To simulate a Spitzer/IRS observation, we degraded the MIRI spectrum of XX~Cha to the resolving power and S/N of Spitzer/IRS low-resolution spectroscopy. We produced two versions of the degraded spectrum: one based on the original MIRI spectrum, which includes all molecular line emission, and another based on the line-removed spectrum that was also used for our MIRI dust compositional fits. We then applied our spectral decomposition routine to the degraded spectra, and the resulting best fits are shown in Fig.~\ref{fig:app:MIRI_degraded_to_Spitzer}. In general, the derived dust mass fractions are broadly consistent between the two cases, although the effects of the degeneracy between large amorphous Mg-silicate grains are noticeable in the mass fractions. In addition, the constraints on annealed silica are relatively weak, as its characteristic features at $12.6$ and $16.1\ \mu$m become rather noisy. Consequently, the fit cannot efficiently distinguish between amorphous and annealed silica.
Across our sample, the impact of molecular lines on the dust features is minor outside a few narrow wavelength regions known to be strongly affected by molecular emission (e.g., around $6.5\ \mu$m and $14.5\ \mu$m). The differences between the original MIRI spectra degraded to the resolving power of Spitzer and the line-removed versions are typically only $0.5$--$1\%$.
This finding is expected to remain valid for other T Tauri and Herbig objects with relatively strong silicate features, but is expected to break down for VLMSs, which tend to exhibit weaker silicate features and stronger molecular line emission \citep{Jang2025}. 

\begin{figure}
	\centering
	\begin{tikzpicture}[
image/.style = {text width=0.48\textwidth, 
                 inner sep=0pt, outer sep=0pt},
node distance = 1mm and 1mm 
                        ] 
\node [image] (frame1)
    {\includegraphics[width=0.99\hsize,trim={0.0cm 0.3cm 0.0cm 0.25cm},clip]{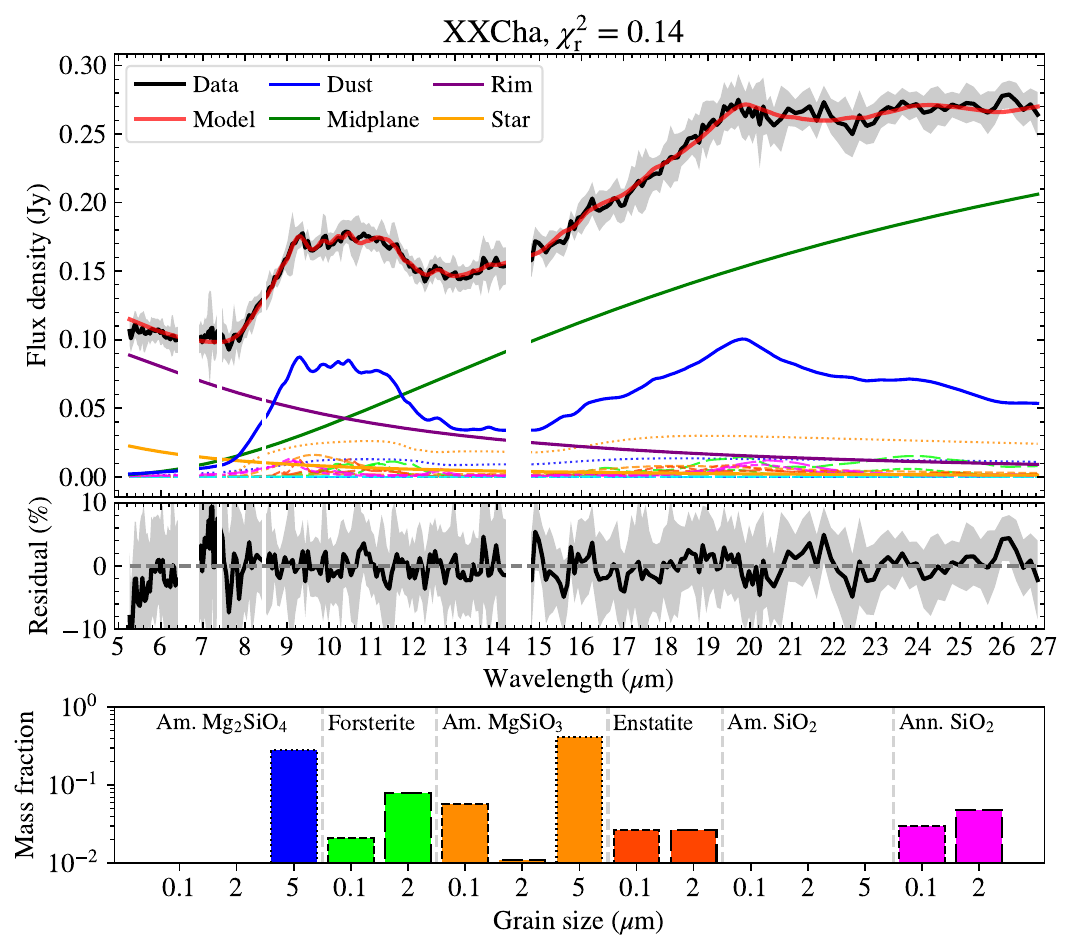}};
\node at (0.2,4.0) {\textbf{Original MIRI spectrum degraded}};
\node [below=of frame1] (text1)  {\textbf{Line-removed MIRI spectrum degraded}};
\node [image,below=of text1] (frame2) 
    {\includegraphics[width=0.99\hsize,trim={0.0cm 0.3cm 0.0cm 0.25cm},clip]{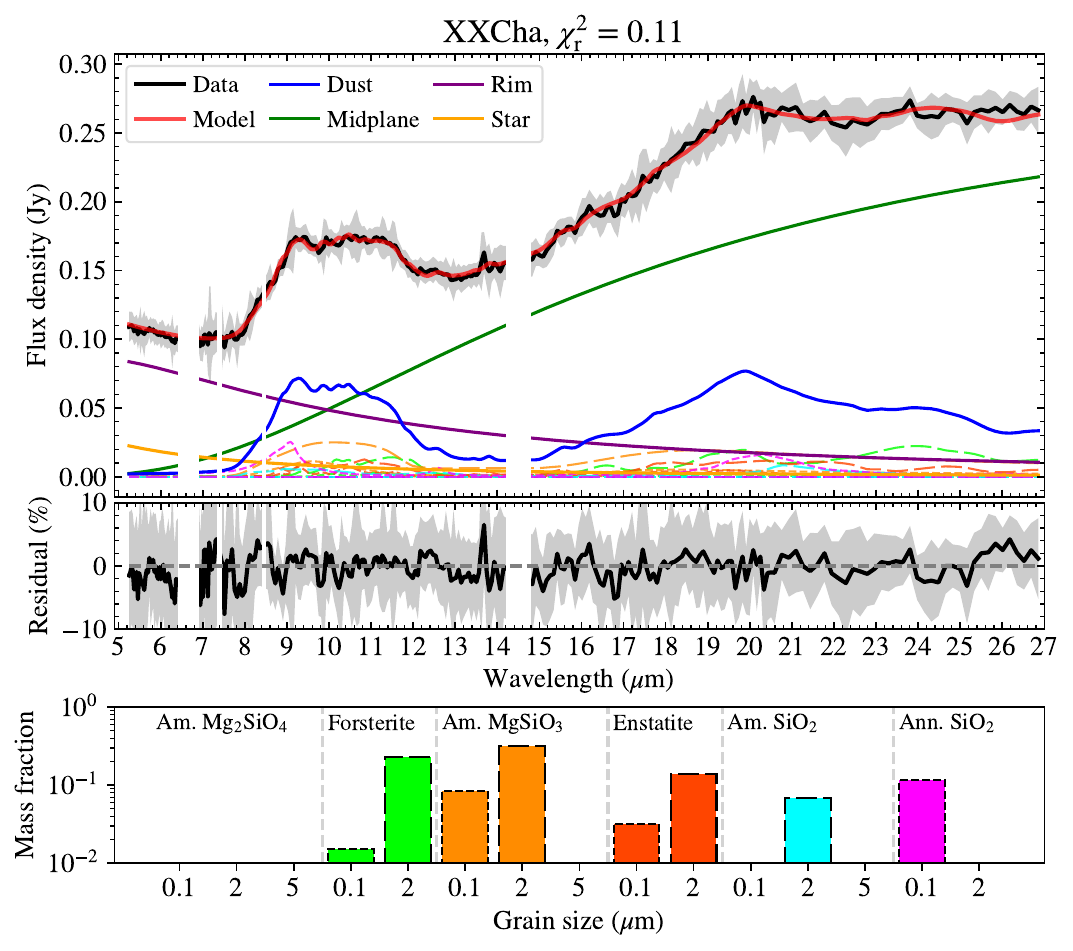}};

\end{tikzpicture}

	\caption{Fits to the MIRI spectra of XX~Cha degraded to the resolving power and S/N of Spitzer/IRS low-resolution spectroscopy. The upper panel shows the result obtained using the original MIRI spectrum of XX~Cha as input, which includes all molecular line emission. The lower panel shows the same analysis performed on the line-removed MIRI spectrum.}
         	\label{fig:app:MIRI_degraded_to_Spitzer}
\end{figure}

\end{appendix}

\end{document}